\documentclass{these}
\usepackage{graphicx}
\usepackage{float}
\usepackage{vmargin}
\usepackage{amsmath}
\usepackage{amssymb}
\usepackage{amsthm}
\usepackage{amsfonts}
\usepackage{indentfirst}
\usepackage[font=small,labelfont=bf,up]{caption}
\usepackage{pstricks}
\usepackage[loose,bf]{subfigure}
\usepackage{multirow}
\usepackage{array}
\usepackage{ae}
\usepackage{bbm}
\usepackage{bm}
\usepackage[toc]{appendix}
\usepackage{cite}     
\usepackage{lscape}

\usepackage{url}

\newcommand{\di}{\mathrm{\; d}}
\newcommand{\moy}[1]{\left\langle #1 \right\rangle}
\newcommand{\moyen}[1]{\left\langle #1 \right\rangle}
\newcommand{\derivb}[1]{\frac{\partial #1}{\partial \beta}}
\newcommand{\derivmb}[2]{\frac{\partial^{#2} #1}{\partial \beta^{#2}}}

\DeclareMathOperator*{\Tr}{Tr}
\newcommand{\iii}{i}
\newcommand{\iip}{p}
\newcommand{\iih}{h}

\newcommand{\diag}[1]{
  \hspace{-1.5mm}
  \begin{tabular}{c}
    \includegraphics[scale=.4]{#1}
  \end{tabular}
  \hspace{-1.5mm}
}

\newcommand{\inlinediag}[1]{
  \hspace{-1.5mm}
  \begin{tabular}{c}
    \includegraphics[scale=.2]{#1}
  \end{tabular}
  \hspace{-1.5mm}
}

\title{Probl\`emes inverses dans les mod\`eles de spin}
\author{Vitor SESSAK}
\date{16/09/2010}
\labo{Laboratoire de Physique Th\'eorique de l'\'Ecole Normale Sup\'erieure}
\school{\'Ecole Doctorale de Physique la R\'egion Parisienne --- ED 107}
\speciality{Physique}
\university{l'Universit\'e Pierre et Marie CURIE}{\Large{UNIVERSITE PIERRE ET MARIE CURIE}}
\advisor[M]{R\'emi Monasson et Nicolas Sourlas} 
\coadvisor[M]{} 
\jury{  
\jurymember[Mme]{Leticia Cugliandolo}{Pr\'esident du jury}\\
\jurymember[M]{John Hertz}{Rapporteur}\\
\jurymember[M]{Silvio Franz}{Rapporteur}\\
\jurymember[M]{Florent Krzakala}{Examinateur}\\
\jurymember[M]{R\'emi Monasson}{Directeur de Th\`ese}\\
\jurymember[M]{Nicolas Sourlas}{Invit\'e}\\

}

\graphicspath{{./pictures/}{./graphs/}{paper/diagrams/}{paper/graphics/}}
\begin{document}
\frontmatter
\maketitle

\resume{
Un bon nombre d'expériences récentes en biologie mesurent des systèmes
composés de plusieurs composants en interactions, comme par exemple les réseaux
de neurones. 
Normalement, on a expérimentalement accès qu'au comportement collectif du 
système, même si on s'intéresse souvent à la caractérisation des interactions entre ses différentes composants.
Cette thèse a pour but d'extraire des informations sur les interactions microscopiques du système à partir de son comportement collectif dans deux cas distincts.
Premièrement, on étudie un système décrit par un modèle d'Ising plus général. On trouve
des formules explicites pour les couplages en fonction des corrélations et
magnétisations. 
Ensuite, on s'intéresse à un système décrit par un modèle de Hopfield. Dans ce
cas, on obtient non seulement une formule explicite pour inférer les patterns, mais
aussi un résultat qui permet d'estimer le nombre de mesures nécessaires pour 
avoir une inférence précise.
}{
Several recent experiments in biology study systems composed of several
interacting elements, for example neuron networks. Normally, measurements
describe only the collective behavior of the system, even if in most
cases we would like to characterize how its different parts interact.
The goal of this thesis is to extract information about the microscopic
interactions as a function of their collective behavior for two different cases.
First, we will study a system described by a generalized Ising model. We
find explicit formulas for the couplings as a function of the correlations
and magnetizations. In the following, we will study a system described by a
Hopfield model. In this case, we find not only explicit formula for
inferring the patterns, but also an analytical result that allows one to
estimate how much data is necessary for a good inference.
}

\chapter*{Remerciements}

Si je devais remercier tous ceux dont j'ai envie, cette liste serait
beaucoup trop longue. Je fais alors le compromis de me restreindre à ceux
que j'ai croisé quotidiennement pendant ma thèse (ou pendant une période).

Tout d'abord, je voudrais remercier Rémi Monasson de m'avoir fait découvrir
tous les domaines forts intéressants dans lesquels j'ai travaillé pendant ma
thèse. Merci également d'avoir été aussi efficace afin de me débloquer quand je n'arrivais pas au bout d'un
calcul. Finalement, je lui suis très reconnaissant de m'avoir donné l'occasion
de travailler dans le cadre très exceptionnel de l'IAS et pour toutes les
démarches faites pour faciliter mon installation aux USA.

J'aimerais aussi remercier Simona Cocco pour toutes les discussions très
fructueuses et de m'avoir aider sur ma thèse à plusieurs reprises.

Une autre personne que je ne peux pas oublier est Stan
Leibler. Merci de m'avoir si bien accueilli à l'IAS et de m'avoir donner l'occasion
d'exposer mon travail à son groupe.

Un merci également très particulier à Carlo pour toute son aide et sa patience pendant notre
collocation à Princeton et d'avoir été un ami de toute heure
aussi bien à Paris qu'à Princeton. Merci aussi à Laeticia pour sa compagnie.

J'en profite pour remercier Stan, Rémi, Simona, Carlo, Laetitia et Arvind pour
tous les déjeuners relaxants à Princeton\footnote{Je ne peux pas m'empêcher
de remercier le cuisiner de la cantine de l'IAS pour tous les repas
délicieux.}.

Un merci particulier également à John Hertz, de m'avoir invité pour faire un
séminaire et de son accueil chaleureux à Stockholm.

Je n'oublie pas mes collègues de bureau et d'étage : merci à Florent d'être toujours
de bonne humeur et d'avoir toujours des choses intéressantes à
dire; merci à tous les thésards de la géophysique, pour les déjeuners dans une
très bonne ambiance; et merci à Sebastien d'être un collègue de bureau très
sympathique.

Je voudrais aussi remercier ma famille d'avoir été si patiente pendant toutes ces
longues années d'absence et de m'avoir toujours encouragé.
Finalement, je veux remercier de tout mon c\oe{ur} Camilla, d'avoir
toujours été de mon côté pendant toutes ces années ensemble, d'avoir toujours
été si patiente dans mes moments de mauvaise humeur et d'avoir accepté si patiemment
mes longues absences à l'étranger.

\tableofcontents

\mainmatter
\part{Introduction}
\label{part1}
\chapter{Biological motivation and related models}
\label{chap_motiv}

In the last years, we have seen a remarkable growth in the number of experiments
in biology that generate an overwhelming quantity of data. In several cases,
like in neuron assemblies, proteins and gene networks, most of the data
analysis focuses on identifying correlations between different parts of the
system.
Unfortunately, identifying the correlations on their own is only of limited
scientific value:
most of the underlying properties of the system can only be understood by
describing the interaction between their different parts.
This work finds its place in developing statistical mechanics tools to
derive these interactions from measured correlations.

In this introductory chapter, we present two biological problems that inspired
this thesis.
First, in section~\ref{sec_neuron} we give a brief introduction to neurons and
how they exchange information in a network.
We discuss some experiments where the individual activity of up to a 
hundred interacting neurons is measured. For this example, the neurons are the
interacting parts and they interact via synapses, whose details are very hard
to extract experimentally.

In a second part, we discuss some recent works on the
analysis of families of homologous proteins, i. e.,
proteins that share an evolutionary ancestry and function.
The variation of the amino acids inside these families are highly correlated
which is deeply related to the biological function of the proteins.
In general terms,
we can say thus that individual amino acid variations play the role of
interacting parts with very complicated interactions, as we will see in
section~\ref{sec_prot}.


\section{Neuron networks}
\label{sec_neuron}

One of the most important scientific questions of the 21st century is the
understanding of the brain. It is widely accepted that its complexity is due to 
the organization of neurons in complex networks. If we consider, for example, 
the human brain, we can count about $10^{11}$ neurons connected by about 
$10^{14}$ connections. Even much simpler organisms like the 
\textit{Drosophila melanogaster} fruit fly counts about 100,000 neurons.

\begin{figure}[H]
\begin{center}
\includegraphics[width=0.85\columnwidth]{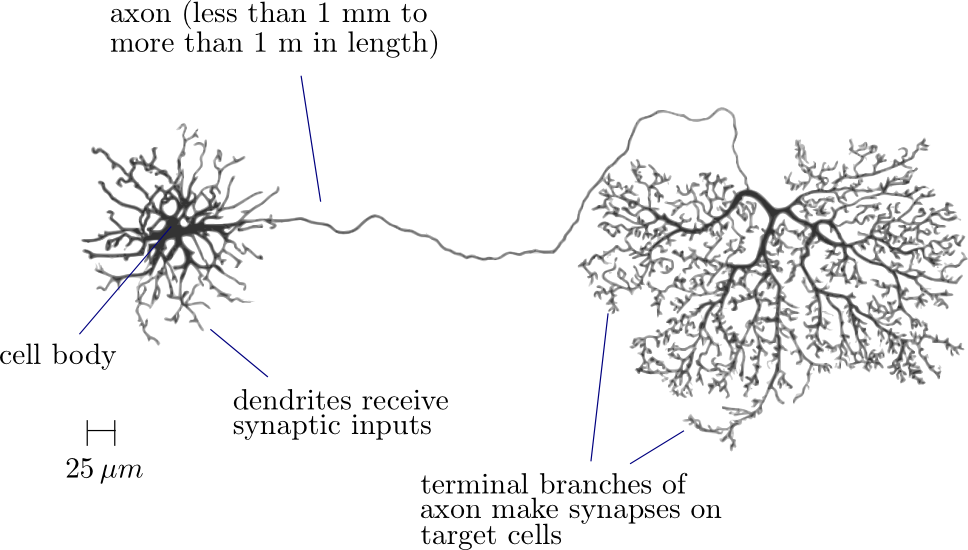}
\end{center}
\caption{Schema of a neuron \cite{alberts-cell}. The diameter of the cell
body is typically of the order of $10 \, \mu m$, while the length of 
dendrites and axons varies considerably with the neuron's function.}
\label{schema_neu}
\end{figure}

A typical neuron can be schematized as a cell composed of three parts: the cell 
body, dendrites and one axon (see Fig.~\ref{schema_neu}). A dendrite is composed
of several branches in a tree-like structure and is responsible for receiving 
electric signals from other neurons. The axon is a longer, ramified single 
filament, responsible for sending electrical signals to other neurons. 
A connexion between two neurons in most cases happens between an axon and a 
dendrite\footnote{As common in biology, such simplified description of 
a neuron and synapses has exceptions. Some axons transmit signals while some 
dendrites receive them. We also find axon-axon and dendrite-dendrite 
synapses\cite{churchland1989neurophilosophy}.}. We call such connections
\textit{synapses}.

Like most cells, neurons have an electrical potential difference between their
cytoplasm and the extracellular medium. This potential difference
is regulated by the exchange of ions (such as $\text{Na}^+$ and $\text{K}^+$) 
through the cell membrane, which can be done in two ways: passively, by proteins
called \textit{ion channels}
that selectively allow the passage of a certain ion from the most concentrated 
medium to the least and,
conversely, actively by
proteins called \textit{ion pumps} that consume energy to increase the ion
concentration difference.

\begin{figure}[H]
\begin{center}
\includegraphics[width=0.4\columnwidth]{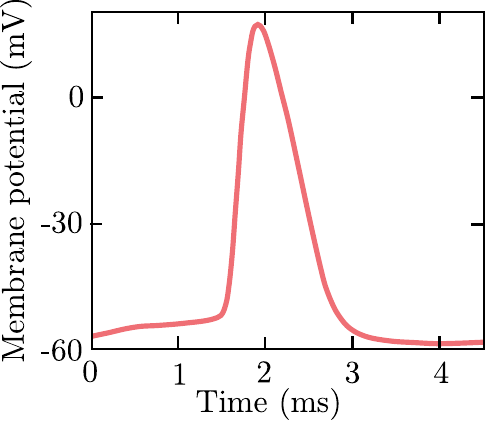}
\end{center}
\caption{Typical voltage as a function of time graph for a firing neuron
\cite{naundorf2006unique}.}
\label{fig_fire}
\end{figure}

A typical neuron has a voltage difference of
about $-70 \,  mV$ when it is not receiving any signal from other neurons. We
call this voltage the \textit{resting potential} of the neuron. 
If the voltage of a neuron reaches a threshold (typically about $-50 \, mV$),
a feedback mechanism makes ion channels of the membrane to open, 
making the voltage increase rapidly up to $100 \, mV$ (depending on the neuron
type), after which it reaches
saturation and decreases quickly, recovering the resting potential 
after a few $ms$ (see Fig.~\ref{fig_fire}). We call this process
\textit{firing} or \textit{spiking}. One important characteristic of 
the spikes is that once the voltage reaches the threshold, its shape and its 
intensity do not depend on the details of how the threshold was attained.

When a neuron fires, its axon releases neurotransmitters at every synapse. 
Those neurotransmitters make ions channels in the dendrites open, changing 
the membrane potential of the neighboring neurons.
Different neurotransmitters cause the opening of different ion channels, allowing 
for both excitatory synapses, which increase the neuron potential, and inhibitory 
synapses which decrease it. Since synapses can be excitatory or inhibitory
to different degrees, most models define a \textit{synaptic weight} with the
convention that excitatory synapses have a positive synaptic weight and 
inhibitory synapses have a negative one, as we will see in
section~\ref{sec_models}.
Another important feature of synapses is that they are \emph{directional}: if a
neuron $A$ can excite a neuron $B$, the converse is not necessarily true:
neuron $B$ might inhibit neuron $A$, or simply not be connected to it at all.

\subsection{Multi-neuron recording experiments}
\label{sec_multineu}

While much progress has been done in describing individual neurons, 
understanding their complex interaction in a network is 
still an unsolved problem. One of the most promising advances in this area was 
the development of techniques for recording simultaneously
the electrical activity of several 
cells individually
\cite{meister1994multi}.

\begin{figure}[H]
\begin{center}
\includegraphics[width=0.65\columnwidth]{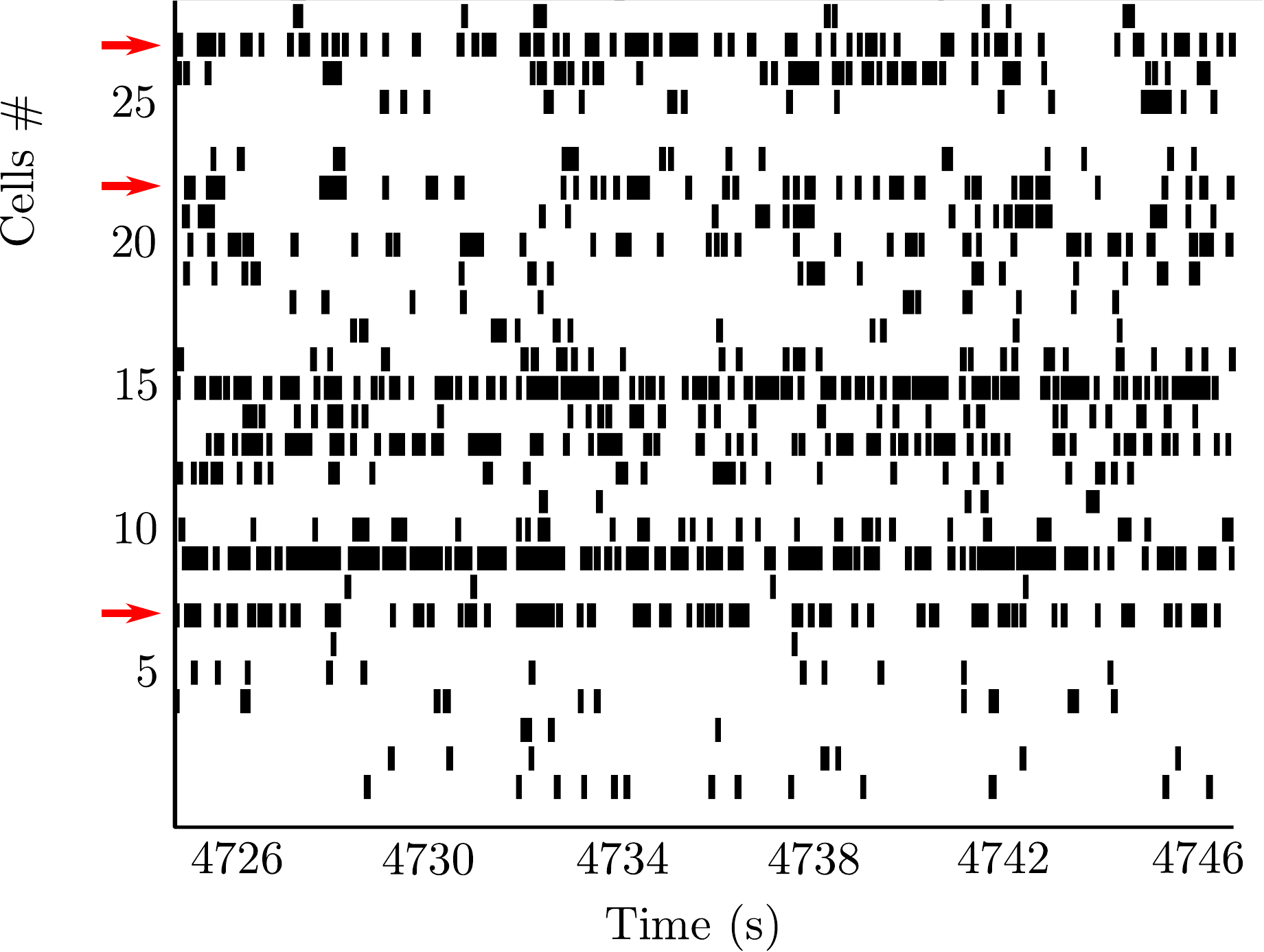}
\end{center}
\caption{Typical measurement of spike trains\cite{peyrache}. Each line corresponds to a single 
neuron. Black vertical bars correspond to spikes.}
\label{fig_spike}
\end{figure}

In these experiments, a microarray counting as many as 250 electrodes
is placed in contact 
with the brain tissue. The potential of each electrode is recorded for up 
to a few hours. Each one of the electrodes might be affected by the activity of
more than
one neuron and, conversely, a single neuron might affect more than one 
electrode. Thus, a computational-intensive calculation is needed to factorize
the signal as the sum of the influence of several different neurons. This 
procedure is known as \textit{Spike Sorting} \cite{peyrache} and its results
are \textit{spike trains}, i.e., time sequences of the state of each cell:
firing or
at rest.
An example of a set of spike trains can be seen in Fig.~\ref{fig_spike}.

\begin{figure}[H]
\begin{center}
\includegraphics[width=0.5\columnwidth]{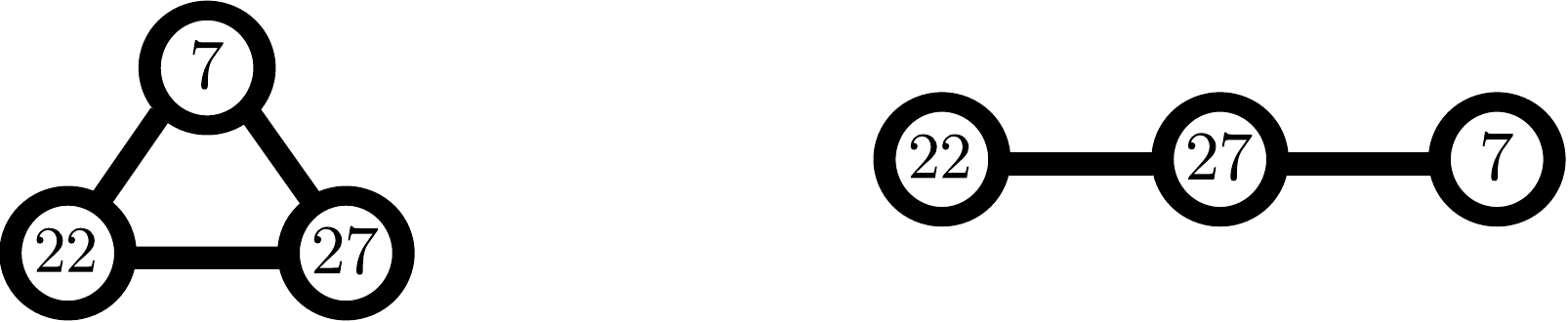}
\end{center}
\caption{Two different possible configurations for three positively 
correlated neurons.}
\label{fig_connec}
\end{figure}

In principle, one should be able to pinpoint the synapses and the synaptic 
weights from the spike trains. However, extracting this information is a 
considerable challenge. First of all, it is not possible with current 
technology to measure every neuron of a network. Thus, experiments 
measure just a small fraction of the system even if the network is small.
That means that all that we can expect to find are \emph{effective} 
interactions that depend on all links between the cells that are not measured.
Secondly, one can not naively state that if the activity of two neurons is
correlated then they are connected by a synapse.
Consider, for example, neurons 7, 22 and 27 of Fig.~\ref{fig_spike}, indicated
by the red arrows. We can
clearly see that there is a tendency for all three of firing at the same time, but
distinguishing between the two possible connections shown in
Fig.~\ref{fig_connec} is not trivial.

\subsection{Models for neuron networks}
\label{sec_models}

Before talking about what have already been done to solve this problem and
our contribution to it, we present some models for neural
networks. We will proceed by first introducing a model that describes rather
faithfully real
biological networks, the leaky Integrate-and-Fire model. Afterwards, we
introduce the Ising model, which is much more tractable analytically.
Finally, we will look at the Ising model from a different point of view by
studying one particular case of it: the Hopfield model.

\subsubsection{Leaky integrate-and-fire model}

The leaky integrate-and-fire model, first proposed by Lapicque in 1907
\cite{lapicque1907,abbott1999lapicque,gerstner2002spiking,burkitt2006review},
 is a straightforward modelization of the firing process
presented in section~\ref{sec_neuron}. It supposes that neurons behave like
capacitors with a small leakage term to account for the fact that the membrane is not a 
perfect insulator. Posing $V(t)$ as the function representing the difference of
potential between the inside and the outside of the membrane:
\begin{equation}
\label{eq_if}
C \frac{\di V}{\di t} = - \frac{V(t)}{R} + I(t) \, ,
\end{equation}
\noindent where $C$ is the capacitance of the neuron, $R$ is the resistance of
the cell membrane and $I(t)$ is the total current due to the synapses of neighboring 
neurons. If we introduce the characteristic time of leakage $\tau = RC$, we can 
rewrite this equation as
\begin{equation}
\label{eq_if2}
\tau \frac{\di V}{\di t} = - V(t) + R I(t) \, .
\end{equation}

Spikes are modelized solely by their ``firing time'' 
$t_f$. This firing time is defined as the moment where the neuron's potential
reaches a firing threshold value $V_{tr}$. Implicitly, it is given by the
equation
%
\begin{equation}
V(t_f) = V_{tr} \, .
\end{equation}
%
Every time a neuron spikes, its potential is reset to zero and its synapses
produce a signal in the form of some function $f(t)$. We can thus describe
the signal $S(t)$ that this neuron send to its neighbors as a function only of the
set of firing times $\{t_f^r\}_{r=1,\ldots,N_\text{spikes}}$
\begin{equation}
S(t) = \sum_{r=1}^{N_\text{spikes}} f(t - t_f^{r}) \, .
\end{equation}
The choice of the function $f(t)$ can be based on biological measures, mimicking
the behavior shown in Fig.~\ref{fig_fire} or can
be a simple Dirac-delta function to make calculations easier.

Finally, we can introduce the synaptic weights to model a neuron network
with the following equations:
\begin{eqnarray}
\label{eq_IF3}
\tau_i \frac{\di V_i}{\di t} &=& - V_i(t) + \sum_j J_{ij} S_j(t) \, ,\\
\label{eq_IF4}
V_i(t_f^{i,r}) &=& 1 \, ,\\
\nonumber\\
\label{eq_IF5}
S_i(t) &=& \sum_{r=1}^{N_\text{spikes}^i} f(t - t_f^{i,r})\, ,
\end{eqnarray}
where $V_i$ is the potential of the neuron $i$ rescaled so that 
$V_{tr}=1$, $t_f^{i,r}$ is the time of the $r$-th spike of the neuron $i$
and $J_{ij}$ is the matrix of the synaptic weights. Note that it is usually
assumed as an approximation that the function $f(t)$ is identical for all
neurons.

This model is very popular due to its balance between biological accuracy
and relative simplicity. It is also very well-suited for computer simulations by the direct
integration of its differential equations. On the other hand, while some numerical work has
been done on the inference of synaptic weights from spike trains using this model
\cite{monas_pnas}, it is not very practical for analytical results.

\subsubsection{Generalized Ising model / Boltzmann Machine}
The Boltzmann Machine  \cite{mcculloch1943logical} is a model of neuron networks 
that mimics less well real biological systems than the leaky Integrate-and-Fire 
model. It is however considerably simpler, being even exactly solvable for some
special networks. In this model, the state of a neuron is 
fully described by a spin variable $\sigma = \pm 1$ with the convention of 
$\sigma=+1$ if the neuron is firing and $\sigma = -1$ if it is 
not\footnote{The convention of $\sigma=1$ for a firing neuron and $\sigma=0$
for resting is also common.}.
The dynamics of the system is ignored\footnote{It is possible to define a time
  evolution in this model using the Glauber dynamics
  if needed.}
and we describe only
the probability $P(\{\sigma_1,\ldots,\sigma_N\})$  of finding the network of $N$ neurons in a state 
$\{\sigma_1,...,\sigma_N\}$, which is  given by the Boltzmann weight of a
generalized Ising model
\begin{eqnarray}
P(\{\sigma_1,...,\sigma_N\}) &=& 
\frac{1}{Z} e^{- \beta H(\{\sigma_1,...,\sigma_N\})}\, ,
\end{eqnarray}
with
\begin{eqnarray}
Z &=& \sum_{\{\sigma\}} e^{-\beta H(\{\sigma_1,...,\sigma_N\})}\, ,
\end{eqnarray}
where $Z$ is the partition function of the model, $\beta$ is a parameter of
the model, that in the context of spins represents the inverse temperature and
we introduced the notation
\begin{equation}
\sum_{\{\sigma\}} \equiv \sum_{\sigma_1=\pm 1}\, \sum_{\sigma_2=\pm 1}\,\, \cdots\,\, \sum_{\sigma_N=\pm 1} \, .
\end{equation}

The Hamiltonian should take into account the connection between
neurons and the fact that some minimum input is needed for the neuron to reach
the threshold and fire. The widely used expression is
\begin{eqnarray}
\label{eq_hamil}
H(\{\sigma_1,...,\sigma_N\}) &=&
- \frac{1}{2 N}\sum_{i,j} J_{ij} \sigma_i \sigma_j 
- \sum_i h_i \sigma_i\, ,
\end{eqnarray}
where $J_{ij}$ corresponds to the synaptic weight and $h_i$ is a term
that models the threshold as a ``field'' favoring the neuron to be in the rest
position.

Two features of this model are particularly pertinent for what
follows. First, it is directly defined in the language of statistical 
physics and allows the use of its framework with no additional complications. Secondly, if one
measures the averages $\langle \sigma_i \rangle$ and the correlations
$\langle \sigma_i \sigma_j \rangle$ of a spike train, the Boltzmann
machine arises naturally as a model consistent with
these measurements, as we will discuss in more detail in section~\ref{subsec_max_entr}.
On the other hand, a significant shortcoming of this model is that synapses are
\emph{symmetric}, i. e., $J_{ij}=J_{ji}$, which is not necessarily true in
biological systems.

It is important to note that the Hamiltonian shown in Eq.~(\ref{eq_hamil}) can give
rise to a rich diversity of behaviors depending on the choice of $J_{ij}$:
ferromagnetism, frustration, glassy systems, etc, as we will see
in chapter~\ref{chap_ising}.

\subsubsection{Particular case: Hopfield model}
\label{subsec_intro_hop}
Until now, we have presented models for neurons in completely arbitrary neuron
networks.
In this section we will describe a model that uses the same modelization for
neurons we presented in the last section but restricts the
synaptic weights $J_{ij}$ to a particular form:
\begin{equation}
J_{ij} = \sum_{\mu=1}^p \xi_i^\mu \xi_j^\mu\, ,
\label{eqeqeq}
\end{equation}
where $\xi_i^\mu$ are real values that we will discuss in the following.
This particular case of the generalized Ising model is called the Hopfield
model and was proposed to describe a system that stores a given number $p$
of memories and is capable to retrieve them when given a suitable input.
The form shown in Eq.~(\ref{eqeqeq}) was chosen so that the Hamiltonian we saw
in Eq.~(\ref{eq_hamil}) can be rewritten as
\begin{equation}
\label{eq_hop}
H = -  \frac{1}{2N} \sum_{\alpha=1}^p \left(\sum_i \xi_i^\alpha \sigma_i \right)^2
\end{equation}
which has the property that $\sigma_i = \mathrm{sign}(\xi_i^\alpha)$ is a local energy minimum for 
every $\alpha$ if $p \ll N$ and the patterns are more or less orthogonal.

The interpretation of this model as a model for associative memory comes from
the fact that, under certain conditions, if our system has an initial
configuration similar to one of the vectors $\xi^\mu$ it will evolve to the
configuration
$\sigma_i = \mathrm{sign}(\xi_i^\mu)$. In this context, we normally call the
$p$ vectors  $\{\xi^1,...,\xi^p\}$ memories (or \textit{patterns}).
More rigorously, we will see in section~\ref{sec_hop} that in the
limit $N \gg 1$,
the system can retrieve up to
$\alpha_c N$ stored binary patterns with $\alpha_c \simeq 0.138$.

The Hopfield model can also be seen as an approximation of the 
general Boltzmann Machine for a finite-rank $J$ matrix. Indeed, let's write the
eigenvector decomposition of the matrix $J$,
\begin{equation}
\label{eq_1133}
J_{ij} = \sum_{\alpha=1}^N \lambda_\alpha v_{\alpha,i} v_{\alpha,j} \, ,
\end{equation}
with $\{\lambda_\alpha\}$ and $\{v_{\alpha,i}\}$ being respectively the eigenvalues and
eigenvectors 
of the matrix $J$. If we truncate this summation up to the first $p$ highest 
eigenvalues and pose $\xi_i^\alpha = \sqrt{\lambda_\alpha}v_{\alpha,i}$, we find exactly the same equation as Eq.~(\ref{eq_hop}). On the
other hand, the
limit of $p=N$ does \emph{not} make this approximation exact, since
Eq.~(\ref{eq_hop}) cannot account for negative eigenvalues
of the matrix  $J_{ij}$.


This model got a renewed interest when experimentalists started looking for
patterns in spike train recording data.
A recent
experiment with rats made by
Peyrache et al. \cite{peyrache}
compared the spike activity of neurons in
two different moments: when the rat was looking for food in a maze and when
it was sleeping.
The main statistical tool used by the authors was the 
\textit{Principal Component Analysis (PCA)}, i.e., finding the eigenvalues and 
eigenvectors of the correlation matrix of the measured neuron activity. They 
showed that the eigenvectors that were the most strongly correlated with
neuron activity when the 
rat was choosing a direction in the maze were revisited during his next sleep. 
The authors interpreted this finding as the well-known process of memory 
consolidation during sleep.
In part~\ref{part3}, we will show that the author's proceeding of extracting
patterns from neural data using the PCA is closely 
related to fitting spike trains with a Hopfield model.

\section{Homologous proteins}
\label{sec_prot}

We say that two different proteins are homologous if they have both a common
evolutionary origin\cite{muddy} and a similar sequence, which normally also imply
a similar function. The comparison of the proteins of a homologous group gives some
valuable insight of which features are really essential for their biological function.

\begin{figure}[H]
\begin{center}
\includegraphics[width=0.9\columnwidth]{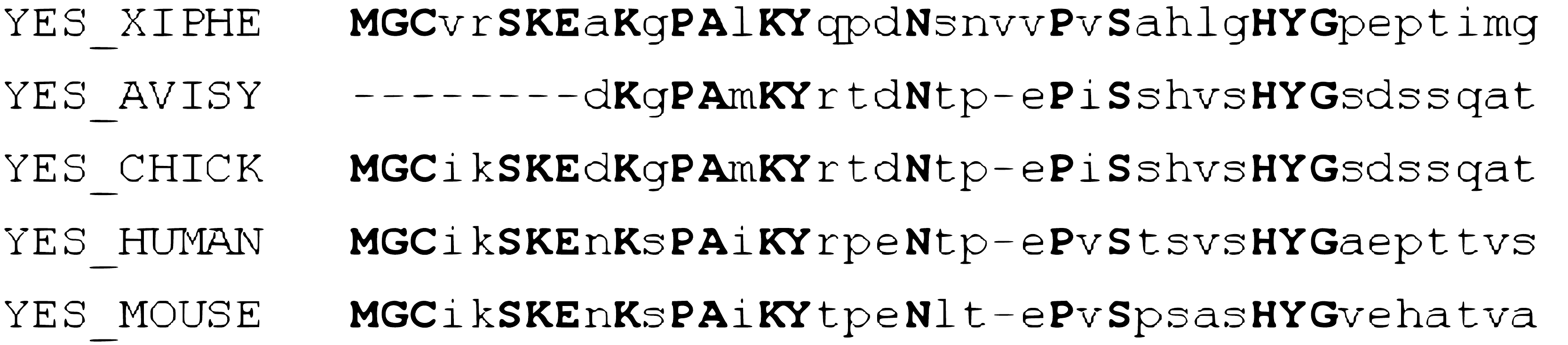}
\end{center}
\caption{A set of 41 sequences containing SH2 domain. Each line correspond to a different
  protein and each letter correspond to an amino acid, with conserved ones in
  bold. These sequences were matched using a multiple sequence
  alignment software\cite{muscle}.}
\label{fig_homol}
\end{figure}

The first step while comparing two or more homologous proteins is to align their
sequences in a way that maximizes the number of identical basis
(see Fig.~\ref{fig_homol}). This procedure is known as
\textit{Multiple Sequence Alignment (MSA)} \cite{lockless1999MSA}.
Since during evolution there could have been insertion or deletion of basis,
the optimal alignment will involve adding empty spaces to the alignment
in an optimal way, what makes the MSA problem NP-complete, i.e., solving it
needs a number of operations that grows exponentially with the number of
sequences.

It is natural to suppose that the most important parts of a protein
should vary significantly less than the least important ones, since most
mutations in important parts yield non-functional proteins.
Consequently, the most straightforward analysis one can do with aligned
sequences is to evaluate how the distribution of amino acids in a given
position deviates from a random uniform distribution\cite{capra2007predicting}.

While considering each position separately was proved to be
useful for identifying functional groups, a much richer behavior was found by
considering pairwise correlation between sites. First of all it has been
shown that by
taking into account both conservation and correlation one can describe more
accurately which sites of proteins are essential for its function
than by just considering conservation alone\cite{lichtarge1996evolutionary}.

Secondly, a remarkable experiment by Russ et al. \cite{func_prot} created
artificial proteins by randomly picking amino acids with a probability
distribution that
reproduced the averages and the pairwise correlations of a group of 
homologous natural proteins. 
He showed that these new proteins fold to a native tertiary
structure similar to that of the natural proteins of the group.
Conversely, he showed that random proteins that were generated \emph{without}
taking into account correlations do not fold into a well-defined
three-dimensional structure, an essential step for a protein to be functional.

Moreover, an interesting paper by Halabi et al. \cite{sectors} showed that the
correlation matrix has a particular structure: the amino acids can be
separated in disjoint groups (or \textit{sectors}) that are only
correlated to other amino acids inside the same sector. Each sector has a
distinct functional role and has evolved practically independently from
the others.

Finally, studying the two-basis correlations was shown to be a very good
way to infer which pairs of amino acids are spatially close in the
three-dimensional structure of the protein \cite{burger2010disentangling}.
Yet, some non-trivial work is needed to know if two basis are
correlated because they are spatially close one to another or because they are
spatially close to a third base, a problem very similar to the one presented
in section~\ref{sec_multineu} for neurons. To solve such a problem, a
paper published in 2009 \cite{weigt2009identification} proposed a very
simplified model to describe a family of proteins composed by
$N$ amino acids: it supposes that the proteins
that constitute the family are randomly chosen among all possible proteins
with length $N$ and that the probability of a given protein is given by:
\begin{equation}
  P(A_1,\ldots,A_N) = \frac{1}{Z} \exp\left[\sum_{i<j}J_{ij}(A_i,A_j)
    + \sum_i h_i(A_i)\right] \, ,
\end{equation}
where $A_i \in \{1,\ldots,22\}$ describe the $i$-th amino acid of the protein
and $J_{ij}(A_i,A_j)$ and $h_i(A_i)$ are real-valued functions.
This modelization is very similar to the Ising model we saw above and
reduces the problem of finding which basis are actually close in the
three-dimensional structure of the protein to the problem of finding which
functions $J_{ij}$ and $h_i$ of the Hamiltonian best describe a set of measured
two-site correlations.

To sum up, in the same way we saw in section~\ref{sec_neuron} for neurons,
we are dealing with a large number of
correlated data where the pairwise correlation plays a special role. While for
neurons we wanted to infer a synaptic network, in this case we would be
interested in extracting an expression for the \textit{effective fitness} of the
proteins of the group, i.e., a quantity that would say how well a
protein performs its biological role as a function of its amino acids sequence.

\chapter{Some classical results on Ising-like models}
\label{chap_ising}

As we have seen in chapter~\ref{chap_motiv}, Ising-like models are
modelizations of neural networks which are particularly suitable for
analytical calculations. In this chapter, we present some classical
results for some of these models, such as the Sherrington-Kirkpatrick and
the Hopfield model.
Since normally most of the behavior of
the system can be deduced from the partition function $Z$, it is normally said
that a model is ``solved'' when one evaluates this quantity explicitly.
We start by reminding some results for the Ising model as it was originally
defined. In the sequence, we will present
for both the Hopfield and the Sherrington-Kirkpatrick models
the procedure for evaluating $Z$ in general lines, since it will be
useful later in chapter~\ref{ch_inf_mat}. Indeed, as we will do similar
calculations, the comparison with these classical results will be
enlightening.

\section{The Ising model}
The original Ising Model was proposed by Wilhelm Lenz and first
studied by
Lenz's PhD student Ernst Ising as a simple model for
ferromagnetism and phase transitions. This model supposes that the atoms 
of
a magnet are arranged in a lattice and the spin of each atom $i$ is described 
by a binary variable $\sigma_i = \pm 1$.
In addition, it assume that each atom interacts only with its closest
neighbors, so we
can write the energy of the system as
\begin{equation}
\label{eq_hami_ising}
  H = - J \sum_{<i,j>} \sigma_i \sigma_j
  + h \sum_i \sigma_i \, ,
\end{equation}
\noindent where $J$ is the energy of the interaction between neighbors, favoring
spins to be aligned
and $h$ corresponds to an external magnetic field. The notation
$\sum_{<i,j>}$ means summing over all the pairs $i,j$ where $i$ and
$j$ are closest neighbors.

We suppose that the probability of the different states of the system is given 
by the Boltzmann distribution
\begin{eqnarray}
P(\{\sigma_1,...,\sigma_N\}) &=& 
\frac{1}{Z} e^{- \beta H(\{\sigma_1,...,\sigma_N\})}\, ,
\label{eq_boltpro}
\end{eqnarray}
with
\begin{eqnarray}
Z &=& \sum_{\{\sigma\}} e^{-\beta H(\{\sigma_1,...,\sigma_N\})}\, ,
\end{eqnarray}
where $\beta = \frac{1}{k_\mathrm{B} T}$, $k_\mathrm{B}$ is the Boltzmann
constant and $T$ is the temperature. In the following, unless explicitly
stated, we will absorb
the constant $\beta$ in the Hamiltonian to make notations lighter, but we
might still use the terms ``high temperature'' and ``low temperature'' to
refer to the magnitude of $J$ and $h$ in temperature units.
The \textit{thermal average}
of a quantity $f(\{\sigma_1,...\sigma_N\})$ is given by
\begin{eqnarray}
\moy{f(\{\sigma_1,...\sigma_N\})} =
\frac{1}{Z} \sum_{\{\sigma\}}
f(\{\sigma_1,...\sigma_N\}) e^{-H(\{\sigma_1,...,\sigma_N\})} \, .
\end{eqnarray}

The concept of ``closest neighbors'' depends both on the form of the lattice 
and its dimension. The one-dimensional case, where spins are arranged on a line,
was solved right after the model was proposed and shown to present no phase
transitions. With a brief calculation\cite{lebellac}, one can also find the
two-site correlation in the $h=0$ case, 

\begin{equation}
\moy{\sigma_i \sigma_j} =\left(\tanh J\right)^{|i-j|}\, .
  \label{eq_corr_isi1d}
\end{equation}

In two dimensions,
the Ising model was solved
after a mathematical
\textit{tour de force} \cite{onsager1944crystal} and shown to
have a second-order phase transition that separates a \textit{ferromagnetic
phase} (where magnetizations -- given by $m=\langle \sigma_i \rangle$ -- are non
zero)
from a \textit{paramagnetic phase} of zero magnetization.

Another case that shows a phase transition is the \emph{infinite dimension}
limit of the model, where the lattice is a complete graph, i.e., each spin is 
neighbor of every other one. In this case the Hamiltonian is given by
\begin{equation}
\label{eq_cw}
  H = - \frac{J}{N} \sum_{\text{$i<j$} } \sigma_i \sigma_j
  - h \sum_i \sigma_i \, ,
\end{equation}
where we did a rescaling of $J\rightarrow J/N$ to keep the Hamiltonian
extensive.
It is a classical calculation to show that in this case the magnetization
is given by the implicit equation
\begin{equation}
m = \tanh(J m + h)\, ,
\end{equation}
which presents a ferromagnetic/paramagnetic phase transition on $J=1$.
We can also obtain the connected correlation of the model:
\begin{equation}
\moy{\sigma_i \sigma_j} - \moy{\sigma_i}\moy{\sigma_j} = \frac{1}{N} \frac{J (1-m^2)^2}{1-J(1-m^2)}\,.
\end{equation}

Besides the different choices of lattice, there are several possible 
generalizations of the model expressed by small changes in the
Hamiltonian (\ref{eq_hami_ising}). For example, on can add interactions between
three sites with a term $J \sum_{i,j,k} \sigma_i \sigma_j \sigma_k$.
Of particular interest for this work is the
generalization of the lattice by defining
arbitrary two-site interactions and making the external field site-dependant:
\begin{equation}
  \label{eq_hami_is2}
  H = - \sum_{i < j}J_{ij} \sigma_i \sigma_j  - \sum_i h_i \sigma_i \, ,
\end{equation}
as we have already seen in section~\ref{sec_models}. In this case,
Ising models are also a privileged ground for the modeling of disordered
systems for two main reasons: first, they are specially convenient for
obtaining exact results and, secondly, the Ising model and all its
generalizations are particularly suitable for computer simulations using
Monte-Carlo methods \cite{krauth2006statistical}, with systems of up to a
thousand spins being
tractable.

We will now look at some particular cases.

\section{Sherrington-Kirkpatrick model}
The Sherrington-Kirkpatrick model (or \textit{SK} model) is  a simplified model
of disordered systems\cite{SK}. In this model, the Hamiltonian is given by
\begin{equation}
  \label{eq_hami_sk}
  H = - \sum_{i < j}\left(J_{ij} + \frac{J_0}{N}\right) \sigma_i \sigma_j\, ,
\end{equation}
where $J_0$ corresponds to a ferromagnetic component of the system. Each
$J_{ij}$ is chosen randomly with a Gaussian distribution
\begin{equation}
  P(J_{ij}) = \frac{1}{\sqrt{2 \pi J^2}}
  e^{-\frac{J_{ij}^2}{2J^2}} \, ,
\end{equation}
where $J$ represents the typical magnitude of couplings. To work with
intensive quantities,
we pose $J = \tilde{J}/\sqrt{N}$, $\tilde{J}$ being $O(1)$. We will denote the 
average of a value $f$
according to the distribution of $J_{ij}$ by $\overline{f}$, not to 
confound with the thermal average $\moy{f}$.

In the following we will look into the technical details of the solution
of this model for two reasons. First, there are some interesting concepts that 
emerge and secondly, we will do a similar calculation in part~\ref{part3}.

\subsection{Replica solution of the SK model}

As we discussed in the beginning of the chapter, to
solve this model we need to evaluate the free-energy $F = -\log Z$, a
quantity that depends on the particular sampling of $J_{ij}$.
Since the free-energy is extensive, we expect it to be self-averaging, i. e.,
to converge to its average value in respect to $J_{ij}$ when one increases
the size of the system.
We would like thus to evaluate $\overline{\log Z}$ to find the typical behavior
of the system.
Since evaluating
$\overline{Z^n}$ is much easier than evaluating $\overline{\log Z}$,
we will first evaluate $\overline{Z^n}$ for every integer $n$ and consider that
\begin{equation}
  \overline{\log Z} = \lim_{n \rightarrow 0} \frac{\overline{Z^n}-1}{n}\, .
\end{equation}
This procedure, known as the \textit{replica trick} \cite{edwards_anderson},
is useful for correctly solving several
statistical mechanics problems but is not  mathematically rigorous:
the limit depends on the behavior of $Z^n$ for $n \ll 1$ which is not
unambiguously defined as an analytic continuation of the integer
values of $Z^n$.

Initially, we have
\begin{eqnarray}
Z^n &=& \sum_{\{\sigma\}} \exp \left[\sum_{\alpha=1}^n 
\sum_{i < j}\left(J_{ij} + \frac{J_0}{N}\right) \sigma_i^\alpha \sigma_j^\alpha \right] \, ,
\end{eqnarray}
which is just the partition function of $n$ identical, non-interacting copies
of the system. Evaluating its average, we obtain
\begin{eqnarray}
\overline{Z^n} &=&  \sum_{\{\sigma\}} \exp \left[\sum_{\alpha=1}^n 
\sum_{i < j}\frac{J_0}{N} \sigma_i^\alpha \sigma_j^\alpha
+\frac{1}{2}\sum_{i<j}
\left(J \sum_{\alpha=1}^n  \sigma_i^\alpha \sigma_j^\alpha\right)^2
\right] \nonumber\\
 &=&  e^{N^2 J^2 n /4} \sum_{\{\sigma\}} \exp \left[\sum_{\alpha=1}^n 
\sum_{i < j}\frac{J_0}{N} \sigma_i^\alpha \sigma_j^\alpha
+\sum_{i<j}
J^2 \sum_{1 \leq \alpha < \gamma \leq n}  \sigma_i^\alpha \sigma_j^\alpha \sigma_i^\gamma \sigma_j^\gamma
\right] \nonumber\\
 &\cong&  e^{N^2 J^2 n /4} \sum_{\{\sigma\}} \exp \left[ \frac{J_0}{2N}
\sum_{\alpha=1}^n  \left(\sum_{i}
\sigma_i^\alpha \right)^2
+\frac{J^2}{2} \sum_{1 \leq \alpha < \gamma \leq n} \left(\sum_{i}
 \sigma_i^\alpha  \sigma_i^\gamma 
\right)^2 
\right] \, , \nonumber\\
\label{eq_001}
\end{eqnarray}
where in the last passage we neglected a term subdominant in $N$.

Using an integral transform, Eq.~(\ref{eq_001}) can be written as
\begin{eqnarray}
  \label{eq_int}
\overline{Z^n} &\cong&
e^{N^2 J^2 n /4}\int \prod_{1 \leq \alpha < \gamma \leq n}
\frac{\di q_{\alpha \gamma}}{\sqrt{2 \pi N^{-1}}}
\prod_{\alpha=1}^n \frac{\di m_{\alpha}}{\sqrt{2 \pi N^{-1}}}
\sum_{\{\sigma\}} e^U \, ,
\end{eqnarray}
where $U$ is given by
\begin{eqnarray}
U&=&
-\frac{N}{2} \sum_{1 \leq \alpha < \gamma \leq n} q_{\alpha \gamma}^2
-\frac{N}{2} \sum_{\alpha=1}^n m_\alpha^2
+ \sqrt{J_0} \sum_\alpha m_\alpha \sum_i \sigma_i^\alpha
\nonumber\\
&& +\tilde{J} \sum_{1 \leq \alpha < \gamma \leq n}
q_{\alpha \gamma} \sum_i \sigma_i^\alpha \sigma_i^\gamma \, .
\label{eq_int2}
\end{eqnarray}
Note that with this writing the sites are decoupled.
Consequently we have
\begin{equation}
\begin{split}
\sum_{\{\sigma\}} e^U =&
\left\{\sum_{\{\sigma\}} \exp\left[
-\frac{1}{2} \sum_{1 \leq \alpha < \gamma \leq n} q_{\alpha \gamma}^2
-\frac{1}{2} \sum_{\alpha=1}^n m_\alpha^2 \right. \right.
\\
&+ 
\left.\left.
 \sqrt{J_0} \sum_\alpha m_\alpha \sigma^\alpha
 +\tilde{J} \sum_{1 \leq \alpha < \gamma \leq n}
q_{\alpha \gamma}  \sigma^\alpha \sigma^\gamma
\right]\right\}^N \, .
\end{split}
  \label{eq_bla1}
\end{equation}

Finally we could in principle evaluate the integrals in Eq.~(\ref{eq_int}) using the saddle-point
approximation. Yet, finding the set of $q_{\alpha \gamma}$ and $m_\alpha$ that
constitute the saddle point for an arbitrary $n$ is non trivial.

\begin{figure}[H]
\begin{center}
\includegraphics[width=0.75\columnwidth]{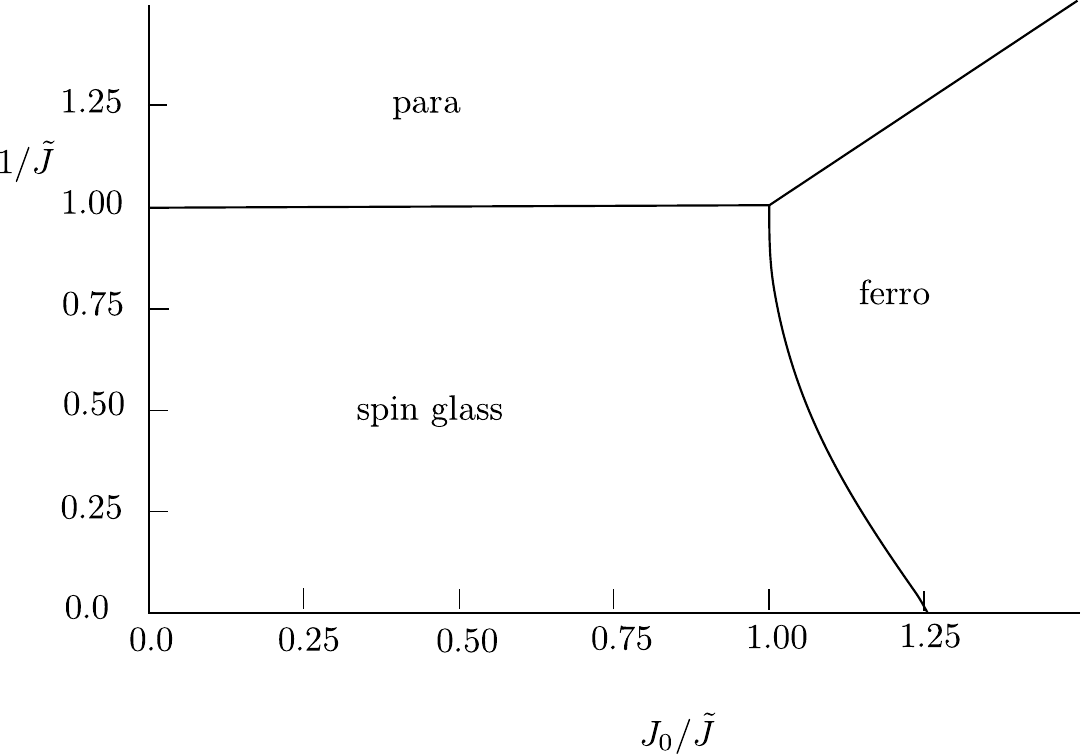}
\end{center}
\caption{Phase diagram of the Sherrington-Kirkpatrick model\cite{SK}.}
\label{fig_sk}
\end{figure}

To find the maximum of Eq.~(\ref{eq_bla1}),
one classically assumes
that all the different copies of the system have
identical statistical properties.  This is known as the \textit{replica symmetric} ansatz.
Mathematically, it corresponds to setting $q_{\alpha \gamma} = q$ and $m_\alpha = m$.
This hypothesis can be shown to yield a good approximation of the free-energy
and to correctly find the phase diagram of the model (Fig.~\ref{fig_sk}), 
but for very low temperatures it yields a negative entropy and hence
this supposition is clearly unjustified in this regime.

The parameters $m$ and $q$ have straightforward physical meanings in
the replica-symmetric case: $m = \overline{\moy{\sigma_i}}$, 
which means that if $m \neq 0$, 
the system has a preferred magnetization that does not vanish after
averaging with respect to the disorder. We say the system is in a ferromagnetic
phase. The other parameter $q$ can be written as
$q = \overline{\frac{1}{N}\sum_i\moy{\sigma_i}^2}$.
When $m = 0$ and $q \neq 0$, the system has a non-zero
magnetization for a giving sampling of $J_{ij}$, but this magnetization
vanishes when averaging with respect to $J_{ij}$. In this case, we say our system is in a \emph{spin glass} 
phase, 
where the system is frozen in one of the several (random) local minima of the 
energy. Finally, the case $q= m=0$ correspond to the paramagnetic phase.

\begin{figure}[H]
\begin{center}
\includegraphics[width=0.3\columnwidth]{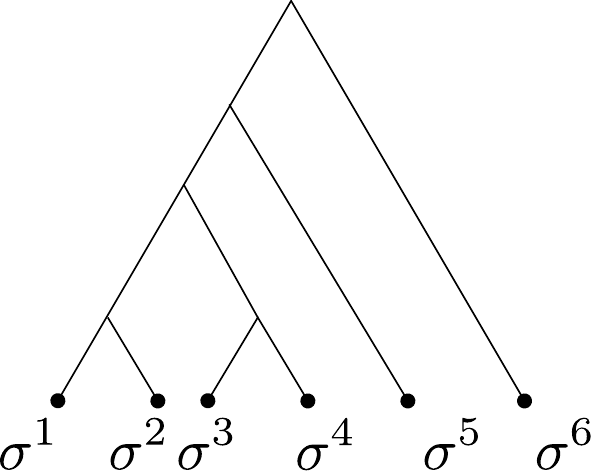}
\end{center}
\caption{Topology of the distance between the different replicas for the Parisi
solution.}
\label{fig_ultra}
\end{figure}

The correct saddle-point of equation~(\ref{eq_bla1}) was found in 
the late 70's by G. Parisi by defining the value of the matrix 
$q_{\alpha \gamma}$ at the saddle-point through an iterative procedure.
Note that in the general case, $q_{\alpha \gamma}$ is the overlap between
the replicas $\alpha$ and $\gamma$:
\begin{equation}
q_{\alpha \beta} =
\overline{\frac{1}{N} \sum_i \sigma_i^\alpha \sigma_j^\gamma}\, .
\end{equation}
His solution have a very interesting property:
if we consider any three replicas $\alpha$, $\gamma$ and $\rho$ and their
overlaps $q_{\alpha \gamma}$, $q_{\gamma \rho}$ and $q_{\alpha \rho}$ we will
have two identical overlaps and one that is strictly larger than the other
two. We can represent the replicas as the leaves of a three where the length
of the path from one leaf to another is the overlap between the replicas
(see Fig.~\ref{fig_ultra}). This distance defines an
\emph{ultrametric} structure for the replicas. More precisely,
we say that a metric space is ultrametric if
for any three points $x,y,z$ we have $d(x, z) \leq \max\{d(x, y), d(y, z)\}$
\cite{ultrametricity}.

The details of the Parisi solution can be found on \cite{parisi}.

\section{TAP Equations}
\label{sec_TAP}

We will now present another way of solving the SK model that remain correct
in the low-temperature regime: the TAP equations. This solution is of
particular interest to this work since it shares some common points
with our procedure for solving the inverse Ising model presented in
part~\ref{part2}.
In this section, we will derive these results following the work of Georges
and Yedidia \cite{G-Y}, since this formulation will be particularly useful for
what follows.

The TAP equations are a mean-field approximation for the SK model
derived by Thouless, Anderson and Palmer \cite{TAP1977}.
Its starting point is the same Hamiltonian as Eq.~(\ref{eq_hami_sk}) with
$J_0=0$:
\begin{equation}
\label{eq_hamil_start}
H = - \sum_{i<j} J_{ij} \sigma_i \sigma_j \, .
\end{equation}
%
To be able to make a small-coupling expansion, we introduce an inverse
temperature $\beta$ in our Hamiltonian.
We add also a Lagrange multiplier $\lambda(\beta)$ fixing $\moy{\sigma_i}=m_i$
\begin{equation}
H = - \beta \sum_{i<j} J_{ij} \sigma_i \sigma_j - \sum_i \lambda_i(\beta) (\sigma_i -m_i) \, ,
\end{equation}
and the corresponding partition function is
\begin{equation}
Z = \sum_{\{\sigma\}} \exp\left[  \beta \sum_{i<j} J_{ij} \sigma_i \sigma_j + \sum_i \lambda_i(\beta) (\sigma_i -m_i)\right] \, .
\end{equation}

For $\beta=0$, the Hamiltonian is trivial since it describes decoupled spins.
In this case,

\begin{equation}
\tanh \lambda_i(0) = m_i
\end{equation}

\noindent and
\begin{eqnarray}
\left. \log Z \right|_{\beta=0} &=& \sum_i \log \left[2\cosh(\lambda_i(0))\right]
- \lambda_i(0)m_i \nonumber\\
 &=& -\sum_i \frac{1+m_i}{2} \log\frac{1+m_i}{2}
-\sum_i \frac{1-m_i}{2} \log\frac{1-m_i}{2} \, .
\end{eqnarray}
For a general $\beta$, we do a Taylor expansion around $\beta=0$

\begin{eqnarray}
\lambda_i(\beta) &=& \lambda_i(0) + \left. \frac{\partial \lambda_i(\beta)}{\partial \beta}\right|_{\beta = 0} \beta +  \left. \frac{\partial^2 \lambda_i(\beta)}{\partial \beta^2}\right|_{\beta = 0}\frac{\beta^2}{2}  + ...
\end{eqnarray}
and
\begin{eqnarray}
F(\beta) = \log Z &=& F(0) +  \left. \frac{\partial F(\beta)}{\partial \beta}\right|_{\beta = 0} \beta +  \left. \frac{\partial^2 F(\beta)}{\partial \beta^2}\right|_{\beta = 0} \frac{\beta^2}{2}  + ... \, .
\end{eqnarray}
Each one of the derivatives of this series can be written as thermal averages
of decoupled spins. For example

\begin{eqnarray}
\left. \frac{\partial F(\beta)}{\partial \beta}\right|_{\beta = 0} &=&
\frac{1}{Z} \sum_{\{\sigma\}} \left[\sum_{i<j} J_{ij} \sigma_i \sigma_j+
\sum_i \left.\frac{\partial \lambda_i(\beta)}{\partial \beta}\right|_{\beta=0}(\sigma_i - m_i) \right] \times \nonumber\\
&& \times \exp\left[\sum_i\lambda_i(0) (\sigma_i - m_i)\right] \nonumber\\
&=& \sum_{i<j} J_{ij} m_i m_j \, ,
\end{eqnarray}
and
\begin{eqnarray}
\left. \frac{\partial \lambda_i(\beta)}{\partial \beta}\right|_{\beta = 0} &=& -\left. \frac{\partial^2 F(\beta)}{\partial \beta \partial m_i}\right|_{\beta = 0}
= - \sum_{j (\neq i)} J_{ij} m_j  \, .
\end{eqnarray}
Continuing this expansion with respect to $\beta$ up to the next order, we get

\begin{eqnarray}
F(\beta = 1) &=& -\sum_i \frac{1+m_i}{2} \log\left(\frac{1+m_i}{2}\right)
-\sum_i \frac{1-m_i}{2} \log\left(\frac{1-m_i}{2}\right)
\nonumber\\
&&+ \sum_{i<j} J_{ij} m_i m_j + \frac{1}{2} \sum_{i<j} J_{ij}^2(1-m_i^2)(1-m_j^2)
\end{eqnarray}
and
\begin{eqnarray}
\lambda_i(\beta=1) &=& \tanh^{-1} m_i - \sum_{j\, (\neq i)} J_{ij} m_j
+ m_i \sum_{j\,(\neq i)} J_{ij}^2 (1-m_j^2) \, .
\end{eqnarray}
Finally, to get back to our original Hamiltonian~(\ref{eq_hamil_start}), we set $\lambda_i=0$, obtaining:
\begin{eqnarray}
\log Z &=& -\sum_i \frac{1+m_i}{2} \log\left(\frac{1+m_i}{2}\right)
-\sum_i \frac{1-m_i}{2} \log\left(\frac{1-m_i}{2}\right)
\nonumber\\
&&+ \sum_{i<j} J_{ij} m_i m_j + \frac{1}{2} \sum_{i<j} J_{ij}^2(1-m_i^2)(1-m_j^2)
\label{eq_TAP_GY}
\end{eqnarray}
and
\begin{eqnarray}
\label{eq_dificil}
\tanh^{-1} m_i &=& \sum_{j\, (\neq i)} J_{ij} m_j
- m_i \sum_{j\,(\neq i)} J_{ij}^2 (1-m_j^2) \, ,
\end{eqnarray}
which are the original TAP equations. Note that the next terms of this expansion
are on higher powers of $J_{ij}$, that are defined in the SK model to be
$O(N^{-1/2})$ and thus negligible in the $N \rightarrow \infty$ limit.
Remark also that solving the $N$ coupled equations (\ref{eq_dificil}) is a
hard problem in general,  but feasible in the limit $J \gg 1$ or
close to the spin-glass phase transition \cite{TAP1977}.

\section{Hopfield model}
\label{sec_hop}

In this section, we will discuss in more detail the Hopfield model that
we have already presented in section~\ref{subsec_intro_hop}, based on the
work of Amit et al. \cite{amit1}.
We will consider a more general Hamiltonian than the one presented previously,
with the addition of local external fields:

\begin{equation}
  H =
  - \frac{1}{2N} \sum_{\mu=1}^p  \sum_{ij} \xi_i^\mu \xi_j^\mu \sigma_i \sigma_j
  - \sum_i h_i \sigma_i \, .
  \label{eq_hop_usual}
\end{equation}

\noindent The corresponding partition function is
\begin{eqnarray}
  Z &=& \int \prod_{\mu=1}^p \frac{\di m_\mu}{\sqrt{2 \pi \beta^{-1} N^{-1}}}
  \sum_{\{\sigma\}}
  \exp\left[ -\frac{\beta N}{2} \sum_{\mu=1}^p m_\mu^2 +
 \right.\nonumber\\
&&+\left.
   \beta \sum_{\mu=1}^p m_\mu \sum_{i} \xi_i^\mu \sigma_i +
\beta \sum_i h_i \sigma_i
   \right] \label{eq_hop_int00} \\
\label{eq_hop_int0}
  &=& \int \prod_{\mu=1}^p \frac{\di m_\mu}{\sqrt{2 \pi \beta^{-1} N^{-1}}}
   \nonumber\\
&& \exp \left\{ -\frac{\beta N}{2} \sum_{\mu=1}^p m_\mu^2 + \sum_i  \log\left[2 \cosh\left(\beta \sum_{\mu=1}^p m_\mu \xi_i^\mu + \beta h_i\right)\right]\right\}
   \, . \nonumber\\
\label{eq_hop_int}
\end{eqnarray}
If the number of patterns $p$ remains finite when $N \rightarrow \infty$,
we can solve this
integral using the saddle-point approximation

\begin{eqnarray}
  \log Z &=& -\frac{\beta N}{2} \sum_{\mu=1}^p m_\mu^2 + \sum_i  \log\left[2 \cosh\left(\beta \sum_{\mu=1}^p m_\mu \xi_i^\mu + \beta h_i\right)\right] \, ,
\label{eq_hop1234}
\end{eqnarray}
where
\begin{eqnarray}
  \label{eq_1234}
m_\mu &=& \frac{1}{N} \sum_i \xi_i^\mu \tanh \left(\beta \sum_\nu m_\nu \xi_i^\nu+ 
\beta h_i\right)\, .
\end{eqnarray}
The solutions of the equation~(\ref{eq_1234}) depend on the
details of the patterns and on the external fields. For instance,
let's
consider the case where $h_i=0$ and $\xi_i^\mu$ taken
randomly according to a Bernoulli distribution
$P(\xi_i^\mu = 1) = P(\xi_i^\mu = -1) = 1/2$.
In this case, we can show that if
$\beta < 1$, the only solution to the saddle-point equations is $m_\mu=0$,
which corresponds to a paramagnetic phase, while
for $\beta > 1$, non-trivial solutions of Eq.~(\ref{eq_1234}) do exist.
The solutions that are global minima of the free-energy are the states
where the magnetization over one pattern $m_\mu$ is non-zero while the others are zero, which correspond exactly
to the thermodynamic states where one retrieves the $\mu$-th pattern.

\begin{figure}[H]
\begin{center}
\includegraphics[width=0.55\columnwidth]{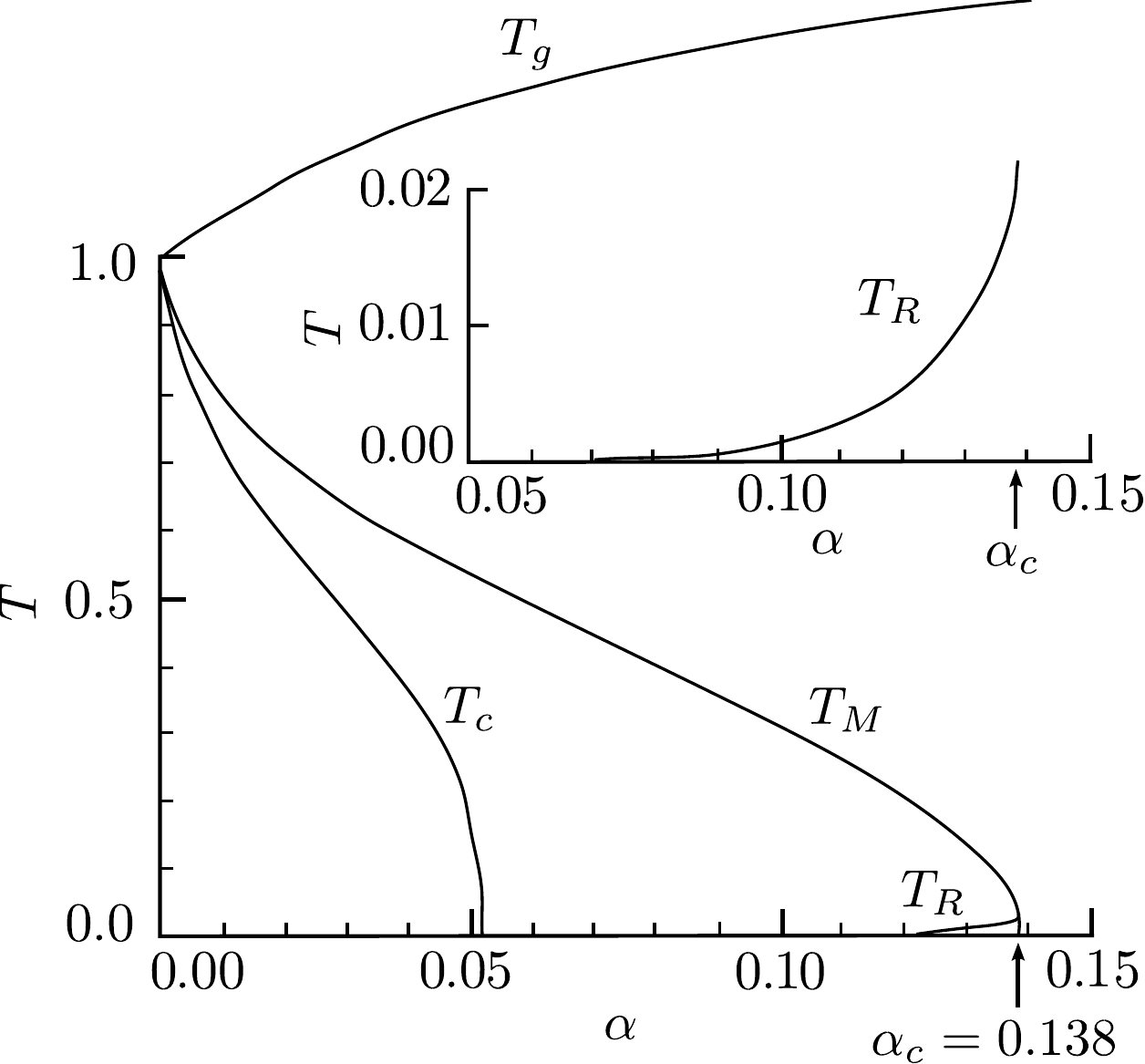}
\end{center}
\caption{Phase diagram of the Hopfield model\cite{amit1987statistical} of
  $p=\alpha N$ patterns.
The temperature $T_g$ corresponds to a transition from a paramagnetic phase 
to a spin-glass phase.
For $T < T_M$ the patterns are a \emph{local} minima of the free-energy and
for $T<T_c$ these minima are global. The temperature $T_R$ is the one below which
the replica-symmetric solution is false (see inset).}
\label{fig_hop}
\end{figure}

Dealing with the case of $p = \alpha N$, for $\alpha$ finite is considerably 
harder. It can be however treated through a calculation similar to that we will see in part~\ref{part3}
using the replica trick \cite{amit1992modeling}. In this 
case, the system has a ferromagnetic phase, where it retrieves
one of the patterns, a paramagnetic phase and a spin glass phase.
In the solution of Amit et al.,
as in the Sherrington-Kirkpatrick model, one needs to make a replica-symmetric
hypothesis
to solve the saddle-point equations of the problem. In the case of the
Hopfield model, for all but the very lowest temperatures the replica-symmetric
solution yields the correct expression of the free energy. The phase diagram
of the model is depicted in Fig.~\ref{fig_hop}.

\section{Graphical models}

The Ising problem is a particular case of a class of problems known in the
statistics community as \textit{undirected graphical models}
\cite{graphical}, which are 
statistical models
where the probability distribution can be factorized over the \textit{cliques}
of a certain graph $G$.

\begin{figure}[H]
\begin{center}
\includegraphics[width=0.5\columnwidth]{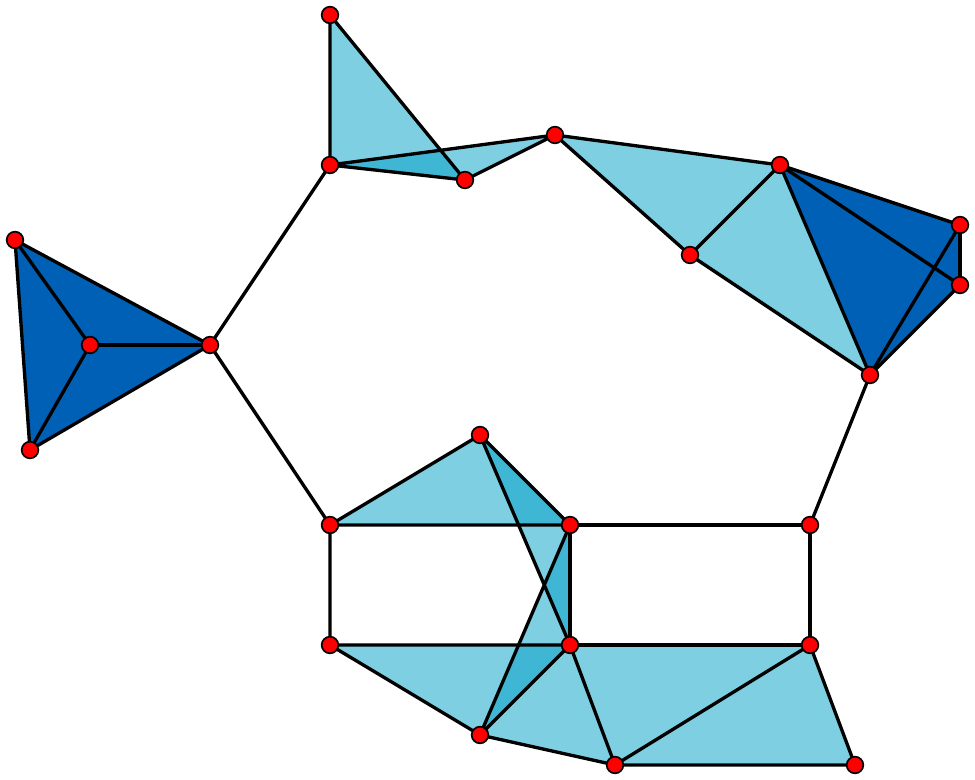}
\end{center}
\caption{In this graph the maximal cliques are the two dark-blue
colored subgraphs, each one of the 11 light blue colored triangles and all the edges that
are not part of any of them \cite{ep_wiki}.}
\label{cliques}
\end{figure}

A clique of a graph $G$ is a subgraph $C \subset G$ that is fully connected
(see Fig~\ref{cliques}). We pose $\mathcal{C}$ as the set of maximal cliques 
of a graph $G$, i.e., cliques that are not contained in any other clique. We
say a probability distribution over $N$ variables $x_1, \ldots, x_N$ is a graphical model if it 
can be factorized as
\begin{equation}
P(x_1,...,x_N) = \frac{1}{Z} \prod_{C \in \mathcal{C}} p_{C}(\{x_k\}_{k\, \text{vertex of $C$}}) \, ,
\end{equation}
where the underlying graph $G$ has $N$ vertices and $Z$ is a normalizing
constant of the probability.

In the case of the Ising model described in the
beginning of this chapter, the underlying graph is the lattice and the maximal
cliques are the edges that connect two neighbors. The probability of
a configuration on the most general graphical
model on a lattice is then
\begin{equation}
  \label{eq_general_clique}
P(x_1,...,x_N) = \frac{1}{Z} \exp\left[
  \sum_{<i,j>} \log(p_{ij}(x_i,x_j))
  \right] \, .
\end{equation}
If we assume that our variables $x_i$ can take binary values $\pm 1$,
the function $p_{ij}(x_i,x_j)$ is a function of $\{0,1\}\times\{0,1\}\rightarrow \mathbb{R}$, i. e., it can only take four different values:
$p_{ij}(+1,+1)$, $p_{ij}(+1,-1)$, $p_{ij}(-1,+1)$ and $p_{ij}(-1,-1)$.
There are an infinity of ways to express such a function with simple
operations. We will choose the one that resembles the most with the
Hamiltonian of a generalized Ising model:
\begin{equation}
\log(p_{ij}(x_i,x_j))= J_{ij}\, x_i\, x_j + \tilde{h}_{i j} \,x_i + \tilde{h}_{ji}\, x_j + \kappa_{ij}\,.
\end{equation}
Indeed, we can easily solve the linear system to find the four unknown values
($J_{ij}$, $\tilde{h}_{ij}$, $\tilde{h}_{ji}$ and $\kappa_{ij}$) as a function
of the four different values of $p_{ij}(x_i, x_j)$.

Absorbing all the constants $\kappa_{ij}$ in the normalization $Z$ and posing
$h_i = \sum_j \tilde{h}_{ij}$,
our probability is

\begin{equation}
  \label{eq_general_clique2}
P(x_1,...,x_N) = \frac{1}{Z} \exp\left[
  \sum_{<i,j>} J_{ij} x_i x_j + \sum_i h_i x_i
  \right] \, ,
\end{equation}

\noindent which correspond to the probability of a Ising-like model
 with couplings between closest neighbors and where both the
 local fields and the coupling
between the neighbors are site-dependent\footnote{Note that allowing non-uniform
  coupling between neighbors
  allows for systems with much more complex behaviors than just a simple
  ferromagnetic-paramagnetic transition. For an example, see
  the Edwards-Anderson model\cite{edwards_anderson}.}.

\subsection{Message-passing algorithms}
\label{subsec_message}

The expectation propagation is an interesting approximation for the general
problem of evaluating averages according to a graphical model. The starting
point of this method is the fact that when the underlying graph is a tree,
we can evaluate these averages exactly.
Suppose that we want to evaluate
\begin{equation}
P(x_s) = \sum_{\{x_1,...,x_{s-1},x_{s+1},...,x_N\}} P(x_1,...,x_N) \, .
\end{equation}
We choose to represent our tree with $s$ as its root.
In this case, we can write
\begin{equation}
\label{eq_asewq}
  P(x_s) = \sum_{\{x_1,...,x_{s-1},x_{s+1},...,x_N\}}
\frac{1}{Z}  \prod_{(r,s) \in E(G)} p_{r,s}(x_r, x_s) \, ,
\end{equation}
where $E(G)$ is the set of edges of the graph $G$.
We can decompose this expression on each branches starting
on $s$.

\begin{figure}[H]
\begin{center}
\includegraphics[width=0.35\columnwidth]{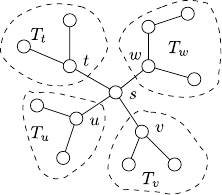}
\end{center}
\caption{Example tree. Note the branches $T_t, T_u, T_v$ and $T_w$ starting
  on its root.}
\label{fig_graphe}
\end{figure}

\noindent For the tree shown in Fig.~\ref{fig_graphe}, for example, it
will be

\begin{equation}
\begin{split}
  P(x_s) = \frac{1}{Z} 
  & \left[ \sum_{\{x_r\}_{r \in V(T_t)}} p_{s,t}(x_s,x_t)
  \prod_{(r,d) \in E(T_t)} p_{r,d}(x_r, x_d) \right] \times \\
   \times &\left[ \sum_{\{x_r\}_{r \in V(T_u)}} p_{s,u}(x_s,x_u)
  \prod_{(r,d) \in E(T_u)} p_{r,d}(x_r, x_d) \right] \times \\
   \times &\left[ \sum_{\{x_r\}_{r \in V(T_v)}} p_{s,v}(x_s,x_v)
  \prod_{(r,d) \in E(T_v)} p_{r,d}(x_r, x_d) \right] \times \\
   \times &\left[ \sum_{\{x_r\}_{r \in V(T_w)}} p_{s,w}(x_s,x_w)
  \prod_{(r,d) \in E(T_w)} p_{r,d}(x_r, x_d) \right]
\end{split}
\label{eq_propa}
\end{equation}
  
\noindent where each term in the product represents the contribution of one
branch.
As we can see, we transformed the Eq.~(\ref{eq_asewq}) in four independent problems
defined in each branch which can be solved separately.
By repeating the procedure recursively, it is possible to solve the problem
with a small number of operations. Note that the same divide-and-conquer method
can be used also for evaluating $Z$.

We now would like to reformulate this solution as an algorithm that would also
be well defined in graphs with cycles, even if not to give an exact solution
nor being guaranteed to converge. 
The algorithm work by passing in each iteration messages $M_{st}$ from every two 
vertex $s$ and $t$ connected by an edge, corresponding to an iterative 
relation

\begin{equation}
\label{eq_mesprop}
  M_{tr}(x_r) \leftarrow \kappa \sum_{x_t'} \left[p_{rt}(x_r, x_t')
  \prod_{u \in N(t), u \neq r} M_{ut} (x_t')\right] \, ,
\end{equation}

\noindent where $N(t)$ is the set of neighbors of $t$ and $\kappa$ is a
normalization constant fixing $\sum_{x_r} \prod_{t \in N(r)} M_{tr}(x_r) = 1$.

We can recover $P(x_s)$ with the formula
\begin{equation}
P(x_s) = \kappa  \prod_{t \in N(s)} M_{ts}(x_s) \, .
\end{equation}
Note that the fixed point of this algorithm is the solution
of~(\ref{eq_propa}).
This algorithm is the simplest message-passing algorithm and is known as
\textit{belief propagation}. Several variations of this algorithm
can be found in the literature\cite{graphical}.

\chapter{Inverse Problems}
\label{ch_inverse}

As exemplified in the previous chapters, most of the problems in statistical mechanics
consist of describing the collective behavior of a large number of interacting parts. In general, the
individual behavior of parts and how they interact are either described by
first principles or can be very accurately measured.
Unfortunately, as we have seen in chapter~\ref{chap_motiv}, for a few problems
like neuron networks, the behavior of the parts and/or how they interact is
not known, even if we can measure their collective behavior.
In these cases, we would like to deduce the behavior and interactions
of the parts from the available data.
We talk then of \textit{inverse problems}.



Inverse problems are often \textit{ill posed}, i.e., there is more than one
possible set of laws or parameters that can describe the observed data.
To give an example,
suppose all we know about a real-valued random variable $x$ is
that $\moy{x}=0$ and $\moy{x^2}=1$. Even if we restrain ourselves to
Bernoulli
distributions, there is an infinity of distributions satisfying our
conditions: for any real $a$, $P(x=a)=1/(1+a^2)$ and
$P(x=-1/a)=1-1/(1+a^2)$ meet our requirements.

Intuitively, a possible criterion for choosing one among all these distributions
is to look for the least ``restrictive'' one, i. e., the one which allows as many different values as possible. To formalize this criterion, we need to define the
\textit{Shannon entropy} of a statistical distribution. We will start thus
this chapter by defining this entropy and presenting how to optimize it to put
inverse problems in a well-defined framework. In the sequence, we will present
the Bayesian inference, which is a complementary approach to the entropy
optimization.
Finally, we will define and present some known
results for the inverse problem of most interest for this work: the inverse
Ising problem.

\section{Maximal entropy distribution}
\label{subsec_max_entr}

The Shannon entropy of a random variable is a measure of the quantity
of information unknown about it. It is defined by the sum
\begin{equation}
  S = - \sum_\Omega P(\Omega) \log P(\Omega) \, ,
  \label{eq_def_S}
\end{equation}
where $P(\Omega)$ is the probability of the configuration $\Omega$ of the
system.
Its interpretation as the quantity of information comes from the Shannon's
source coding theorem, which states that the best theoretically possible
compression algorithm can encode a sampling of $N$ values taken with the
distribution $P$ using $NS$ bits in the $N \rightarrow \infty$ limit.

If we are looking for the most general distribution that reproduces a set
of averages $f_i = \moy{f_i(X)}$, it is reasonable to look for the one that
maximizes $S$. This is known as the \textit{principle of maximum
entropy}. It can be interpreted as the model that satisfies our constraints,
i. e., reproducing the prescribed set of averages,
while
imposing as few extra conditions as possible.

Let us now consider the
interesting case of random binary variables $\sigma_i = \pm 1$, constraint to
satisfy a set of local averages $\moy{\sigma_i}=m_i$ and correlations
$\moy{\sigma_i \sigma_j}=C_{ij}$\cite{tkacik2006ising}.
In principle, one could also consider imposing higher order correlations, like
the three-site ones $C_{ijk}= \moy{\sigma_i \sigma_j \sigma_k}$,
but doing so would only be useful in situations where one knows such
high-order couplings precisely.
Unfortunately, to extract such data from an experimental system one needs to
measure a very large number of configurations of the system, which is
rarely possible. We choose then to deal with only one and two-site correlations.
In this case, we define generically the probability 
$P(\{\sigma_i\}) = p_{\sigma_1,...,\sigma_N}$
of a configuration $\{\sigma_1,\ldots,\sigma_N\}$ and
we can write the entropy as
\begin{equation}
S= -\sum_{\{\sigma\}} p_{\sigma_1,...,\sigma_N} \log p_{\sigma_1,...,\sigma_N} \, .
\end{equation}
In order to impose the constraints on the averages and correlations, we add
the Lagrange multipliers $h_i$, $J_{ij}$ and $\lambda$ respectively associated
to $m_i= \langle \sigma_i\rangle$, $C_{ij} = \langle \sigma_i \sigma_j \rangle$
and to the normalization of the probability
$\sum_{\{\sigma\}} p_{\sigma_1,...,\sigma_N} = 1$. We obtain
\begin{equation}
\begin{split}
S =& -\sum_{\{\sigma\}} p_{\sigma_1,...,\sigma_N} \log p_{\sigma_1,...,\sigma_N}
+\sum_i h_i \left(m_i -  \sum_{\{\sigma\}} p_{\sigma_1,...,\sigma_N} \sigma_i\right)
\\
& + \sum_{i,j} J_{ij} \left(C_{ij} -  \sum_{\{\sigma\}} p_{\sigma_1,...,\sigma_N} \sigma_i \sigma_j\right) + \lambda \left( \sum_{\{\sigma\}} p_{\sigma_1,...,\sigma_N} - 1\right) \, .
\end{split}
\end{equation}
Optimizing $S$ on $p_{\sigma_1,...,\sigma_N}$ we obtain
\begin{equation}
0 = \lambda - 1 - \log p_{\sigma_1,...,\sigma_N} + \sum_i h_i \sigma_i
+\sum_{ij} J_{ij} \sigma_i \sigma_j \, .
\label{eq_opt1}
\end{equation}
Solving Eq.~(\ref{eq_opt1}), we derive the probability distribution
\begin{equation}
P(\{\sigma\}) = e^{\lambda-1}
\exp\left(\sum_{ij} J_{ij} \sigma_i \sigma_j + \sum_i h_i \sigma_i\right) \, ,
\end{equation}
which corresponds exactly to the Boltzmann distribution for the 
generalized Ising model we presented in the beginning of chapter~\ref{chap_ising} (Eqs.~(\ref{eq_boltpro}) and~(\ref{eq_hami_is2}))
for $\lambda-1 = -\log Z$.

At this point, we know how the probability distribution depends on $J_{ij}$
and on $h_i$. To solve completely this problem, we need to express $J_{ij}$
and $h_i$ in terms of the imposed averages and correlations.
Since $J_{ij}$ and $h_i$ are Lagrange multipliers,
the values they should take to reproduce our averages and
correlations correspond to an extrema of the entropy. To know whether it
is a maximum or a minimum, we can do an explicit calculation
of $\frac{\partial S}{\partial J_{ij}}$ and $\frac{\partial S}{\partial h_i}$
to see that the entropy is a \emph{convex} function of the parameters
$J_{ij}$ and $h_i$.
Thus, we need to look for the set of parameters that \emph{minimizes}
the entropy. Moreover, the convexity assures that if a minimum exists, it is
unique.

To illustrate, let's examine the case of a two-spin system with
$\moy{\sigma_1}=m_1$, $\moy{\sigma_2}=m_2$ and $\moy{\sigma_1 \sigma_2}=C$.
The entropy is given by
\begin{eqnarray}
  S &=& \log\left\{\sum_{\sigma_{1=\pm 1}} \sum_{\sigma_2=\pm 1}
  \exp\left[J \sigma_1 \sigma_2 + h_1 \sigma_1 + h_2\sigma_2\right]\right\}
  - J C - h_1 m_1 - h_2 m_2 \, ,\nonumber\\
  &=& \log\left\{e^{J + h_1 + h_2} + e^{J - h_1 - h_2} + e^{-J + h_1 - h_2}
  + e^{-J - h_1 + h_2}\right\} \nonumber\\
  &&  - J C - h_1 m_1 - h_2 m_2 \, .
  \label{eq_bb11}
\end{eqnarray}
Since we have only two spins, the optimization of $S$ with respect to
$h_1$, $h_2$ and $J$ can be done explicitly and we obtain
\begin{eqnarray}
  \label{eq_J2spi}
J &=& \frac{1}{4}\log \left[
  \frac{(1+C-m_1-m_2)(1+C+m_1+m_2)}{(1-C-m_1+m_2)(1-C+m_1-m_2)}\right] \\
h_1 &=& \frac{1}{4} \log \left[
  \frac{(1-C+m_1-m_2)(1+C+m_1+m_2)}{(1+C-m_1-m_2)(1-C-m_1+m_2)}\right]\\
h_2 &=& \frac{1}{4} \log \left[
  \frac{(1-C-m_1+m_2)(1+C+m_1+m_2)}{(1+C-m_1-m_2)(1-C+m_1-m_2)}\right] \, .
  \label{eq_bb12}
\end{eqnarray}

\section{Bayesian inference}

Suppose now that we are not only interested in finding the best
parameters to fit some data, but also in attributing a probability
distribution to the set of these possible parameters.
If our model depends on a set of unknown parameters 
$\{\lambda_i\}$ we could
in principle write the probability $P(\{X_i\}|\{\lambda_i\})$ of measuring
any set of configurations $\{X_i\}$. The Bayes theorem states that the probability of
a set of \emph{parameters} $\{\lambda_i\}$ as a function of a set of measures $\{X_i\}$ is
\begin{equation}
  \label{eq_bayes}
  P(\{\lambda_i\}|\{X_i\}) = \frac{  P(\{X_i\}|\{\lambda_i\}) P_0(\{\lambda_i\})}{P(\{X_i\})} \, ,
\end{equation}
where $P_0(\{\lambda_i\})$ is the \textit{a priori} probability of the
parameters $\{\lambda_i\}$
and $P(\{X_i\})$ is the marginal probability of $\{X_i\}$, which can also be
interpreted as a normalization constant:
\begin{equation}
P(\{X_i\}) = \sum_{\{\lambda_i\}}  P(\{X_i\}|\{\lambda_i\}) P_0(\{\lambda_i\})
\, .
\label{eq_bayesnorm}
\end{equation}

If one is looking to the set of parameters $\{\lambda_i\}$ that best describes
the measures $\{X_i\}$,
a natural choice is the one that maximizes
Eq.~(\ref{eq_bayes}). Such choice is known as the
\textit{maximum a posteriori (MAP) estimator}. 

On the other hand,
when the prior is not known, a common procedure is to look for the set of 
$\{\lambda_i\}$ that maximizes $P(\{X_i\}|\{\lambda_i\})$, which is the same
as setting the prior to $P_0(\{\lambda_i\}) = 1$. We call such procedure
the \textit{maximum likelihood estimation}.

To illustrate, suppose that we have a coin that we know is biased in
the following way:
$P(\text{favored side}) = 1/2 + \epsilon$ and 
$P(\text{unfavored side}) = 1/2 - \epsilon$, but we do not know if head or
tails is favored. We toss this coin three times and get three heads. Using
the Bayes theorem, we have
\begin{equation}
P(\text{tails is favored}|\text{3 heads}) = 
\frac{P(\text{3 heads}|\text{tails favored}) P_0(\text{tails favored})}
{P(\text{3 heads})} \, .
\label{eq_3_12}
\end{equation}
Since we have no \emph{prior} knowledge whether head or tails is favored,
we have $P_0(\text{tails favored}) = P_0(\text{heads favored}) = 1/2$,
which leads to

\begin{eqnarray}
P(\text{tails are favored}|\text{3 heads}) &=& 
\frac{\left(\frac{1}{2}-\epsilon\right)^3}{\left(\frac{1}{2}-\epsilon\right)^3
+\left(\frac{1}{2}+\epsilon\right)^3} \, ,
\end{eqnarray}
and
\begin{eqnarray}
P(\text{heads are favored}|\text{3 heads}) &=& 
\frac{\left(\frac{1}{2}+\epsilon\right)^3}{\left(\frac{1}{2}-\epsilon\right)^3
+\left(\frac{1}{2}+\epsilon\right)^3} \, .
\end{eqnarray}
Unsurprisingly, we conclude that it is more likely that the coin's favored
side is heads.

Let us now look at a slightly different situation where the bias $\epsilon$ is
unknown.
As before, there is an unknown favored side and heads are obtained three times.
We would like to determine $\epsilon$.
From Eq.~(\ref{eq_bayes}), we obtain an expression similar to
Eq.~(\ref{eq_3_12})
\begin{eqnarray}
P(\epsilon|\text{3 heads}) &=& 
\frac{\left(\frac{1}{2}+\epsilon\right)^3 P_0(\epsilon)}{\int_{-1/2}^{1/2} \left(\frac{1}{2}+\epsilon'\right)^3 P_0(\epsilon') \di \epsilon'} \, .
\label{eq_315}
\end{eqnarray}

Remark in the denominator the normalization according to
Eq.~(\ref{eq_bayesnorm}).
In this case the probability has a strong dependence on the 
prior, which is unknown in the majority of inference problems. Fortunately,
this problem gets less and less important when one increases the amount of
data. For example, suppose that instead of doing just three coin tosses, we
toss it a large number of times, getting $N$ heads and $M$ tails. In this case,
Eq.~(\ref{eq_315}) becomes

\begin{eqnarray}
P(\epsilon|\text{$N$ heads, $M$ tails}) &=& 
\frac{
\left(\frac{1}{2}+\epsilon\right)^N
\left(\frac{1}{2}-\epsilon\right)^M
 P_0(\epsilon)}
{
\int_{-1/2}^{1/2} 
\left(\frac{1}{2}+\epsilon'\right)^N
\left(\frac{1}{2}-\epsilon'\right)^M
 P_0(\epsilon') \di \epsilon'} \, ,
\label{eq_316}
\end{eqnarray}
\noindent where the binomial term 
$\left(\begin{array}{c} N+M \\ N \end{array} \right)$ gets canceled out with 
the normalization. Since the function 
$\left(\frac{1}{2}+\epsilon\right)^N \left(\frac{1}{2}-\epsilon\right)^M$
has a very sharp peak around $(N-M)/(2M+2N)$, we can do the following
approximation:
\begin{equation}
  \left(\frac{1}{2}+\epsilon\right)^N
\left(\frac{1}{2}-\epsilon\right)^M
 P_0(\epsilon) \, \simeq  \, 
  \left(\frac{1}{2}+\epsilon\right)^N
\left(\frac{1}{2}-\epsilon\right)^M
 P_0\left(\frac{N-M}{2M+2N} \right) \, .
 \end{equation}
If we apply this approximation to Eq.~(\ref{eq_315}), 
the value
$P_0\left(\frac{N-M}{2M+2N} \right)$
appears both in the numerator and in the denominator and will cancel out.
The probability is thus independent of the unknown function $P_0(\epsilon)$.

In other situations, the prior might be useful to make an inference procedure
more robust. Suppose for example that we are measuring a system composed
by a large 
number of spins $\{\sigma_i\}$. By pure coincidence (or lack of data), two
particular sites, $i$ and $j$, have identical spin values in all the 
measured configurations. Without defining a prior (i.e, setting $P_0(\alpha)=1$),
the algorithm will
infer an infinite-valued coupling between the two sites to account for this,
which is non-physical and numerically problematic. On the other hand, if we
suppose that $P_0(\alpha)$ is a Gaussian distribution, the prior will skew the inferred
values away from very large values, avoiding the problematic solutions.

\subsection{Relationship with entropy maximization}

Suppose that we make $L$ independent measurements $\{\sigma_l\}$ of a system
we would like to describe using a set of parameters $\alpha$. Since
the measures are independent, we can write
\begin{equation}
\log P(\{\sigma\}|\alpha) = \sum_{l=1}^L \log P(\sigma_l|\alpha) \, .
\end{equation}
Using the maximum a posteriori principle and the Bayes theorem, the set
of $\alpha$ that best describes the data is
\begin{equation}
  \label{eq_eqeq}
\alpha = 
\mathop{\text{argmax}}_{\alpha'} \left[ \log P_0(\alpha') + \sum_{l=1}^L \log P(\sigma_l|\alpha')\right] \, ,
\end{equation}
where $P_0(\alpha)$ is the prior probability of $\alpha$. 
If we want to use the principle of the maximization of the entropy, one should
estimate the entropy from the data as
\begin{equation}
S(\alpha) = -\frac{1}{L} \sum_{l=1}^L \log P(\sigma_l|\alpha)
\, \simeq
- \moy{\log P(\sigma|\alpha)}_{\sigma} \, ,
\end{equation}
using the definition of an average, we can show that $S(\alpha)$ corresponds
to the usual definition of the entropy
\begin{equation}
S(\alpha) \, \simeq  \,
-\sum_{\{\sigma\}} P(\sigma|\alpha) \log P(\sigma|\alpha) \, .
\end{equation}
As we saw in the last section, we should then minimize $S(\alpha)$ with respect
to the parameters $\alpha$, what corresponds exactly to maximizing
$P(\{\sigma\}|\alpha)$, as one would do using the maximum likelihood method.

\section[The I.I.P.: some results from the literature]{The inverse Ising problem: some results from the literature}
\label{sec_inv_ising}
We call the problem of finding the set of couplings $\{J_{ij}\}$ and local
fields $\{h_i\}$ from the set of magnetizations $\{m_i\}$ and correlations
$\{C_{ij}\}$ of a generalized Ising model
the \textit{inverse generalized Ising problem} \cite{bialek-berry}.
In the following, we will omit the mention ``generalized'' for simplicity.
We expect
this problem to be particularly hard, since as we have
seen in chapter~\ref{chap_ising}, the direct problem of finding the
magnetizations from the model's parameters is already a non-trivial one.

\subsection{Monte Carlo optimization}

One can use the fact that the direct problem is numerically solvable with Monte
Carlo (\textit{MC}) methods \cite{krauth2006statistical} to solve the inverse
problem with the following algorithm \cite{ackley1985learning}:
\begin{enumerate}
  \item start with an initial guess for the parameters $J_{ij}^0$ and $h_i^0$.
  \item do a Monte-Carlo simulation to find the set of magnetization
    $m_i^\text{est}$ and
    correlations  $C_{ij}^\text{est}$ corresponding to these parameters.
    \label{step2}
  \item update $J_{ij}$ according to
    $J_{ij}^{t+1} = J_{ij}^{t} + \eta(t)(C_{ij} - C_{ij}^\text{est}) +
    \alpha (J_{ij}^{t} - J_{ij}^{t-1})$ for some chosen function $\eta$ and
    constant $\alpha$.
    \label{step3}
  \item update $h_i$ analogously to $J_{ij}$.
    \label{step4}
  \item repeat steps~\ref{step2} -- \ref{step4} until
    $\max_{ij} (C_{ij} - C_{ij}^\text{est}) < \epsilon$.
\end{enumerate}
The number of steps necessary to reach a certain accuracy depends
both on the initial parameters, the function $\eta$ and the parameter $\alpha$.
An important drawback of this algorithm is that it is very inefficient: at
each step one must do a Monte-Carlo simulation that is very time-consuming
if one needs an accurate result. There are others modified versions
of this algorithm that improve the number of necessary
steps \cite{inference_mc}, but they all involve doing a MC simulation at
every step and thus have the same drawbacks.

\subsection{Susceptibility propagation}
\label{sec_susc_prop}

In 2008, M. M\'ezard and T. Mora had the interesting idea of
modifying the message passing algorithm we have seen in
section~\ref{subsec_message} to solve the inverse Ising
problem \cite{mora,valerie_susc_prop}.

In their paper, the authors first write the Belief Propagation equations to 
find
the values of $C_{ij}$ and $m_i$. They reinterpret these
equations by identifying $\{J_{ij},h_i\}$ as the unknowns and
$\{C_{ij},m_i\}$ as the input data and describe a message passing procedure
that converges to the right fixed-point in trees.
The details can be found in \cite{valerie_susc_prop}.

This procedure, in the same way as the belief propagation for the direct
problem, is exact on trees and an approximation for graphs that contain loops.
If it converges (which is not guaranteed in graphs with loops), it do so
in polynomial time, which makes it much faster than
the Monte Carlo optimization. The main drawback of this method is that 
for graphs with loops the resulting approximated solution might be very far 
from the optimal solution of the problem.

\subsection{Inversion of TAP equations}
Another approach for solving the inverse Ising problem was proposed by
Roudi et al.\cite{hertz_tap}. Their starting point are the TAP equations we
already saw in section~\ref{sec_TAP}:
\begin{equation}
  \tanh^{-1} m_i = h_i +  \sum_{j\, (\neq i)} J_{ij} m_j
- m_i \sum_{j\,(\neq i)} J_{ij}^2 (1-m_j^2) \, .
\end{equation}
Taking the derivative of this expression with respect to $m_j$ and
noting that $(C^{-1})_{ij}=\partial h_i / \partial m_j$, we have
\begin{equation}
 (C^{-1})_{ij}=-J_{ij} - 2 J_{ij}^2 m_i m_j + (1-m_i^2)^{-1} \delta_{ij} \, ,
  \label{eq_tap1}
\end{equation}
\noindent
which is easily solvable for $J_{ij}$.

Note that if $J_{ij}$ is small, we can neglect the $J_{ij}^2$ term, yielding
an explicit solution for the couplings
\begin{equation}
J_{ij}  =  (C^{-1})_{ij} -  (1-m_i^2)^{-1} \delta_{ij}  \, .
  \label{eq_tap_mf}
\end{equation}

The inversion of Eq.~(\ref{eq_tap1}) has the same strengths and drawbacks of
the use of the TAP
equations in the direct problem: it is exact in the large size limit for the
SK model and we might expect it to work well only in models where the
couplings are small.

\subsection{Auto-consistent equations}

Recently, a novel approach for finding the parameters of an Ising 
model was proposed by the statistics community\cite{wainwright-high}.
Its main idea reposes on the fact that for a Ising system, the magnetization
respects
\begin{eqnarray}
m_i &=& \sum_{\{\sigma_k\}_{k \neq i}} \sum_{\sigma_i = \pm 1} \sigma_i
\exp\left[
\sigma_i \sum_k J_{ij} \sigma_j +
\mathop{\sum_{j<k}}_{j\neq i\neq k} J_{jk}\sigma_j \sigma_k + \sum_j h_j \sigma_j + h_i \sigma_i \right]
\nonumber\\
 &=& \moy{\tanh\left(\sum_{j\,(\neq i)} J_{ij} \sigma_j + h_i \right)} \, ,
\label{eq_moy1}
\end{eqnarray}
\noindent where we used the fact that
$\sum_{\sigma = \pm 1} \sigma e^{\sigma A} = \tanh(A) \sum_{\sigma = \pm 1} e^{\sigma A}$.
An analogous expression can be derived for the correlations:
\begin{eqnarray}
C_{ij} = 
\moy{
\frac{
 {A_{ij}(\sigma,+,+)} 
-{A_{ij}(\sigma,+,-)} 
-{A_{ij}(\sigma,-,+)} 
+{A_{ij}(\sigma,-,-)} 
}{
 {A_{ij}(\sigma,+,+)} 
+{A_{ij}(\sigma,+,-)} 
+{A_{ij}(\sigma,-,+)} 
+{A_{ij}(\sigma,-,-)} 
}} \, ,
\label{eq_corr1}
\end{eqnarray}
where
\begin{equation}
A_{ij}(\sigma,\tau,\rho) = \exp\left[ \rho \tau J_{ij} + \rho \sum_k J_{ik} \sigma_k + \tau \sum_k J_{jk} \sigma_k + \tau h_i + \rho h_j\right] \, .
\end{equation}

Suppose now that we have a set of $L$ independent measures of the
full spin configuration of our system $\{\sigma^1,...,\sigma^L\}$, with
$\sigma^k = \{\sigma^k_1,... \sigma^k_N\}$. We can then estimate Eqs.~(\ref{eq_moy1}) 
and~(\ref{eq_corr1}):
\begin{equation}
\frac{1}{L} \sum_{l=1}^L \sigma_i^l =
\frac{1}{L} \sum_{l=1}^L \tanh\left(\sum_{j\,(\neq i)} J_{ij} \sigma_j^l + h_i \right) \, ,
\end{equation}
and, respectively,
\begin{equation}
\begin{split}
\frac{1}{L} \sum_{l=1}^L \sigma_i^l \sigma_j^l =&\\
=\frac{1}{L} \sum_{l=1}^L &\frac{
 {A_{ij}(\sigma^l,+,+)} 
-{A_{ij}(\sigma^l,+,-)} 
-{A_{ij}(\sigma^l,-,+)} 
+{A_{ij}(\sigma^l,-,-)} 
}{
 {A_{ij}(\sigma^l,+,+)} 
+{A_{ij}(\sigma^l,+,-)} 
+{A_{ij}(\sigma^l,-,+)} 
+{A_{ij}(\sigma^l,-,-)} 
} \, .
\end{split}
\end{equation}
We have thus a 
system of coupled non-linear equations for
$J_{ij}$ and $h_i$ which can be solved without the need to evaluate
the partition function $Z$.

This procedure allows one to find the couplings from the
measured data in polynomial time, but it has a few drawbacks. First
of all, this procedure is not optimal according to the Bayes theorem.
It depends on all high-order correlations while the optimal Bayes
inference depends only on magnetizations and correlations. 
Accordingly, this method does not work if the Hamiltonian used to generate
the data has any  three or higher order couplings. This is 
particularly awkward for the case of inferring neural synapses where
the hypothesis of the Ising model is just an approximation.
Finally, solving the set of equations for $J_{ij}$ is a non-trivial
problem. The original paper\cite{wainwright-high} proposes an algorithm
to solve it that unfortunately does not work in the low-temperature regime.





\part{Some results on the inverse Ising problem}
\label{part2}
\chapter{The inverse Ising problem in the small-correlation limit}
\label{sec_expansion}

In section~\ref{sec_inv_ising}, we have introduced the inverse Ising problem
and discussed what has been done in the literature to solve it.
In this chapter, we propose a small-correlation expansion procedure that
allows one to find the couplings and magnetizations up to any given power on
the correlations\cite{sessak}. We will find an explicit expression for the couplings
and magnetizations that is correct up to $O((\text{largest connected correlation})^3)$.

We consider the generalized Ising model for a system composed of $N$ spins $\sigma_i=\pm 1$, 
$i=1,\ldots ,N$, whose Hamiltonian is given by
\begin{equation}
H(\{\sigma_i\}) = -\sum_{i<j} J_{ij}\, \sigma_i\, \sigma_j - \sum_i h_i \, \sigma_i \, ,
\end{equation}
as we have already introduced in chap.~\ref{chap_ising}, in Eq.~(\ref{eq_hami_is2}).
We want to find the values of couplings $J_{ij}^*$ and fields
$h^*_i$ such that
the average values of the spins and of the spin-spin correlations
match the prescribed magnetizations $m_i$, given by
\begin{equation}\label{eqmc}
m_i = \moy{\sigma_i} \, ,
\end{equation}
and connected correlations 
$c_{ij}$, defined by
\begin{equation}
c_{ij} = \moy{\sigma_i \sigma_j} -m_i\; m_j \, .
\end{equation}
For given fixed magnetizations and correlations, the entropy of the generalized
Ising model, obtained in section~\ref{subsec_max_entr}, is given by

\begin{equation*}
S(\{J_{ij}\},\{\lambda_{i}\};\{m_i\},\{ c_{ij}\} ) =
\log Z(\{J_{ij}\}, \{h_i\}) -
\sum_{i<j} J_{ij} \, (c_{ij} +m _i \, m_j)- \sum_i h_i \, m_i  \, ,\\
\end{equation*}
\vspace{-0.5cm}
\begin{equation}
  \label{eqG2}
  \begin{split}
\phantom{xxxx}&= \log \sum_{\{\sigma_i\}} \exp\left\{
\sum_{i<j} J_{ij} \left[ \sigma_i\sigma_j  -
  c_{ij} -m_i m_j\right] + \sum_i h_i (\sigma_i - m_i)\right\}\, ,\\ 
  &= \log \sum_{\{\sigma_i\}} \exp\left\{
\sum_{i<j} J_{ij} \left[ (\sigma_i - m_i)(\sigma_j - m_j) -
  c_{ij}\right] + \sum_i \lambda_i (\sigma_i - m_i)\right\} \, ,
\end{split}
\end{equation}
where the new fields $\lambda_i$ are simply related to the physical fields
$h_i$ through $\lambda _i= h_i + \sum_j J_{ij} \, m_j$. In this same section,
we have seen that the couplings $J_{ij}^*$ and fields $h_i^*$ are the ones
that minimize the entropy.
As discussed in chap.~\ref{chap_ising},
the exact evaluation of the entropy shown in Eq.~(\ref{eqG2}) for a given set of
$J_{ij}$ and $\lambda _i$ is, in general, a computationally
challenging task, not to say about its minimization. To obtain a
tractable expression we multiply all connected correlations $c_{ij}$ 
in Eq.~(\ref{eqG2}) by the same small parameter $\beta$, which can be
interpreted as a fictitious inverse temperature. Our entropy is thus
\begin{multline}
S(\{J_{ij}\},\{\lambda_{i}\};\{m_i\},\{\beta\, c_{ij}\} ) =\\
=   \log \sum_{\{\sigma_i\}} \exp\left\{
\sum_{i<j} J_{ij} \left[ (\sigma_i - m_i)(\sigma_j - m_j) -
  \beta c_{ij}\right] + \sum_i \lambda_i (\sigma_i - m_i)\right\} \, ,
\end{multline}
In this chapter we want to expand the entropy in powers of $\beta$ as a
function of the magnetizations and correlations:
\begin{equation}
  S(\{m_i\},\{\beta \, c_{ij}\}) = S^0 + \beta S^1 + \beta^2 S^2 + \ldots\, ,
\end{equation}
Accordingly, $J_{ij}^*$ and $\lambda_i^*$ can also be written as series on $\beta$:
\begin{eqnarray}
  J_{ij}^*(\{m_i\},\{\beta \, c_{ij}\}) &=& J_{ij}^0 + \beta J_{ij}^1 + \beta^2 J_{ij}^2 + \ldots\, , \\
  \lambda_{i}^*(\{m_i\},\{\beta \, c_{ij}\}) &=& \lambda_{i}^0 + \beta \lambda_{i}^1 + \beta^2 \lambda_{i}^2 + \ldots\, ,
\end{eqnarray}
where we omit the dependency of the terms on $\{m_i\}$ and $\{c_{ij}\}$ to
make notations lighter.
The entropy we are looking for will be obtained when setting $\beta =1$
in the expansion.
Since the parameter $\beta$ multiply every
value of $c_{ij}$, we have that $S^k = O(c_{ij}^k)$. We can thus deduce that
our expansion for $S$ will be convergent for small enough couplings.
Note that once we have expressed the entropy as a series on $\beta$, we can
retrieve an expansion for couplings and fields using the following
identities, that follow from the definition of the entropy:
\begin{equation}
  \frac{\partial S(\{m_i\},\{\beta \;c_{ij}\} )}
       {\partial c_{ij}} = -\beta J_{ij}^*(\beta) \, ,
       \label{eq_derivcij}
\end{equation}
and
\begin{equation}
   \frac{\partial S(\{m_i\},\{\beta \;c_{ij}\} )}
        {\partial m_{i}} = -\lambda_{i}^*(\beta) \, .
       \label{eq_derivmi}
\end{equation}
Thus, once we have found an expansion for $S$, it is trivial to deduce from it
an expansion for $J_{ij}^*$ and $\lambda_i^*$.

The calculation of the entropy
$S(\{m_i\},\{\beta\, c_{ij}\} )$
is straightforward for $\beta =0$ since
spins are uncoupled in this limit. In this case, the values of the couplings
and fields minimizing the entropy are thus
\begin{equation}
J^0_{ij}  =0 \qquad  \text{and}\qquad \lambda^0_i = \tanh^{-1} (m_i) \ .
  \label{eq_jij0}
\end{equation}
Accordingly, the entropy for $\beta=0$ is
\begin{equation}
  \label{eq_s0}
  S^0 =
  -\sum_i \left[\frac{1+m_i}{2}\ln\frac{1+m_i}{2} + 
\frac{1-m_i}{2}\ln\frac{1-m_i}{2}\right]\, .
\end{equation}

To find the non-trivial terms of the entropy
we proceed in the following way: first we
define a potential $U$ over the spin configurations at inverse
temperature $\beta$ through (note the new last term)
\begin{equation}
\begin{split}
  U (\{\sigma_i\})&= 
\sum_{i<j} J_{ij}^*(\beta) \left[ (\sigma_i - m_i)(\sigma_j - m_j) -
  \beta\; c_{ij}\right]
  + \sum_i \lambda_i^*(\beta) (\sigma_i - m_i)
  \\
  &
  + \sum_{i<j} c_{ij} \int_0^\beta \di \beta' J_{ij}^*(\beta') \, ,
\end{split}
  \label{eqU}
\end{equation}
and a modified entropy (compare to Eq.~(\ref{eqG2}))
\begin{equation}  \label{eeerr}
\tilde S(\{m_i\},\{c_{ij}\}, \beta ) = 
\log \sum_{\{\sigma_i\}} e^{U(\{\sigma_i\})} \ .
\end{equation}
Notice that $U$ depends on the coupling values $J^*_{ij}(\beta')$
at all inverse temperatures $\beta' < \beta$.
The true entropy (at its minimum) and the modified entropy 
are simply related to each other through
\begin{equation}
S(\{m_i\},\{c_{ij}\}, \beta ) = \tilde S(\{m_i\},\{c_{ij}\}, \beta ) - \sum_{i<j} c_{ij} \int_0^\beta \di \beta' J_{ij}^*(\beta')
\label{realG} \ .
\end{equation}
The modified entropy $\tilde S$ in Eq.~(\ref{eeerr}) was chosen to be
independent of $\beta$. Indeed, it
has an explicit
dependence on $\beta$ through the potential $U$ (Eq.~(\ref{eqU})), and
an implicit dependence through the couplings and the fields. As the
latter are chosen to minimize $S$, the full derivative of $\tilde S$
with respect to $\beta$ coincides with its partial derivative, and we
get  
\begin{equation}
\label{gconst}
\frac{d\tilde S}{d\beta}= \frac{\partial \tilde{S}}{\partial \beta} =- \sum_{i<j}  c_{ij}\; J_{ij}^*(\beta) + 
\sum_{i<j} c_{ij} \; J_{ij}^* (\beta)=0 \ .
\end{equation}
The above equality is true for any $\beta$.
Consequently, $\tilde S$ is constant and equal to
its value at $\beta=0$, $S^0$, given in Eq.~(\ref{eq_s0}).

In the following, we will use the fact that $\tilde{S}$ does not depend on
$\beta$ to write self-consistency equations from which we will deduce our
expansion. We will start by presenting $S^1$ and $S^2$
since their calculations differ from those used for higher orders.
Afterwards,
we will present the calculations for $S^3$ as a generalizable example
of the general method, which will be presented in the sequence.

\section{Evaluation of \textit{S}\textsuperscript{1} and \textit{S}\textsuperscript{2}}
To find $S^1$, we derive Eq.~(\ref{realG})
with respect to $\beta$:
\begin{eqnarray}
S^1 &=& \left.  \derivb{S}\right|_0 =
\left.\derivb{\tilde{S}}\right|_0 - \sum_{i<j} c_{ij} J_{ij}^*(0) \, ,\nonumber\\
&=& 0 \, ,
\end{eqnarray}
since $\tilde{S}$ does not depend on $\beta$ (Eq.~(\ref{gconst})) and $J_{ij}^*(0) = J_{ij}^0= 0$
(Eq.~(\ref{eq_jij0})). A direct
consequence of this, deriving from Eq.~(\ref{eq_derivmi}), is
\begin{equation}
 \left.\derivb{\lambda_i}\right|_0 = 0 = \lambda^1\, .
\label{eq_lambda1_0}
\end{equation}

To evaluate the next term $S^2$, we note that since $
\frac{\di \tilde{S}(\beta)}{\di \beta}=0$ for any $\beta$,
we have in particular that
\begin{equation}
  \derivb{\tilde{S}} = 0 =
\frac{1}{\sum_{\{\sigma\}} e^{U(\{\sigma\},\beta)}}
\sum_{\{\sigma\}} \derivb{U(\{\sigma\})} e^{U(\{\sigma\})} = \moyen{\derivb{U}}\, .
\label{eq_b1b1}
\end{equation}
Evaluating Eq.~(\ref{eq_b1b1}) explicitly for $\beta=0$ yields
\begin{eqnarray}
  0 =
-\sum_{i<j}c_{ij} J_{ij}^*(0)
+\sum_{i<j} \left.\derivb{J_{ij}^*}\right|_0 \moyen{(\sigma_i - m_i)(\sigma_j - m_j)}_0 \, ,
\end{eqnarray}
This equality is trivial, since we know that $J_{ij}^*(0)=0$ and the averages
also vanish since the spins are uncoupled for $\beta=0$.
We must thus look at the second derivative of $\tilde{S}$:
\begin{equation}
  \derivmb{\tilde{S}}{2} = 0 = \moyen{\derivmb{U}{2}}
  + \moyen{\left(\derivb{U}\right)^2}   - \moyen{\derivb{U}}^2 \, .
\label{eq_b1b2}
\end{equation}
Explicitly, the first term corresponds to
\begin{eqnarray}
\derivmb{U}{2} &=& \sum_{i<j} \derivmb{J_{ij}^*}{2}(\sigma_i-m_i)(\sigma_j-m_j) +
\sum_i \derivmb{\lambda_i^*}{2} (\sigma_i-m_i)
\nonumber\\
&&- \sum_{i<j} \derivb{J_{ij}^*} c_{ij} -
\beta \sum_{i<j} \derivmb{J_{ij}^*}{2} c_{ij} \, ,
\end{eqnarray}
which for $\beta=0$ yields
\begin{eqnarray}
\left.\derivmb{U}{2}\right|_0 &=&
- \sum_{i<j} c_{ij} \left.\derivb{J_{ij}^*}\right|_0 \, ,
\end{eqnarray}

The next one is given by
\begin{eqnarray}
\left . \left( \derivb{U} \right)^2 \right|_0 &=& 
\sum_{i<j} \sum_{k<l} \derivb{J_{ij}^*} \derivb{J_{kl}^*} 
\left(\sigma_i - m_i\right) \left(\sigma_j - m_j\right)
\left(\sigma_k - m_k\right) \left(\sigma_l - m_l\right)
\nonumber\\
&&+ \sum_{i,j}
\derivb{\lambda_i^*}  \derivb{\lambda_j^*} (\sigma_i-m_i)(\sigma_j-m_j) 
\nonumber\\
&&+ 2 \sum_{i<j} \sum_k \derivb{J_{ij}^*} \derivb{\lambda_i^*} (\sigma_i-m_i)
(\sigma_j-m_j)(\sigma_k-m_k) \, ,
\end{eqnarray}
which for $\beta=0$ reduces to
\begin{eqnarray}
\moyen{\left( \derivb{U} \right)^2}_0
&=&\sum_{i<j} \left(\left. \derivb{J_{ij}^*} \right|_0
\right)^2 (1 - m_i^2)(1 - m_j^2) \, ,
\end{eqnarray}
where we used in the last equation the fact that $\left.\derivb{\lambda_i^*}\right|_0=\lambda_i^1 = 0$ (Eq.~(\ref{eq_lambda1_0})). The last term vanishes as consequence of
Eq.~(\ref{eq_b1b1}).

Finally, we can rewrite Eq.~(\ref{eq_b1b2}) as
\begin{equation}
\label{moyennes}
\sum_{i<j} c_{ij} \left. \derivb{J_{ij}^*}\right|_0 =
\sum_{i<j} \left(\left. \derivb{J_{ij}^*} \right|_0
\right)^2 (1 - m_i^2)(1 - m_j^2)  \, ,
\end{equation}
whose simpler solution is given by
\begin{equation}
 \left. \derivb{J_{ij}^*}\right|_0 = \frac{c_{ij}}{(1-m_i^2)(1-m_j^2)} = J_{ij}^1\, .
 \label{eq_jij2}
\end{equation}
Using Eqs.~(\ref{realG}) and~(\ref{eq_derivmi}) we can now deduce
\begin{equation}
  S^2 = -\frac{1}{2}\sum_{i<j} \frac{c_{ij}^2}{(1-m_i^2)(1-m_j^2)} \, ,
\end{equation}
and
\begin{equation}
  \left.\derivmb{\lambda_i^*}{2}\right|_0 = 2 m_i \sum_j \frac{c_{ij}^2}{(1-m_i^2)(1-m_j^2)^2} = 2 \lambda_i^2\, .
\end{equation}
Finally, we have the value of $\lambda_i^2$, $S^2$ and our first non-trivial
estimation of $J_{ij}^*$.
We can verify the correctness of Eq.~(\ref{eq_jij2}) by noting it is the
first-order approximation of Eq.~(\ref{eq_J2spi}) for small $c$.

\section{Evaluation of \textit{S}\textsuperscript{3}}
\label{sec_evals3}
Like in previous section, we calculate the third derivative of $\tilde{S}$ with
respect to $\beta$:
\begin{equation}
0= \derivmb{\tilde{S}}{3}=
\moyen{\derivmb{U}{3}} + 3\moyen{\derivmb{U}{2}\derivb{U}} + \moyen{\left(\derivb{U}\right)^3} \, ,
\end{equation}
which yields, after evaluating the averages (see appendix~\ref{ap_A}):
\begin{equation}
  \sum_{i<j} c_{ij}\left. \derivmb{J_{ij}^*}{2}\right|_0=
-4 \sum_{i<j} \frac{c_{ij}^3 m_i m_j}{(1-m_i^2)^2(1-m_j^2)^2} -
6 \sum_{i<j<k} \frac{c_{ij} c_{jk} c_{ki}}{(1-m_i^2)(1-m_j^2)(1-m_k^2)}\, .
\end{equation}
Taking the third derivative of Eq.~(\ref{realG}), we can show that
\begin{equation}
  -\sum_{i<j} c_{ij}\left. \derivmb{J_{ij}^*}{2}\right|_0=
  \left.\derivmb{S}{3}\right|_0 = 6  S^3 \, .
\end{equation}  
Comparing the two last equations, we
finally find the expression for $S^3$:
\begin{equation}
  S^3 = 
\frac{2}{3} \sum_{i<j} \frac{c_{ij}^3 m_i m_j}{(1-m_i^2)^2(1-m_j^2)^2}
+ \sum_{i<j<k} \frac{c_{ij} c_{jk} c_{ki}}{(1-m_i^2)(1-m_j^2)(1-m_k^2)} \, .
\end{equation}

\section{Higher orders}

The expansion procedure can be continued order by order using the same procedure
as in section~\ref{sec_evals3}. To evaluate $S^k$ having already evaluated
all $S^n$ for $n<k$, one must evaluate
$\left. \derivmb{\tilde S}{k}\right|_0$ as a sum of
averages with respect to uncoupled spins. After evaluating explicitly the
averages, one will have
\begin{equation} 
\left. \derivmb{\tilde S}{k}\right|_0 =- \sum_{i<j} c_{ij} 
\left. \derivmb{J_{ij}^*}{k-1}\right|_0+ Q_k \, ,
\label{anyorder}
\end{equation}
where $Q_k$ is a (known) function of the magnetizations, correlations, 
and of the
derivatives in $\beta=0$ of the couplings $J_{ij}^*$ and fields
$\lambda_i^*$ of order $\le \max (1,k-2)$. See Appendices \ref{ap_A} and \ref{app_mag}.

Finally, as  $\tilde S$ is constant by
virtue of Eq.~(\ref{gconst}), both sides of Eq.~(\ref{anyorder}) vanish. Using
Eq.~(\ref{realG}), we have
\begin{equation}
  S^k = -\frac{Q_k}{k!} \, ,
\end{equation}
which allows then to find the fields and couplings using Eqs.~(\ref{eq_derivcij}-\ref{eq_derivmi}).

Using this procedure, we could go up to $S^4$ (details are on
Appendix~\ref{ap_A}). Using the notations
\begin{equation}
L_i = \moy{\left(\sigma_i - m_i\right)^2}_0 = 1-m_i^2 \, ,
\end{equation}
which is basically the variance of an independent spin of average $m_i$ and
\begin{equation}
K_{ij} = (1-\delta_{ij}) \frac{\moy{\left(\sigma_i - m_i\right)\left(\sigma_j - m_j\right)}_0}
{\moy{\left(\sigma_i - m_i\right)^2}_0\moy{\left(\sigma_j - m_j\right)^2}_0} =
(1-\delta_{ij}) \frac{c_{ij}}{L_i L_j} \, ,
\end{equation}
where we have multiplied our definition of $K_{ij}$ by one minus a Kronecker
symbol so
that $K_{ii}= 0$, what makes our notations simpler. With these definitions,
we have
\begin{eqnarray}
S &=& 
-\sum_i \left[\frac{1+m_i}{2}\ln\frac{1+m_i}{2} + 
\frac{1-m_i}{2}\ln\frac{1-m_i}{2}\right]
\nonumber\\
&&-
\frac{\beta^2}{2} \sum_{i<j} K_{ij}^2 L_iL_j +
\frac{2}{3} \beta^3 \sum_{i<j}K_{ij}^3 m_i m_j L_i L_j
+ \beta^3 \sum_{i<j<k} K_{ij} K_{jk}K_{ki} L_i L_j L_k\nonumber\\
&&- \frac{\beta^4}{12} \sum_{i<j} K_{ij}^4 \left[1 + 3m_i^2 +3m_j^2 + 9 m_i^2m_j^2\right] L_i L_j
-\frac{\beta^4}{2} \sum_{i<j} \sum_k K_{ik}^2 K_{kj}^2 L_k^2 L_i L_j
\nonumber\\
&&- \beta^4 \sum_{i<j<k<l}
(K_{ij} K_{jk} K_{kl} K_{li} +
  K_{ik} K_{kj} K_{lj} K_{il} + K_{ij} K_{jl} K_{lk} K_{ki})
  L_i L_j L_k L_l
\nonumber\\
&&+ O(\beta^5) \, .
\label{eqG}
\end{eqnarray}
The result for $J_{ij}^*$ is
\begin{eqnarray}
J_{ij}^*(\{c_{kl}\}, \{m_i\},\beta) &=&
\beta K_{ij} -2 \beta^2 m_i m_j K_{ij}^2
- \beta^2 \sum_k K_{jk}K_{ki} L_k\nonumber\\
&&+ \frac{1}{3} \beta^3 K_{ij}^3 \left[1 + 3m_i^2 +3m_j^2 + 9 m_i^2m_j^2\right]
\nonumber\\
&&
+ \beta^3 \sum_{\substack{k \\ (\neq i,\,\neq j)}} K_{ij}
(K_{jk}^2 L_j +K_{ki}^2 L_i)  L_k \nonumber\\
\label{jij}
&&+ \beta^3 \sum_{\substack{k, l \\ (k\neq i,\, l\neq j)}} K_{jk} K_{kl} K_{li}L_k L_l
+
 O(\beta^4) \, ,
\end{eqnarray}  
and the physical field is given by
\begin{eqnarray}
h_l^*(\{c_{ij}\}, \{m_i\},\beta) &=&
\frac{1}{2} \ln \left( \frac{1+m_l}{1-m_l}\right)
-\sum_j J_{lj}^* m_j
+ \beta^2 \sum_{j (\neq l)} K_{lj}^2 m_l L_j
\nonumber\\
&&-\frac{2}{3} \beta^3 (1+3m_l^2) \sum_{j (\neq l)} K_{lj}^3 m_j L_j
-2 \beta^3 m_l \sum_{j<k} K_{lj} K_{jk} K_{kl} L_j L_k \nonumber\\
&&+ 2 \beta^4 m_l \sum_{i<j} \sum_k K_{ik} K_{kj} K_{jl} K_{li} L_i L_j L_k\nonumber\\
&&+
\beta^4 m_l \sum_{j} K_{lj}^4 L_j \left[1 + m_l^2 + 3 m_j^2 +3m_l^2m_j^2\right]
\nonumber\\
&&+ \beta^4 m_l \sum_{i\, (\neq l)} \sum_j K_{ij}^2 K_{jl}^2 L_i L_j^2 
+ O(\beta^5) \, .
\label{eq_hi}
\end{eqnarray}

\section{Checking the correctness of the expansion}
As we can see in Appendix~\ref{ap_A}, the calculations for getting to
Eq.~(\ref{eqG}) are long and error-prone. In this section, we will look at the
different methods used to verify the correctness of these calculations.

\subsection{Comparing the values of the external field with TAP equations}
\label{sec_compar}
In section~\ref{sec_TAP}, we presented an expansion of the free energy of the
direct Ising model for small couplings. The first two orders were developed by
Thouless et al. to solve the SK model, and are given in Eq.~(\ref{eq_TAP_GY}).
In 1991 A. Georges and J. Yedidia
\cite{G-Y} published the next two orders of this expansion. They found
\begin{eqnarray}
- \beta F(\{\beta J_{ij}\},\{m_i\})
&=& -\sum_i \left[\frac{1+m_i}{2}\ln\frac{1+m_i}{2} + \frac{1-m_i}{2}\ln\frac{1-m_i}{2}\right] + \beta \sum_{i<j} J_{ij} m_i m_j
\nonumber \\ 
&+& \frac{\beta^2}{2} \sum_{i<j} J_{ij}^2 L_iL_j + 
\frac{2 \beta^3}{3}  \sum_{i<j} J_{ij}^3 m_i m_j L_i L_j 
+ \beta^3 \sum_{i<j<k} J_{ij} J_{jk} J_{ki} L_i L_j L_k
\nonumber \\ 
&-& \frac{\beta^4}{12} \sum_{i<j} J_{ij}^4 (1+3m_i^2+3m_j^2-15m_i^2m_j^2) L_i L_j
\nonumber \\ 
&+& 2 \beta^4  \sum_{i<j} \sum_k J_{ij}^2 J_{jk} J_{ki} m_i m_j L_i L_j L_k
\nonumber \\ 
&+&  \beta^4 \sum_{i<j<k<l}
(J_{ij} J_{jk} J_{kl} J_{li} +
  J_{ik} J_{kj} J_{lj} J_{il} + J_{ij} J_{jl} J_{lk} J_{ki})
L_i L_j L_k L_l\, . \nonumber\\
\label{eq_gy}
\end{eqnarray}

From this result, we can derive the external fields as a function of $J_{ij}$ and $m_i$ through:
\begin{equation}
h_i(\{J_{ij}\},\{m_i\}) = \frac{\partial F}{\partial m_i} \, .
\label{eq_gy_hi}
\end{equation}
For example, up to $J^2$, we have
\begin{equation}
  \label{eq_hij2}
  h_i = \frac{1}{2} \ln\left(\frac{1-m_i}{1-m_j}\right) - \sum_j J_{ij} m_j + \sum_j J_{ij}^2 m_i L_j + O(J^3) \, .
\end{equation}

We would like to compare this equation to our result for $h_i(\{c_{ij}\},\{m_i\})$,
given in Eq.~(\ref{eq_hi}), in order to check the correctness of our expansion.
To rewrite Eq.~(\ref{eq_hij2}) as a function of $\{c_{ij}\}$, we use the 
expansion for $J_{ij}(\{c\},\{m\})$ obtained by us, $J_{ij} = K_{ij} + O(c^2)$.
We rewrite then Eq.~(\ref{eq_hij2}) as
\begin{equation}
  h_i = \frac{1}{2} \ln\left(\frac{1-m_i}{1-m_j}\right) - \sum_j J_{ij} m_j + \sum_j K_{ij}^2 m_i L_j + O(c^3) \, ,
\end{equation}
which corresponds exactly to the first three terms of Eq.~(\ref{eq_hi}).
We followed the same procedure using all the terms of Eq.~(\ref{eq_gy}) and
the expansion of $J_{ij}$. We could verify then all the
orders of Eq.~(\ref{eq_hi}). The details are in Appendix~\ref{ap_feio}.


\subsection{Numerical minimum-squares fit}
\label{sec_numeric}

In this section, we present a method to verify our expansion for $S$
given in Eq.~(\ref{eqG}) numerically.
For that, we rewrite our result in a slightly more general way, introducing
the coefficients $\{a_1,\ldots,a_6\}$:
\begin{equation}
  \begin{split}
S^\text{dev} =& 
-\sum_i \left[\frac{1+m_i}{2}\ln\frac{1+m_i}{2} + 
\frac{1-m_i}{2}\ln\frac{1-m_i}{2}\right]
\\
&+
a_1  \sum_{i<j} K_{ij}^2 L_iL_j +
a_2  \sum_{i<j}K_{ij}^3 m_i m_j L_i L_j
+ a_3 \sum_{i<j<k} K_{ij} K_{jk}K_{ki} L_i L_j L_k\\
&+ a_4 \sum_{i<j} K_{ij}^4 \left[1 + 3m_i^2 +3m_j^2 + 9 m_i^2m_j^2\right] L_i L_j
+a_5 \sum_{i<j} \sum_k K_{ik}^2 K_{kj}^2 L_k^2 L_i L_j
\\
&+ a_6 \sum_{i<j<k<l}
(K_{ij} K_{jk} K_{kl} K_{li} +
  K_{ik} K_{kj} K_{lj} K_{il} + K_{ij} K_{jl} K_{lk} K_{ki})
  L_i L_j L_k L_l\, .
  \end{split}
\end{equation}
We would like to obtain these coefficients through a numerical fit using data
generated by exact enumeration. Afterwards, we can verify if these results
match those derived formerly in this chapter.

We proceeded in the following way:
\begin{enumerate}
  \item We choose randomly $\{c_{ij}\}$ and $\{m_i\}$ for a system with $N=5$ spins.
    Both the correlations $\{c_{ij}\}$ and the magnetizations $\{m_i\}$ are chosen
    randomly with an uniform distribution in the interval
    $[-10^{-12},10^{-12}]$ and $[-1,1]$, respectively.
    The values of $c_{ij}$ are
    very small so that the terms on $c^{k+1}$ in the expansion of $S$ are
    negligeable with respect to those in $c^k$.\label{stepaa}

  \item We find numerically the minimum $S^\text{num}$ of the entropy $S$ with
    respect to $J_{ij}$ and
    $h_i$. This calculation has to be done with a very large numerical
    precision to
    account for the very small values of $c_{ij}$. We used 400 decimal units.
    \label{stepbb}

  \item We repeat steps~\ref{stepaa} and~\ref{stepbb} for different samplings
    of  $\{c_{ij}\}$ and $\{m_i\}$ to evaluate
    $D = \moy{(S^\text{dev} - S^\text{num})^2}_{\{c_{ij}\},\{m_i\}}$. In our case,
    we used 60 different random values of $\{c_{ij}\}$ and $\{m_i\}$.

  \item We find the set of $a=\{a_1,\ldots,a_6\}$ that minimizes $D$. Note that
    since $D$ is a quadratic function of the coefficients $a_i$, this method
    can still be done efficiently if we go further in the expansion and have
    a much larger set $a$.
\end{enumerate}

The obtained values of $\{a_1,\ldots,a_6\}$ (see table~\ref{tab1}) show a very
good agreement with Eq.~(\ref{eqG}), giving support to our derivation.

\begin{table}[H]
\begin{center}
\begin{tabular}{|l|c|c|c|c|c|c|}
  \hline
  Constant&$a_1$ & $a_2$ & $a_3$ & $a_4$ & $a_5$ & $a_6$ \\
  \hline
  Error $\phantom{a^{a^{a^a}}}$ & $3.7\cdot 10^{-32}$ & $5.4 \cdot 10^{-20}$ & $2.1\cdot 10^{-18}$ &
  $8.2\cdot 10^{-12}$ & $6.7\cdot 10^{-8}$ & $3.8\cdot 10^{-6}$ \\
 \hline
\end{tabular}
\end{center}
\caption{Agreement between theoretical and numerical values of $\{a_1,\ldots,a_6\}$ }
\label{tab1}
\end{table}



\chapter{Further results based on our expansion for the inverse Ising model}
\label{chap_sums}
In this chapter, we will see some useful results that follow from the expansion
made in the last chapter. In particular, we will sum some infinite subsets of the
expansion, what will make the expansion more robust.

To make some results in the following more visual,
we will introduce a diagrammatical notation. A point
in a diagram represents a spin and a line represents a $\beta K_{ij}$ link. 
We do not represent the polynomial in the variables $\{m_i\}$ that
multiplies each link. Summation over the indices is implicit.
Using these conventions, we can write our entropy as:
\begin{eqnarray}
S(\{c_{kl}\}, \{m_i\},\beta)  &=&
  -\diag{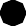} -
  \frac{1}{2} \diag{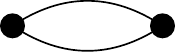}
  +\frac{2}{3}\diag{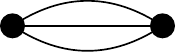}
  +\diag{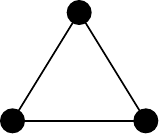}
  \nonumber \\
  &&-
  \frac{1}{12}\diag{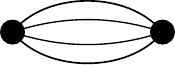}
  -\frac{1}{2}\diag{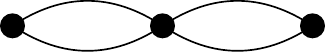} -
  \diag{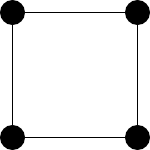} \, .
\end{eqnarray}

We can also represent $J_{ij}^*$ diagrammatically, with the difference that we
connect the $i$ and $j$ sites with a dashed line that do not represent any term
in the expansion. The summation over indices are only done in sites that are
not connected by a dashed line. We obtain
\begin{eqnarray}
J_{ij}^*  &=&
  \diag{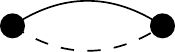}
  -2 \diag{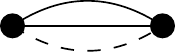}
  -\diag{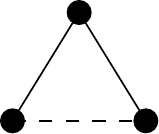}
 \nonumber \\
  &&+
  \frac{1}{3}\diag{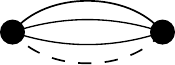}
  +\diag{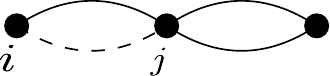} +\diag{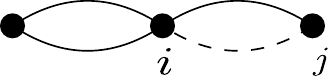} +
  \diag{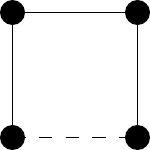} \, .
  \label{diagsJ}
\end{eqnarray}

\section{Loop summation}
\label{sec_loop}

If we rewrite Eq.~(\ref{eqG}) in a slightly different form, a particular
subset of the terms in the expansion seems to follow a regular pattern
(mind the last three terms):
\begin{eqnarray}
S &=& 
-\sum_i \left[\frac{1+m_i}{2}\ln\frac{1+m_i}{2} + 
\frac{1-m_i}{2}\ln\frac{1-m_i}{2}\right]
+\frac{2}{3} \beta^3 \sum_{i<j}K_{ij}^3 m_i m_j L_i L_j
\nonumber\\
&&
+ \frac{\beta^4}{6} \sum_{i<j} K_{ij}^4 \left[1 - 3m_i^2 -3m_j^2 - 3 m_i^2m_j^2\right] L_i L_j
\nonumber\\
&&
-\frac{\beta^2}{2} \sum_{i<j} K_{ij}^2 L_iL_j
+ \beta^3 \sum_{i<j<k} K_{ij} K_{jk}K_{ki} L_i L_j L_k\nonumber\\
&&- \frac{\beta^4}{8} \sum_{i,j,k,l} K_{ij} K_{jk} K_{kl} K_{li}  L_i L_j L_k L_l
\nonumber\\
&&+ O(\beta^5) \, ,
\label{eq_12345}
\end{eqnarray}
where we have used the identity
\begin{equation}
\begin{split}
- \frac{\beta^4}{8} &\sum_{i,j,k,l} K_{ij} K_{jk} K_{kl} K_{li}  L_i L_j L_k L_l
= -\frac{\beta^4}{2} \sum_{i<j} \sum_k K_{ik}^2 K_{kj}^2 L_k^2 L_i L_j
- \frac{\beta^4}{4} \sum_{i<j} K_{ij}^4 L_i^2 L_j^2
 \\
& - \beta^4 \sum_{i<j<k<l}
(K_{ij} K_{jk} K_{kl} K_{li} +
  K_{ik} K_{kj} K_{lj} K_{il} + K_{ij} K_{jl} K_{lk} K_{ki})
  L_i L_j L_k L_l\, .
\end{split}
\label{eq_contractions}
\end{equation}
The last three terms of Eq.~(\ref{eq_12345}) can be written in a different form:
\begin{eqnarray}
S^\text{loop} &= &-\frac{\beta^2}{4} \sum_{i,j} K_{ij}^2 L_iL_j
+ \frac{\beta^3}{6} \sum_{i,j,k} K_{ij} K_{jk}K_{ki} L_i L_j L_k
\nonumber\\
&&- \frac{\beta^4}{8} \sum_{i,j,k,l} K_{ij} K_{jk} K_{kl} K_{li}  L_i L_j L_k L_l
\, ,
\nonumber\\
&=&
-\frac{\beta^2}{4} \Tr(M^2) + \frac{\beta^3}{6} \Tr(M^3) -
\frac{\beta^4}{8} \Tr(M^4) \, ,
\label{eq_53}
\end{eqnarray}
where $M$ is the matrix defined by $M_{ij} = K_{ij} \sqrt{L_i L_j}$ and we will
justify in the following the notation $S^\text{loop}$.
Since $K_{ii}=0$, we have $\Tr{M}=0$, which implies that Eq.~(\ref{eq_53}) can be
rewritten as
\begin{eqnarray}
S^\text{loop} 
&=&
\frac{1}{2} \Tr\left(\beta M - \frac{\beta^2}{2} M^2 + \frac{\beta^3}{3} M^3 -
\frac{\beta^4}{4} M^4\right)\, .
\end{eqnarray}
We now make the hypothesis that if we continue this expansion to higher orders
on $\beta$ we will found all the other terms on $(-M)^k/k$. Thus,
\begin{eqnarray}
S^\text{loop} 
&=&
\frac{1}{2} \Tr\left(\beta M - \frac{\beta^2}{2} M^2 + \frac{\beta^3}{3} M^3 -
\frac{\beta^4}{4} M^4 + \cdots\right)
\nonumber\\
&=& \Tr \left[\log(1+\beta M)\right] = \log\left[ \det(1+\beta M)\right] \, .
  \label{eq_sloop}
\end{eqnarray}
In diagrammatic terms, Eq.~(\ref{eq_sloop}) corresponds to summing all single-loop diagrams
and their possible contractions (see Eq.~(\ref{eq_contractions}) for example):
\begin{eqnarray}
S^\text{loop}  &=&
  \frac{1}{2} \diag{loop1}
  +\diag{loop2}
  - \diag{loop3} + \diag{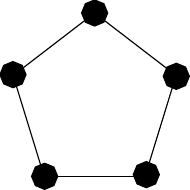} + \ldots \nonumber\\
  &&-\frac{1}{2}\diag{3spins_lin} -  \frac{1}{4}\diag{loop_d1}
  -  \frac{1}{3}\diag{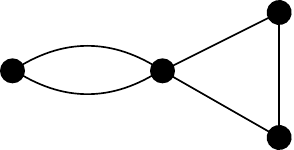} + \ldots \, .
\label{eq_diag_sloop}
\end{eqnarray}

From Eqs.~(\ref{eq_sloop}) and~(\ref{eq_derivcij}), we can also derive a
formula for the contribution $J_{ij}^\text{loop}$ of $S^\text{loop}$ to
$J_{ij}^*$:
\begin{eqnarray}
  \label{eq_jijloop}
J_{ij}^\text{loop}
&=& \frac{1}{\sqrt{(1-m_i^2)(1-m_j^2)}} \left[M (M+1)^{-1}\right]_{ij} \, ,
\end{eqnarray}
which is exactly the formula for the mean-field approximation found in
Eq.~(\ref{eq_tap_mf}). This is not only a very good evidence that the hypothesis
we used on Eq.~(\ref{eq_sloop}) is correct but also gives a physical
interpretation to $S^\text{loop}$.
Finally, we can combine this result with the previous
ones, yielding:

\begin{eqnarray}
S &=& S^\text{loop}
-\sum_i \left[\frac{1+m_i}{2}\ln\frac{1+m_i}{2} + 
\frac{1-m_i}{2}\ln\frac{1-m_i}{2}\right]
 +\frac{2}{3} \sum_{i<j} K_{ij}^3 m_i m_j L_i L_j
\nonumber\\
&&+ \frac{\beta^4}{6} \sum_{i<j} K_{ij}^4 \left[1 - 3m_i^2 -3m_j^2 - 3 m_i^2m_j^2\right] L_i L_j + O(\beta^5) \, ,
\end{eqnarray}
and
\begin{eqnarray}
J_{ij}^* &=& J_{ij}^{\text{loop}}
 - 2 \beta^2 m_i m_j K_{ij}^2
 \nonumber\\
&& - \frac{2}{3} \beta^3 K_{ij}^3 \left[1 - 3m_i^2 -3m_j^2 - 3 m_i^2m_j^2\right]
+ O(\beta^4)  \, .
\label{resum2}
\end{eqnarray}

Note that  the infinite series shown in Eq.~(\ref{eq_sloop})
is divergent when one of the eigenvectors of $M$
is greater than one, while Eqs.~(\ref{eq_sloop}) and~(\ref{eq_jijloop})
remain stable for all positive eigenvalues of $M$.
In practical terms, the loop summation is much more robust for inferring
the couplings than the
simple power expansion in Eq.~(\ref{eqG}).
In the next section, we propose a simple numerical verification of our
hypothesis that confirm this assertion.

\subsubsection{Numerical verification of our series expansion and the loop sum}

\begin{figure}[H]
  \begin{center}
    \includegraphics[width=.6\columnwidth]{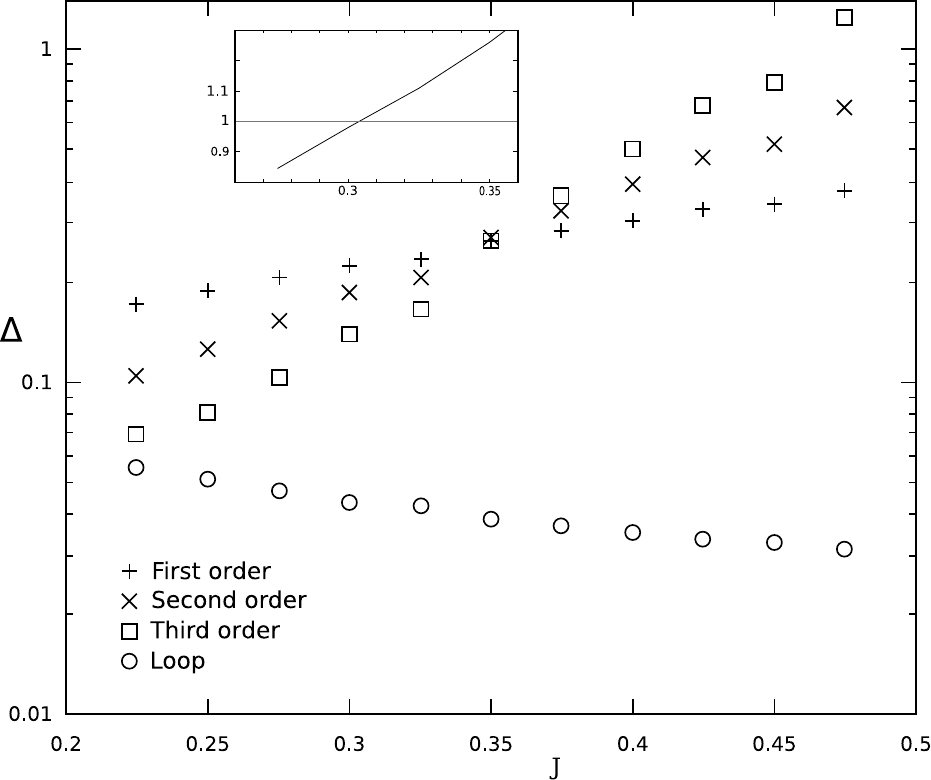}
  \end{center}
\caption{Relative error $\Delta$ given in Eq.~(\ref{delta2}) on the inferred
  couplings as a function of the parameter $J$ of the
Sherrington-Kirkpatrick model with $N=200$ spins. Monte Carlo
simulations are run over 100 steps. Averages and error bars 
are computed over 100 samples. Top: orders $\beta$, $\beta^2$ and $\beta^3$ of the expansion.
Bottom: expression (\ref{resum2}) which includes the sum over all
loop diagrams. Inset: largest eigenvalue $\Lambda$ of matrix $M$ as a
function of $J$.}
\label{orders3}
\end{figure}

We have tested the behavior of the series on the 
Sherrington-Kirkpatrick model in the paramagnetic phase. We 
randomly drew a set of $N\times(N-1)/2$ couplings $J_{ij}^{true}$ 
from uncorrelated normal 
distributions of variance $J^2/N$. From Monte-Carlo simulations, we
calculated the correlations and
magnetizations, inferring the couplings 
$J_{ij}^*$ from Eqs.~(\ref{jij}) and~(\ref{resum2}) and compared the outcome to
the true couplings through the estimator
\begin{equation}\label{delta2}
\Delta = \sqrt{\frac2{N(N-1)J^2}\sum_{i<j} \left(J_{ij}^*-
    J_{ij}^{true}\right)^2} \ .
\end{equation}
The quality of inference can be seen in Figure 
\ref{orders3} for orders (powers of $\beta$)  1,2, and 3 (corresponding
respectively to the symbols $+$, $\times$ and $\square$).
For large couplings the inference gets
worse as the order of the expansion increases due to the presence of terms
with alternating signs in the expansion as discussed. Indeed, in the inset
we show that for $J\approx 0.3$ the highest eigenvalue approaches 1. In this
figure, we plot also the value for $J_{ij}^*$ obtained from
Eq.~(\ref{resum2}) (as circles) and we can clearly see that it outperforms
the other formulas.

If we try to use the same procedure to test the performance of our
loop summation formula, we get results like the ones shown in Figure~\ref{mc22}:

\begin{figure}[H]
  \begin{center}
      \includegraphics[width=.6\columnwidth]{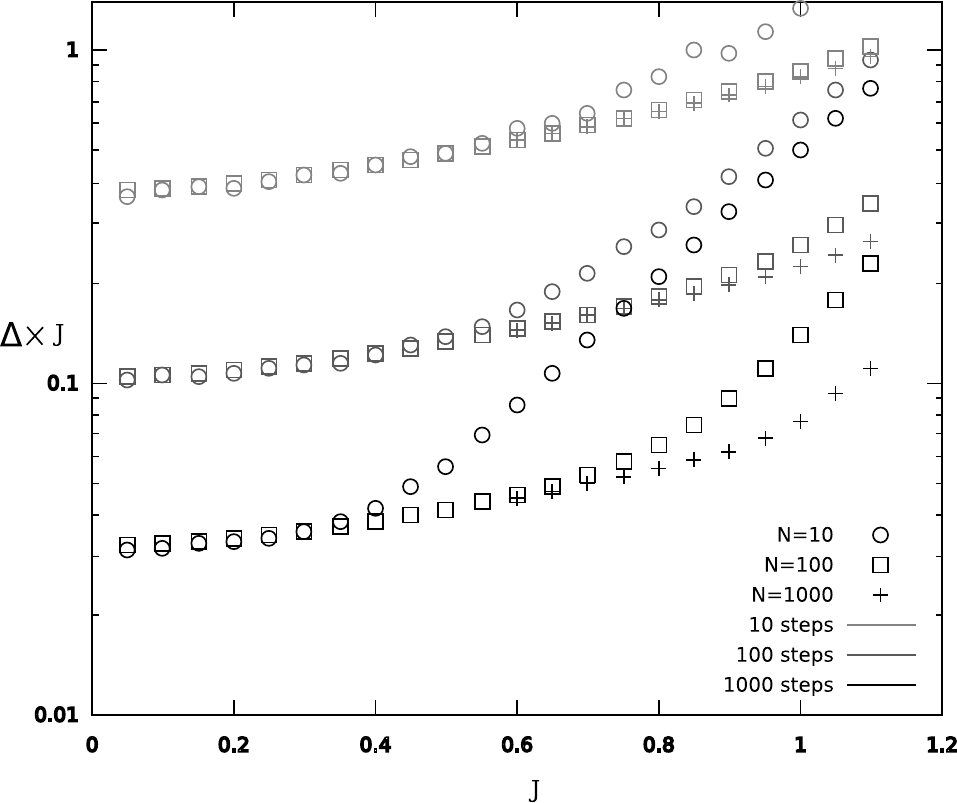}
  \end{center}
\caption{Absolute error $J\times \Delta$ on the inferred couplings 
as a function of the parameter $J$ of the
Sherrington-Kirkpatrick model. Inference is done through formula
(\ref{resum2}), which takes into account all loop diagrams. 
The error decreases with the number of spins and the number of Monte
Carlo steps (shown on the figure).}
      \label{mc22}
  \end{figure}

We can clearly see above that the error on the inferred couplings for the
Sherrington-Kirkpatrick model is essentially due to the noise
in the MC estimates of the correlations and magnetizations, since it decreases
with the number of steps.

\section{Combining the two-spin expansion and the loop diagrams}

In the previous section, we have identified a set of diagrams whose sum yields
the mean-field approximation we saw in chapter~\ref{chap_ising}.
In section~\ref{subsec_max_entr}, we have inferred exactly the value of
$J$ for a system composed of only two spins
(Eqs.~(\ref{eq_bb11}-\ref{eq_bb12})).
We will see that it is easy to identify summable diagrams also in this
two-spin case.
Our system is composed only by two spins $i$ and $j$, thus there can be no
diagrams of more than two vertices in the expansion
of $S$. Moreover, since the formula is exact, the expansion in the case of two
spins
contains \emph{all} the two-spin diagrams.
Indeed, the first four terms of the Taylor expansion of Eq.~(\ref{eq_J2spi}) on
small $c_{ij}$ are
\begin{eqnarray}
J_{ij}
&=&
\beta K_{ij} -2  m_i m_j K_{ij}^2
+ \frac{1}{3} K_{ij}^3 \left[1 + 3m_i^2 +3m_j^2 + 9 m_i^2m_j^2\right]
  + \ldots \, , \nonumber\\
  &=&
  \diag{loop1_p}
  -2 \diag{2spins_3_p}
  +\frac{1}{3}\diag{loop_d1_p} \, .
\end{eqnarray}
It is easy to identify this formula as the two-spin diagrams in
Eq.~(\ref{jij}). Using the
explicit formula for $S^\text{2-spin}$ given in Eq.~(\ref{eq_bb11}) and
applying Eqs.~(\ref{eq_J2spi}--\ref{eq_bb12}), we obtain

\begin{eqnarray}
S^\text{2-spin}_{ij} &=& S^\text{1-spin}_i+S^\text{1-spin}_j
\nonumber\\
&&
+ \frac{1}{4}
\log\left[1+\frac{c_{ij}}{(1-m_i)(1-m_j)}\right][c_{ij} + (1-m_i)(1-m_j)]
\nonumber\\
&&
+ \frac{1}{4}
\log\left[1-\frac{c_{ij}}{(1-m_i)(1+m_j)}\right][c_{ij} - (1-m_i)(1+m_j)]
\nonumber\\
&&
+ \frac{1}{4}
\log\left[1-\frac{c_{ij}}{(1+m_i)(1-m_j)}\right][c_{ij} - (1+m_i)(1-m_j)]
\nonumber\\
&&
+ \frac{1}{4}
\log\left[1+\frac{c_{ij}}{(1+m_i)(1+m_j)}\right][c_{ij} + (1+m_i)(1+m_j)]\,,
\label{eq_sss}
\end{eqnarray}
where
\begin{eqnarray}
  S^\text{1-spin}_i &=& - \left[\frac{1+m_i}{2}\ln\frac{1+m_i}{2} + 
\frac{1-m_i}{2}\ln\frac{1-m_i}{2}\right] \,.
\end{eqnarray}
We have now an explicit formula for the sum of all two-spin diagrams.
To go as further as possible in our expansion, we would like to sum both
all the loop and 2-spin diagrams.
To combine Eq.~(\ref{eq_sss}) with Eq.~(\ref{eq_sloop}), we need
to remove the diagrams that are counted twice, since the loop expansion
contains two-spin diagrams (see for ex. the first diagram in Eq.~(\ref{eq_diag_sloop})).
To evaluate the two-spin diagrams of $S^\text{loop}$, we can simply
evaluate it for the particular case of $N=2$, where all the diagrams involving
three or more spins are zero. Thus,
\begin{equation}
  S^\text{loop and 2-spin} = \frac{1}{2}\log\left[\det\left(
    \begin{array}{cc}
      1 & K_{ij} \sqrt{L_i L_j}\\
      K_{ij} \sqrt{L_i L_j} & 1
      \end{array}
    \right)\right] = \frac{1}{2} \log \left(1-K_{ij}^2 L_i L_j\right) \, .
\end{equation}
Finally, we can write an equation combining both sums:
\begin{eqnarray}
S^\text{2-spin + loop} &=& 
\sum_i S^\text{1-spin}_i
+ \sum_{i<j} \left[S_{ij}^\text{2-spin}-S^\text{1-spin}_i-S^\text{1-spin}_j
  \right]
\nonumber\\
&&
+S^\text{loop}
-\frac{1}{2} \sum_{i<j}\log(1-K_{ij}^2 L_i L_j) \, .
\label{eq_s_2sp_loop}
\end{eqnarray}
Note that this formula contains all diagrams shown in Eq.~(\ref{eqG}).
The corresponding formula for $J_{ij}^*$ is
\begin{eqnarray}
J_{ij}^{*(\text{2-spin+loop})} &=& J_{ij}^{*\text{loop}}
+ J_{ij}^{*\text{2-spin}}- \frac{K_{ij}}{1-K_{ij}^2 L_i L_j} \, ,
\label{eq_2spiloop}
\end{eqnarray}
where, as we have already seen in Eq.~(\ref{eq_J2spi}),
\begin{eqnarray}
 J_{ij}^{*(\text{2-spin})}
&=&
\phantom{+}\frac{1}{4}\ln\left[ 1 + K_{ij}(1+m_i)(1+m_j) \right]\nonumber\\
&&+ \frac{1}{4}\ln\left[ 1 + K_{ij} (1-m_i)(1-m_j) \right] \nonumber\\
&&-\frac{1}{4}\ln\left[ 1 - K_{ij}(1-m_i)(1+m_j) \right]  \nonumber\\
&&-\frac{1}{4}\ln\left[ 1 - K_{ij}(1+m_i)(1-m_j) \right] \label{2spi} \, .
\end{eqnarray}

\subsection{Three spin diagrams}
In the case of a system with a zero local magnetization, we can find a rather
simple expression for couplings $J_{ij}$ of a system composed of only three
spins $\sigma_i,\sigma_j, \sigma_k$ :
\begin{equation}
  \begin{split}
    J_{ij;k}^{* \text{3-spin}} = &\frac{1}{4}
\log\left[\frac{1 + c_{ij} - c_{ik} -
    c_{jk}}{1 - c_{ij} - c_{ik} + c_{jk}} \right]  \\
     -  & \frac{1}{4}\log \left[\frac{1 - c_{ij} + c_{ik} - c_{jk}}{1 -
     c_{ij} - 
     c_{ik} + c_{jk}}\right] + 
         \frac{1}{4}\log \left[\frac{1 + c_{ij} + c_{ik} + c_{jk}}{1 - c_{ij} -
	 c_{ik} + c_{jk}}\right] \, .
  \end{split}
\end{equation}
Proceeding in the same way we did with the two-spin diagrams,
we can combine this formula with the previous results:
\begin{equation}\label{all2}
  \begin{split}
	 J_{ij}^{\text{2-spin+loop+3-spin}} = 
 &J_{ij}^{*\text{2-spin+loop}} + \sum_{k\, (k\neq i,k\neq j)} J_{ij;k}^{* \text{3-spin}} 
		       - \sum_{k\, (k\neq i,k\neq j)} \Bigg\{
J_{ij}^{\text{2 spins}}\\
&\phantom{xyzxx}+\frac{c_{ij} - c_{ik} c_{jk}}{
   1-c_{ij}^2-c_{ik}^2-c_{jk}^2 +
			       2 c_{ij}c_{jk}c_{ki}} -
\frac{c_{ij}}{1-c_{ij}^2}
			       \Bigg\}\, .
  \end{split}
\end{equation}

\section{Quality of the inference after summing the loops and 2-3 spin diagrams}

In this section, we will look at how our results perform for two different
well-known models. First we will look analytically at the one-dimensional Ising
model and
afterwards we will see numerical results for the Sherrington-Kirkpatrick model.
We will test both the inference using just the loop diagrams we saw in
Eq.~(\ref{eq_jijloop}), the combination of loops and two-spin diagrams
we saw in Eq.~(\ref{eq_2spiloop}) and the combination of loops, 2-spins and 3-spins
diagrams we saw in Eq.~(\ref{all2}).

\subsection{One-dimensional Ising}
For the one-dimensional Ising model, we can evaluate exactly the coupling
as a function of the correlations (see Eq.~(\ref{eq_corr_isi1d})):
\begin{equation}
  J_{kl} =  (\delta_{k+1,l}+\delta_{k-1,l}) \tanh^{-1}\left(c_{ij}^{\frac{1}{|i-j|}}\right) \, ,
\end{equation}
where $\delta_{ij}$ is the Kronecker symbol.
Note that the obtained value of $J_{kl}$ should not depend on the pair of sites
$i,j$ chosen.
Using our formula for $J_{ij}^\text{2-spin}$ given in Eq.~(\ref{2spi}), we have
\begin{equation}
  J_{ij}^\text{2-spin}= \tanh^{-1} c_{ij}
  = \tanh^{-1}\left[ \left(\tanh J\right)^{|i-j|}\right] \, ,
\end{equation}
which predicts correctly the values of the couplings between closest neighbors
$J_{i,i+1}=J$, but gives an non-zero result for the other couplings.
On the other hand, using the loop summation formula from Eq.~(\ref{eq_jijloop}), we get
\begin{equation}
  J_{ij}^\text{loop}= \frac{c}{1-c^2}\left(\delta_{i,i+1}+\delta_{i,i-1}\right)
  \, ,
\end{equation}
where $c=c_{i,i+1} = \tanh J$ and $\delta_{i,j}$ is the Kronecker function.
The loop sum correctly predicts that the model has only
closest-neighbor couplings but does not predict correctly its value.

Finally, using both the 2-spin diagrams and the loop sum
(see Eq.~(\ref{eq_2spiloop})), we have
\begin{eqnarray}
J_{ij}^\text{2-spin+loop} &=& 
 J (\delta_{i,i+1} + \delta_{i,i-1}) +
\left[\tanh c_{ij} - \frac{c_{ij}}{1 - c_{ij}^2}\right](1-\delta_{i,i+1})
(1-\delta_{i,i-1}) \nonumber \\
&=& J (\delta_{i,i+1} + \delta_{i,i-1}) + O(c^6) \ , 
\end{eqnarray}
which is correct to the order $O(c^6)$. As we will see in the following, the
next contribution to the
couplings coming from the
expansion in Eq.~(\ref{eq_66}) corresponds to $\inlinediag{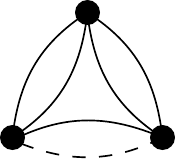}$, whose
leading term is indeed proportional to
$c_{i,i+2} \cdot c_{i,i+1}^2  \cdot c_{i+1,i+2}^2 = c^6$.

\subsection{Sherrington-Kirkpatrick model}

In section~\ref{sec_loop}, we saw that when we used Monte-Carlo simulations to
evaluate
the quality of the inference for the SK model we were limited mostly by
the numerical errors of the MC simulation.
To have more precise values, we now evaluate the error due to our
truncated expansion using a program that calculates $c_{ij}$ through
an exact enumeration of all $2^N$ spin configurations. We are limited
to small values of $N$ (10, 15 and 20). However the case of a small 
number of spins is particularly interesting since, for the SK model, 
the summation of loop diagrams is exact in the
limit $N \rightarrow \infty$, as we discussed in section~\ref{sec_TAP}. The importance of terms
not included in the loop summation is thus better 
studied at small $N$.

\begin{figure}[H]
 \begin{center}
   \includegraphics[width=.6\columnwidth]{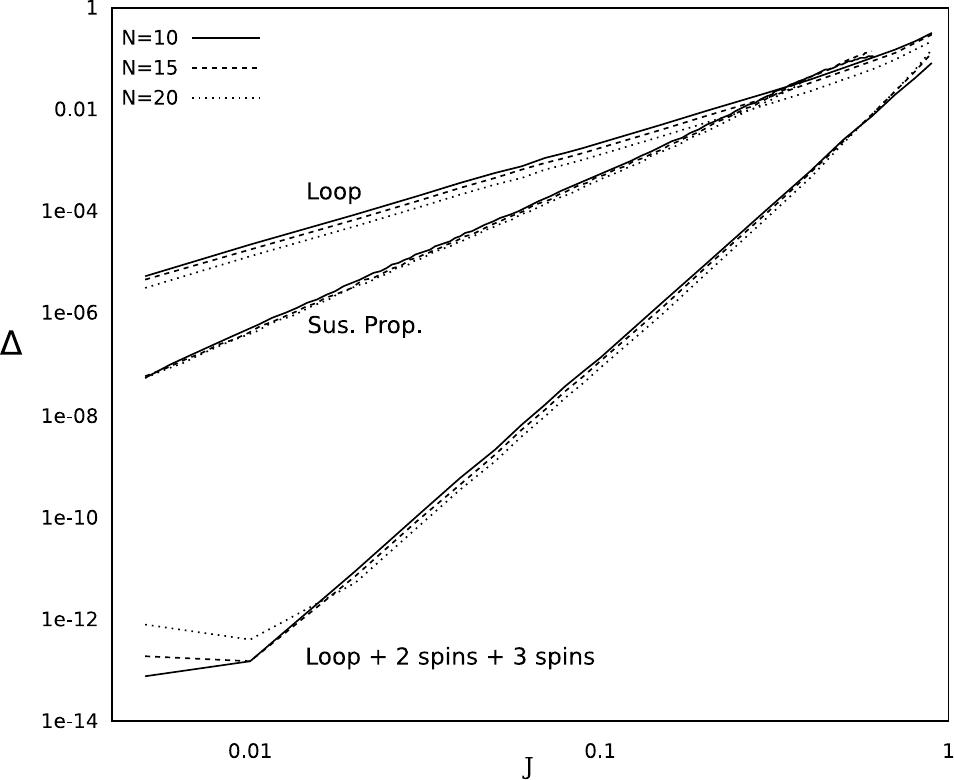}
 \end{center}
\caption{Relative error $\Delta$ (Eq.~(\ref{delta2})) as a function of $J$ for the SK
  model for our summation $J_{ij}^\text{2 spin + loop + 3 spin}$ 
(Eq.~(\ref{all2})) compared to the Susceptibility Propagation method of M\'ezard and 
Mora \cite{mora} and loop resummation $J_{ij}^\text{loop}$ 
(Eq.~(\ref{eq_jijloop})).}
 \label{orders2aa}
\end{figure}

We compared the quality of the inference using the loop summation
(Eq.~(\ref{eq_jijloop})), the combination of loop summation and all diagrams up
to three spins (Eq.~(\ref{all2})) and the method of susceptibility propagation
we discussed in section~\ref{sec_susc_prop}.
Results are shown in Figure \ref{orders2aa}. The error is remarkably
small for weak couplings (small $J$), and is dominated by finite-digit accuracy 
($10^{-13}$) in this limit. Not surprisingly it behaves better than
simple loop summation, and also outperforms the
susceptibility propagation algorithm.

\section{Numerical evaluation of high-order diagrams}

In the two last sections we saw that summing all loop diagrams improved
considerably the robustness of the inference. In this section, we will try to
find some
more terms numerically, in the hope that we might find other classes of exactly
summable diagrams.
We will proceed in a similar way as
in section~\ref{sec_numeric}, where we used a numerical fit to validate our
expansion
in small $\beta$. We will use the same method to find new diagrams in the
expansion by guessing a general form of the lowest-order missing terms
and finding numerically their coefficients.

We started by defining a list of several possible corrections to
Eq.~(\ref{eq_s_2sp_loop}) (since Eq.~(\ref{all2}) is only valid for $m_i=0$) such as
%
%
\begin{equation}
\begin{split}
  S = S^\text{2-spin + loop}&\\
+ \sum_{i,j,k} K_{ij} K_{jk}^2 K_{ki}^2
\large[&a_1 + a_2 (m_i +  m_j) + a_3 (m_j + m_i)m_k +
  a_4 m_i m_j\\
  &+ a_5 (m_i^2 + m_j^2) + a_6 m_k^2 + a_7 (m_i^3 + m_j^3)
  \\
  &
  + a_8 m_k^3 + a_9 m_i m_j m_k + a_{10} m_i m_j m_k^2\large]\, .
\end{split}
\label{eq_fit}
\end{equation}
Note that in the same way as in Eq.~(\ref{eqG}), the numerical coefficients
in the expansion must be a fraction of small integers, as a consequence
of our expansion procedure.

We followed the same method described in section~\ref{sec_numeric}
for each one of our guesses.
We found that for all of them, with the exception of the one shown in Eq.~(\ref{eq_fit}),
the fitted coefficients did not correspond to a fraction of small integers
as required.
Unlikewise, for the guess shown in Eq.~(\ref{eq_fit}) all the
%
coefficients were zero except $a_4=-1$ and $a_{10}=1$. If
eventually
there was one extra term missing in Eq.~(\ref{eq_fit}) (for example a term
on $K_{ij}^3K_{jk} K_{ki}$), the parameters other than $a_4$ and $a_{10}$ would
have some bogus value to compensate for the missing term. Accordingly, finding
only two non-zero coefficients is a strong evidence of the lack of additional
corrections other than those shown in Eq.~(\ref{eq_fit}).
Indeed, this is corroborated by the value of the squared mean deviation, which is of
the same order as $c^6$, as we would expect in an expansion
with no supplementary term on $c^5$  missing.
Finally, the expansion up to order $O(c^5)$ is given by
\begin{equation}
S = S^\text{2-spin + loop}
- \sum_{i,j,k} K_{ij} K_{jk}^2 K_{ki}^2 m_i m_j L_i L_j L_k^2 + O(c^6)\, .
\end{equation}

In the particular case of a system with zero magnetization, the guesses of the
corrections are much simpler since there is no arbitrary polynomial on $m$
multiplying each term. We could then find all the terms of the expansion
up to $O(c^8)$:
\begin{eqnarray}
S  &=& S^{\text{2-spin+loop}} -\frac{1}{6} \diag{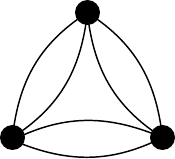}
  + \frac{2}{3}\diag{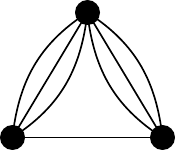}
  - \frac{1}{4} \diag{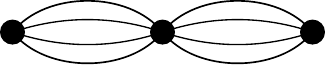}
  \nonumber\\
  &&
  - 2 \diag{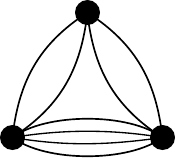}
  +\frac{1}{8} \diag{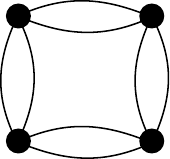}
  +\frac{1}{2} \diag{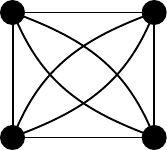}
  + O(c^9) \, .
\label{eq_66}
\end{eqnarray}



\section{Expansion in n-spin diagrams}

All the results seen up to now were only
valid on a small correlation limit. Unfortunately, for actual neuron data,
there might be two or more neurons with very strongly correlated activity.
Consequently, here we will try another approach, based on the fact that
neurons spend the most of their time at rest. Their magnetization is thus
very close to $-1$ (or $+1$, depending on which convention one chooses for
the rest state). We derive thus an expansion of the couplings valid
for values of magnetization close to $\pm 1$. The technical details can
be found on appendix~\ref{app_mag}. Our final result is
\begin{eqnarray}
  J_{ij}^\text{k-spin diagrams} &=& J_{ij}^\text{2-spin}+J_{ij}^\text{3-spin}+ \ldots
  J_{ij}^\text{$k$-spin}
  -\text{(repeated diagrams)} \, ,
\end{eqnarray}
and the error is given by
\begin{eqnarray}
J_{ij}^* &=& 
J_{ij}^\text{k-spin diagrams} 
+  O\left[(1-m_i^2)^{k-2}\right] \, .
\label{eq_errmag}
\end{eqnarray}
where $J_{ij}^\text{k-spin}$ is the sum of all diagrams in the expansion
of $J_{ij}$ involving $k$ spins.

The results seen previously in this work (see Eq.~(\ref{eqG})) suggest that $J_{ij}^\text{k-spins}$
is of order $O(c^k)$, with the lowest order diagram being the loop over
$k$ spins, thus
\begin{equation}
J_{ij}^* =
J_{ij}^\text{k-spin diagrams}
+ O\left(c^{k+1}\right) \, .
\label{eq_errc}
\end{equation}
We can then expect that summing all diagrams up to $k$ spins might be a very
good approximation both in the strong magnetization regime
(see Eq.~(\ref{eq_errmag})) and in the week
correlation one (see Eq.~(\ref{eq_errc})).

Unfortunately, even for values of $k$ as small as $k=4$, we cannot find an
exact expression for $J_{ij}^\text{k-spin}$
as we did for $k=2$ in Eq.~(\ref{eq_J2spi}).
In their
PNAS paper, Cocco et al. \cite{monas_pnas} note that $J_{ij}^\text{k-spin}$
can be obtained \emph{numerically}, by exact enumeration of all $2^k$ possible
states of a $k$-spin system. Using this method, they could sum all diagrams
up to 7 spins. They could also combine this method with summing all loop
diagrams, which has improved the performance of their inference.

\part{Inference of Hopfield patterns}
\label{part3}
In part~\ref{part2}, we have seen a very general treatment of the inverse
Ising problem. In this part, we are interested in a particular case of
the same problem: inferring the patterns of a Hopfield model, introduced
in section~\ref{sec_hop}.

In chap.~\ref{ch_inf_mat},
we deal with the problem of inferring a set of $p$ patterns from the measured
data under the supposition that the number of patterns is a non-extensive
quantity.
We derive explicit formulas
for the patterns as a function of the magnetizations and correlations in both
the paramagnetic and ferromagnetic phases of the model in the limit of large
system size. Interestingly, for the paramagnetic case we find in the leading
order the same formula found in section~\ref{sec_loop} for the loop summation.

The goal of chapter~\ref{chap_hop} is to find an estimation of how many times
one needs to measure a Hopfield system to be able to have a good estimate of
its patterns. To this end, we use the concept of Shannon entropy introduced in
chapter~\ref{ch_inverse} to
estimate the quantity of information we lack about the system.
We evaluate explicitly the entropy for a typical realization of the system
as a function of the number of measurements. We find that when the system
is magnetized according to one of the patterns, we can find this pattern
using just a non-extensive number of measures. On the other hand, to find
the patterns that were not visited in any of our measures one needs an extensive
number of measures.

\chapter{Pattern inference for the Hopfield model}
\label{ch_inf_mat}

Up to this point we have dealt with the problem of inferring a coupling
matrix $\{J_{ij}\}$ of a generalized Ising model. In this chapter, we will look
at the particular case of a Hopfield model. There are several potential
advantages of this model: first, as we saw in
section~\ref{subsec_intro_hop}, in some experiments one needs to infer a
set of patterns from the measured data. Secondly, we might expect that reducing
the number of degrees of freedom might make the inference procedure more
stable. Finally, the Hopfield model can be solved analytically and thus we
expect to have a better control of the inference errors.

In principle, one could proceed by first inferring the matrix
$\{J_{ij}\}$ from the data, as we have done in part~\ref{part2}, and then
diagonalizing it to extract a set of patterns.
The problem with this approach is that it is not optimal from the
Bayes point of view: the inferred patterns are not the ones that maximize
the \textit{a posteriori} probability.
This is particularly relevant when the assumption that the underlying system
is governed by a Hopfield model is just an
approximation, as will almost always be the case
in biological data. In this case, 
we cannot guarantee that the patterns obtained by diagonalizing
the $\{J_{ij}\}$ matrix are the ones that best describe the data.

This method was carried out in a paper recently published by
Haiping Huang ~\cite{huang2010reconstructing}, where several different methods
for solving the inverse Ising model was used to find the couplings
of a Hopfield model. In their
paper, they show that the method presented in chapters~\ref{sec_expansion}
and~\ref{chap_sums} does not perform significantly better than the naive
mean-field method for a Hopfield model, which gives yet another reason to
look for a method specific to this model.

In this chapter, we suppose that we have measured $L$ configurations of a
system that is governed by a
Hopfield model.
We would like to deduce both the sign and the magnitude of the patterns
from the data using a Bayesian inference, as defined in chapter~\ref{ch_inverse}.
We suppose that we wait long enough between two successive measures so that
they show no temporal correlation, i.e., that  our configurations constitute
an independent and identically distributed sampling of a Boltzmann
distribution.

We will first start with the simple, albeit not very
useful, case of Hopfield model
with a single pattern. In this case, the calculations can be done in a few lines
and allow one to get an idea of the structure of the solution for the
general case. Afterwards,
we deal with the case of any number of patterns $p$, which is the main result
of this chapter.

\section{A simpler case: inference of a single pattern}
\label{sec_inf_mat}
In this section we suppose that we have measured $L$ independent configurations
of system
$\sigma^1, \ldots, \sigma^L$, where each configuration are given by
$\sigma^l =\{\sigma^l_1, \ldots, \sigma_N^l\}$ and $\sigma_i^l = \pm 1$. From this data,
we can, for example, evaluate the measured correlations and magnetizations
\begin{eqnarray}
m_i = \frac{1}{L} \sum_{l=1}^L \sigma_i^l \, , \quad \quad
C_{ij} = \frac{1}{L} \sum_{l=1}^L \sigma_i^l \sigma_j^l \, .
\label{eq_666}
\end{eqnarray}
We suppose that our system can be described by a Hopfield pattern with a single
pattern.
Its Hamiltonian is thus given by
\begin{equation}
H(\{\sigma_i\}) = - \frac{1}{N}\sum_{i<j} \xi_i \xi_j \sigma_i \sigma_j \, ,
\end{equation}
where $\xi_i$ are real values that describe the pattern, $\sigma_i$ are the spin variables and
$N$ is the number of spins of the system. The partition function is given by
\begin{eqnarray}
Z(\beta,\{\xi_i\}) &=&  \sum_{\{\sigma\}} \exp\left[\frac{\beta}{N}\sum_{i<j} \xi_i \xi_j \sigma_i \sigma_j\right] \, .
\label{eq_part_p1}
\end{eqnarray}
For the rest of this chapter we will avoid the explicit dependence on $\beta$ by
performing the
change of variables $\xi_i \rightarrow \xi_i/\sqrt{\beta}$.
We will also omit the dependence of the
partition function on $\{\xi_i\}$ to simplify notations, posing
$Z(\beta,\{\xi_i\}) \equiv Z$.

We would like to infer both the sign and magnitude of $\xi_i$ from the
measured configurations $\{\sigma^l\}$. Using the Bayes theorem as announced in
Eq.~(\ref{eq_bayes}), the 
likelihood of the patterns is given by
\begin{equation}
P(\{\xi_i\}|\{\sigma^l\}) = \frac{P_0(\{\xi_i\}) }{Z^L P(\{\sigma^l\})} \prod_{l=1}^L 
\exp\left[\frac{1}{N}\sum_{i<j} \xi_i \xi_j \sigma_i^l \sigma_j^l\right] \, .
\label{eq_likeli0}
\end{equation}
Using Eq.~(\ref{eq_666}), we can rewrite this expression as
\begin{equation}
\begin{split}
\frac{1}{L} \log P(\{\xi_i\}|\{\sigma^l\}) = &- \log Z - \log P(\{\sigma^l\}) \\
&+ \frac{1}{L}\log P_0(\{\xi_i\})
+ \exp\left[\frac{1}{N}\sum_{i<j} C_{ij} \xi_i \xi_j\right] \, .
\end{split}
\label{eq_likeli}
\end{equation}
To maximize this expression, we need to evaluate $\log Z$ explicitly. Using
an integral transform, Eq.~(\ref{eq_part_p1}) becomes
\begin{eqnarray}
Z &=& \int_{-\infty}^\infty \frac{\di x}{\sqrt{2 \pi N^{-1}}} \sum_{\{\sigma\}}
\exp\left[-\frac{N}{2} x^2 + x \sum_i \xi_i \sigma_i \right] \, ,\\
&=& \int_{-\infty}^\infty \frac{\di x}{\sqrt{2 \pi N^{-1}}} 
\exp\left\{-\frac{N}{2} x^2 + \sum_i \log\left[2\cosh (x \xi_i) \right]
\right\} \, .
\label{eq_Z11}
\end{eqnarray}
In the following, we will set the prior to $P_0(\{\xi_i\})=1$, since we can see
from Eq.~(\ref{eq_likeli}) that it is irrelevant for large values of $L$.
In the following we will treat separately the ferromagnetic case ($m \neq 0$)
and the paramagnetic case ($m=0$).

\subsection{Ferromagnetic case}
In the ferromagnetic phase, we can use the saddle-point method to evaluate
the integral in Eq.~(\ref{eq_Z11}), yielding
\begin{equation}
\log Z = -\frac{N}{2} x^2 + \sum_i \log\left[2\cosh (x \xi_i) \right] + O(1/N)
\end{equation}
with $x$ given by
\begin{equation}
x = \frac{1}{N} \sum_i \xi_i \tanh(x \xi_i)\, .
\end{equation}

To infer $\{\xi_i\}$ from the data, we follow the maximum likelihood principle
and maximize Eq.~(\ref{eq_likeli}) with
respect to $\{\xi_i\}$, obtaining
\begin{equation}
\frac{1}{N} \sum_j C_{ij} \xi_j = x \tanh(m \xi_i) \,.
\label{eq_cc}
\end{equation}
It is easy to see that for such a system the correlation is dominated by its
non-connected part:
$C_{ij} = m_i m_j + O(1/N)$. Applying this result to Eq.~(\ref{eq_cc}),
we obtain
\begin{eqnarray}
\xi_i &=& \frac{1}{x} \tanh^{-1} m_i \, ,
\end{eqnarray}
and
\begin{eqnarray}
x^2 &=& \frac{1}{N} \sum_i m_i \tanh^{-1} m_i \, .
\end{eqnarray}
We found that the pattern is simply a function of the local magnetization,
what is not very surprisingly knowing we are dealing with the ferromagnetic
phase.

\subsection{Paramagnetic case}

In the paramagnetic case, the saddle point is $x=0$. Consequently, we need to
evaluate
the next term in the large $N$ limit to find a non-trivial partition
function:
\begin{eqnarray}
Z &=&  \int_{-\infty}^\infty \frac{\di m}{\sqrt{2 \pi N^{-1}}} 
\exp\left\{-\frac{N}{2} m^2 + \sum_i \log\left[2\cosh (m \xi_i) \right]
\right\}  \, ,\nonumber\\
&=&  \int_{-\infty}^\infty \frac{\di m}{\sqrt{2 \pi N^{-1}}} 
\exp\left\{-\frac{N}{2} m^2 + \frac{m^2}{2} \sum_i   \xi_i^2\right\}
\, , \nonumber\\
&=& \left(1 - \frac{1}{N}\sum_i \xi_i^2\right)^{-\frac{1}{2}} \, .
\end{eqnarray}
Maximizing with respect to $\xi_i$, we obtain

\begin{equation}
\sum_j C_{ij} \xi_j = \frac{1}{1-\frac{1}{N} \sum_j \xi_j^2} \xi_i \, .
\end{equation}
\noindent We conclude thus that $\xi_i$ is proportional to an 
eigenvector $v_i$ of the matrix $C$.
The corresponding eigenvalue $\lambda$ allows one to find the
proportionality constant
\begin{equation}
\xi_i = \sqrt{1-\frac{1}{\lambda}} \, v_i \, ,
\label{eq_inf11}
\end{equation}
where the eigenvectors are normalized in the following way
\begin{equation}
\frac{1}{N} \sum_j v_i^2 = 1 \, .
\label{eq_inf12}
\end{equation}
It still remain to be decided which pair of eigenvalue/eigenvector we
should pick to obtain our pattern. Since we suppose our patterns are
real-valued, Eq.~(\ref{eq_inf11}) implies that we should choose an
eigenvector greater than one. Moreover, if more than one eigenvalue satisfy
this condition, we can easily deduce from Eqs.~(\ref{eq_inf11}),
(\ref{eq_inf12}) and~(\ref{eq_likeli0}) that the greatest eigenvalue is the
one that maximizes the likelihood.


We verified this formula using Monte-Carlo simulations and the inferred
patterns showed a good agreement with the real ones used to make the simulation.

\section{Inference of continuous patterns for \textit{\protect\iip} > 1}

In this section, we look at the more general and interesting case
of a system with several patterns. In the same way as in the previous section,
we will treat the ferro- and paramagnetic case separately. But first, however,
we show how the problem is theoretically harder to define in this case.

\subsection{Discussion on the gauge}
\label{subsec_gauge}
A major issue in the inference of the patterns of the Hopfield model is that
if the patterns can take real values
the problem is ill-defined: there are many patterns that could describe equally
well the data. Suppose for example the case $p=2$:
\begin{equation} 
H = N \left( \frac{1}{N}\sum_i \xi_i^1 \sigma_i \right)^2 + N \left( \frac{1}{N}\sum_i \xi_i^2 \sigma_i \right)^2 \,.
\end{equation} 
If we define alternative patterns 
$\tilde{\xi_i^1} =  \xi_i^1 \cos \theta + \xi_i^2 \sin \theta$ 
and
$\tilde{\xi_i^2} =  -\xi_i^1 \sin \theta + \xi_i^2 \cos \theta$, our new 
Hamiltonian is
\begin{eqnarray} 
\tilde{H} &=& N \left[ \frac{1}{N}\sum_i (\xi_i^1 \cos \theta + \xi_i^2 \sin \theta)
 \sigma_i \right]^2 
\nonumber\\
&&+ N \left[ \frac{1}{N}\sum_i
 ( -\xi_i^1 \sin \theta + \xi_i^2 \cos \theta) \sigma_i \right]^2 \, ,
\\
&=& H \, .
\end{eqnarray}

More generally, if one has $p$ patterns, doing a rotation in $p$ dimensions
will not change the Hamiltonian of the system. We say that the model has
a \textit{gauge invariance} in respect to such rotations.
Thus, for inferring the patterns
one need either to add additional constraints to the remove the
$p(p-1)/2$ degrees of freedom or
to add a prior probability to the patterns, which would select a particular
preferred rotation. In the following, we will choose the
former solution, since it makes our calculations simpler.

\subsection{Ferromagnetic phase}

Suppose that we have a sample of $L$ measures of our system, containing $l_1$
measures where the system was magnetized 
according to the first patterns, $l_2$ measures with the system magnetized
according to the second and so on. In this case, we have
\begin{eqnarray}
C_{ij} &=& \frac{1}{L} \sum_{l} \sigma_i^l \sigma_j^l \nonumber\\
&=& \sum_{k=1}^p \frac{1}{L} \sum_{l\in l_k} \sigma_i^l \sigma_j^l \nonumber \\
&=& \frac{1}{L} \sum_{k=1}^p  l_k m_i^k m_j^k + O(1/N) \, ,
\label{eq_hopferro}
\end{eqnarray}
where
\begin{eqnarray}
m^k &=& \frac{1}{N} \sum_i \xi_i^k \tanh(m_k \xi_i^k) \, ,
\end{eqnarray}
and
\begin{eqnarray}
m^k_i &=& \tanh(m^k \xi_i^k) \, .
\label{eq_seila43}
\end{eqnarray}
A consequence of Eq.~(\ref{eq_hopferro}) is that the matrix $C_{ij}$ has
exactly $p$ eigenvalues that are extensive and their corresponding eigenvectors
are proportional to $m_i^k$. We can thus easily solve
Eqs.~(\ref{eq_hopferro}-\ref{eq_seila43}) by diagonalizing the matrix $C_{ij}$.
Note that we did not have fixed a gauge when writing these equations, but
considering that $m_i^k$ are proportional to the eigenvectors of the matrix
$C_{ij}$ implies that they are orthogonal. Thus, this procedure is equivalent
of choosing the gauge that satisfies:
\begin{equation}
\sum_i \tanh(m^k \xi_i^k) \tanh(m^{k'} \xi_i^{k'}) = 0, \quad \text{for $k\neq k'$} \, .
\end{equation}

\subsection{High external field case}
\label{sec_hop_para}

In this section, we will suppose that our external field is strong enough
so that the spins are not magnetized
according to any of the patterns but only according to the external field.
We will be interested in inferrin both the patterns of our model but also
the value of the external fields, which might be site-dependent.
The calculations that will follow can be considerably simplified by replacing
our usual 
Hamiltonian (Eq.~(\ref{eq_hop_usual})) by a slightly different one:
\begin{equation}
H = -\frac{1}{N} \sum_{\mu=1}^p \sum_{i<j} \xi_{i}^\mu \xi_j^\mu (\sigma_i - \tanh h_i) (\sigma_j - \tanh h_j)
- \sum_i h_i \sigma_i \, .
\label{eq_4411}
\end{equation}
This Hamiltonian can be related to the usual one by a translation on the local
fields:
\begin{equation}
h_i \rightarrow h_i - \sum_\mu \sum_{j \, (\neq i)} \xi_i^\mu \xi_j^\mu \tanh h_j
\, .
\end{equation}
From Eq.~(\ref{eq_bayes}),
the \textit{a posteriori} probability for the inference is then given by
%
\begin{equation}
\begin{split}
P(\{\xi_i^\mu\}&|\{\sigma^l\}) = \frac{P_0(\{\xi_i^\mu\}) }{Z(\{\xi_i^\mu\})^L P(\{\sigma^l\})} 
\times \\
&\times\prod_{l=1}^L\exp\left[
\frac{\beta}{N} \sum_{\mu=1}^p \sum_{i<j} \xi_i^\mu \xi_j^\mu (\sigma_i^l -\tanh h_i) (\sigma_j^l - \tanh h_j) + \beta \sum_i h_i \sigma_i^l\right] \, .
\end{split}
\end{equation}
Introducing the measured values of the magnetizations and of the
connected correlation
\begin{equation}
m_i = \frac{1}{L} \sum_l \sigma_i^l \, , \quad \quad
c_{ij} = \frac{1}{L} \sum_l \sigma_i^l \sigma_j^l - m_i m_j \, ,
\end{equation}
we can rewrite our probability as
\begin{equation}
\begin{split}
P(\{\xi_i^\mu\}|&\{\sigma^l\}) = \frac{P_0(\{\xi_i^\mu\}) }{Z(\{\xi_i^\mu\})^L P(\{\sigma^l\})}
\exp\Bigg[
\frac{\beta L}{N} \sum_{\mu=1}^p \sum_{i<j} \xi_i^\mu \xi_j^\mu c_{ij} \\
&+ \frac{\beta L}{N} \sum_\mu \sum_{i<j} \xi_i^\mu\xi_j^\mu(m_i -\tanh h_i)(m_j - \tanh h_j)
+ \beta L \sum_i h_i m_i\Bigg] \, .
\end{split}
\label{eq_p_sigma}
\end{equation}
To follow our Bayesian approach of maximizing this probability in respect to
the patterns and external fields, we need first to evaluate $\log Z$
explicitly. Since this calculation is straightforward
its details can be found in appendix~\ref{ap_logZ}. We obtain
\begin{eqnarray}
\label{logZ}
\log Z = 
\sum_i \log( 2 \cosh h_i) - 
\frac{1}{2} \sum_\mu \log \chi_\mu
+ \frac{1}{4 N}\sum_{\mu, \nu} \frac{s_{\mu \nu}^2 - r_{\mu\nu}}{\chi_\mu \chi_\nu}
+ O(1/N^{3/2}) \, , \nonumber\\
\end{eqnarray}
where
\begin{eqnarray}
r_{\mu \nu} &=& \frac{1}{N} 
\sum_i (\xi_i^\mu)^2 (\xi_i^\nu)^2 (1 - 3 \tanh^2 h_i)(1- \tanh^2 h_i) \, ,\\
\chi_\mu &=& 1 - \frac{1}{N} \sum_i (\xi_i^\mu)^2 (1-\tanh^2 h_i) \, ,\\
s_{\mu \nu} &=& \frac{1}{\sqrt{N}} \sum_i \xi_i^\mu \xi_i^\nu (1-\tanh^2 h_i) \, .
\end{eqnarray}

\subsubsection{Optimization of the probability}

In the following, we will ignore the prior, which is justified in the limit
$L \rightarrow \infty$.
To maximize Eq.~(\ref{eq_p_sigma}) one could just maximize
the following quantity
\begin{eqnarray}
\log P &=& \frac{1}{2N} \sum_{i,j,\mu} \xi_i^\mu \xi_j^\mu c_{ij}
+\frac{1}{2N} \sum_{i,j,\mu}\xi_i^\mu \xi_j^\mu (m_i - \tanh h_i)(m_j - \tanh h_j)
\nonumber\\
&&+ \sum_i h_i m_i - \log Z \, .
\label{eq_543}
\end{eqnarray}
As discussed in section~\ref{subsec_gauge}, we need to choose a gauge to make
our problem well-defined. From Eq.~(\ref{logZ}), a natural choice to
simplify our equations is
adding a Lagrange multiplier $ x_{\mu\nu}$ to fix the gauge $s_{\mu \nu}= 0$.
We obtain thus
\begin{equation}
\begin{split}
\log P = \frac{1}{2N} \sum_{i,j,\mu} \xi_i^\mu \xi_j^\mu c_{ij}
+\frac{1}{2N} \sum_{i,j,\mu}\xi_i^\mu \xi_j^\mu (m_i - \tanh h_i)(m_j - \tanh h_j)
+ \sum_i h_i m_i
\\
- \sum_i \log( 2 \cosh h_i) +
\frac{1}{2} \sum_\mu \log \chi_\mu
- \frac{1}{4 N}\sum_{\mu, \nu} \frac{r_{\mu\nu}}{\chi_\mu \chi_\nu}
- \frac{1}{N \sqrt{N}} \sum_{\mu \neq \nu} x_{\mu\nu} s_{\mu\nu} \, .
\end{split}
\end{equation}
Optimizing with respect to $h_i$, we obtain
\begin{equation}
\begin{split}
\frac{\partial \log P}{\partial h_i} = 0 = &
m_i - \tanh h_i 
- \sum_\mu (1-\tanh^2 h_i) \xi_i^\mu \frac{1}{N} \sum_j \xi_j^\mu (m_j-\tanh h_j)
\\
&+ \sum_\mu \frac{(\xi_i^\mu)^2}{ N \chi_\mu} \tanh h_i (1-\tanh^2 h_i)
+ O(1/N^{3/2}) \, .
\end{split}
\label{eq12}
\end{equation}
Posing
\begin{equation}
\alpha_\mu = \sum_i \xi_i^\mu (m_i - \tanh h_i) \, ,
\end{equation}
we can multiply Eq.~(\ref{eq12}) by $\xi_i^\mu$ and sum over $i$, yielding
%
\begin{eqnarray}
\label{eigen}
\chi_\nu \alpha_\nu &=& 
- \frac{1}{\sqrt{N}}\sum_\mu \frac{1}{\chi_\mu} 
\frac{1}{\sqrt{N}}\sum_i (\xi_i^\mu)^2\xi_i^\nu \tanh h_i (1-\tanh^2 h_i) \, .
\end{eqnarray}
Eq.~(\ref{eigen}) shows that unless the matrix $s_{\mu\nu}$ happens to have an 
eigenvalue equal to $\chi_\mu$, $\alpha_\mu$ it is of order $1/\sqrt{N}$.
We can then rewrite Eq.~(\ref{eq12}) as
\begin{eqnarray}
\label{e16}
m_i - \tanh h_i &=& - \frac{1}{N} \sum_\mu \frac{(\xi_i^\mu)^2}{\chi_\mu} \tanh h_i (1-\tanh^2 h_i) + O(1/N^{3/2}) \, .
\end{eqnarray}
which shows clearly that $m_i = \tanh h_i + O(1/N)$.

We will now optimize our probability with respect to $\xi_i^\mu$:
%
\begin{eqnarray}
\frac{1}{N} \sum_j c_{ij} \xi_j^\mu &=& \frac{1}{N} \frac{1}{\chi_\mu} \xi_i^\mu (1-\tanh^2 h_i) 
- \frac{1}{N^2} \sum_\nu \frac{\xi_i^\mu (\xi_i^\nu)^2}{\chi_\mu \chi_\nu}
(1-3 \tanh^2 h_i)(1-\tanh^2 h_i)
\nonumber\\
&&- \frac{1}{N^2} \sum_\mu \frac{r_{\mu \nu}}{\chi_\mu^2 \chi_\nu} \xi_i^\mu
(1-\tanh^2 h_i) - \frac{2}{N^2} \sum_{\nu(\neq \mu)} x_{\mu \nu} \xi_i^\nu
(1-\tanh^2 h_i)
\nonumber\\
&&+ O(1/N^{5/2}) \, ,
\label{e17}
\end{eqnarray}
Where we used the fact that
$\frac{1}{N} \alpha_\mu (m_i - \tanh h_i) = O(1/N^{5/2})$ to ignore subleading
terms. In the following, we will start by solving this equation in the
leading order on $N$.

\subsubsection{Solution in the leading order}

We want to find the solutions $\xi_i^{\mu,0}$ and $h_i^0$ of
Eqs.~(\ref{e16}) and~(\ref{e17}) up to the leading order on $N$.
Ignoring sub-leading terms, this equations are respectively given by
\begin{eqnarray}
\label{eq_domin_xi}
\frac{1}{N} \sum_j c_{ij} \xi_j^{\mu,0} &=& \frac{1}{N} \frac{1}{\chi_\mu} \xi_i^{\mu,0} (1-\tanh^2 h_i^0) \, ,  
\end{eqnarray}
and
\begin{eqnarray}
h_i^0 &=& \tanh^{-1} m_i \, .
\end{eqnarray}
If we define $v_i^\alpha$ as the eigenvector associated to the eigenvalue $\lambda_\alpha$ of the matrix 
\begin{equation}
\label{matrice}
M_{ij} = \frac{c_{ij}}{\sqrt{1-\tanh^2 h_i^0}\sqrt{1-\tanh^2 h_j^0}}\, ,
\end{equation}
 we have
\begin{eqnarray}
\xi_i^{\mu,0} = \sqrt{1-\frac{1}{\lambda_\mu}} \frac{v_i^\mu}{\sqrt{1-\tanh^2 h_i^0}} \, .
\label{matrice2}
\end{eqnarray}
Note that the orthogonality of the eigenvectors assures that our gauge
$s_{\mu \nu}=0$ is respected.

The procedure of diagonalizing the correlation matrix to extract underlying
information is known in the statistics literature by the name of Principal 
Component Analysis. We already mentioned in the end of section \ref{sec_models}
how this method is currently used to find patterns in neuronal data.
It was also used by Ranganathan et al. \cite{sectors} to find functional 
groups in proteins. Our approach gives a Bayesian justification for using the
PCA for neuron data and allows us to write the probability distribution for
the measured system: it is just the Boltzmann distribution for the obtained
Hopfield model.

Note that Eq.~(\ref{matrice}) implies
\begin{eqnarray}
J_{ij} &=& \sum_\alpha \frac{\lambda_\alpha -1}{\lambda_\alpha} 
\frac{v_i^\alpha v_j^\alpha}{\sqrt{(1-m_i^2)(1-m_j^2)}} 
\nonumber \\
&=& \frac{1}{\sqrt{(1-m_i^2)(1-m_j^2)}} \left[M^{-1} (M-1)\right]_{ij} \, ,
\end{eqnarray}
which is exactly the same formula found in section~\ref{sec_loop} for
the loop summation, but using the convention of $M_{ii}=1$ instead of
$M_{ii} = 0$. While this is an encouraging sign that our calculations are
correct, it does not bring any new result. We will thus now try to go
beyond the leading terms in $N$ and look for the first correction to this
expression.

\subsubsection{Sub-dominant corrections}

Applying our first approximation of the patterns $\xi^{\mu,0}$ to
Eq.~(\ref{e16}), we can evaluate the first correction to the field $h_i$.
Posing $h_i = h_i^0 + h_i^1$, we have
\begin{eqnarray}
h_i^1 = \frac{1}{1-\tanh^2 h_i^0}\frac{1}{N} \sum_\mu \frac{(\xi_i^{\mu,0})^2}{\chi_\mu^0} \tanh h_i^0 (1-\tanh^2 h_i^0)
\end{eqnarray}
Since now we have a more precise value of $h_i$, we can redo the procedure
of solving Eq.~(\ref{eq_domin_xi}) we used for the leading order using
this new value. We will denote the obtained patterns $\xi_i^{\mu,1}$. Note
that $\xi_i^{\mu,1}$ is correct up to the leading order in $N$, in
the same way as $\xi_i^{\mu,0}$, since it neglects the subdominant terms
in Eq.~(\ref{e17}).
Finally, to find an expression for the patterns that is correct up to the first
leading order, $\xi_i^{\mu,2} = \xi_i^{\mu,1} + \delta_i^\mu$, we do
a Taylor expansion of the dominant orders of Eq.~(\ref{e17}) around
$\xi_i^\mu = \xi_i^{\mu,1}$:

\begin{eqnarray}
&&\phantom{=}
\frac{1}{N} \sum_i c_{ij} \delta_i^\mu - \frac{1}{N} \frac{1}{\chi_\mu} \delta_i^\mu (1-\tanh^2 h_i) 
+ \frac{2}{N^2} \sum_{\nu (\neq \mu)} x_{\mu \nu} \xi_i^{\nu,1} (1-\tanh^2 h_i)
\nonumber\\
&&\phantom{=}
- \frac{2}{N^2 \chi_\mu^2} \sum_j \xi_i^{\mu,1} \xi_j^{\mu,1} \delta_j^\mu
(1-\tanh^2 h_i)(1-\tanh^2 h_j) =
\nonumber\\
&&=
- \frac{1}{N^2} \sum_\nu \frac{\xi_i^{\mu,1} (\xi_i^{\nu,1})^2}{\chi_\mu \chi_\nu}
(1-3 \tanh^2 h_i)(1-\tanh^2 h_i) \nonumber \\ 
&&\phantom{=}- \frac{1}{N^2} \sum_\mu \frac{r_{\mu \nu}}{\chi_\mu^2 \chi_\nu} \xi_i^{\mu,1}
(1-\tanh^2 h_i)\, .
\label{eq_feia}
\end{eqnarray}

Similarly, our gauge equations $s_{\mu \nu} = 0$ yields
\begin{eqnarray}
\label{eq_gauge}
\frac{1}{\sqrt{N}}
\sum_i (\xi_i^{\mu,1} \delta_i^\nu +\xi_i^{\nu,1} \delta_i^\mu) 
(1-\tanh^2 h_i^1)= 0\, .
\end{eqnarray}

At last, Eqs.~(\ref{eq_feia}) and~(\ref{eq_gauge}) form a non-homogeneous
linear system in the
variables $\delta_i^\mu$ and $x_{\mu \nu}$ ($\mu <\nu$) which can be solved
numerically by a simple matrix inversion.

\subsection{Numerical verification}

In this section we will verify numerically the correctness of both the
inference up to the dominant order shown in Eqs.~(\ref{matrice})
and~(\ref{matrice2}) and its
subdominant corrections shown in Eqs.~(\ref{eq_feia}) and~(\ref{eq_gauge}).
We will proceed in the following way: first, we will choose a set of patterns
$\{\xi_i^\mu\}$ and fields $\{h_i\}$ . Then, we will use a numerical method
(that we will explain in the following) to compute the correlations and
local magnetizations of the corresponding Hopfield model. Using these
quantities, we will
use our inference procedure to find the the inferred patterns, both
in the dominant order $\{\xi_i^{\mu,0}\}$ and with the subdominant corrections
$\{\xi_i^{\mu,2}\}$. Finally, we will compare the obtained patterns with our
initially chosen ones to estimate the quality of our inference.

\subsubsection{Numerical evaluation of the correlations of the Hopfield model}

To verify the validity of our inference, we needed a numerical method that
produced very precise values for the correlations, so we could be sure that any
disagreement between the real patterns and the inferred ones were due to
shortcomings of the inference procedure and not due to numerical errors.
Moreover, we wanted a numerical simulation that
scales well for increasing $N$, since our equations are correct in the
large $N$ limit.

To be able to satisfy these requirements, we restricted our input patterns
and fields to a four-block configuration:
\begin{eqnarray}
\begin{array}{cccccccccccccc}
&&\multicolumn{3}{c}{N/4}
&\multicolumn{3}{c}{N/4}
&\multicolumn{3}{c}{N/4}
&\multicolumn{3}{c}{N/4}
\\
\vspace{-.4cm}
&
&\multicolumn{3}{c}{\overbrace{\phantom{\begin{array}{ccc}a_1&\cdots&a_1\end{array}}}}
&\multicolumn{3}{c}{\overbrace{\phantom{\begin{array}{ccc}a_1&\cdots&a_1\end{array}}}}
&\multicolumn{3}{c}{\overbrace{\phantom{\begin{array}{ccc}a_1&\cdots&a_1\end{array}}}}
&\multicolumn{3}{c}{\overbrace{\phantom{\begin{array}{ccc}a_1&\cdots&a_1\end{array}}}}
\\
\xi^1 &= &\{ a_1, & \cdots, & a_1, & b_1, & \cdots, & b_1, &
             c_1, & \cdots, & c_1, & d_1, & \cdots, & d_1 \} \, , \\
\xi^2 &= &\{ a_2, & \cdots, & a_2, & b_2, & \cdots, & b_2, &
             c_2, & \cdots, & c_2, & d_2, & \cdots, & d_2 \} \, , \\
\xi^3 &= &\{ a_3, & \cdots, & a_3, & b_3, & \cdots, & b_3, &
             c_3, & \cdots, & c_3, & d_3, & \cdots, & d_3 \} \, , \\
h     &= &\{ h_1, & \cdots, & h_1, & h_2, & \cdots, & h_2, &
             h_3, & \cdots, & h_3, & h_4, & \cdots, & h_4 \} \, , \\
\end{array}
\label{eq_4blocks}
\end{eqnarray}
where $\{a_i\}, \{b_i\}$, etc are real values.
Since in our calculations we suppose that
$\sum_i\xi^\mu_i \xi^\nu_i = O(\sqrt{N})$ and
$\sum_i \xi_i^\mu \tanh h_i = O(\sqrt{N})$,
we have restricted our number of patterns to three since it is
impossible using four blocks to satisfy these conditions for more patterns.
In our simulation we have chosen values of $\{a_i\}$, $\{b_i\}$, $\ldots\,$ so that
the patterns obey our orthogonality condition and that the magnitude of the
patterns are small enough not to be too close to the phase transition at
$|\xi_i^\mu| =1$.

With the choice of patterns shown in Eq.~(\ref{eq_4blocks}), the partition
function is
\begin{equation}
\begin{split}
 Z =
\sum_{\{\sigma\}} \exp\Bigg\{&
  N \sum_{\mu=1}^3\Bigg[
  a_\mu^2 \left(\frac{1}{N}\sum_{i=     1}^{N/4 } \sigma_i\right)^2
+ b_\mu^2 \left(\frac{1}{N}\sum_{i= N/4+1}^{N/2 } \sigma_i\right)^2
\\
+&c_\mu^2 \left(\frac{1}{N}\sum_{i= N/2+1}^{3N/4} \sigma_i\right)^2
+ d_\mu^2 \left(\frac{1}{N} \sum_{i=3N/4+1}^{N  } \sigma_i\right)^2
\Bigg]
\Bigg\}\, ,
\end{split}
\end{equation}
We can see that the Hamiltonian depends only on
$m_1= \sum_{i=1}^{N/4 } \sigma_i$, $m_2= \sum_{i=N/4+1}^{N/2 } \sigma_i$, etc.
Now we replace the sum over all the spin configurations by a sum over all
possible values of $m_1$, $m_2$, $m_3$ and $m_4$, which allows one to
evaluate the partition function, the correlations and the magnetizations
with a complexity of $O(N^4)$.

\subsubsection{Comparison of the inferred with the real patterns}

We cannot directly compare the inferred patterns with the real ones to evaluate
the error of our procedure since they might differ in gauge. Thus, we used in
our comparison a value that does not depend on the gauge:
$J_{ij} = \sum_\mu \xi_i^\mu \xi_j^\mu$.
The results of the comparison of the real patterns
with the inferred ones can be seen in the following graph

\begin{figure}[H]
\begin{center}
\includegraphics[width=0.75\columnwidth]{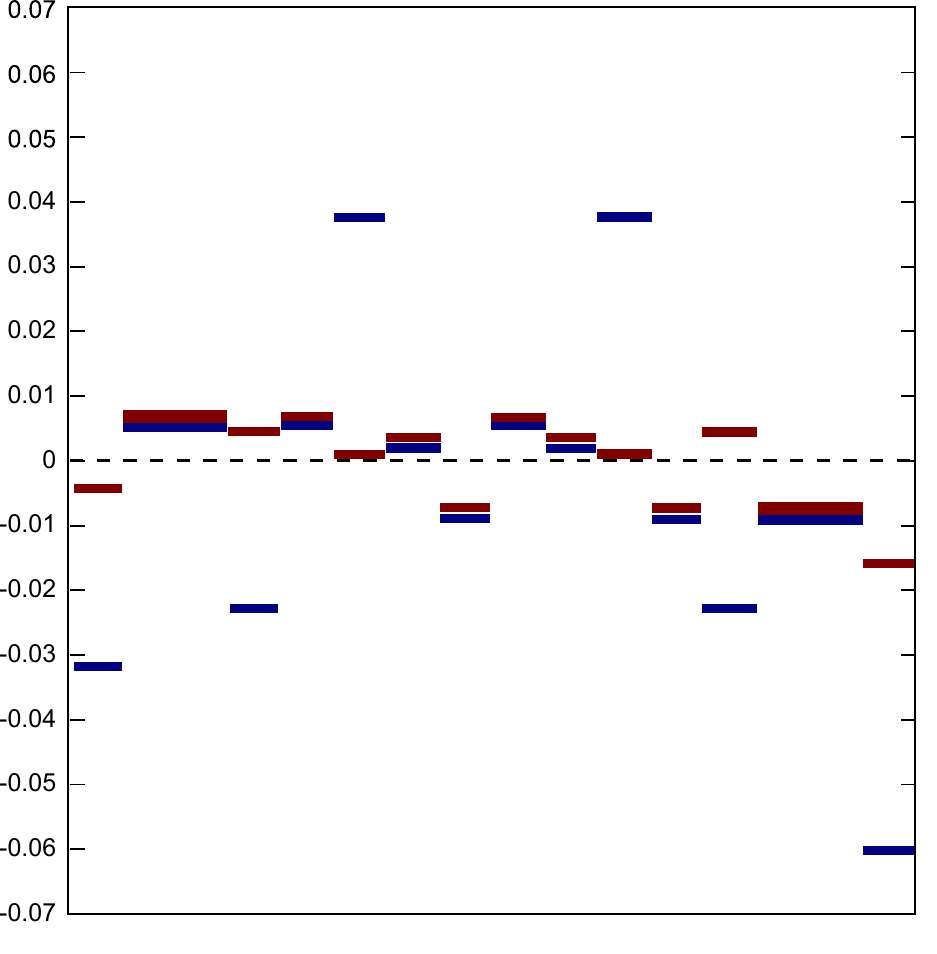}
\caption{Inference error for the different elements of the matrix $J_{ij}$ for
$N=100$.
In blue we have $J_{ij}^\text{real} - \sum_\mu \xi_i^{\mu,0} \xi_j^{\mu,0}$ and
in red we have $J_{ij}^\text{real} - \sum_\mu \xi_i^{\mu,2} \xi_j^{\mu,2}$.
In this graph, we have
$\{a_1,b_1,c_1,d_1\} = \{0.693,0.4, -0.8,0.4\}$,
$\{a_2,b_2,c_2,d_2\} = \{0.693,-0.8, 0.4,0.4\}$,
$\{a_3,b_3,c_3,d_3\} = \{0,0.693, 0.693,0.693\}$
and
$\{h_1,h_2,h_3,h_4\} = \{0.254, -0.283, 0.416, -0.380\}$.
}
\label{grblah}
\end{center}
\end{figure}

We can see clearly in Figure~\ref{grblah} that the subdominant corrections
improve the inference quality. Unfortunately,
these corrections are very sensible to noisy data, and we could not
see an improvement of the inference for both
Monte-Carlo and real neuron data.

\chapter{Evaluation of the inference entropy for the Hopfield model}
\label{chap_hop}

As introduced in section~\ref{subsec_max_entr},
a good estimate of how much data is needed to infer a pattern is given by
the information-theoretical entropy (Eq.~(\ref{eq_def_S})) of the a posteriori
probability of the patterns, given by
\begin{equation}
S[\{\sigma_i^l\}] = - \sum_{\{\xi\}} P[\{\xi_i^\mu\}|\{\sigma_i^l\}] \log P[\{\xi_i^\mu\}|\{\sigma_i^l\}] \, ,
\label{eq_S22}
\end{equation}
where $\{\xi_i^\mu\}$ are the patterns of our Hopfield model (see
section~\ref{sec_hop}) and $\{\sigma_i^l\}$ are a set of $L$ measured
configurations of the system.

The entropy can be interpreted as the quantity of information that is missing
about our system. Thus, when $S \ll 1$, we can say that we have enough data to
infer with very little error the patterns.
In this chapter we will evaluate the entropy for the inference of
the Hopfield model. As before, we will first treat the simpler case of a single
pattern before dealing with the more general and interesting case of an
arbitrary number of patterns $p$.

\section{Case of a single binary pattern}
\label{sec_mattis}

Let us recall the usual partition function of the Hopfield model
(Eq.~(\ref{eq_hop_usual})) for $p=1$
\begin{eqnarray}
Z(\beta,\{\xi_i\}) &=&  \sum_{\{\sigma\}} \exp\left[\frac{\beta}{N}\sum_{i<j} \xi_i \xi_j \sigma_i \sigma_j\right] \, .
\end{eqnarray}
In the particular case where $\xi_i=\pm 1$, we can
pose $\sigma_i'=\xi_i \sigma_i$ and rewrite the partition
function as
\begin{equation}
Z(\beta) =  \sum_{\{\sigma'\}} \exp\left[\frac{\beta}{N}\sum_{i<j} \sigma_i' \sigma_j'\right] \, .
\label{eq_mattis_cw}
\end{equation}
This
equation is exactly the same as Eq.~(\ref{eq_cw}) for zero external field, 
which means that the thermodynamics of this model is identical to the
infinite-dimensional Ising model presented in chapter~\ref{chap_ising}. From a 
more
technical point of view, contrary to the usual Hopfield model with $p>1$, 
the partition function is independent of the pattern, which make
the calculations considerably simpler.

In this case, the Bayes a posteriori probability is given by
Eq.~(\ref{eq_likeli0}), which we recall here:
\begin{equation}
P(\{\xi_i\}|\{\sigma_i^l\}) = \frac{P_0(\{\xi_i\}) }{Z(\beta)^L \mathcal{N}(\beta,\{\sigma_i^l\})} \prod_{l=1}^L 
\exp\left[\frac{1}{N}\sum_{i<j} \xi_i \xi_j \sigma_i^l \sigma_j^l\right] \, ,
\end{equation}
where we used a different notation for the normalization $\mathcal{N}$.
Applying this result to Eq.~(\ref{eq_S22}) and setting the prior
$P_0(\{\xi_i\})=1$, we obtain
\begin{eqnarray}
S[\{\sigma_i^l\}] &=& -  \frac{1}{\mathcal{N}[\{ \sigma_i^l\}] Z(\beta)^L}  \sum_{\{\xi\}} \exp 
\left(\frac{\beta}{N} \sum_{l=0}^L \sum_{i<j} \xi_i \xi_j \sigma_i^l  \sigma_j^l \right) \times \nonumber\\
&&\times\left[- \log \left(\mathcal{N}[\{\sigma_i^l\}] Z(\beta)^L\right) +\left(\frac{\beta}{N} \sum_{l=0}^L \sum_{i<j} \xi_i \xi_j \sigma_i^l  \sigma_j^l \right)\right] \, .
\label{eq_S123}
\end{eqnarray}
Introducing the variable
\begin{equation}
\tilde{N}[\{\sigma_i^l\}] \equiv \mathcal{N}[\{\sigma_i^l\}] Z(\beta)^L= \sum_{\{\xi\}}\exp \left(\frac{\beta}{N} \sum_{l=0}^L \sum_{i<j} \xi_i \xi_j \sigma_i^l  \sigma_j^l \right)\, ,
\label{eq_ntil}
\end{equation}
we can rewrite Eq.~(\ref{eq_S123}) as
\begin{eqnarray}
S[\{\sigma_i^l\}] &=&  \frac{\log \tilde{N}[\{\sigma_i^l\}]}{\tilde{N}[\{ \sigma_i^l\}]}  \sum_{\{\xi\}}
 \exp \left(\frac{\beta}{N} \sum_{l=0}^L \sum_{i<j} \xi_i \xi_j \sigma_i^l  \sigma_j^l \right) \nonumber\\
&&-  \sum_{\{\xi\}} \frac{\beta}{N} \sum_{l=0}^L \sum_{i<j} \xi_i \xi_j \sigma_i^l  \sigma_j^l 
 \exp \left(\frac{\beta}{N} \sum_{l=0}^L \sum_{i<j} \xi_i \xi_j \sigma_i^l  \sigma_j^l \right) \nonumber \\
&=& \log \tilde{N}[\{\sigma_i^l\}] - \beta \frac{\partial \log \tilde{N}[\{\sigma_i^l\}]}{\partial \beta} \, .
\label{eq_entro11}
\end{eqnarray}
Thus, the entropy can be trivially evaluated from $\tilde{N}[\{\sigma_i^l\}]$. 

We can see in Eq.~(\ref{eq_ntil}) that
the expression for $\tilde{N}$ is formally identical to the partition function
of a Hopfield model where the $L$ measured configurations $\{\sigma_i^l\}$ 
play the role of the patterns and $\{\xi_i\}$ replace the spin variables:
\begin{equation}
Z_\text{Hop} = \sum_{\{\sigma\}} \exp \left(\frac{\beta}{N} \sum_{\mu=1}^L \sum_{i<j} \xi_i^\mu \xi_j^\mu \sigma_i  \sigma_j \right)\, .
\label{eq_logZ_not}
\end{equation}
For such analogy, the inference entropy shown in Eq.~(\ref{eq_entro11}) has a 
thermodynamic meaning: it is the thermodynamic entropy of the model.

Eqs.~(\ref{eq_ntil}) and~(\ref{eq_entro11}) give the entropy
of the system for a particular set of measures $\{\sigma_i^l\}$. Now,
it is natural to expect the entropy to be very reproducible across
different sets of measurements. In this context, we are interested in evaluating
the average of the entropy with respect to all possible measurements.
Supposing that the data is produced measuring an actual Hopfield model with
a pattern $\{\tilde{\xi}_i\}$, we have
\begin{eqnarray}
\moy{S} &=& \left.\moy{\log \tilde{N}[\{\sigma_i^l\}]}\right|_{\tilde{\beta}=\beta} 
- \beta \left(\frac{\partial}{\partial \beta} \moy{\log \tilde{N}[\{\sigma_i^l\}]}
\right)_{\tilde{\beta}=\beta} \, ,
\label{eq_seilaqnome}
\end{eqnarray}
with
\begin{eqnarray}
\moy{\log \tilde{N}} &=& \frac{1}{Z\left(\tilde{\beta}\right)^L} \sum_{\{\sigma\}}
\exp\left(\frac{\tilde{\beta}}{N} 
 \sum_{l=0}^L \sum_{i<j} \tilde{\xi_i} \tilde{\xi_j} \sigma_i^l  \sigma_j^l \right) 
\log \tilde{N}[\{\sigma_i^l\}] \, ,
\end{eqnarray}
where we replaced our usual inverse temperature $\beta$ by a new variable
$\tilde{\beta}$ since we should not take the
derivative in respect to it in Eq.~(\ref{eq_seilaqnome}).

The entropy of the system is very different if $\beta<1$ and the 
system
is in the paramagnetic phase or if $\beta>1$ in which case the system is in the ferromagnetic phase. We
will see these two cases separately in the following sections.

\subsection{Ferromagnetic case}

In the ferromagnetic case, if we want to evaluate a thermal average of
some quantity $X$, we have
\begin{eqnarray}
\moy{X} &=& \frac{1}{Z^L(\tilde{\beta})} \sum_{\{\sigma\}} X(\{\sigma_i^l\}) \exp\left(\frac{\tilde{\beta}}{N} 
 \sum_{l=0}^L \sum_{i<j} \tilde{\xi_i} \tilde{\xi_j} \sigma_i^l  \sigma_j^l \right) 
 \, , \nonumber\\
&=& \sum_{\{\sigma\}} X(\{\sigma_i^l\}) \prod_{l=1}^L \prod_{i=1}^N
\frac{e^{\tilde{\beta} m^* \sigma^l_i \tilde{\xi}_i}}{2 \cosh \left(\tilde{\beta} m^*\right)} \, ,
\end{eqnarray}
where $m^*$ is the solution of the equation $m^* = \tanh (\tilde{\beta} m^*)$.

Since $\log \tilde{N}$ is formally identical to the free-energy of the
Hopfield model, we start with the saddle-point solution obtained in 
chap.~\ref{chap_ising}, given in Eqs.~(\ref{eq_hop1234}) and~(\ref{eq_1234}):
\begin{eqnarray}
  \log \tilde{N} &=& -\frac{\beta N}{2} \sum_{l=1}^L m_l^2 + \sum_i  \log\left[2 \cosh\left(\beta \sum_{l=1}^L m_l \sigma_i^l \right)\right] \, ,
  \label{eq_7p12}
\end{eqnarray}
and
\begin{eqnarray}
m_l &=& \frac{1}{N} \sum_i \sigma_i^l \tanh \left(\beta \sum_s m_s \sigma_i^s\right) \, .
\label{eq_m_isi}
\end{eqnarray}
We recall that the physical meaning of $m_\mu$ in the Hopfield model is the 
magnetization of the spins according to the pattern $\mu$. In our case, it
represents the overlap between the pattern we are inferring and the $l$-th
configuration:
$m_l = \frac{1}{N}\sum_i \xi_i \sigma_i^l$. 

Eq.~(\ref{eq_m_isi}) has always a solution in the form 
$\{m_l\}=\{{m,m,...,m\}}$
\cite{amit1}.
Since $\{\sigma_i^l\}$ are
measured configurations of an Ising system in the ferromagnetic phase, we expect
them to be very similar: they are all close to the same minimum of the free
energy of the Ising model. We expect thus that the solution
$\{m_l\}=\{{m,m,...,m\}}$
to be the minimum of our modified 
free-energy of Eq.~(\ref{eq_7p12}). Under this hypothesis, the entropy reads
\begin{equation}
\frac{\moy{S}}{N} = \moy{\ln \left[ 2 \cosh \left(\beta m \sum_l \sigma^l \right)\right] }
-\moy{\left(\beta m \sum_l \sigma^l \right) \tanh \left(\beta m \sum_l \sigma^l \right) } \, ,
\label{eq_S_ferro}
\end{equation}
where the value of $m$ that satisfies Eq.~(\ref{eq_m_isi}) is $m=m^*$,
fact proven in Appendix~\ref{ap_m}.

\subsubsection{Asymptotic behavior}

One can note that the entropy of this system is the same of the system composed
of a single spin $\rho =\pm 1$ connected with $L$ independent, magnetized
spins. The partition function for this system of a single spin is
\begin{equation}
Z = 2 \cosh \left(\beta m \sum_i \sigma_i \right) \, ,
\end{equation}
and the free-energy reads
\begin{equation}
F = \moy{\log \left[2 \cosh \left(\beta m \sum_i \sigma_i \right)\right]}_{\sigma_i} \, ,
\end{equation}
where the average is done with respect to independent spins $\sigma$ 
with magnetization $m$. If we want to calculate the average entropy 
of this system, we note that for large $x$,
$\log\left(2 \cosh x\right) - x \tanh x \approx (1+2 |x|)e^{-2|x|}$.
Consequently
\begin{equation}
\moy{S} \approx \sum_{k=0}^L \left(\begin{array}{c} L \\ k \end{array} \right) 
\frac{e^{\beta m (L-2k)}}{\left[2 \cosh(\beta m)\right]^L}
e^{-2 \beta m |L-2k|} \, .
\end{equation}
One may remark that the probability that our single spin $\rho$ is not
aligned with its partners is exponentially small on $L$. There are two 
extreme cases that contributes to this probability

\begin{figure}[H]
\begin{center}
\includegraphics[width=0.75\columnwidth]{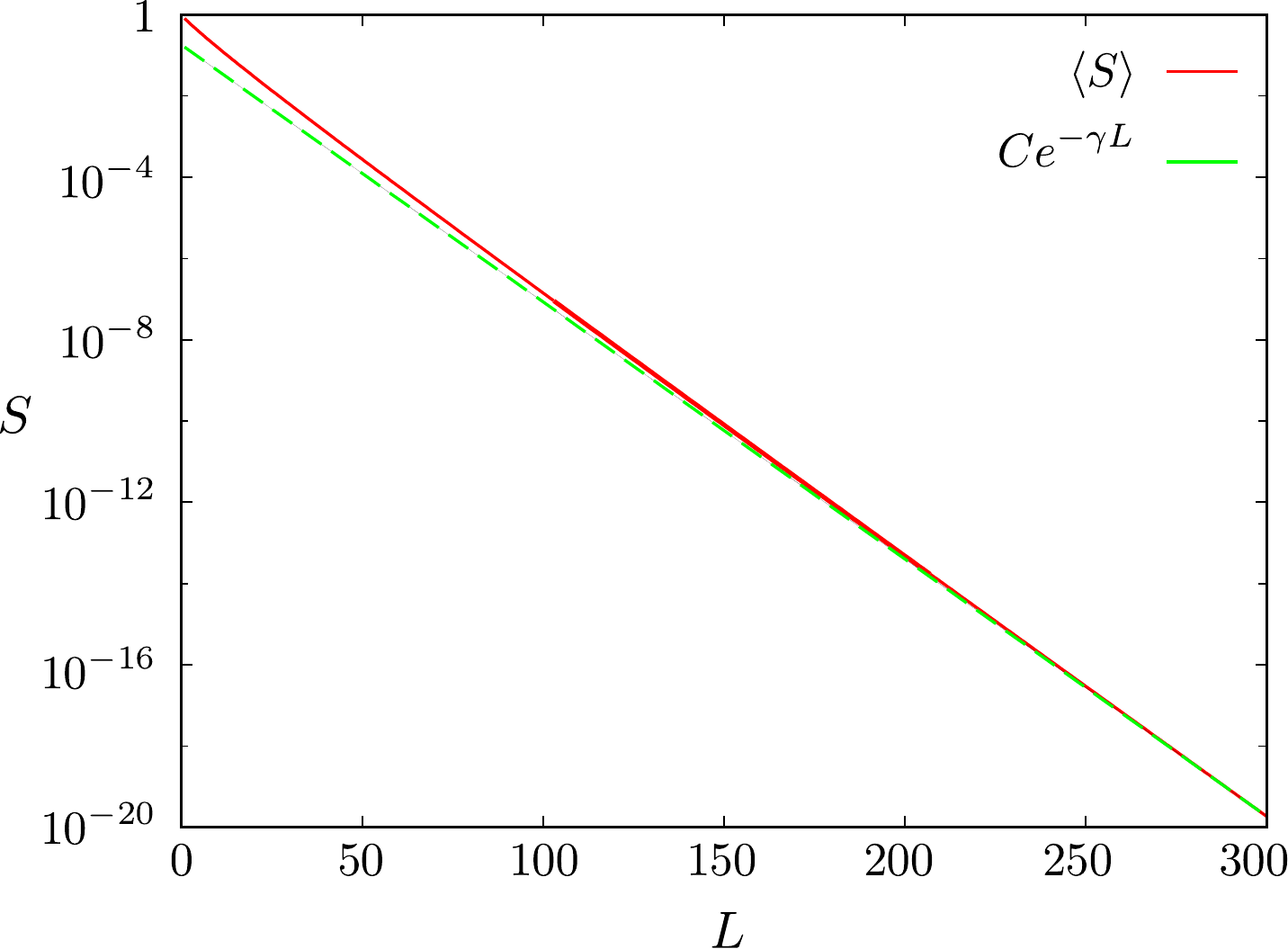}
\caption{Entropy per spin as a function of $L$ for $\beta=1.1$ according
to Eq.~(\ref{eq_S_ferro}).}
\label{grEntr1}
\end{center}
\end{figure}

\begin{enumerate}
\item
\label{item1}
 The spins $\sigma_i$ obeys $\sum_i \sigma_i \approx 0$  (which is very
  unlikely) and consequently our spin $\rho$ is random.

\item 
\label{item2}
We have the highly probable situation of
$\sum_i \sigma_i \approx m L$, but our spin $\rho$ is misaligned with
the others,
which is very unlikely.
\end{enumerate}
The case \ref{item1} has a probability $\approx \left[\cosh (\beta m) \right]^{-L}$
and gives a $\log 2$ contribution to the entropy, while the case \ref{item2}
has a probability
O(1) but gives a contribution of order $e^{-2 \beta m^2 L}$ to the entropy.
Since for all $\beta$, $\log \cosh (\beta m) < 2 \beta m^2$ (note that $m$
is an implicit function of $\beta$)
, case \ref{item1} 
dominates the behavior of the system for $L \rightarrow \infty$. So, for
large $L$, we have the general behavior
\begin{equation}
S \propto e^{-\gamma L},
\end{equation}
with
\begin{equation}
 \gamma = \log \cosh\left(\beta m\right) \, .
\end{equation}
We can see how well the entropy converges to this asymptotic behavior in
Fig.~\ref{grEntr1}.

We can thus conclude that for $L \gg 1/\gamma$ we can infer the unknown pattern
with a very small probability of error.

\subsection{Paramagnetic case}
\label{sec_mattis_para}
In the paramagnetic phase we have $m=0$. According to 
Eq.~(\ref{eq_S_ferro}), the entropy is thus equal to $\log 2$ for all values 
of $L$.
This is correct under the hypothesis that $L$ remains \emph{finite} when
$N \rightarrow \infty$. For $L = \alpha N$, the results of the previous section
do not hold since we cannot use the saddle-point approximation for $\log Z$.

We start with Eq.~(\ref{eq_hop_int00})
\begin{equation}
  \tilde{N} = \int \prod_{l=1}^L \frac{\di m_l}{\sqrt{2 \pi \beta^{-1} N^{-1}}}
  \sum_{\{\xi\}}
  \exp\left[ -\frac{\beta N}{2} \sum_{l=1}^L m_l^2
   +\beta \sum_{l=1}^L m_l \sum_{i} \sigma_i^l \xi_i 
   \right]
   \label{eq_720}
\end{equation}
and we would like to find the average $\moy{\tilde{N}}_\sigma$ with respect to all possible
realizations of the measures $\{\sigma_i^l\}$, like we have done  in the
ferromagnetic case. The main difference is that now the saddle-point values of
$m_l$ are of order $O(1/\sqrt{N})$.

To find the correct solution under these conditions, we need to use the replica
trick as explained in section~\ref{sec_hop}.
The calculations are very similar to those done in the solution of the
Hopfield model by Amit et al. \cite{amit2} and the details can be found in
Appendix~\ref{ap_entro_para}.
The obtained value for the entropy is given by
\begin{eqnarray}
\moy{S}_\sigma &=& -\frac{\alpha \beta (1-q) 
\left\{1 - \beta\left[2 - 2 q(1-\beta) - (1+t^2)\beta\right]\right\}}
{2 (1-\beta) \left[ 1- (1-q) \beta\right]^2}
- \frac{\alpha}{2} \ln \left[1-(1-q)\beta\right] \nonumber\\
&&+\moy{\ln \left[ 2 \cosh \left( \frac{\alpha}{2} \hat{t} + z \sqrt{\alpha \hat{q}} \right) \right] }_z
- \frac{\alpha}{2} \left[(1-q)\hat{q} + t \hat{t}\right]  \, ,
\label{entrop12}
\end{eqnarray}
where $\moy{f}_z = \int_{-\infty}^{\infty} f(z) e^{-z^2/2} \di z$,
we remind that $\alpha = L/N$ and
\begin{eqnarray}
\label{col1}
t &=& \moy{\sigma_i^l \xi_i}_\sigma = \moy{\tanh\left(\frac{\alpha}{2} \hat{t} + z \sqrt{\alpha \hat{q}}  \right)}_z \, , \\
\label{col2}
q &=& \moy{\left(\frac{1}{N} \sum_i \xi_i \sigma_i^l\right)^2}_\sigma =\moy{\left[\tanh\left(\frac{\alpha}{2} \hat{t} +  z \sqrt{\alpha \hat{q}}  \right)\right]^2}_z \, , \\
\label{col3}
\hat{t} &=& \frac{2 t \beta^2}{(1-\beta)[1-(1-q)\beta]} \, , \\
\label{col4}
\hat{q} &=& \beta^2 \frac{q (1 - \beta) + t^2 \beta}{(1-\beta)\left[1 - (1-q)\beta\right]^2} \, ,
\end{eqnarray}
where we remind that $t$ is the overlap between the real and the inferred
pattern.
One can verify that we only have
a non zero solution to these equations when $\alpha > \alpha_c$, with 
$\alpha_c = \left(1-\frac{1}{\beta}\right)^2$. In this case both
$q$ and $t$ are non-zero.

When $\alpha < \alpha_c$, $t$ is zero which implies that the inferred pattern
has no resemblance to the real one.
Surprisingly, the entropy decreases linearly in this regime for increasing
$\alpha$, which can be
possibly interpreted as a growth of
the set of patterns known to be \emph{incompatible} with the data.

Note that in contradistinction with the ferromagnetic case, to infer the
patterns in the paramagnetic phase it takes
a number of measures that is proportional to the size of the system, implying
that is much harder to extract information from it.

\subsubsection{Numerical verification}

\begin{figure}[H]
\begin{center}
\includegraphics[width=0.85\columnwidth]{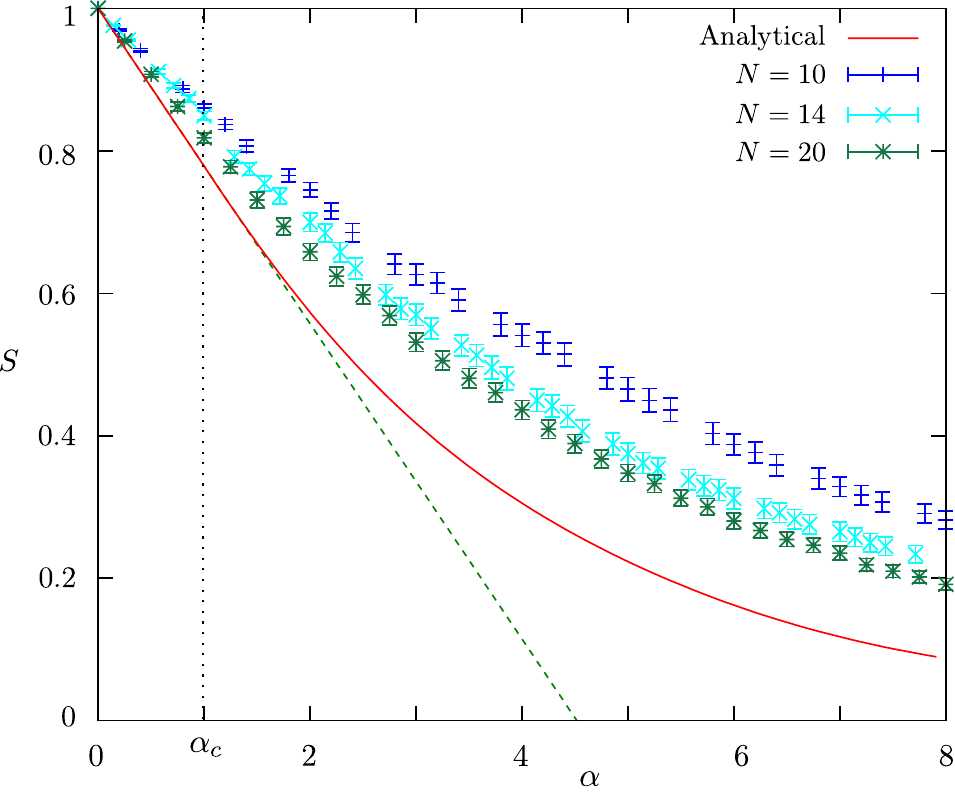}
\caption{Entropy per spin as a function of $\alpha$ for $\beta=0.5$. The
solid line correspond to Eqs.~(\ref{entrop12}--\ref{col4}) while the points
correspond to numerical values obtained by exact enumeration. The dashed line
corresponds to the behavior of $\moy{S}$ in the regime of $q=t=0$ while the
vertical line indicates $\alpha_c$.}
\label{grEntr2}
\end{center}
\end{figure}

In Figure~\ref{grEntr2}, one can see the behavior of the entropy as a function
of $\alpha$ for a fixed inverse temperature $\beta=0.5$.
To evaluate numerically $\moy{S}$ as shown in Fig.~\ref{grEntr2}, we used the
following algorithm:

\begin{enumerate}
\item Evaluate $Z$ by exact enumeration;
\item Generate $L = \alpha N$ configurations $\{\sigma_i^l\}$ according to the Boltzmann weight
      by rejection sampling;\label{step2_ch5}
\item Evaluate $\mathcal{N}[\{\sigma_i^l\}]$ by exact enumeration;
\item Evaluate $S[\{\sigma_i^l\}]$ by exact enumeration.
\end{enumerate}

For every $\alpha$, we repeated this procedure one hundred times with different
random seeds, which gives different configurations $\{\sigma_i^l\}$ in
step~\ref{step2_ch5}. The
points in the graph correspond to the averages of the set of obtained values
of $S$ and the error bars were
calculated using the standard deviation. The result of this procedure
supports our analytical results. Indeed, in Figure~\ref{grEntr2} we notice that
the bigger values of $N$ are much closer to the analytical curve, which suggests
that the difference between the analytical and numerical results is due to
the small value of $N$ used in the numerical calculations.

\section[Case of a single pattern $\xi_{\protect\iii} \in \mathbb{R}$]{Case of a single pattern $\bm{\xi_{\protect\iii} \in \mathbb{R}}$}

The results of the previous section are valid only if our pattern takes bimodal
values, what might be a very particular case.
To verify if the inference of a real-valued pattern presents any qualitative
difference with respect to the bimodal case, we propose ourselves to 
evaluate explicitly the entropy for the real-valued case in this section.

The calculations are similar to those of the last
section, but with the supplementary complication that the
partition function depends on the exact values of the pattern whereas in the
bimodal case it depended only on the temperature. The partition function
is given by
\begin{eqnarray}
\log Z\left[\{\xi_i\}\right] &=& -\frac{N}{2} m[\{\xi_i\}]^2 + 
\sum_i \log\left[ 2 \cosh (m \xi_i)\right] \, ,
\label{eq_logZ1111}
\end{eqnarray}
with
\begin{eqnarray}
\label{eq_m11111}
m[\{\xi_i\}] &=& \frac{1}{N} \sum_i \xi_i \tanh\left(\xi_i m[\{\xi\}] \right) \, .
\end{eqnarray}
Since $\xi_i$ is continuous,
we must modify the definition of the entropy replacing the sum by an integral
\begin{eqnarray}
S[\{\sigma_i^l\}] &=& - \int \prod_i \di\xi_i P[\{\xi_i\}|\{\sigma_i^l\}] \log P[\{\xi_i\}|\{\sigma_i^l\}]\, .
\end{eqnarray}
Using Bayes' theorem, we obtain
\begin{equation}
\begin{split}
S[\{\sigma_i^l\}] &= -  \frac{1}{\mathcal{N}[\{ \sigma_i^l\}]}
\int \prod_i \di\xi_i \, 
\frac{P_0(\{\xi_i\})}{Z^L(\{\xi_i\})}  
\exp 
\left(\frac{1}{N} \sum_{l=0}^L \sum_{i<j} \xi_i \xi_j \sigma_i^l  \sigma_j^l \right) \times \\
 &
\times\Bigg[- \log \mathcal{N}[\{\sigma_i^l\}] - L \log Z(\{\xi_i\}) + \log  P_0(\{\xi_i\}) 
\phantom{\frac{A}{B}\sum_a^b}    \\
&
\left. \phantom{\Bigg [ +}
+\frac{1}{N} \sum_{l=0}^L \sum_{i<j} \xi_i \xi_j \sigma_i^l  \sigma_j^l \right]
\, ,
\end{split}
\end{equation}
where $\mathcal{N}[\{\sigma_i^l\}]$ is given by
\begin{eqnarray}
\mathcal{N}[\{\sigma_i^l\}] &=& \int \prod_i \di\xi_i \, P[\{\xi_i\}|\{\sigma_i^l\}] \\
 &=&   \int \prod_i \di\xi_i \,
\frac{P_0(\{\xi_i\})}{Z^L(\{\xi_i\})}  
\exp 
\left(\frac{1}{N} \sum_{l=0}^L \sum_{i<j} \xi_i \xi_j \sigma_i^l  \sigma_j^l \right) \, .
\end{eqnarray}

As in the previous section,
we would like to write $S$ as a derivative of the normalization.
For that, we define a modified normalization $\tilde{N}$
by introducing a parameter $\beta$ to $\mathcal{N}$:
\begin{equation}
\begin{split}
\tilde{N}[\{\sigma_i^l\},\beta] &=\\
= \int \prod_i& \di\xi_i
\exp 
\left(\frac{\beta}{N} \sum_{l=0}^L \sum_{i<j} \xi_i \xi_j \sigma_i^l  \sigma_j^l
+ \beta \log P_0(\{\xi_i\}) - \beta L \log Z(\{\xi_i\})\right) \, .
\end{split}
\label{eqn123}
\end{equation}
Note that while $\beta$ is similar to a inverse temperature, it is not strictly
one, since it multiplies also the partition function. Using this definition,
we can rewrite our entropy as
\begin{equation}
\label{eq_s_n_deriv}
S[\{\sigma_i^l\}] = \log \tilde{N}[\{\sigma_i^l\},1] - \left. \frac{\partial \log \tilde{N}[\{\sigma_i^l\},\beta]}{\partial \beta} \right|_{\beta=1} \, .
\end{equation}

A supplementary difficulty of the continuous case is the dependence
of $\log Z$ on $m[\{\xi\}]$, which depends on the patterns implicitly according
to Eq.~(\ref{eq_m11111}). To make that dependence explicit,
we introduce the following identity
\begin{equation}
f(m[\xi]) = \int \di m \int \di x  f(m) \exp \left[-i N x m + 
i \sum_i \xi_i \tanh(m \xi_i) \right] \, ,
\end{equation}
which is just Dirac's delta written in the integral form. We can thus write the
$e^{-\beta L \log Z}$ term in Eq.~(\ref{eqn123}) as
\begin{equation}
\label{eq_seila1123}
\begin{split}
e^{-\beta L \log Z} = \int \di m \int \di x
\exp\Bigg\{\frac{\beta L N}{2} m^2
&- \beta L \sum_i \log\left[ 2 \cosh (m \xi_i)\right]
\\&
-i N x m
+ i \sum_i \xi_i \tanh(m \xi_i) \Bigg\} \, .
\end{split}
\end{equation}

Like in the previous section, we would like to evaluate the average of the
entropy with respect to the different possible measures and to the different
real pattern $\{\tilde{\xi}_i\}$:
\begin{equation}
\moy{S} = \int \prod_i \di \tilde{\xi}_i\, P_0(\{\tilde{\xi}_i\})
\sum_{\sigma} \frac{1}{Z(\{\tilde{\xi_i}\})} S(\{\sigma_i^l\})
\exp\left[ \frac{1}{N} \sum_{l=1}^L \sum_{i<j} \tilde{\xi}_i \tilde{\xi}_j
\sigma_i^l \sigma_j^l \right]
\end{equation}
where we impose that the distribution of the unknown real patterns
$\{\tilde{\xi}_i\}$ is the same of our prior distribution $P_0(\{\xi_i\})$.

The details of the replica calculation, similar to the last section,
can be found in Appendix~\ref{ap_continu}. It yields

\begin{equation}
\begin{split}
\frac{1}{N}& \moy{\log \tilde{N}} = \frac{\beta}{2} \sum_{l=1}^L Q_l^2
+ \frac{\beta L m^2}{2}
\\
&+ \int \di \tilde{\xi} \, p_0(\tilde{\xi}) \sum_{\{\sigma^l\}} \exp \left[
\tilde{m} \tilde{\xi} \sum_{l=1}^L \sigma^l - L \log (2 \cosh(\tilde{m} \tilde{\xi}))
\right] \times \\
& 
\phantom{\int}
\times \log\left\{ \int \di \xi  \exp \left[ 
\beta \sum_{l=1}^L Q_l \xi \sigma^l - \beta L \log(2 \cosh(m \xi)) 
+ \beta \log p_0(\xi) \right]
\right\} \, ,
\end{split}
\end{equation}
where
\begin{eqnarray}
\tilde{m} &=& \int \di \tilde{\xi} \, p_0(\tilde{\xi}) \, \tilde{\xi}
\tanh(\tilde{m} \tilde{\xi})
= \moy{\tilde{\xi} \tanh(\tilde{m}\tilde{\xi})} \, ,\\
m &=& \int \di \tilde{\xi} \, \sum_{\{\sigma\}} \exp \left[
\sum_{l=1}^L \tilde{m} \tilde{\xi} \sigma^l - L \log 2 \cosh(\tilde{m} \tilde{\xi})
\right] \times \nonumber\\
&&\times \frac{
\int \di \xi \, p_0(\xi) \xi \tanh(m \xi) \exp \left[\sum_{l=1}^L Q^l \sigma^l \xi 
- L \log 2 \cosh (m \xi)\right]
}{
\int \di \xi \, p_0(\xi) \exp \left[\sum_{l=1}^L Q^l \sigma^l \xi 
- L \log 2 \cosh (m \xi)\right]
} \, , \\
Q_k &=& \int \di \tilde{\xi} \, \sum_{\{\sigma\}} \exp \left[
\sum_{l=1}^L \tilde{m} \tilde{\xi} \sigma^l - L \log 2 \cosh(\tilde{m} \tilde{\xi})
\right] \times\nonumber\\
&&\times \frac{
\int \di \xi \, p_0(\xi) \xi \sigma^l \exp \left[\sum_{l=1}^L Q^l \sigma^l \xi 
- L \log 2 \cosh (m \xi)\right]
}{
\int \di \xi \, p_0(\xi) \exp \left[\sum_{l=1}^L Q^l \sigma^l \xi 
- L \log 2 \cosh (m \xi)\right] \, ,
}
\end{eqnarray}
and $\prod_i e^{p_0(\xi_i)}= P_0(\{\xi\})$ is the prior probability of the
pattern and the probability distribution we used to average over the real
pattern, supposed to be identical.

From this last equation it is straightforward to evaluate the entropy using
Eq.~(\ref{eq_s_n_deriv}).


\section{Case of a system with \textit{p} patterns, magnetized following a strong external field}

In this section we consider the same conditions as we had in
section~\ref{sec_hop_para} for the inference: we have a Hopfield model with
$p$ patterns and we introduce a local external field strong
enough so that the system is not magnetized according to any of the patterns.
The Hamiltonian is thus the same as Eq.~(\ref{eq_hop_usual}).
The calculations are very similar to these of section~\ref{sec_seila1}, so
they can be found in Appendix~\ref{app_B}.
We find that the entropy associated with each one of the patterns is described
by exactly the same equations
as single pattern in the paramagnetic phase, seen in
section~\ref{sec_mattis_para}. We have thus the same behavior:
we need thus a number of measures that is proportional to the size of the
system and the exact form of the entropy is given by
Eqs.~(\ref{entrop12}--\ref{col4}).

\section[Two patterns, magnetized according to the first]{A particular case: a system with two patterns, magnetized according to the first}
\label{sec_seila1}

In last section, we saw that when our system does not visit any one of the
patterns the inference of each one of them is as hard as describing a system
composed of a single pattern 
in the paramagnetic phase. In this section, we will see if there
is any qualitative change in the situation 
where the system is magnetized according to one
pattern but we would like to infer a second, non-visited one.

We suppose that our system's Hamiltonian is given by
\[
H = - \beta \sum_{i<j} (\xi_i^1 \xi_j^1 + \xi_i^2 \xi_j^2) \sigma_i \sigma_j \, ,
\]
where the patterns $\xi_i^\mu = \pm 1$ are binary.
We suppose we did $L = \alpha N$ measures of the configuration of this system,
all of them while the system was magnetized
following the first pattern. Supposing that the first pattern is perfectly 
known (what is reasonable since our entropy on it goes with
$e^{- \gamma L} = e^{O(N)}$ as the inference of the first pattern is very
similar to the ferromagnetic case we saw previously), 
we study the entropy of the a posteriori distribution of the second pattern
\begin{equation}
S = - \sum_{\{\xi^2_i\}} P[\{\xi^2_i\}|\{\xi^1_i\},\{\sigma^l_i\}] \log P[\{\xi^2_i\}|\{\xi^1_i\},\{\sigma^l_i\}]\, .
\end{equation}
In the same way we did with the single pattern case in section \ref{sec_mattis},
we average this entropy with respect both to the first pattern and to the
measured configurations. The probability is given by

\begin{eqnarray}
P[\{\xi^2_i\}|\{\xi^1_i\},\{\sigma^l_i\}] &=& 
\frac{P[\{\sigma^l_i\}|\{\xi^1_i\},\{\xi^2_i\}] P[\{\xi^2_i\}]}{\mathcal{N}[\{\xi^1_i\},\{\sigma^l_i\}]} \, ,\\
P[\{\sigma^l_i\}|\{\xi^1_i\},\{\xi^2_i\}] &=&
\frac{1}{Z_\text{Hop}[\beta, \{\xi^1_i\},\{\xi^2_i\}]^L} 
\exp\left[\frac{\beta}{N}\sum_{\mu=1,2} \sum_{l=1}^L \sum_{i<j}
\sigma_i^l \sigma_j^l \xi_i^\mu \xi_j^\mu \right] \, ,  \nonumber\\
\end{eqnarray}
where the normalization of the probability $\mathcal{N}$ is given by
\begin{eqnarray}
\mathcal{N}[\{\xi^1\},\{\sigma^l\}] &=& \sum_{\{\xi^2\}} 
\frac{1}{Z_\text{Hop}[\beta, \{\xi^1\},\{\xi^2\}]^L}
\exp\left[
\frac{\beta}{N} \sum_{\mu=1,2} \sum_{l=1}^L \sum_{i<j} \sigma_i^l \sigma_j^l \xi_i^\mu \xi_j^\mu
\right] P[\xi^2] \, . \nonumber\\
\end{eqnarray}
As done in Section~\ref{sec_mattis}, we write our entropy as a 
derivative of a modified normalization $\tilde{N}$ given by
\begin{eqnarray}
\label{eqNtil}
\tilde{N}[\{\xi^1\},\{\sigma^l\}] &=& \sum_{\{\xi^2\}} 
\exp\Bigg[
\frac{\beta}{N} \sum_{\mu=1,2} \sum_{l=1}^L \sum_{i<j} \sigma_i^l \sigma_j^l \xi_i^\mu \xi_j^\mu
\nonumber\\
&&- \frac{\beta}{\tilde{\beta}} L 
\log Z_\text{Hop}[\tilde{\beta},\{\xi^1\},\{\sigma^l\}]
\Bigg] \, ,
\end{eqnarray}
so that we can easily deduce the entropy as
\begin{eqnarray}
\label{eqS00}
S &=& \left. \log \tilde{N} \right|_{\beta=\tilde{\beta}} 
- \left.\frac{\partial \log \tilde{N}}{\partial \beta}\right|_{\beta=\tilde{\beta}}  \, ,
\end{eqnarray}
where we have supposed $P[\xi^2]=2^{-N}$. We will now evaluate this quantity
explicitly.


Doing a calculation very similar to the derivation of Eq.~(\ref{logZ}), we
can write the partition function of the Hopfield model for
$p=2$ for a system magnetized in the first pattern, obtaining:
\begin{eqnarray}
\log Z_\text{Hop} &=&  - \frac{\beta N}{2} m^{*2} + N \log \left[ 2 \cosh 
(\beta m^* ) \right] 
 - \log \left[1 - \beta (1 - m^{*2})\right] \nonumber\\
&&+ \frac{1}{2} \frac{ \beta m^{*2}}{1 - \beta (1-m^{*2})}\left[ \frac{1}{\sqrt{N}} \sum_i \xi^1_i \xi^2_i\right]^2 \, ,
\end{eqnarray}
with $m^*$ given by $m^* = \tanh(\beta m^*)$.

\subsection{Evaluation of $\bm{\langle}$log \textit{\~N}$\bm\rangle$}

To evaluate $\moy{\log \tilde{N}}$, we use again the replica
trick, obtaining
\begin{equation}
\begin{split}
\moy{N^n} = &\frac{1}{2^N} \sum_{\{\xi^1\}} \frac{1}{2^N} \sum_{\{\tilde{\xi}^2\}}
\sum_{\{\sigma\}} \sum_{\{\xi^{2,\nu}\}} \exp \Bigg\{ \\
&\frac{\beta}{N} \sum_{\nu=1}^n \sum_{l=1}^L \sum_{i,j} \sigma_i^l \sigma_j^l (\xi_i^1 \xi_j^1+
\xi_i^{2,\nu} \xi_j^{2,\nu})
-  \frac{\beta}{\tilde{\beta}} L \sum_{\nu=1}^n
 \log Z_\text{Hop}\left[\tilde{\beta}, \{\xi^1\}, \{\xi^{2,\nu}\}\right]
 \\
&+ \tilde{\beta} \sum_{l=1}^L \sum_{i,j} \sigma_i^l \sigma_j^l (\xi_i^1 \xi_j^1+
\tilde{\xi}^2_i \tilde{\xi}^2_j)
- L  \log Z_\text{Hop}\left[\tilde{\beta}, \{\tilde{\xi}^1\}, \{\tilde{\xi}^2\}\right]
\Bigg\} \, ,
\end{split}
\label{eqnn}
\end{equation}
where we denote by $\tilde{\xi^2}$ the real second pattern and by 
$\xi^2$ the inferred one.

That expression can be simplified to (see Appendix~\ref{ap_entrohop})

\begin{eqnarray*}
\moy{N^n} &=& 
\sum_{\{\xi^1\}} \sum_{\{\tilde{\xi}^2\}}
\sum_{\{\xi^{2,\nu}\}}
\exp \left[
C_1 - \frac{L}{2} \frac{\tilde{\beta} \tilde{m^*}^2 u^2}{1 - \tilde{\beta}(1-\tilde{m^*}^2)}
 - \frac{L}{2} \frac{\beta \tilde{m^*}^2}{1 - \tilde{\beta}(1-\tilde{m^*}^2)}
\sum_\nu s_\nu^2
\right. \\
&&- \left.  \frac{L}{2} \log \det M + \frac{L}{2} A^t M^{-1} A
\right] \, ,
\end{eqnarray*}
where
\begin{eqnarray*}
u = \frac{1}{\sqrt{N}} \sum_i \xi_i^1 \tilde{\xi}_i^2 \, ,
& \quad \quad \quad &
t_\mu = \frac{1}{N} \sum_i \tilde{\xi}_i^2 \xi_i^{2,\mu} \, ,\\
s_\mu = \frac{1}{\sqrt{N}} \sum_i \xi_i^1 \xi_i^{2,\mu} \, ,&&
q_{\mu, \nu} = \frac{1}{N} \sum_i \xi_i^{2,\mu} \xi_i^{2,\nu} \, ,
\end{eqnarray*}
%
%
\[
M= \left(
\begin{array}{ccccccc}
d_1 & \tilde{u} & \tilde{s}_1 & \tilde{s}_2 & \tilde{s}_3 & \cdots & \tilde{s}_n\\
\tilde{u} & d_2 & \tilde{t}_1 & \tilde{t}_2 & \tilde{t}_3 & \cdots & \tilde{t}_n\\
\tilde{s}_1 & \tilde{t}_1 & d_3 & \tilde{q}_{1,2} & \tilde{q}_{1,3} & \cdots & \tilde{q}_{1,n}\\
\tilde{s}_2 & \tilde{t}_2 & \tilde{q}_{1,2} & d_3  & \tilde{q}_{2,3} & \cdots & \tilde{q}_{2,n} \\
\vdots \raisebox{-3ex}{} & \vdots & \vdots & &\multicolumn{2}{l}{\hfill \raisebox{.8ex}{.} \hfill . \hfill \raisebox{-.8ex}{.}} \hfill  & \vdots \\
\tilde{s}_n & \tilde{t}_n & \tilde{q}_{1,n} &   \multicolumn{2}{l}{\dotfill} & \tilde{q}_{n-1,n} & d3 \\
\end{array}
\right) \, ,
\]
\[
A = \left(
\begin{array}{cccccc}
\sqrt{N} (\tilde{m^*}-m^*)(\beta n + \tilde{\beta}) &
\tilde{m^*} \tilde{\beta} u &
\tilde{m^*} \beta s_1 &
\tilde{m^*} \beta s_2 &
\cdots &
\tilde{m^*} \beta s_n
\end{array}
\right) \, ,
\]
and
\begin{eqnarray*}
d_1 &=& (\beta n + \tilde{\beta}) \left[ 1 -(\beta n + \tilde{\beta})
(1-\tilde{m^*}^2)\right]\, ,  \\
d_2 &=& \tilde{\beta} \left[ 1 - \tilde{\beta}(1-\tilde{m^*}^2)\right] \, ,\\
d_3 &=& \beta \left[ 1 - \beta(1-\tilde{m^*}^2)\right] \, ,\\
\tilde{u}_\nu &=& -(1-\tilde{m^*}^2)\tilde{\beta}(\beta n+\tilde{\beta})u_\nu \, ,\\
\tilde{t}_\nu &=& - (1-\tilde{m^*}^2) \beta \tilde{\beta} t_\nu \, ,\\
\tilde{s}_\nu &=& - (1-\tilde{m^*}^2) \beta(\beta n+\tilde{\beta})s_\nu \, ,\\
\tilde{q}_{\nu,\sigma} &=& - (1-\tilde{m^*}^2) \beta^2 q_{\nu,\sigma} \, .\\
\end{eqnarray*}


Interestingly, the value of the entropy for the replica-symmetric case
of this system is exactly the same of section~\ref{sec_mattis_para} 
(Eqs.~(\ref{entrop12})--(\ref{col4})), but with a equivalent
inverse temperature of $\beta' = \beta (1-m^2)$. We show thus that the behavior
saw in the last section remains valid in this case.


\chapter{Conclusion}

In this work we have studied the general problem of extracting information
from the measured activity of interacting parts.
This problem is a recurring one in several domains, like biology
(with the study of neuron networks and proteins), physics, social science
and economics.

To solve such a problem, one needs to make a supposition about the underlying
system that has generated the data. Our results can thus be separated in two
main parts:
in the first, we supposed that our system is described by a generalized
Ising model, i.e., an Ising model with a Hamiltonian given by
$H(\{\sigma_i\}) = \sum_{i<j} J_{ij} \sigma_i \sigma_j + \sum_i h_i \sigma_i$.
In this case, the underlying information to be extracted are the couplings
$J_{ij}$ and external fields $h_i$ used in the Hamiltonian. A second part
of our results deals with the case where the underlying system is described by
the Hopfield model of $p$ patterns, with an Hamiltonian given by
$H(\{\sigma_i\}) =\sum_{i<j}\sum_{\mu=1}^p \xi_i^\mu \xi_j^\mu \sigma_i \sigma_j
+ \sum_i h_i \sigma_i$. In this case, the information to be obtained
are the values of the patterns $\{\xi_i^p\}$ and of the external fields
$h_i$.

In part~\ref{part2} of this thesis, we derived an explicit formula for
the couplings $J_{ij}$ and external fields $h_i$ as a function of the
magnetizations $m_i = \moy{\sigma_i}$ and connected correlations
$c_{ij} = \moy{\sigma_i \sigma_j} - m_i m_j$. That formula was obtained
through a small-correlation expansion and was evaluated up to order three
on the correlations. We developed also a general method through which one
could continue the expansion up to any desired order.

Unfortunately, the performance of our approximation up to order three degraded
very quickly when increasing the values of the correlations. To workaround this
limitation, we identified some terms of our expansion that, once grouped
together, corresponded exactly to the first terms of a mean-field approximation
already known in the literature. We could then replace these terms by the
full mean-field approximation to find a formula that was much more robust
against large values of correlation. Moreover, we could identify a second
set of terms easy to interpret: all the terms that involve
only two sites can be shown to represent the exact solution of a system
composed of only two spins. Thus, in the same way, we replaced these terms
by their corresponding sum. Finally, we obtained a fairly simple formula
that was very stable numerically and had a straightforward interpretation:
it unifies the contribution of a mean-field approximation (which works very
well in systems with several small couplings) with an independent-pair
approximation (which
performs well in systems with few, but strong, couplings). We verified that
our formula outperforms existing methods for the inverse Ising model.

A possible future extension of this work might be to find out if other
methods of solving the inverse Ising problem can be interpreted in the
framework of our expansion. For instance, the susceptibility propagation
method is exact on trees, and an interesting perspective would be to associate
this method with the exact sum of
all terms that are not zero in this case (in the diagrammatic notation
we introduced in chapter~\ref{chap_sums} it should correspond to 
diagrams that do \emph{not} contain loops). It might be interesting then
to add to our formula the exact sum of all those terms. Another promising
venue of research would be to understand what dominates the error of our
formula for systems with small couplings, beyond the loop summation we
discussed in
chapter~\ref{chap_sums}.

Unfortunately, our results have a few limitations. First of all, they cannot
work for a system that is magnetized in a ferromagnetic or glassy state or
any other system with high correlations. Secondly, while our
method of expanding on small-correlation is valid up to any given order,
the calculations are impractically long beyond order four.

In the last part this thesis we dealt with the inference of the patterns
of a Hopfield model.
This approach has several potential advantages: first of
all, the Hopfield model can be solved analytically, which should make our
calculations more precise and simpler. Moreover, for a fixed number of patterns
$p$, we have only $p N$ real values to infer while for the inverse
Ising problem we had all the $N(N-1)/2$ elements of the matrix $J_{ij}$.
Having a smaller number of degrees of freedom can potentially improve the
quality of our inference when the input data is noisy,
since it can avoid ``fitting the noise''.
Finally, this model is potentially valid also in the ferromagnetic
phase.

To infer the patterns from the data, we used a Bayesian approach: we looked for
the set of patterns that maximized the a posteriori probability. We found
an explicit result exact in the large system size limit for the patterns in
terms of the measured correlations and magnetizations. Since the Hopfield model
is a particular case of the generalized Ising model, we could compare
this formula with our previous results to show they correspond to the mean-field
approximation of part~\ref{part2}.
The formula corresponds also a the well-known method for extracting
patterns from data: the
Principal Component Analysis (\textit{PCA}). Thus, our calculations
provide a rigorous justification for this method.

To find a formula that provides a better inference than the mean-field formula,
we evaluated the first subleading correction to the patterns. We found a formula
that works well with synthetic data obtained by exact enumeration but quickly
falls apart for noisy input.

Besides finding an explicit formula for the patterns, we wanted to evaluate
how much data is needed for finding a precise estimation of the patterns.
For that, we computed the information-theoretical entropy of the inference
for a typical realization of the measurements. We find that if the system is
magnetized according to a given pattern, the quality of our estimation
grows exponentially with the number of measures. Interestingly, this number
does not depend of the size of the system in the limit of a very large system.
On the other hand, if we are looking for a pattern where none of the
measurements were magnetized, one needs a number of measures that is
proportional to the size of the system.

An interesting perspective for future work would be to try to apply the Hopfield
results for the data analysis of groups of homologous proteins, where both the
PCA and a Hopfield-like Hamiltonian have shown to yield interesting results.

A final remark is that our results can only be used to extract information
from data constituting independent samples from a Gibbs distribution, which
is not normally true for biological experiments.
More precisely, a fundamental premise of our derivations is that the
probability of $L$ meaured configurations is just the product of
$L$ Boltzmann weights.

Concerning experimental data,
neuron activity usually present strong temporal correlations. As the
free-energy landscape has potentially many minima, it might be that
the measured configurations correspond to a very particular subset of all the
states and thus the measures are not independent.
In the case of protein families, the situation is particularly bad: due to
their common evolutionary origin, there is a strong bias favoring sampling
proteins similar to their common ancestor. In future investigations, it would be
interesting to work around this problem or at least see to which extent it
interferes with the inference procedure.
A possibility would be to take into consideration two contributions to the
probability: a term associated to the fitness of the protein, similar to what
was done here, and a term that accounts for the evolutionary history of the
family.

\chapter*{Résumé détaillé}
Récemment, un grand nombre d'expériences en biologie qui génèrent une quantité
très importante de données ont vu le jour. Dans une partie considérable de ces
expériences, dont on peut citer les réseaux de neurones, l'analyse de données
consiste grosso modo d'identifier les corrélations entre les différentes parties
du système. Malheureusement, les corrélations en soi n'ont qu'une valeur
scientifique limité~: la plupart des propriétés intéressantes du système sont
décrites plutôt par l'interaction entre ses différentes parties. Le but de ce
travail est de créer des outils pour permettre de déterminer les interactions
entre les différentes parties d'un système en fonction de ses corrélations.

Dans ce résumé en langue française, on va commencer par une introduction
où on exposera des résultats classiques sur le principal
système qui a motivé ce travail~: les réseaux des neurones. Ensuite, on
va parler brièvement de la modélisation qu'on a choisi pour ce travail et des
résultats connus sur des systèmes similaires.

Dans une deuxième partie, on présentera un développement en petites corrélations
du problème d'Ising inverse.
Finalement, dans une dernière partie on traitera le problème d'Hopfield
inverse, ie., trouver les patterns du modèle à partir des champs et corrélations
locales.

\section*{Introduction}
Une des questions scientifiques plus importantes du 21ème siècle est la
compréhension du cerveau. Aujourd'hui, il est bien connu que la complexité
du cerveau est un produit de l'organisation des ses cellules (les neurones)
en des réseaux complexes.
Un neurone typique est composé de trois parties~: un corps cellulaire, qui
contient le noyau de la cellule, des dendrites, responsables pour la réception
des signaux des autres cellules et un axone, qui envoie des signaux à des
autres neurones (voir image). Les connexions entre neurones sont appelées
synapses et ont lieu typiquement entre un axone et un dendrite.

\begin{figure}[H]
\begin{center}
\includegraphics[width=0.85\columnwidth]{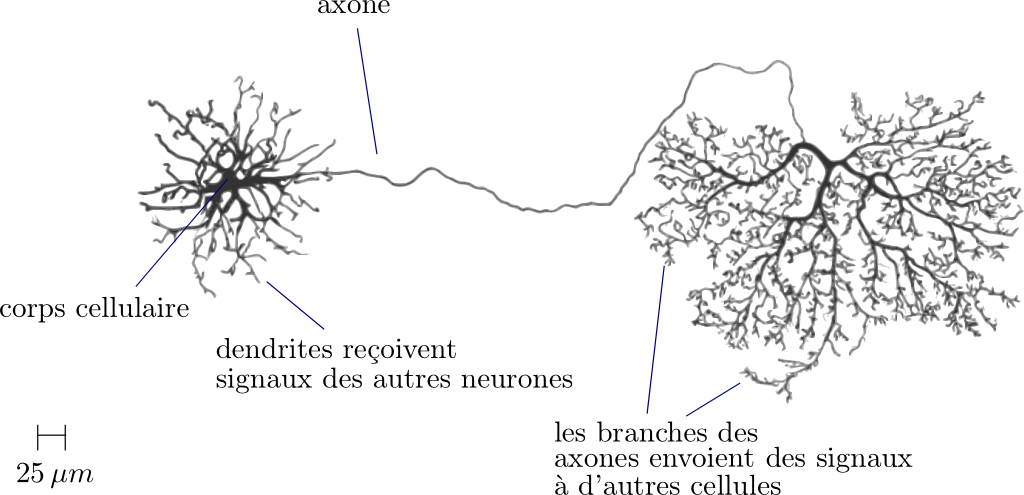}
\end{center}
\caption{Schéma d'un neurone \cite{alberts-cell}. Le diamètre du corps
cellulaire est typiquement de $10 \, \mu m$, pendant que la taille des 
dendrites et des axones varie considérablement avec la fonction du neurone.}
\label{schema_neu_fr}
\end{figure}

Comme toutes les cellules, les neurones possèdent une différence de potentiel
entre son cytoplasme et le milieu inter-cellulaire. Cette différence de
potentiel est contrôlée par des mécanismes de pompes d'ions, qui peuvent augmenter
ou diminuer ce potentiel. Quand la différence de potentiel d'un neurone
atteint un certain seuil, un mécanisme de \textit{feedback} active les pompes,
faisant le
potentiel croître rapidement jusqu'à environ 100 mV (en dépendant du type de
neurone), après quoi il atteint la saturation et décroît, en revenant au
potentiel de repos de la cellule. On appelle ce processus un \textit{spike}.

À chaque fois qu'un neurone émet un spike, son axone libère des
neurotransmisseurs dans les cellules à lesquelles il est connecté, en changent
leur différence de potentiel. Comme les synapses peuvent être
excitatrices ou inhibitrices, les modèles normalement définissent un poids pour
les synapses, avec la convention que un poids positive correspond à une
synapse qui augmentent le potentiel des neurones auquel elle est connecté
(et donc favorise les spikes) et un poids négatif au cas où la synapse décroît
ce potentiel.

Une nouvelle venue de recherche très prometteuse dans le domaine des
neurosciences a été le développement des techniques d'enregistrement
multi-neurones. Dans ces expériences, une matrice de micro-électrodes
(contant jusqu'à 250 électrodes) est mise en contact avec le tissu cérébral
et le potentiel à chaque électrode est mesuré pendant quelques heures. Un
procédé sophistiqué d'analyse de données permet alors d'identifier
la activité individuelle de chaque neurone en contact avec les
électrodes. La sortie typique de un enregistrement multi-neurones est montrée
dans la figure suivante.

\begin{figure}[H]
\begin{center}
\includegraphics[width=0.65\columnwidth]{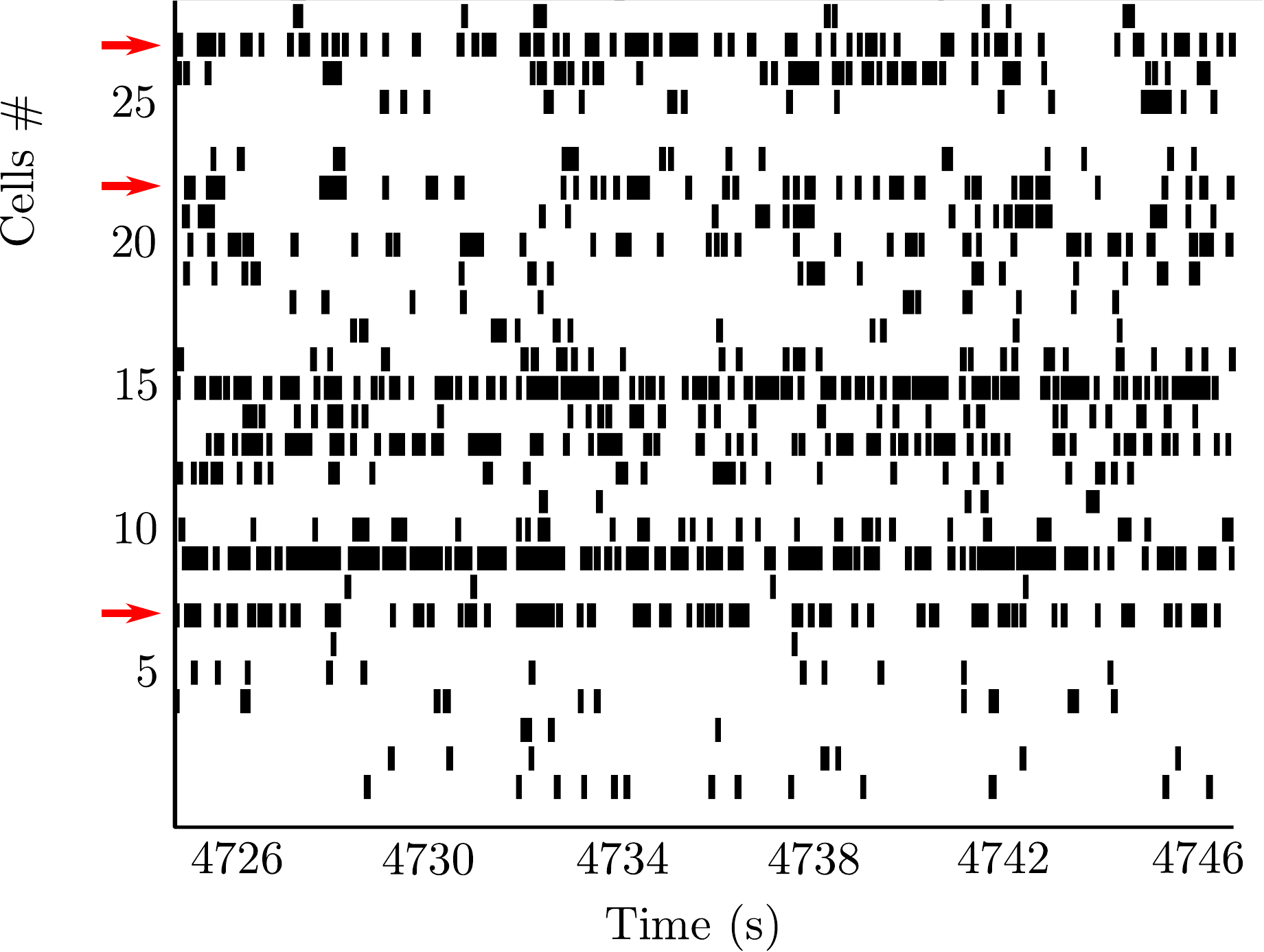}
\end{center}
\caption{Résultat typique d'une expérience d'enregistrement multi-électrodes\cite{peyrache}. Chaque ligne correspond à un seul neurone, pendant que les barres verticales correspondent à des spikes.}
\label{fig_spike_fr}
\end{figure}

En principe, on pourrait trouver les synapses à partir des enregistrements
multi-électrodes, mais extraire cet information n'est pas trivial. Naïvement,
on aurait envie de dire que si l'activité de deux neurones sont corrélées ils
sont connectés par une synapse, mais considérons trois neurones dont l'activité
est corrélé~: on voit aisément que toutes les deux configurations montrées dans
l'image suivante peut rendre compte des ces corrélations.

\begin{figure}[H]
\begin{center}
\includegraphics[width=0.5\columnwidth]{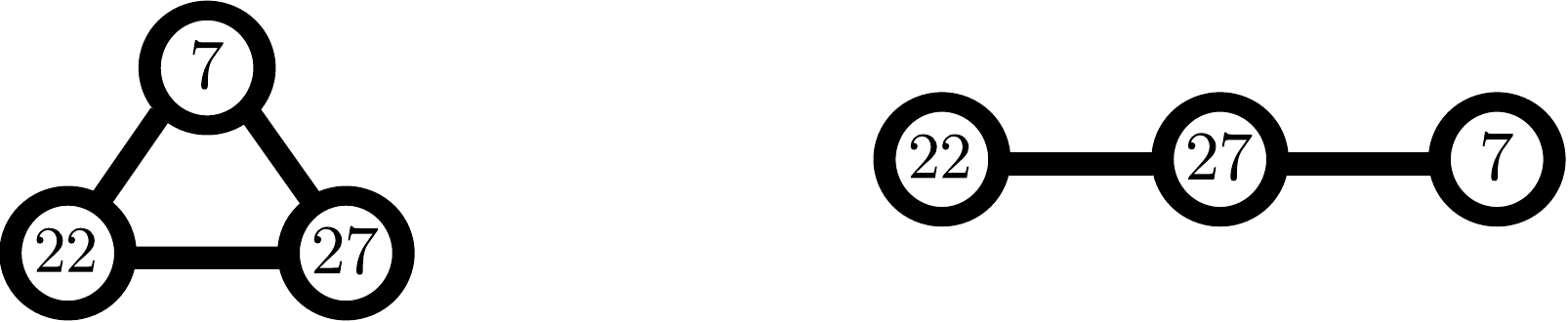}
\end{center}
\caption{Deux configurations possibles pour trois neurones corrélés.}
\label{fig_connec_fr}
\end{figure}

Pour pouvoir donner une contribution à ce problème, il est nécessaire d'abord
de choisir un modèle pour les réseaux de neurones. Le modèle d'intéresse pour
cette thèse est la machine de Boltzmann \cite{mcculloch1943logical}.
Dans ce modèle, on modélise l'état
d'un neurone par une variable binaire~: $\sigma=+1$ si il est en train d'émettre
un spike, $\sigma=-1$ sinon. Le réseau de N neurones est alors décrit par
un vecteur $\{\sigma_1, \ldots, \sigma_N\}$.
Pour simplifier, on ignore totalement la dynamique
du système et le modèle décrit seulement la probabilité
$P(\{\sigma_1,\ldots,\sigma_N\})$
de trouver le système dans un état $\{\sigma_1,\ldots,\sigma_N\}$, qui est
donnée par le poids de Boltzmann de un modèle d'Ising généralisé~:
\begin{eqnarray}
P(\{\sigma_1,...,\sigma_N\}) &=& 
\frac{1}{Z} e^{- \beta H(\{\sigma_1,...,\sigma_N\})}\, ,
\end{eqnarray}
avec
\begin{eqnarray}
Z &=& \sum_{\{\sigma\}} e^{-\beta H(\{\sigma_1,...,\sigma_N\})}\, ,
\end{eqnarray}
où $Z$ est la fonction de partition du modèle, $\beta$ est un paramètre du
modèle, qui dans le contexte de spins correspond à la température inverse et
on introduit la notation
\begin{equation}
\sum_{\{\sigma\}} \equiv \sum_{\sigma_1=\pm 1}\, \sum_{\sigma_2=\pm 1}\,\, \cdots\,\, \sum_{\sigma_N=\pm 1} \, .
\end{equation}

Le Hamiltonien doit prendre en compte les synapses entre les neurones et qu'il
doit recevoir une certaine quantité minimum des signaux pour émettre un
spike. L'expression la plus utilisée est
\begin{eqnarray}
\label{eq_hamil_fr}
H(\{\sigma_1,...,\sigma_N\}) &=&
- \frac{1}{2 N}\sum_{i,j} J_{ij} \sigma_i \sigma_j 
- \sum_i h_i \sigma_i\, ,
\end{eqnarray}
où $J_{ij}$ correspond au poids des synapses et $h_i$ est un terme qui modélise
le seuil de spike comme un champs qui attire le neurone vers son état de repos.
Ce modèle est intéressante pour ce travail pour deux raisons~: d'abord, il
permet de mettre en \oe{uvre} directement les outils développés dans le contexte
de la mécanique statistique et des systèmes désordonnés. En outre, ce modèle
est émerge naturellement quand on cherche un modèle qui peut rendre compte
d'un ensemble de moyennes $\moy{\sigma_i}$ et de corrélations
$\moy{\sigma_i \sigma_j}$.
C'est important à noter que dans ce modèle les couplages sont symétriques,
ie, $J_{ij} = J_{ji}$, ce qui n'est pas forcement vrai dans des systèmes
biologiques.

Le modèle d'Ising généralisé contient comme cas particulier des différents
modèles classiques de la littérature. On peut citer le modèle d'Ising ordinaire,
le modèle de Sherrington-Kirkpatrick et, d'un intérêt particulier pour cette
thèse, le modèle de Hopfield. Dans ce modèle, le Hamiltonien est donné par
\begin{equation}
\label{eq_hop_fr}
H =
-  \frac{1}{2N} \sum_{\alpha=1}^p \left(\sum_i \xi_i^\alpha \sigma_i \right)^2
\, ,
\end{equation}
où $\xi_i^\alpha$ sont des valeurs réels et en développant le carré on voit
bien qu'il correspond au modèle d'Ising généralisé par
\begin{equation}
J_{ij} = \sum_{\mu=1}^p \xi_i^\mu \xi_j^\mu\, .
\label{eqeqeq_fr}
\end{equation}

L'idée derrière le modèle d'Hopfield est de rendre compte d'un système qui
garde un nombre $p$ de mémoires et peut les retrouver à partir d'un état
initial similaire à la mémoire recherché. Effectivement, dans
l'équation~(\ref{eq_hop_fr}) on voit bien que si les mémoires $\xi^p$ sont
à peu près orthogonaux les unes avec les autres, l'état
$\sigma_i = \xi_i^\alpha$ correspond à un minimum local de l'énergie. Un
résultat classique (voir \cite{amit1}) est que dans certaines conditions,
ces états sont aussi des minimums locales de l'énergie libre, ce qui permet
effectivement de dire que le système peut retrouver ces mémoires.

Déterminer le comportement d'un système décrit par un modèle d'Ising généralisé
est normalement très difficile et il n'existe pas de solution générale pour
ce problème. On peut, par contre, trouver des solutions approchés dans un
certain limite de validité. Un cas particulier que nous intéresse est le
développement en petites couplages introduite par Thouless, Anderson and Palmer
(dites équations TAP) \cite{TAP1977}. Le résultat qu'ils ont obtenu dit que
pour un système avec un Hamiltonien donné par
\begin{equation}
H = - \sum_{i<j} J_{ij} \sigma_i \sigma_j
\end{equation}
l'énergie libre est donnée par
\begin{eqnarray}
\log Z &=& -\sum_i \frac{1+m_i}{2} \log\left(\frac{1+m_i}{2}\right)
-\sum_i \frac{1-m_i}{2} \log\left(\frac{1-m_i}{2}\right)
\nonumber\\
&&+ \sum_{i<j} J_{ij} m_i m_j + \frac{1}{2} \sum_{i<j} J_{ij}^2(1-m_i^2)(1-m_j^2)
+ O(J^3) \, ,
\label{eq_TAP_GY_fr}
\end{eqnarray}
où $m_i = \moy{\sigma_i}$ est la magnétisation locale du système, qui selon
les équations TAP obéissent
\begin{eqnarray}
\label{eq_dificil_fr}
\tanh^{-1} m_i &=& \sum_{j\, (\neq i)} J_{ij} m_j
- m_i \sum_{j\,(\neq i)} J_{ij}^2 (1-m_j^2) \, .
\end{eqnarray}

Malheureusement, dans le cadre des expériences avec les neurones, on n'a
pas besoin d'avoir une méthode pour trouver le comportement d'un système en
fonction des paramètres, mais faire le contraire~: trouver les
paramètres qui rendent compte au mieux des résultats observés. On parle alors de
``problèmes inverses''.

Travailler avec des problèmes inverses entraîne deux complications
supplémentaires~: d'abord, même pour les cas où on peut résoudre le problème
directe, trouver les paramètres qui décrivent les données est normalement
un problème difficile. En outre, donner une signification mathématique à la
expression ``mieux décrit le système'' est aussi un point délicat. Une cas
possible est que plusieurs différents choix de paramètres peuvent décrire
exactement les mesures. La situation contraire, où il n'existe pas d'ensemble
de paramètres qui rendent compte des données (à cause des erreurs
expérimentaux), est tout aussi possible.

Pour rendre compte de ce problème, il est utile de postuler que les paramètres
qui décrivent le problème suivent eux aussi une loi statistique. On peut alors
appliquer le théorème de Bayes qui dit que la probabilité que les paramètres
valent $\{\lambda_i\}$ en fonction des mesures $\{X_i\}$ est donnée par
\begin{equation}
  \label{eq_bayes_fr}
  P(\{\lambda_i\}|\{X_i\}) = \frac{  P(\{X_i\}|\{\lambda_i\}) P_0(\{\lambda_i\})}{P(\{X_i\})} \, ,
\end{equation}
où $P_0(\{\lambda_i\})$ est la probabilité \textit{a priori} des paramètres et
$P(\{X_i\})$ est la probabilité marginale de $\{X_i\}$, qui peut aussi être
interprétée comme une normalisation de la probabilité
$P(\{\lambda_i\}|\{X_i\})$~:
\begin{equation}
P(\{X_i\}) = \sum_{\{\lambda_i\}}  P(\{X_i\}|\{\lambda_i\}) P_0(\{\lambda_i\})
\, .
\label{eq_bayesnorm_fr}
\end{equation}
Ce théorème permet de donner une définition précise du ``meilleur'' ensemble
de paramètres pour décrire le système comme ceux qui maximisent
$P(\{\lambda_i\}|\{X_i\})$. À part cela, s'il existe plus d'un ensemble de
paramètres qui rendent compte des données, un choix judicieux de
$P_0(\{\lambda_i\})$ permet de ``choisir'' parmi ces solutions et de rendre
le problème bien définit.

\section*{Le problème d'Ising inverse}

On considère le modèle d'Ising généralisé avec un Hamiltonien donné par
\begin{equation}
H(\{\sigma_i\}) = -\sum_{i<j} J_{ij}\, \sigma_i\, \sigma_j - \sum_i h_i \, \sigma_i \, ,
\end{equation}
qui définit alors une probabilité sur les états suivant la loi de Boltzmann
\begin{equation}
P(\{\sigma_i\}) = \frac{1}{Z} e^{-H(\{\sigma_i\})} \, ,
\end{equation}
où $Z$ est la fonction de partition, donnée par
\begin{equation}
Z = \sum_{\{\sigma_i\}} e^{-H(\{\sigma_i\})} \, .
\end{equation}
Normalement, quand on étudie un tel système on cherche à déterminer les
magnétisations locales $m_i = \moy{\sigma_i}$ et les corrélations à deux sites
$c_{ij}= \moy{\sigma_i \sigma_j} - m_i m_j$ en fonction des couplages $\{J_{ij}\}$ et des champs
$\{h_i\}$. Par contre, on parle de problème d'Ising inverse quand on cherche à
déterminer $J_{ij}$ et $h_i$ en fonction des corrélations et magnétisations.

Notre point de départ pour résoudre ce problème est l'entropie de Shannon du
problème\footnote{Pour plus de détails, voir Chapitre~\ref{ch_inverse}
et~\ref{sec_expansion} de la version en anglais.}
\begin{equation*}
S(\{J_{ij}\},\{\lambda_{i}\};\{m_i\},\{ c_{ij}\} ) =
\log Z(\{J_{ij}\}, \{h_i\}) -
\sum_{i<j} J_{ij} \, (c_{ij} +m _i \, m_j)- \sum_i h_i \, m_i  \, ,\\
\end{equation*}
\vspace{-0.5cm}
\begin{equation}
  \label{eqG2_fr}
  \begin{split}
\phantom{xxxx}&= \log \sum_{\{\sigma_i\}} \exp\left\{
\sum_{i<j} J_{ij} \left[ \sigma_i\sigma_j  -
  c_{ij} -m_i m_j\right] + \sum_i h_i (\sigma_i - m_i)\right\}\, ,\\ 
  &= \log \sum_{\{\sigma_i\}} \exp\left\{
\sum_{i<j} J_{ij} \left[ (\sigma_i - m_i)(\sigma_j - m_j) -
  c_{ij}\right] + \sum_i \lambda_i (\sigma_i - m_i)\right\} \, ,
\end{split}
\end{equation}
où on a introduit des nouveaux champs externes $\lambda_i$ qui sont liés
avec le vrai champs $h_i$ par $\lambda _i= h_i + \sum_j J_{ij} \, m_j$. On voit
aisément calculant $\partial S/\partial J_{ij}$ et $\partial S/\partial h_i$
qui l'ensemble des $J_{ij}^*$ et $\lambda_i^*$ qui reproduisent les
magnétisations $m_i$ et corrélations $c_{ij}$ est celui qui \emph{minimise} $S$.
On va aussi s'intéresser à la valeur de $S$ au minimum
\begin{equation}
\label{aa_fr}
S(\{m_i\},\{ c_{ij}\} ) = \min_{\{J_{ij}\}} \min_{\{\lambda_i\}}
S(\{J_{ij}\},\{\lambda_{i}\};\{m_i\},\{ c_{ij}\} ) \, ,
\end{equation}
car à partir de cet expression il est possible d'extraire les valeurs de
$J_{ij}^*$ et $\lambda_i^*$ utilisant
\begin{equation}
  \frac{\partial S(\{m_i\},\{\beta \;c_{ij}\} )}
       {\partial c_{ij}} = -\beta J_{ij}^*(\beta) \, ,
       \label{eq_derivcij_fr}
\end{equation}
et
\begin{equation}
   \frac{\partial S(\{m_i\},\{\beta \;c_{ij}\} )}
        {\partial m_{i}} = -\lambda_{i}^*(\beta) \, .
       \label{eq_derivmi_fr}
\end{equation}

Comme le problème d'Ising inverse est très difficile à résoudre en toute
généralité, on va chercher une expansion pour des valeurs petites des
corrélations. On les multiplie alors toutes pour un paramètre $\beta$ qui
on introduit, de façon à qu'un expansion en série autour de $\beta=0$ correspond
à une série en petites corrélations. L'équation~(\ref{eqG2_fr}) s'écrit alors
\begin{multline}
S(\{J_{ij}\},\{\lambda_{i}\};\{m_i\},\{\beta\, c_{ij}\} ) =\\
=   \log \sum_{\{\sigma_i\}} \exp\left\{
\sum_{i<j} J_{ij} \left[ (\sigma_i - m_i)(\sigma_j - m_j) -
  \beta c_{ij}\right] + \sum_i \lambda_i (\sigma_i - m_i)\right\} \, .
\end{multline}
On cherche maintenant à trouver une expansion de $S$ au minimum
(voir Eq.~(\ref{aa_fr})) pour $\beta$ petit~:
\begin{equation}
  S(\{m_i\},\{\beta \, c_{ij}\}) = S^0 + \beta S^1 + \beta^2 S^2 + \ldots\, ,
\end{equation}
d'où on pourra extraire aussi des séries pour $J_{ij}^*$ et $\lambda_i^*$
utilisant l'éqs.~(\ref{eq_derivcij_fr}) et~(\ref{eq_derivmi_fr}).

La détermination du premier terme de l'expansion de $S$ est trivial, car quand
$\beta=0$ on a des spins décorrélés (qui correspond à des spins indépendants)~:
\begin{equation}
  \label{eq_s0_fr}
  S^0 =
  -\sum_i \left[\frac{1+m_i}{2}\ln\frac{1+m_i}{2} + 
\frac{1-m_i}{2}\ln\frac{1-m_i}{2}\right]\, .
\end{equation}

Pour trouver les termes non-triviaux de l'entropie, on procède de la manière
suivante~: d'abord, on définie un potentiel $U$ sur les configurations de spins
par (noter le nouveau terme à la fin)
\begin{equation}
\begin{split}
  U (\{\sigma_i\})&= 
\sum_{i<j} J_{ij}^*(\beta) \left[ (\sigma_i - m_i)(\sigma_j - m_j) -
  \beta\; c_{ij}\right]
  + \sum_i \lambda_i^*(\beta) (\sigma_i - m_i)
  \\
  &
  + \sum_{i<j} c_{ij} \int_0^\beta \di \beta' J_{ij}^*(\beta') \, ,
\end{split}
  \label{eqU_fr}
\end{equation}
et une nouvelle entropie (à comparer avec éq.~(\ref{eqG2_fr}))
\begin{equation}  \label{eeerr_fr}
\tilde S(\{m_i\},\{c_{ij}\}, \beta ) = 
\log \sum_{\{\sigma_i\}} e^{U(\{\sigma_i\})} \ .
\end{equation}
Notez que la valeur de $U$ dépends de la valeur des couplages $J_{ij}^*(\beta')$
à tous les valeurs de $\beta' < \beta$ et pas seulement pour la valeur de
$\beta$ pour laquelle on veux calculer $U$. $S$ et $\tilde{S}$ se relient
par
\begin{equation}
S(\{m_i\},\{c_{ij}\}, \beta ) = \tilde S(\{m_i\},\{c_{ij}\}, \beta ) - \sum_{i<j} c_{ij} \int_0^\beta \di \beta' J_{ij}^*(\beta')
\label{realG_fr} \ .
\end{equation}

L'expression de $\tilde{S}$ a été choisi de sorte qu'elle soit indépendante
de $\beta$, comme un petit calcul permet de le vérifier
\begin{equation}
\label{gconst_fr}
\frac{d\tilde S}{d\beta}= \frac{\partial \tilde{S}}{\partial \beta} =- \sum_{i<j}  c_{ij}\; J_{ij}^*(\beta) + 
\sum_{i<j} c_{ij} \; J_{ij}^* (\beta)=0 \ .
\end{equation}
Notez que cela est valable pour toute valeur de $\beta$, d'où $\tilde{S}$ est
constante et égal à sa valeur en $\beta=0$, $S^0$, donnée par
l'éq.~(\ref{eq_s0_fr}).

On utilise alors le fait que $\tilde{S}$ est indépendant de $\beta$ pour écrire
des équations d'auto-consistence pour les dérivés de $S$. Par exemple, pour
déterminer $S^1$, on écrit
\begin{eqnarray}
S^1 &=& \left.  \derivb{S}\right|_0 =
\left.\derivb{\tilde{S}}\right|_0 - \sum_{i<j} c_{ij} J_{ij}^*(0) \, ,\nonumber\\
&=& 0 \, ,
\end{eqnarray}
où on a utilisé que $J_{ij}^*(0)=0$ qui découle du fait que à
température nulle les spins sont découplés.

Pour des ordres plus élevés,
on doit procéder ordre par ordre. Pour trouver $S^k$ en ayant déjà calculé
$S^p$ pour tout $p<k$,
on démarre avec l'équation $\left. \derivmb{\tilde S}{k}\right|_0 = 0$, égalité
conséquence de l'éq.~(\ref{gconst_fr}). Après un calcul explicité de la dérivé,
on se retrouvera avec une équation du type
\begin{equation} 
\left. \derivmb{\tilde S}{k}\right|_0 =- S^k
+ Q_k  = 0\, ,
\label{anyorder_fr}
\end{equation}
où $Q_k$ est une expression qui dépend des magnétisations, corrélations et
des dérivés à l'ordre $p < k-2$ en zéro des couplages $J_{ij}^*$ et des
champs $\lambda_i^*$. En utilisant les éqs.~(\ref{eq_derivcij_fr})
et~(\ref{eq_derivmi_fr}), on peut trouver ces dérivés des couplages et des
champs en fonction des valeurs déjà connus de $S^p$ pour $p<k$ et avoir
une expression explicite pour $Q_k$ et donc pour $S^k$. Le résultat final
pour $S$ est donc
\begin{eqnarray}
S &=& 
-\sum_i \left[\frac{1+m_i}{2}\ln\frac{1+m_i}{2} + 
\frac{1-m_i}{2}\ln\frac{1-m_i}{2}\right]
+\frac{2}{3} \beta^3 \sum_{i<j}K_{ij}^3 m_i m_j L_i L_j
\nonumber\\
&&
+ \frac{\beta^4}{6} \sum_{i<j} K_{ij}^4 \left[1 - 3m_i^2 -3m_j^2 - 3 m_i^2m_j^2\right] L_i L_j
\nonumber\\
&&
-\frac{\beta^2}{2} \sum_{i<j} K_{ij}^2 L_iL_j
+ \beta^3 \sum_{i<j<k} K_{ij} K_{jk}K_{ki} L_i L_j L_k\nonumber\\
&&- \frac{\beta^4}{8} \sum_{i,j,k,l} K_{ij} K_{jk} K_{kl} K_{li}  L_i L_j L_k L_l
\nonumber\\
&&+ O(\beta^5) \, .
\label{eqG_fr}
\end{eqnarray}
Les couplages sont données par
\begin{eqnarray}
J_{ij}^*(\{c_{kl}\}, \{m_i\},\beta) &=&
\beta K_{ij} -2 \beta^2 m_i m_j K_{ij}^2
- \beta^2 \sum_k K_{jk}K_{ki} L_k\nonumber\\
&&+ \frac{1}{3} \beta^3 K_{ij}^3 \left[1 + 3m_i^2 +3m_j^2 + 9 m_i^2m_j^2\right]
\nonumber\\
&&
+ \beta^3 \sum_{\substack{k \\ (\neq i,\,\neq j)}} K_{ij}
(K_{jk}^2 L_j +K_{ki}^2 L_i)  L_k \nonumber\\
\label{jij_fr}
&&+ \beta^3 \sum_{\substack{k, l \\ (k\neq i,\, l\neq j)}} K_{jk} K_{kl} K_{li}L_k L_l
+
 O(\beta^4) \, ,
\end{eqnarray}  
et le champs $h$ par
\begin{eqnarray}
h_l^*(\{c_{ij}\}, \{m_i\},\beta) &=&
\frac{1}{2} \ln \left( \frac{1+m_l}{1-m_l}\right)
-\sum_j J_{lj}^* m_j
+ \beta^2 \sum_{j (\neq l)} K_{lj}^2 m_l L_j
\nonumber\\
&&-\frac{2}{3} \beta^3 (1+3m_l^2) \sum_{j (\neq l)} K_{lj}^3 m_j L_j
-2 \beta^3 m_l \sum_{j<k} K_{lj} K_{jk} K_{kl} L_j L_k \nonumber\\
&&+ 2 \beta^4 m_l \sum_{i<j} \sum_k K_{ik} K_{kj} K_{jl} K_{li} L_i L_j L_k\nonumber\\
&&+
\beta^4 m_l \sum_{j} K_{lj}^4 L_j \left[1 + m_l^2 + 3 m_j^2 +3m_l^2m_j^2\right]
\nonumber\\
&&+ \beta^4 m_l \sum_{i\, (\neq l)} \sum_j K_{ij}^2 K_{jl}^2 L_i L_j^2 
+ O(\beta^5) \, .
\label{eq_hi_fr}
\end{eqnarray}
Dans ces expressions on a utilisé les notations
\begin{equation}
L_i = \moy{\left(\sigma_i - m_i\right)^2}_0 = 1-m_i^2 \, ,
\end{equation}
qui est la déviation standard d'un spin indépendant $m_i$ et
\begin{equation}
K_{ij} = \delta_{ij} \frac{\moy{\left(\sigma_i - m_i\right)\left(\sigma_j - m_j\right)}_0}
{\moy{\left(\sigma_i - m_i\right)^2}_0\moy{\left(\sigma_j - m_j\right)^2}_0} =
\delta_{ij} \frac{c_{ij}}{L_i L_j} \, ,
\end{equation}
où $\delta_{ij}$ est le symbole de Kronecker.

Malheureusement, on a vérifié que la qualité d'inférence obtenue par cette 
formule se dégrade très rapidement quand on sort de son limite de validité
$c_{ij} \ll 1$. Pour améliorer son stabilité, on a remarqué que les trois
derniers termes de l'éq.~(\ref{eqG_fr}) peuvent s'écrire de la forme d'une
série alterné
\begin{equation}
-\frac{\beta^2}{4} \Tr(M^2) + \frac{\beta^3}{6} \Tr(M^3) -
\frac{\beta^4}{8} \Tr(M^4) \, ,
\end{equation}
où $M$ est la matrice définit par $M_{ij} = K_{ij} \sqrt{L_i L_j}$.
Comme $K_{ii}=0$, on a que $\Tr{M}=0$, donc les trois derniers termes
s'écrivent
\begin{eqnarray}
S^\text{loop} 
&=&
\frac{1}{2} \Tr\left(\beta M - \frac{\beta^2}{2} M^2 + \frac{\beta^3}{3} M^3 -
\frac{\beta^4}{4} M^4\right) 
\nonumber \\
&=&  \Tr \left[\log(1+\beta M)\right] + O(\beta^5)
\nonumber \\
&=& \log\left[ \det(1+\beta M)\right] + O(\beta^5)
\label{eq_111_fr}
\end{eqnarray}

Cette expression peut aussi être retrouvée comme une conséquence de 
l'éq.~(\ref{eq_dificil_fr}), ce qui montre qu'elle correspond à une 
approximation du type ``champs moyen'' de l'entropie. En outre, si on remplace
les trois derniers termes de l'éq.~(\ref{eqG_fr}) par cette expression, on
voit une nette amélioration de la stabilité de l'inférence. Cela s'explique
car très probablement les termes d'ordre supérieure à $O(\beta^5)$ 
du développement de $\log[\det(1+\beta M)]$ sont
contenus dans l'expansion de $S$, et les développer en série correspond à une
série alternée, divergente pour des valeurs modérément grands de $c_{ij}$.

Comme l'expression de champs moyen a pu améliorer considérablement la 
stabilité de l'expansion, on a cherché à intégrer une autre approximation 
possible du problème~: l'approximation de pairs indépendants, dans laquelle on
examine chaque pair de sites $i$ et $j$ comme un système composé de juste deux
spins. Utilisant cette expression et l'approximation de champs moyen décrite
dans le paragraphe précédant, on obtient
\begin{eqnarray}
S^\text{2-spin + loop} &=& 
\sum_i S^\text{1-spin}_i
+ \sum_{i<j} \left[S_{ij}^\text{2-spin}-S^\text{1-spin}_i-S^\text{1-spin}_j
  \right]
\nonumber\\
&&
+S^\text{loop}
-\frac{1}{2} \sum_{i<j}\log(1-K_{ij}^2 L_i L_j) \, .
\label{eq_s_2sp_loop_fr}
\end{eqnarray}
où
\begin{eqnarray}
S^\text{2-spin}_{ij} &=& S^\text{1-spin}_i+S^\text{1-spin}_j
\nonumber\\
&&
+ \frac{1}{4}
\log\left[1+\frac{c_{ij}}{(1-m_i)(1-m_j)}\right][c_{ij} + (1-m_i)(1-m_j)]
\nonumber\\
&&
+ \frac{1}{4}
\log\left[1-\frac{c_{ij}}{(1-m_i)(1+m_j)}\right][c_{ij} - (1-m_i)(1+m_j)]
\nonumber\\
&&
+ \frac{1}{4}
\log\left[1-\frac{c_{ij}}{(1+m_i)(1-m_j)}\right][c_{ij} - (1+m_i)(1-m_j)]
\nonumber\\
&&
+ \frac{1}{4}
\log\left[1+\frac{c_{ij}}{(1+m_i)(1+m_j)}\right][c_{ij} + (1+m_i)(1+m_j)]\,,
\label{eq_sss_fr}
\end{eqnarray}
et
\begin{eqnarray}
  S^\text{1-spin}_i &=& - \left[\frac{1+m_i}{2}\ln\frac{1+m_i}{2} + 
\frac{1-m_i}{2}\ln\frac{1-m_i}{2}\right] \,.
\end{eqnarray}

Cette expression est équivalente à l'éq.~(\ref{eqG_fr}) (ils ne diffèrent 
que en termes d'ordre $O(\beta^5)$ ou plus), mais elle est beaucoup plus
stable numériquement. La formule pour $J_{ij}^*$ qui se déduit de cette
expression est
\begin{eqnarray}
J_{ij}^{*(\text{2-spin+loop})} &=& J_{ij}^{*\text{loop}}
+ J_{ij}^{*\text{2-spin}}- \frac{K_{ij}}{1-K_{ij}^2 L_i L_j} \, ,
\label{eq_2spiloop_fr}
\end{eqnarray}
où
\begin{eqnarray}
 J_{ij}^{*(\text{2-spin})}
&=&
\phantom{+}\frac{1}{4}\ln\left[ 1 + K_{ij}(1+m_i)(1+m_j) \right]\nonumber\\
&&+ \frac{1}{4}\ln\left[ 1 + K_{ij} (1-m_i)(1-m_j) \right] \nonumber\\
&&-\frac{1}{4}\ln\left[ 1 - K_{ij}(1-m_i)(1+m_j) \right]  \nonumber\\
&&-\frac{1}{4}\ln\left[ 1 - K_{ij}(1+m_i)(1-m_j) \right] \label{2spi_fr}
\end{eqnarray}
et
\begin{eqnarray}
  \label{eq_jijloop_fr}
J_{ij}^\text{loop}
&=& \frac{1}{\sqrt{(1-m_i^2)(1-m_j^2)}} \left[M (M+1)^{-1}\right]_{ij} \, .
\end{eqnarray}

\section*{Inférence d'un modèle de Hopfield}

Dans la partie précédente, on s'est intéressé à un modèle d'Ising généralisé
sans préciser la forme des couplages. Maintenant, on s'intéresse au cas 
particulier d'un modèle de Hopfield. Il y a plusieurs avantages d'utiliser
ce modèle~:
d'abord, il existe d'expériences de enregistrement multi-neurones où on
ne cherche pas à trouver des couplages entre les neurones mais d'en extraire des
patterns. Deuxièmement, on s'attend que diminuer le nombre de dégrées de
liberté du problème rends la procédure d'inférence plus stable. Finalement, le
modèle de Hopfield peut être résolu analytiquement, d'où on espère
avoir un meilleur contrôle des erreurs d'inférence.

Notez qu'en principe on pourrais procéder par d'abord inférer la matrice
$\{J_{ij}\}$ des couplages pour ensuite la diagonaliser pour trouver un ensemble
de patterns. Le problème de cette approche est qu'elle n'est pas optimale
d'un point de vue Bayesian~: les patterns trouvés ne seront pas forcement ceux
qui maximisent la probabilité a posteriori. Cela est particulièrement
problématique quand la supposition de que le système qu'on étude est décrit par
un modèle de Hopfield est juste une approximation.

Dans cette partie on va avoir deux buts~: premièrement, on va chercher une
formule permettant d'inférer les patterns en fonction des données. Ensuite,
on va s'intéresser à estimer le nombre de fois qu'on doit mesurer
le système pour avoir une bonne inférence.

Premièrement, on remarque que trouver les patterns qui mieux rendent compte
d'une séquence de données est un problème mal posé, car il y a plusieurs
ensembles de patterns qui peuvent décrire le même système. Supposons pour
exemple le cas de deux patterns ($p=2$)~:

\begin{equation} 
H = N \left( \frac{1}{N}\sum_i \xi_i^1 \sigma_i \right)^2 + N \left( \frac{1}{N}\sum_i \xi_i^2 \sigma_i \right)^2 \,.
\end{equation} 
Si l'on définit un nouveau ensemble de patterns donnée par
$\tilde{\xi_i^1} =  \xi_i^1 \cos \theta + \xi_i^2 \sin \theta$ 
et
$\tilde{\xi_i^2} =  -\xi_i^1 \sin \theta + \xi_i^2 \cos \theta$, le nouveau
Hamiltonien est
\begin{eqnarray} 
\tilde{H} &=& N \left[ \frac{1}{N}\sum_i (\xi_i^1 \cos \theta + \xi_i^2 \sin \theta)
 \sigma_i \right]^2 
\nonumber\\
&&+ N \left[ \frac{1}{N}\sum_i
 ( -\xi_i^1 \sin \theta + \xi_i^2 \cos \theta) \sigma_i \right]^2 \, ,
\\
&=& H \, .
\end{eqnarray}
En générale, si on a $p$ patterns, faire une rotation des patterns en $p$
dimensions ne change pas le Hamiltonien du système. On dit que le système
possède une invariance de gauge. Alors, pour rendre le problème d'inférence
bien posée on doit soit lever cette dégénérescence en additionnant des
contraintes pour lever les $p(p-1)/2$ dégrées de liberté, soit additionner un
prior $P_0$ dans notre probabilité Bayesienne. Pour des raisons techniques,
on va se concentrer sur la première solution.

Pour pouvoir développer un méthode d'inférence, il est nécessaire de traiter
séparément le cas où le système est dans une phase ferromagnétique et le cas
d'une phase paramagnétique. Dans le premier cas, on va supposer que les
données qu'on dispose pour faire l'inférence sont des mesures des
configurations du système. On suppose de plus que ces mesures sont une 
réalisation de la loi de Boltzmann.
Comme les minimums de l'énergie libre correspondent à des configurations 
magnétisés selon un
des patterns\cite{amit1}, on suppose qu'on a mesuré $l_1$ configurations
magnétisés selon le premier pattern, $l_2$ selon le deuxième et ainsi de suite.
Dans ce cas, si on estime à partir des mesures la corrélation entre deux sites,
on a
\begin{eqnarray}
C_{ij} &=& \frac{1}{L} \sum_{l} \sigma_i^l \sigma_j^l \nonumber\\
&=& \sum_{k=1}^p \frac{1}{L} \sum_{l\in l_k} \sigma_i^l \sigma_j^l \nonumber \\
&\approx& \frac{1}{L} \sum_{k=1}^p  l_k m_i^k m_j^k + O(1/N) \, ,
\label{eq_hopferro_fr}
\end{eqnarray}
où
\begin{eqnarray}
m^k &=& \frac{1}{N} \sum_i \xi_i^k \tanh(m_k \xi_i^k) \, ,
\end{eqnarray}
et
\begin{eqnarray}
m^k_i &=& \tanh(m^k \xi_i^k) \, ,
\label{eq_seila43_fr}
\end{eqnarray}
d'où on peut aisément inférer les patterns en diagonalisant la matrice $C_{ij}$.
Notez que cette procédure choisi implicitement un gauge tel que
\begin{equation}
\sum_i \tanh(m^k \xi_i^k) \tanh(m^{k'} \xi_i^{k'}) = 0, \quad \text{avec $k\neq k'$} \, .
\end{equation}

Le cas paramagnétique est plus intéressant, mais plus complexe. D'abord, comme
on cherche un système dans une phase où il n'est pas magnétisé selon aucun
pattern, on suppose qu'il existe un champs externe suffisamment fort pour
que les magnétisations $m_i$ soit dominées par le champs externe locale $h_i$.
Aussi, pour faciliter les calculs qui suivent, on remplace le Hamiltonien
habituel du modèle d'Hopfield (éq.~(\ref{eq_hop_fr})) par
\begin{equation}
H = -\frac{1}{N} \sum_{\mu=1}^p \sum_{i<j} \xi_{i}^\mu \xi_j^\mu (\sigma_i - \tanh h_i) (\sigma_j - \tanh h_j)
- \sum_i h_i \sigma_i \, .
\label{eq_4411_fr}
\end{equation}
Ce Hamiltonien permet de retrouver l'éq.~(\ref{eq_hop_fr}) par une translation
du champs externe~:
\begin{equation}
h_i \rightarrow h_i - \sum_\mu \sum_{j \, (\neq i)} \xi_i^\mu \xi_j^\mu \tanh h_j
\, .
\end{equation}

Utilisant le théorème de Bayes (eq.~(\ref{eq_bayes_fr})), on a alors
\begin{equation}
\begin{split}
P(\{\xi_i^\mu\}&|\{\sigma^l\}) = \frac{P_0(\{\xi_i^\mu\}) }{Z(\{\xi_i^\mu\})^L P(\{\sigma^l\})} 
\times \\
&\times\prod_{l=1}^L\exp\left[
\frac{\beta}{N} \sum_{\mu=1}^p \sum_{i<j} \xi_i^\mu \xi_j^\mu (\sigma_i^l -\tanh h_i) (\sigma_j^l - \tanh h_j) + \beta \sum_i h_i \sigma_i^l\right] \, .
\end{split}
\end{equation}
On voit que cette probabilité ne dépend que des valeurs mesurés des
corrélations et magnétisations
\begin{equation}
m_i = \frac{1}{L} \sum_l \sigma_i^l \, , \quad \quad
c_{ij} = \frac{1}{L} \sum_l \sigma_i^l \sigma_j^l - m_i m_j \, ,
\end{equation}
donnant
\begin{equation}
\begin{split}
P(\{\xi_i^\mu\}|&\{\sigma^l\}) = \frac{P_0(\{\xi_i^\mu\}) }{Z(\{\xi_i^\mu\})^L P(\{\sigma^l\})}
\exp\Bigg[
\frac{\beta L}{N} \sum_{\mu=1}^p \sum_{i<j} \xi_i^\mu \xi_j^\mu c_{ij} \\
&+ \frac{\beta L}{N} \sum_\mu \sum_{i<j} \xi_i^\mu\xi_j^\mu(m_i -\tanh h_i)(m_j - \tanh h_j)
+ \beta L \sum_i h_i m_i\Bigg] \, .
\end{split}
\label{eq_p_sigma_fr}
\end{equation}

Pour optimiser cette quantité par rapport aux patterns $\xi_i^\mu$, il est
nécessaire d'écrire explicitement un développement de $\log Z$ pour $N$ grand,
ce qui est fait dans l'Appendice~\ref{ap_logZ}. La minimisation en soit étant
assez technique, elle peut être consulté au chapitre~\ref{ch_inf_mat}. Le
résultat final, à l'ordre dominante en $N$, est

\begin{eqnarray}
h_i^0 &=& \tanh^{-1} m_i \, .
\end{eqnarray}
et
\begin{eqnarray}
\xi_i^{\mu,0} = \sqrt{1-\frac{1}{\lambda_\mu}} \frac{v_i^\mu}{\sqrt{1-\tanh^2 h_i^0}} \, ,
\label{matrice2_fr}
\end{eqnarray}
où $\lambda_\mu$ et $v^\mu$ sont, respectivement, le $\mu$-ème plus grande
valeur propre de la matrice $M$ et sont vecteur propre associé, où $M$ est donnée
par
\begin{equation}
\label{matrice_fr}
M_{ij} = \frac{c_{ij}}{\sqrt{1-\tanh^2 h_i^0}\sqrt{1-\tanh^2 h_j^0}}\, .
\end{equation}

C'est intéressant de noter que cette expression implique que
\begin{eqnarray}
J_{ij} &=& \sum_\alpha \frac{\lambda_\alpha -1}{\lambda_\alpha} 
\frac{v_i^\alpha v_j^\alpha}{\sqrt{(1-m_i^2)(1-m_j^2)}} 
\nonumber \\
&=& \frac{1}{\sqrt{(1-m_i^2)(1-m_j^2)}} \left[M^{-1} (M-1)\right]_{ij} \, ,
\end{eqnarray}
qui est exactement l'expression déjà trouvée dans l'éq.~(\ref{eq_jijloop_fr})
pour le modèle d'Ising généralisé.

Dans le but de trouver une expression inédite pour l'inférence, on a aussi
déterminé les corrections sous-dominantes correspondant à
l'éq.~(\ref{matrice2_fr}). Les détails se trouvent dans le
chapitre~\ref{ch_inf_mat}. Malheureusement, l'expression trouvée est très
sensible à des erreurs aléatoires sur les magnétisations et corrélations
mesurées, donc peu utile pour des données réels.

Un autre problème intéressant est de savoir combien de mesures on doit faire
d'un système d'Hopfield pour qu'il soit possible d'inférer précisément les
valeurs de ses patterns. Pour donner une réponse à ce problème, on a calculé
l'entropie de Shannon de la procédure d'inférence, car il est raisonnable
de penser que quand $S/N \ll 1$, on peut trouver les patterns avec un erreur
faible. On rappelle que l'entropie de Shannon d'une distribution de
probabilité $P$ défini sur un ensemble $\Omega$ est
\begin{equation}
  S = - \sum_{\omega \in \Omega} P(\omega) \log P(\omega) \, .
  \label{eq_def_S_fr}
\end{equation}
Dans le cas de notre inférence, la probabilité est donné par le théorème
de Bayes (eq.~(\ref{eq_bayes_fr})) et on suppose que nos patterns ne peuvent
valoir que $\pm \sqrt{\beta}$, où $\beta$ est une constante fixe, qu'on
associe à une température inverse. On traite
d'abord le cas d'un seul pattern. On a alors
\begin{eqnarray}
S[\{\sigma_i^l\}] &=& -  \frac{1}{\mathcal{N}[\{ \sigma_i^l\}] Z(\beta)^L}  \sum_{\{\xi\}} \exp 
\left(\frac{\beta}{N} \sum_{l=0}^L \sum_{i<j} \xi_i \xi_j \sigma_i^l  \sigma_j^l \right) \times \nonumber\\
&&\times\left[- \log \left(\mathcal{N}[\{\sigma_i^l\}] Z(\beta)^L\right) +\left(\frac{\beta}{N} \sum_{l=0}^L \sum_{i<j} \xi_i \xi_j \sigma_i^l  \sigma_j^l \right)\right] \, .
\label{eq_S123_fr}
\end{eqnarray}
où
\begin{equation}
\mathcal{N}[\{\sigma_i^l\}] = Z(\beta)^{-L} \sum_{\{\xi\}}\exp \left(\frac{\beta}{N} \sum_{l=0}^L \sum_{i<j} \xi_i \xi_j \sigma_i^l  \sigma_j^l \right)\, .
\label{eq_ntil_fr}
\end{equation}

Notez que $S$ dépend explicitement des mesures $\{\sigma_i\}$. Par contre, il est naturel
d'espérer que pour un système assez grand et pour un nombre pas trop petit
de mesures, l'entropie va dépendre plus des caractéristiques du système que
des détail des mesures effectuées. On s'intéresse alors à calculer $\moy{S}$,
où la moyenne est effectué par rapport à la probabilité de Boltzmann de
mesurer les configurations $\{\sigma_i\}$, en supposant que le système que
les a généré est décrit par un modèle de Hopfield.

Le comportement de $\moy{S}$ est très différent pour les deux phases du système.
S'il est dans une phase ferromagnétique, l'entropie décroit exponentiellement
avec le nombre de mesures $L$. On peut trouver les détails du calcul et 
l'expression analytique de l'entropie dans le chapitre~\ref{chap_hop}. Les
résultats sont représentés dans le graphique suivant~:
\begin{figure}[H]
\begin{center}
\includegraphics[width=0.75\columnwidth]{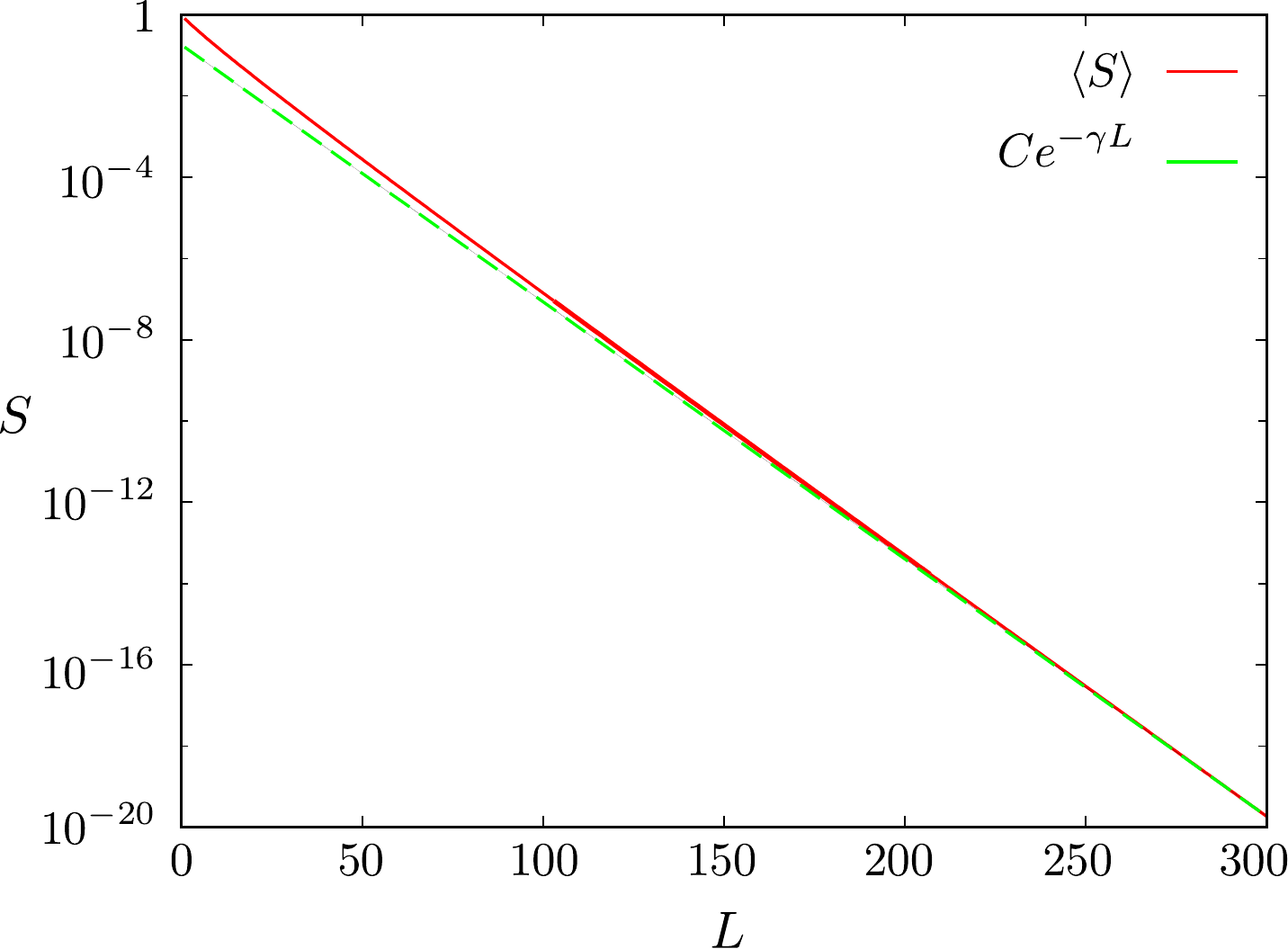}
\caption{Entropie par spin en fonction du nombre de mesures $L$ pour 
$\beta=1.1$. Notez le comportement asymptotique 
$\moy{S} \approx C e^{\gamma L}$ où $\gamma = \log \cosh (\beta m)$.}
\label{grEntr1_fr}
\end{center}
\end{figure}

Ce résultat montre que le nombre de mesures nécessaires pour inférer le
pattern du système est une grandeur intensive du système, ie, il reste finit
quand la taille du système tends vers l'infini.

\begin{figure}[H]
\begin{center}
\includegraphics[width=0.85\columnwidth]{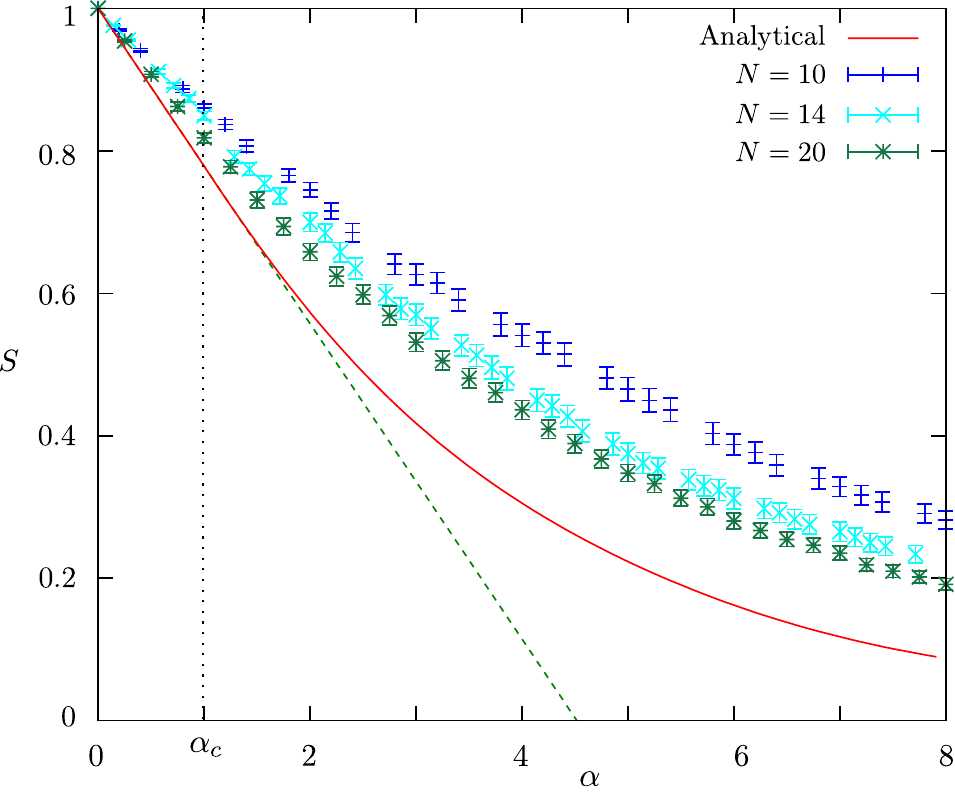}
\caption{Entropie par spin comme fonction de $\alpha$ pour $\beta=0.5$. La
ligne solide correspond à la solution analytique trouvé dans le 
chapitre~\ref{chap_hop} et les points correspondent à des résultats
numériques.}
\label{grEntr2_fr}
\end{center}
\end{figure}

Quand le système est dans la phase paramagnétique, cette situation se modifie.
En fait, il faut un nombre de mesures $L = \alpha N$ proportionnel à la taille
du système pour pouvoir inférer le pattern. Comme on peut voir dans le
chapitre~\ref{chap_hop}, les calculs sont aussi plus complexes, et on est
obligé d'utiliser des méthodes des systèmes désordonnées (notamment la méthode 
des répliques) pour trouver une expression analytique pour l'entropie moyenne.
Un graphique illustrant les résultats obtenus se trouve dans la 
Fig.~\ref{grEntr2_fr}.

Finalement, pour le cas de plusieurs patterns dans la phase paramagnétique,
on retrouve que l'entropie de l'inférence de chaque pattern décroit comme 
l'entropie du système où on n'a qu'un seul pattern (Fig.~\ref{grEntr2_fr}).

\newcounter{paperref}
\appendix
\chapter{Details of the small-$\bm\beta$ expansion}
\label{ap_A}
Let $O$ be an observable of the spin configuration (which can
explicitly depend on the inverse temperature $\beta$), and 
\begin{equation}
\moyen{O} = \frac 1Z \sum_{\{ \sigma\}} O(\{\sigma\}) \;e^{U(\{\sigma\})}
\end{equation} 
its average value, where $U$ is defined in (\ref{eqU}), and
$Z=\exp(\tilde S)$.
The derivative of the average value of $O$ fulfills the following
identity,
\begin{equation}
  \label{eqderiv}
\derivb{\moyen{O}} = \frac{1}{Z} 
\sum_{\{\sigma\}}  \left[\derivb{O}+O\derivb{U} \right] e^U - 
\frac{1}{Z^2} \derivb{Z} \sum_{\{\sigma\}}  Oe^U = \moyen{\derivb{O}} + \moyen{O\derivb{U}}
\end{equation}
where the term in $Z^{-2}$ vanishes as a consequence of (\ref{gconst}).

\section{Second order expansion}

Using (\ref{eqderiv}) and~(\ref{eq_b1b2})
\begin{equation}
  \label{g3}
0= \derivmb{\tilde{S}}{3} = \derivb{}\left[\moyen{\derivmb{U}{2}} + 
\moyen{\left(\derivb{U}\right)^2}\right] = 
\moyen{\derivmb{U}{3}} + 3\moyen{\derivmb{U}{2}\derivb{U}} + \moyen{\left(\derivb{U}\right)^3}
\end{equation}

A straightforward calculation gives (where we omit for clarity the
notation $|_0$ and the ${}^*$ subscript from $J_{ij}$ and $\lambda_i$)

\begin{eqnarray}
\moyen{\derivmb{U}{3}}_0 &=& - 2\sum_{i<j} \derivmb{J_{ij}}{2} c_{ij} \\
\moyen{\derivmb{U}{2}  \derivb{U}}_0 &=&
\sum_{i<j} \derivmb{J_{ij}}{2} \derivb{J_{ij}} L_i L_j +
\sum_{i} \derivmb{\lambda_i}{2} \derivb{\lambda_i} L_i \\
\moyen{\left( \derivb{U}  \right)^3}_0 &=& 
6 \sum_{i<j<k} \derivb{J_{ij}} \derivb{J_{jk}} \derivb{J_{ki}} L_i
L_j L_k +
\nonumber \\ &+&
\sum_{i<j} \left(\derivb{J_{ij}}\right)^3 4 m_i m_j L_i L_j
+6 \sum_{i<j} \derivb{J_{ij}} \derivb{\lambda_i}
\derivb{\lambda_j}L_i L_j
\end{eqnarray}  

Using (\ref{g3}), the
expressions of the derivatives of $\lambda_i$ in $\beta=0$, we obtain
\begin{equation}
Q_2 = -
4 \sum_{i<j} \frac{c_{ij}^3 m_i m_j}{(1-m_i^2)^2(1-m_j^2)^2} -
6 \sum_{i<j<k} \frac{c_{ij} c_{jk} c_{ki}}{(1-m_i^2)(1-m_j^2)(1-m_k^2)}
\end{equation}
from which we deduce
\begin{equation}
\left. \derivmb{S}{3} \right|_0 = 
4 \sum_{i<j} K_{ij}^3 m_i m_j L_i L_j +
6 \sum_{i<j<k} K_{ij} K_{jk} K_{ki} L_i L_j L_k
\end{equation}
and
\begin{equation}
\left. \derivmb{J_{ij}}{2}\right|_0
=
-4 m_i m_j K_{ij}^2 - 2 \sum_{k (\neq i,\, \neq j)} K_{jk} K_{ki} L_k
\ .
\end{equation}

\section{Third order expansion}

The procedure to derive the third order expansion for the coupling is
identical to the second order one. We start from
\begin{equation}
0 = \derivmb{\tilde{S}}{4} = \moyen{\derivmb{U}{4}} + 
3 \moyen{\left(\derivmb{U}{2}\right)^2} + 4 \moyen{\derivmb{U}{3}\derivb{U}}
+ 6\moyen{\left(\derivb{U}\right)^2\derivmb{U}{2}} +
\moyen{\left(\derivb{U}\right)^4}
\end{equation}
and evaluate each term in  the sum:
\begin{eqnarray}
\moyen{\derivmb{U}{4}}_0 &=&
-3 \sum_{i<j} \left. \derivmb{J_{ij}}{3}\right|_0 K_{ij} L_i L_j
\\
\moyen{\left(\derivmb{U}{2}\right)^2}_0 &=&
\sum_{i<j} \left(\derivmb{J_{ij}}{2}\right)^2 L_i L_j +
\sum_i \left(\derivmb{\lambda_i}{2}\right)^2 L_i + 
\left[ \sum_{i<j} K_{ij}^2 L_i L_j \right]^2
\\
\moyen{\derivmb{U}{3}\derivb{U}}_0 &=& 
\sum_{i<j} K_{ij} \derivmb{J_{ij}}{3} L_i L_j
\\
\moyen{\left(\derivb{U}\right)^2\derivmb{U}{2}}_0 &=&
2\sum_{i<k} \sum_j K_{ij}K_{jk}\derivmb{J_{ki}}{2} L_i L_j L_k
+ 4\sum_{i<j} K_{ij}^2 \derivmb{J_{ij}}{2} m_i m_jL_i L_j
\nonumber\\
&+& \sum_i \sum_j K_{ij}^2 \derivmb{\lambda_i}{2}(-2m_i)L_i L_j
- \moyen{\left(\derivb{U}\right)^2}_0 \sum_{i<j} K_{ij}^2 L_i L_j
\\
\moyen{\left(\derivb{U}\right)^4}_0
&=& \sum_{i<j} K_{ij}^4 (3m_i^2+1)L_i (3m_j^2+1)L_j 
+ 3 \sum_{i<j, \,k<l\, (k\neq i,\, l\neq j)}
K_{ij}^2 K_{kl}^2 L_i L_j L_k L_l +
\nonumber\\
&+& 6 \sum_{i<k}\sum_j K_{ij}^2 K_{jk}^2 (3m_j^2+1) L_i L_j L_k +
\nonumber\\
&+& 12 \sum_{i<j<k} K_{ij} K_{jk} K_{ki} L_i L_j L_k \left[
4m_i m_j K_{ij} + 4m_im_k K_{ik} + 4m_km_i K_{ki}\right] +
\nonumber\\
&+& 3\sum_{i,j,k,l \, (\neq)}
K_{ij} K_{jk} K_{kl} K_{li} 
L_i L_j L_k L_l
\end{eqnarray}
Using the results of Eq.~(\ref{eq_derivcij}) and~(\ref{eq_derivmi}) we can write all the terms above in the
same form
\begin{eqnarray}
-3\left[ \sum_{i<j} K_{ij}^2 L_i L_j \right]^2 &=&
- 3 \sum_{i<j,\,k<l\, (k\neq i,\, l\neq j)} K_{ij}^2 K_{kl}^2 L_i L_j L_k L_l
\nonumber\\
&&-6 \sum_{i<j} \sum_k K_{ik}^2 K_{kj}^2 L_i L_j L_k^2 \nonumber\\
&&- 3 \sum_{i<j} K_{ij}^4 L_i^2 L_j^2
\\
12\sum_{i<j}\sum_k K_{ik}K_{kj}\derivmb{J_{ij}}{2} L_i L_j L_k &=&
-48 \sum_{i<j}  \sum_k K_{ij}^2 K_{ik} K_{kj} m_i m_j L_i L_j L_k -
\nonumber\\
&&- 12 \sum_{i,j,k,l \, (\neq)}
K_{ij} K_{jk} K_{kl} K_{li}L_i L_j L_k L_l
\nonumber\\
&&- 24 \sum_{i<j} \sum_k K_{ik}^2 K_{kj}^2 L_i L_j L_k^2
\\
\sum_{i<j} K_{ij}^2\derivmb{J_{ij}}{2} m_i m_j L_i L_j &=&
 -4 \sum_{i<j} K_{ij}^4 m_i^2 m_j^2 L_i L_j \nonumber\\
&&- 2\sum_{i<j} \sum_k K_{ij}^2 K_{jk} K_{ki} m_i m_j L_iL_jL_k
\\
3\sum_{i<j} \left(\derivmb{J_{ij}}{2}\right)^2 L_i L_j &=&
48 \sum_{i<j} K_{ij}^4m_i^2 m_j^2L_i L_j + \nonumber\\
&&48 \sum_{i<j} \sum_k K_{ij}^2 K_{ik} K_{kj} m_i m_j L_i L_j L_k +
\nonumber\\
&&+ 6 \sum_{i,j,k,l \, (\neq)}
K_{ij} K_{jk} K_{kl} K_{li} L_iL_jL_kL_l
\nonumber\\
&&+ 12 \sum_{i<j} \sum_k K_{ik}^2 K_{kj}^2 L_k^2 L_i L_j
\\
6\sum_i \sum_j K_{ij}^2 \derivmb{\lambda_i}{2} (-2m_i)L_iL_j &=&
-24 \sum_i \sum_j K_{ij}^4 m_i^2 (1-m_j^2) L_iL_j
\nonumber \\
&&- 48 \sum_{i<j} \sum_k K_{ik}^2 K_{kj}^2 m_k^2 L_iL_jL_k
\\
3 \sum_k \left(\derivmb{\lambda_k}{2}\right)^2 L_k &=&
24 \sum_{i<j} \sum_k K_{ik}^2 K_{kj}^2 m_k^2 L_iL_jL_k \nonumber\\
&&+ 12 \sum_i \sum_j K_{ij}^4 m_i^2 L_i L_j^2 
\end{eqnarray}
Again we find equation (\ref{anyorder}) with 
\begin{eqnarray}
Q_3 &=& -\sum_{i<j} K_{ij}^4 \left[(3m_i^2+1) (3m_j^2+1) 
-48 m_i^2m_j^2\right]L_iL_j
\nonumber \\
&+& 12 \sum_{i<j}\sum_k K_{ik}^2 K_{jk}^2 L_i L_j L_k^2
+ 3 \sum_{i,j,k,l\,(\neq)}
K_{ij} K_{jk} K_{kl} K_{li}L_i L_j L_k L_l
\nonumber \\
&+& 12 \sum_i \sum_j K_{ij}^4 m_i^2  L_iL_j^2
+ 3 \sum_{i<j} K_{ij}^4 L_i^2 L_j^2
\end{eqnarray}
which gives the fourth order contribution to the entropy,
\begin{eqnarray}
\derivmb{S}{4}  &=& -2\sum_{i<j} K_{ij}^4 \left[1 + 3m_i^2 +3m_j^2 + 9 m_i^2m_j^2\right] L_i L_j
-12 \sum_{i<j} \sum_k K_{ik}^2 K_{kj}^2 L_k^2 L_i L_j
\nonumber\\
&-& 24 \sum_{i<j<k<l}
(K_{ij} K_{jk} K_{kl} K_{li} +
  K_{ik} K_{kj} K_{lj} K_{il} + K_{ij} K_{jl} K_{lk} K_{ki})
  L_i L_j L_k L_l
\end{eqnarray}
and the third order contribution to the coupling,
\begin{eqnarray}
\left. \derivmb{J_{ij}}{3}\right|_0 &=& 
2 K_{ij}^3 \left[1 + 3m_i^2 +3m_j^2 + 9 m_i^2m_j^2\right]
+6 \sum_{k\, (\neq i,\,\neq j)} K_{ij}
(K_{jk}^2 L_j +K_{ki}^2 L_i)  L_k
+
\nonumber\\
&+& 6 \sum_{\substack{k, l \\ (k \neq i, l\neq j)}} K_{jk} K_{kl} K_{li} L_k L_l \, .
\end{eqnarray}

\chapter{Large magnetization expansion}
\label{app_mag}
Equation (\ref{jij}) suggests that to expand $J_{ij}^*$ to the order
of $(L_i)^k$ one has to sum all the diagrams with up to $k+2$ spins. 
This statement is true if the  expansion for $J_{ij}^*$ is of the form
\begin{equation}
\label{forma}
J_{ij}^* = A_{ij} + \sum_k L_k A_{ijk} + \sum_k \sum_l L_k L_l A_{ijkl} + ...
\end{equation}
where the coefficients $A_{i_1 i_2 ... i_n}$ are polynomials in the
couplings $K_{i_\alpha i_\beta}$ and the magnetizations
$m_\alpha$ ($\alpha,\beta < n$).
In the following we will show that the above statement is true to 
any order of the expansion in $\beta$ by recurrence. 
First of all, from (\ref{eq_derivmi}) we see that if $J_{ij}^*$ is of
the form~(\ref{forma}) up to the order $k$, so is $\lambda_i^*$ to the same
order.

As we saw in section~\ref{sec_expansion}, to find an equation
 for $\derivmb{S}{k}$, one must evaluate $  \derivmb{\tilde{S}}{k+1}$. Using
Eq.~\ref{eqderiv}, we can write

\begin{equation}
\label{ss}
  \derivmb{\tilde{S}}{k+1} = \moyen{\left(\derivb{} + \derivb{U}\right)^k \derivb{U}} =
  \sum_{\{\alpha\}} P_\alpha \moyen{\prod_{j=1}^{k+1} \derivmb{U}{\alpha_j}}
\end{equation}
where $\alpha$ is a multi-index with $|\alpha|=k+1$ and $P_\alpha$ a
multiplicity coefficient.
The highest order term of this expression evaluates to
$\sum_{ij} L_i L_j K_{ij} \derivmb{J_{ij}^*}{j} = \derivmb{S}{k}$.

Due to the structure of $U$, spin dependence in (\ref{ss}) will 
come either from the lower derivatives of $J_{ij}^*$ (of the 
form~(\ref{forma}) by hypothesis), from the derivatives of $\lambda_i^*$, or 
explicitly from $U$. In the later case we get a multiplicative 
factor $(\sigma_i-m_i)$. Hence we end up with computing a term, with
$k\ge 1$, of the form
\begin{equation}
  \label{moyes}
\moyen{(\sigma_i-m_i)^k} = (-1)^k (1-m_i^2) \frac{(m+1)^{k-1} - (m-1)^{k-1}}{2}
\end{equation}
Clearly any term including $(\sigma_i-m_i)$ will give a multiplicative
factor $L_i$ after averaging. As spins are decoupled in the $\beta=0$
limit we obtain the product of those factors over the spins in the
diagram as claimed.




\chapter{Verification of \textit{\protect\iih}${}_\protect\iii$(\{\textit{c}\},\{\textit{m}\})}
\label{ap_feio}

In this appendix, we used the procedure described in section~\ref{sec_compar}
to verify all the terms of Eq.~(\ref{eq_hi}). We start by
posing
\begin{equation}
  \label{eq_seila1}
h_i = \frac{1}{2} \ln\left(\frac{1-m_i}{1-m_j}\right) - \sum_j J_{ij} m_j + \sum_j K_{ij}^2 m_i L_j + h_i^{(3)} + h_i^{(4)} + O(c^5) \, ,
\end{equation}
and we will now proceed to evaluate each one of the $h_i^{(k)}$.

\section{Evaluation of \textit{\protect\iih}${}_\protect\iii$\textsuperscript{(3)}(\{\textit{J}\})}

Using Eqs.~(\ref{eq_gy}) and~(\ref{eq_gy_hi}), we have

\begin{eqnarray}
6 h_l^{(3)}
&=& \frac{\partial}{\partial m_l} \derivmb{(\beta F)}{3}
\nonumber\\
&=& -4 \sum_{i<j} J_{ij}^3  \frac{\partial}{\partial m_l} 
\left[m_i(1-m_i^2)m_j(1-m_j^2)\right]
\nonumber\\
&&-6 \sum_{i<j<k} J_{ij} J_{jk} J_{ki}
\frac{\partial}{\partial m_l} 
\left[ (1-m_i^2) (1-m_j^2) (1-m_k^2) \right] \, ,
\end{eqnarray}
which can be simplified to
\begin{equation}
  \begin{split}
6 h_l^{(3)}
= &-4 \sum_{i<j} J_{ij}^3  [
\delta_{il} (1-m_i^2)m_j (1-m_j^2)
- 2 \delta_{il} m_i^2 m_j (1-m_j^2)
\\
&\phantom{-4 \sum_{i<j} J_{ij}^3  [}
+ \delta_{jl} (1-m_j^2)m_i (1-m_i^2)
- 2 \delta_{jl} m_j^2 m_i (1-m_i^2)]
\\
&
+12 \sum_{i<j<k} J_{ij} J_{jk} J_{ki} 
[ \delta_{il} m_i (1-m_j^2) (1-m_k^2)  + 
\delta_{jl} m_j (1-m_i^2) (1-m_k^2) 
\\
&
\phantom{+12 \sum_{i<j<k} J_{ij} J_{jk} J_{ki} [}
+\delta_{kl} m_k (1-m_i^2) (1-m_j^2) 
] \, .
\end{split}
\end{equation}
Finally, we get
\begin{equation}
  \begin{split}
6 h_l^{(3)}
= &-4 (1-m_l^2) \sum_{i (\neq l)} J_{il}^3  m_i (1-m_i^2) +
8 m_l^2 \sum_{i (\neq l)} J_{il}^3 m_i (1-m_i^2)
\\
&+
12 m_l \sum_{i<j\, (i,j \neq l)} J_{li} J_{ij} J_{jl} (1-m_i^2) (1-m_j^2) \, .
\end{split}
  \label{eq_seeila}
  \end{equation}

\section{Evaluation of \textit{\protect\iih}${}_\protect\iii$\textsuperscript{(3)}(\{\textit{c}\})}
\label{sec_h3c}
We will now use Eq.~(\ref{eq_seeila}) to verify the terms of order $\beta^3$
of Eq.~(\ref{eq_hi}).
We start with Eqs.~(\ref{eq_seila1}) and~(\ref{eq_seeila})
\begin{equation}
  \begin{split}
h_i(\{J_{ij}\}) = 
&-\frac{1}{2} \ln \left( \frac{1-m_i}{1+m_i}\right) 
- \sum_{j (\neq i)} J_{ij} m_j 
+ \sum_{j (\neq i)} J_{ij}^2 m_i (1-m_j^2)
\\
&
-\frac{2}{3} (1-m_i^2) \sum_{j (\neq l)} J_{ij}^3  m_j (1-m_j^2) +
\frac{4}{3} m_i^2 \sum_{j (\neq i)} J_{ij}^3 m_j (1-m_j^2)
\\
&
+2 m_i \sum_{j<k (j,k \neq i)} J_{ij} J_{jk} J_{ki} (1-m_j^2) (1-m_k^2) \, .
  \end{split}
\end{equation}
Using Eq.~(\ref{jij}),
we pose
$J_{ij} = K_{ij} - 2K_{ij}^2 m_im_j - \sum_k K_{jk}K_{ki}(1-m_k^2) + O(c^3)$,
yielding
\begin{equation}
  \begin{split}
h_i(\{c\}) = 
&
-\frac{1}{2} \ln \left( \frac{1-m_i}{1+m_i}\right)
- \sum_{j (\neq i)} J_{ij} m_j
\\
&
+ \sum_{j (\neq i)} m_i \left[K_{ij} - 2K_{ij}^2 m_im_j - \sum_k K_{jk}K_{ki}(1-m_k^2)\right]^2 (1-m_j^2)
\\
&-\frac{2}{3} (1-m_i^2) \sum_{j (\neq l)} K_{ij}^3  m_j (1-m_j^2)
+\frac{4}{3} m_i^2 \sum_{j (\neq i)} K_{ij}^3 m_j (1-m_j^2)
\\
&
+2 m_i \sum_{j<k} K_{ij} K_{jk} K_{ki} (1-m_j^2) (1-m_k^2) + O(c^4) \, .
\end{split}
\end{equation}
Simplifying this expression, we get
\begin{equation}
  \begin{split}
h_i(\{J_{ij}\}) =
&
-\frac{1}{2} \ln \left( \frac{1-m_i}{1+m_i}\right)
- \sum_{j (\neq i)} J_{ij} m_j
+ \sum_{j (\neq i)} K_{ij}^2 m_i (1-m_j^2)
\\
&
- 4 \sum_{j (\neq i)} K_{ij}^3 m_i^2 m_j  (1-m_j^2) -
\\
&
-2 \sum_{j,\,k} K_{ij} K_{jk} K_{ki} m_i (1-m_k^2)(1-m_j^2)-
\\
&
-\frac{2}{3} (1-m_i^2) \sum_{j (\neq l)} K_{ij}^3  m_j (1-m_j^2) +
\frac{4}{3} m_i^2 \sum_{j (\neq i)} K_{ij}^3 m_j (1-m_j^2)
\\
&
+2 m_i \sum_{j<k} K_{ij} K_{jk} K_{ki} (1-m_j^2) (1-m_k^2) + O(c^4) \, ,
\end{split}
\end{equation}
and finally
\begin{equation}
  \begin{split}
h_i(\{J_{ij}\}) =
&
-\frac{1}{2} \ln \left( \frac{1-m_i}{1+m_i}\right)
- \sum_{j (\neq i)} J_{ij} m_j
+ \sum_{j (\neq i)} K_{ij}^2 m_i (1-m_j^2)
\\
&
-\frac{2}{3} (1+3m_i^2) \sum_{j (\neq l)} K_{ij}^3  m_j (1-m_j^2)
\\
&
-2 m_i \sum_{j<k} K_{ij} K_{jk} K_{ki} (1-m_j^2) (1-m_k^2) + O(c^4) \, .
\end{split}
\end{equation}
Which matches exactly Eq.~(\ref{eq_hi}).

\begin{landscape}
\section{Evaluation of \textit{\protect\iih}${}_\protect\iii$\textsuperscript{(4)}(\{\textit{J}\})}

Proceeding in the same way as in the last section,
we use that
$24 h^{(4)}_l = \frac{\partial}{\partial m_l} \derivmb{(\beta A)}{4}$ to
evaluate this quantity explicitly:
\begin{equation}
\begin{split}
24 h^{(4)}_l 
=&
\frac{\partial}{\partial m_l} \Bigg[
2 \sum_{i<j} J_{ij}^4 (1-m_i^2)(1-m_j^2) (1+3m_i^2 + 3 m_j^2 - 15 m_i^2m_j^2)
\\
&
\phantom{\frac{\partial}{\partial m_l} \Bigg[}
- 48 \sum_{i<j<k} J_{ij} J_{jk} J_{ki} (J_{ij}m_im_j + J_{jk}m_jm_k + J_{ki} m_km_i) (1-m_i^2) (1-m_j^2)(1-m_k^2)
\\
&
\phantom{\frac{\partial}{\partial m_l} \Bigg[}
- 24 \sum_{i<j<k<l}
(J_{ij} J_{jk} J_{kl} J_{li} +
  J_{ik} J_{kj} J_{lj} J_{il} + J_{ij} J_{jl} J_{lk} J_{ki})
(1-m_i^2)(1-m_j^2)(1-m_k^2)(1-m_l^2)
\Bigg] \, ,
\end{split}
\nonumber
\end{equation}
\begin{equation}
\begin{split}
=&\phantom{+} 8 m_l \sum_{i (\neq l)} J_{il}^4 (1-m_i^2)(1-3m_l^2 -9 m_i^2 + 15m_i^2m_l^2)
+ 48(3m_l^2-1) \sum_i \sum_{j (\neq i)} J_{il}^2 J_{lj} J_{ij} m_i (1-m_i^2)(1-m_j^2)
\\
&
+ 48 m_l \sum_i \sum_j J_{ij}^2 J_{il} J_{jl} m_i m_j (1-m_i^2)(1-m_j^2)
+48 m_l \sum_{i<j} \sum_k J_{ik} J_{kj} J_{jl} J_{li} L_i L_j L_k
- 24 m_l L_l \sum_i \sum_j J_{il}^2 J_{jl}^2 L_i L_j \, .
\end{split}
\end{equation}

Finally, the full value of $h_l(\{J\})$ is
\begin{equation}
\begin{split}
h_l(\{J\}) = &
-\frac{1}{2} \ln \left( \frac{1-m_l}{1+m_l}\right) 
- \sum_{j (\neq l)} J_{lj} m_j
+ \sum_{j (\neq l)} J_{lj}^2 m_l (1-m_j^2)
-\frac{2}{3} (1-m_l^2) \sum_{j (\neq l)} J_{lj}^3  m_j (1-m_j^2)
+\frac{4}{3} m_l^2 \sum_{j (\neq l)} J_{lj}^3 m_j (1-m_j^2)
\\
&
+2 m_l \sum_{j<k} J_{lj} J_{jk} J_{kl} (1-m_j^2) (1-m_k^2)
+ \frac{1}{3} m_l \sum_{i (\neq l)} J_{il}^4 (1-m_i^2)(1-3m_l^2 -9 m_i^2 + 15m_i^2m_l^2)
\\
&
+ 2(3m_l^2-1) \sum_i \sum_{j (\neq i)} J_{il}^2 J_{lj} J_{ij} m_i (1-m_i^2)(1-m_j^2)
+ 2 m_l \sum_i \sum_j J_{ij}^2 J_{il} J_{jl} m_i m_j (1-m_i^2)(1-m_j^2)
\\
&
+2m_l \sum_{i<j} \sum_k J_{ik} J_{kj} J_{jl} J_{li} L_i L_j L_k
-  m_l L_l \sum_i \sum_j J_{il}^2 J_{jl}^2 L_i L_j
\end{split}
\label{eq_mmnn}
\end{equation}

\section{Evaluation of \textit{\protect\iih}${}_\protect\iii$\textsuperscript{(4)}(\{\textit{c}\})}
We will now repeat the same procedure of section~\ref{sec_h3c} up to the order
$\beta^4$. Using Eqs.~(\ref{eq_mmnn}) and (\ref{jij}), we have
\begin{equation}
\begin{split}
h_l(\{c\}) = &
-\frac{1}{2} \ln \left( \frac{1-m_l}{1+m_l}\right)
- \sum_{j (\neq l)} J_{lj} m_j
+ \sum_{j (\neq l)} m_l
\left[K_{lj} - 2K_{lj}^2 m_lm_j - 
\sum_k K_{jk}K_{kl}(1-m_k^2) + \frac{1}{6} \derivmb{J_{lj}}{3}+...\right]^2 
(1-m_j^2)
\\
&
-\frac{2}{3} (1-m_l^2) \sum_{j (\neq l)} 
\left[K_{lj} - 2K_{lj}^2 m_lm_j - \sum_k K_{jk}K_{kl}(1-m_k^2) +...\right]^3  m_j (1-m_j^2)
\\
&
+\frac{4}{3} m_i^2 \sum_{j (\neq l)} \left[K_{lj} - 2K_{lj}^2 m_lm_j - 
\sum_k K_{jk}K_{kl}(1-m_k^2) +...\right]^3 m_j (1-m_j^2)
\\
&
+2 m_l \sum_{j<k, (j,k\neq l)}
\left[K_{lj} - 2K_{lj}^2 m_lm_j - \sum_m K_{lm}K_{mj}(1-m_m^2) +...\right]\times
\\
&
\phantom{2 m_l \sum_{j<k, (j,k\neq l)}}
\times
\left[K_{jk} - 2K_{jk}^2 m_jm_k - \sum_m K_{jm}K_{mk}(1-m_m^2) +...\right]\times
\\
&
\phantom{2 m_l \sum_{j<k, (j,k\neq l)}}
\times
\left[K_{kl} - 2K_{kl}^2 m_km_l - \sum_m K_{km}K_{ml}(1-m_m^2) +...\right]
(1-m_j^2) (1-m_k^2)
\\
&
+ \frac{1}{3} m_l \sum_{i (\neq l)} K_{il}^4 (1-m_i^2)(1-3m_l^2 -9 m_i^2 + 15m_i^2m_l^2)
+ 2(3m_l^2-1) \sum_i \sum_{j (\neq i)} K_{il}^2 K_{lj} K_{ij} m_i (1-m_i^2)(1-m_j^2)
\\
&
+  m_l \sum_i \sum_j K_{ij}^2 K_{il} K_{jl} m_i m_j (1-m_i^2)(1-m_j^2)
+2m_l \sum_{i<j<k} K_{li} K_{ij} K_{jk} K_{ki} L_i L_j L_k
- m_l L_l \sum_i \sum_j K_{il}^2 K_{jl}^2 L_i L_j \, .
\end{split}
\end{equation}
Posing $V_{ij} \equiv - 2K_{ij}^2 m_im_j - \sum_k K_{ik}K_{kj}(1-m_k^2)$ and
ignoring all the terms of order other than $c^4$, we obtain
\begin{equation}
\begin{split}
h^{(4)}_l(\{c\}) =
&\phantom{+}
\frac{1}{3} \sum_{j (\neq l)} m_l \derivmb{J_{lj}}{3} K_{lj} (1-m_j^2)
+\sum_{j (\neq l)}m_l V_{lj}^2 (1-m_j^2)
-2 (1-m_l^2) \sum_{j (\neq l)} K_{lj}^2 V_{lj} m_j (1-m_j^2) 
\\
&+4 m_l^2 \sum_{j (\neq l)} K_{lj}^2 V_{lj} m_j (1-m_j^2) 
+ 2m_l\sum_{j<k} \left(V_{lj} K_{jk} K_{kl}+K_{lj} V_{jk} K_{kl}+
K_{lj} K_{jk} V_{kl}\right)(1-m_j^2)(1-m_k^2)
\\
&
+ \frac{1}{3} m_l \sum_{i (\neq l)} K_{il}^4 (1-m_i^2)(1-3m_l^2 -9 m_i^2 + 15m_i^2m_l^2)
+ 2(3m_l^2-1) \sum_i \sum_{j (\neq i)} K_{il}^2 K_{lj} K_{ij} m_i (1-m_i^2)(1-m_j^2)
\\
&+ 2 m_l \sum_i \sum_j K_{ij}^2 K_{il} K_{jl} m_i m_j (1-m_i^2)(1-m_j^2)
+2m_l \sum_{i<j} \sum_k K_{ik} K_{kj} K_{jl} K_{li} L_i L_j L_k
-  m_l L_l \sum_i \sum_j K_{il}^2 K_{jl}^2 L_i L_j \, .
\end{split}
\end{equation}

We now need to evaluate explicitly the terms in $V_{ij}$, yielding
\begin{eqnarray}
\sum_{j(\neq l)} m_l V_{lj}^2L_j &=&\phantom{+}
4 m_l^3 \sum_{j(\neq l)} K_{lj}^4 m_j^2 L_j
+ m_l \sum_{j} \sum_{k(\neq j)} K_{lk}^2  K_{kj}^2 L_k^2 L_j 
+ 2 m_l \sum_{i<j} \sum_k K_{ik} K_{kj} K_{jl} K_{li} L_i L_j L_k
\nonumber \\
&&- m_l L_l \sum_j \sum_{k(\neq j)} K_{lk}^2 K_{lj}^2 L_j L_k
+ 4 m_l^2 \sum_j \sum_k K_{lj}^2 K_{lk} K_{kj} m_k L_k L_j
-2(1-m_l^2) \sum_j K_{lj}^2 V_{lj} m_j L_j
\nonumber \\
&=&4 m_l L_l \sum_j K_{lj}^4 m_j^2 L_j + 2 L_l \sum_j \sum_k K_{lj}^2 K_{lk} 
K_{kj} m_j L_j L_k \, ,
\end{eqnarray}
and
\begin{eqnarray}
4 m_l^2 \sum_{j (\neq l)} K_{lj}^2 V_{lj} m_j L_j &=&
-8 m_l^3 \sum_j K_{lj}^4 m_j^2 L_j 
-4 m_l^2 \sum_j \sum_k K_{lj}^2 K_{lk} K_{kj} m_j L_j L_k
\nonumber\\
&&
+2m_l\sum_{j<k (j,k\neq l)} \left(V_{lj} K_{jk} K_{kl}+K_{lj} V_{jk} K_{kl}+
K_{lj} K_{jk} V_{kl}\right)(1-m_j^2)(1-m_k^2) 
\nonumber
\\
&=& -4 m_l^2 \sum_j \sum_k  K_{lj}^2 K_{jk}K_{kl}  m_j L_j L_k
 - 2m_l \sum_j \sum_k  K_{jk}^2 K_{jl}K_{lk}  m_j m_k L_j L_k
-
\nonumber
\\
&&
-6 m_l \sum_{i<j} \sum_k K_{ik} K_{kj} K_{jl} K_{li} L_i L_j L_k
- 2 m_l \sum_j \sum_k K_{lk}^2 K_{jk}^2 L_k^2 L_j
+ 2 m_l L_l \sum_j \sum_k K_{lk}^2 K_{jl}^2 L_k L_j
\end{eqnarray}

Using these results, we can write $h^{(4)}_l$ explicitly:
\begin{equation}
\begin{split}
h^{(4)}_l(\{c\}) =
&\phantom{+}
\frac{1}{3} \sum_{j (\neq l)} m_l \derivmb{J_{lj}}{3} K_{lj} (1-m_j^2)
+4 m_l^3 \sum_{j(\neq l)} K_{lj}^4 m_j^2 L_j
+  m_l \sum_{j} \sum_k K_{lk}^2 K_{kj}^2 L_k^2 L_j + 
- m_l L_l \sum_j \sum_k K_{lk}^2 K_{lj}^2 L_j L_k
\\
&
+ 2 m_l \sum_{i<j} \sum_k K_{ik} K_{kj} K_{jl} K_{li} L_i L_j L_k
+ 4 m_l^2 \sum_j \sum_k K_{lj}^2 K_{lk} K_{kj} m_k L_k L_j
+4 m_l L_l \sum_j K_{lj}^4 m_j^2 L_j
\\
&
+2 L_l \sum_j \sum_k K_{lj}^2 K_{lk} 
K_{kj} m_j L_j L_k
-8 m_l^3 \sum_j K_{lj}^4 m_j^2 L_j 
-4 m_l^2 \sum_j \sum_k K_{lj}^2 K_{lk} K_{kj} m_j L_j L_k
\\
&
 -4 m_l^2 \sum_j \sum_k  K_{lj}^2 K_{jk}K_{kl}  m_j L_j L_k
 - 2m_l \sum_j \sum_k  K_{jk}^2 K_{jl}K_{lk}  m_j m_k L_j L_k
-6 m_l \sum_{i<j} \sum_k K_{ik} K_{kj} K_{jl} K_{li} L_i L_j L_k
\\
&
-2 m_l \sum_j \sum_k K_{lk}^2 K_{jk}^2 L_k^2 L_j
+ 2 m_l L_l \sum_j \sum_k K_{lk}^2 K_{jl}^2 L_k L_j
+ \frac{1}{3} m_l \sum_{i (\neq l)} K_{il}^4 (1-m_i^2)(1-3m_l^2 -9 m_i^2 + 15m_i^2m_l^2)
\\
&
+ 2(3m_l^2-1) \sum_i \sum_{j (\neq i)} K_{il}^2 K_{lj} K_{ij} m_i L_iL_j
+ 2 m_l \sum_i \sum_j K_{ij}^2 K_{il} K_{jl} m_i m_j L_i L_j
\\
&
+2m_l \sum_{i<j} \sum_k K_{ik} K_{kj} K_{jl} K_{li} L_i L_j L_k
- m_l L_l \sum_i \sum_j K_{il}^2 K_{jl}^2 L_i L_j \, .
\end{split}
\end{equation}
This expression can be simplified considerably, yielding
\begin{equation}
\begin{split}
h^{(4)}_l(\{c\}) =
&\phantom{+}
\frac{1}{3} \sum_{j (\neq l)} m_l \derivmb{J_{lj}}{3} K_{lj} (1-m_j^2)
- m_l \sum_{j (\neq l)} \sum_k K_{lk}^2 K_{kj}^2 L_k^2 L_j 
-2 m_l \sum_{i<j} \sum_k K_{ik} K_{kj} K_{jl} K_{li} L_i L_j L_k
\\
&
+ \frac{1}{3} m_l \sum_{i (\neq l)} K_{il}^4 (1-3m_l^2 +3 m_i^2 - 9m_i^2m_l^2) L_i
\, .
\end{split}
\end{equation}
Using our result for $\derivmb{J_{ij}}{3}$, we get
\begin{equation}
\begin{split}
h^{(4)}_l(\{c\}) =
&\phantom{+}
\frac{2}{3} m_l \sum_{j (\neq l)} K_{lj}^4 \left[1 + 3m_l^2 +3m_j^2 + 9 m_l^2m_j^2\right] L_j
+2 m_l \sum_{j (\neq l)} \sum_{k\, (\neq l,\,\neq j)} K_{lj}^2
(K_{jk}^2 L_j +K_{kl}^2 L_l) L_j L_k
\\
&
+ 2 m_l \sum_{j (\neq l)} \sum_{k, m (m, j, k, l \text{ distincts})} K_{lj} K_{jk} K_{km} K_{ml} L_j L_k L_m
- m_l \sum_{j (\neq l)} \sum_k K_{lk}^2 K_{kj}^2 L_k^2 L_j 
\\
&
-2 m_l \sum_{i<j} \sum_k K_{ik} K_{kj} K_{jl} K_{li} L_i L_j L_k
+ \frac{1}{3} m_l \sum_{i (\neq l)} K_{il}^4 (1-3m_l^2 +3 m_i^2 - 9m_i^2m_l^2) L_i
\, .
\end{split}
\end{equation}
Finally, using the identity
\begin{eqnarray}
2 m_l \sum_{j (\neq l)} \sum_{k, m (m, j, k, l \text{ distincts})} K_{lj} K_{jk} K_{km} K_{ml} L_j L_k L_m
&=& 4 m_l \sum_{i<j} \sum_k K_{ik} K_{kj} K_{jl} K_{li} L_i L_j L_k
- 2m_l L_l \sum_j\sum_{k (\neq l)} K_{lj}^2 K_{kj}^2 L_i L_j \, ,
\end{eqnarray}
we get to our final result:
\begin{eqnarray}
h^{(4)}_l(\{c\}) &=&
m_l \sum_{j (\neq l)} K_{lj}^4 \left[1 + m_l^2 +3m_j^2 + 3 m_l^2m_j^2\right]L_j
+ m_l \sum_{j (\neq l)} \sum_k K_{lk}^2 K_{kj}^2 L_k^2 L_j
\nonumber\\
&&
+2 m_l \sum_{i<j} \sum_k K_{ik} K_{kj} K_{jl} K_{li} L_i L_j L_k
- 2 m_l L_l \sum_i \sum_l K_{li}^2 K_{lj}^2 L_i L_j
\nonumber\\
&&
+ 2 m_l L_l \sum_{j<k} K_{lk}^2 K_{lj}^2 L_j L_k
\nonumber\\
&=&2 m_l \sum_{i<j} \sum_k K_{ik} K_{kj} K_{jl} K_{li} L_i L_j L_k
+m_l \sum_{j (\neq l)} K_{lj}^4 \left[1 + m_l^2 +3m_j^2 +3 m_l^2m_j^2\right]L_j
+ m_l \sum_{j (\neq l)} \sum_k K_{lk}^2 K_{kj}^2 L_k^2 L_j  \, .
\end{eqnarray}
That confirms, once again, the result found on Eq.~(\ref{eq_hi}).

\end{landscape}

\chapter[Evaluation of $\protect\log Z$]{Evaluation of log \textit{Z}}
\label{ap_logZ}

Using Eq.~(\ref{eq_4411}) and doing an integral transform, we get
\begin{equation}
\begin{split}
\log Z  = \int \prod_{\mu=1}^p \frac{\di m_\mu}{\sqrt{2 \pi}} \sum_{\{\sigma\}}
 \exp \Bigg\{& -\frac{1}{2} \sum_{\mu=1}^p m_\mu^2 + \sum_i h_i \sigma_i
 \\
 &+\frac{1}{\sqrt{N}} \sum_\mu \sum_i m_\mu \xi_i^\mu \left(\sigma_i - \tanh h_i\right)\Bigg\} \, ,
\end{split}
\nonumber
\end{equation}
\begin{equation}
\begin{split}
\phantom{\log Z}  = \int \prod_{\mu=1}^p \frac{\di m_\mu}{\sqrt{2 \pi}} 
 \exp \Bigg\{ &-\frac{1}{2} \sum_{\mu=1}^p m_\mu^2
 -\frac{1}{\sqrt{N}} \sum_\mu \sum_i m_\mu \xi_i^\mu \tanh h_i 
  \\
 &
+\sum_i  \log\left[2 \cosh\left(\frac{1}{\sqrt{N}} \sum_{\mu=1}^p m_\mu \xi_i^\mu +  h_i\right)\right]\Bigg\} \, ,
\end{split}
\end{equation}
which is just Eq.~(\ref{eq_hop_int}) supposing that the magnetizations $m_\mu$
are $O(1/\sqrt{N})$. Doing a Taylor expansion of this equation for large $N$,
we have
\begin{eqnarray}
Z &=& \exp \left[\sum_i \log( 2 \cosh h_i)\right] 
\int \prod_\mu 
 \frac{\di m_\mu}{\sqrt{2 \pi}} \exp\left[
 -\frac{1}{2} \sum_\mu \chi_\mu  m_\mu^2
 \right.
\nonumber\\
&&
\left.
+ \frac{1}{2\sqrt{N}} \sum_{\mu \neq \nu} s_{\mu \nu} m_\mu m_\nu
 \right.
\nonumber\\
&&
\left.
-\frac{1}{12 N} \frac{1}{N} \sum_i \left(\sum_\mu m_\mu \xi_i^\mu \right)^4
(1 - 3 \tanh^2 h_i)(1- \tanh^2 h_i) 
 \right.
\nonumber\\
&&
+ O(1/N^2) 
\Bigg] \, ,
\end{eqnarray}
where
\begin{eqnarray}
\chi_\mu &=& 1 - \frac{1}{N} \sum_i (\xi_i^\mu)^2 (1-\tanh^2 h_i) \,,
\end{eqnarray}
and
\begin{eqnarray}
s_{\mu \nu} &=& \frac{1}{\sqrt{N}} \sum_i \xi_i^\mu \xi_i^\nu (1-\tanh^2 h_i)\, .
\end{eqnarray}

We use the expansion of $e^{x} = 1 + x + \cdots$ to write this equation as a
gaussian integral
\begin{eqnarray}
Z &=& \exp \left[\sum_i \log( 2 \cosh h_i)\right]
\int \prod_\mu 
 \frac{\di m_\mu}{\sqrt{2 \pi}} 
\left[1+
\frac{1}{2\sqrt{N}} \sum_{\mu \neq \nu} s_{\mu \nu} m_\mu m_\nu
 \right.
\nonumber\\
&&
\left.
+ \frac{1}{8 N}\left( \sum_{\mu \neq \nu} s_{\mu \nu} m_\mu m_\nu\right)^2
\right. \nonumber
\\
&&
\left.
-\frac{1}{12 N} \frac{1}{N} \sum_i \left(\sum_\mu m_\mu \xi_i^\mu \right)^4
(1 - 3 \tanh^2 h_i)(1- \tanh^2 h_i) + O(1/N^{3/2}) 
\right]
\nonumber \\
&&
\cdot
\exp\left[ -\frac{1}{2} \sum_\mu \chi_\mu  m_\mu^2 
\right] \, ,
\end{eqnarray}
which can be rewritten as averages in respect to a gaussian distribution
\begin{eqnarray}
Z &=& \exp \left[\sum_i \log( 2 \cosh h_i) - 
\frac{1}{2} \sum_\mu \log \chi_\mu\right] \cdot
\nonumber \\
&&
\left\langle 1
+ \frac{1}{4 N}\sum_{\mu \neq \nu} s_{\mu \nu}^2 m_\mu^2 m_\nu^2
\right.\nonumber\\
&&
\left.
-\frac{1}{12 N} \frac{1}{N} \sum_i \left(\sum_\mu m_\mu \xi_i^\mu \right)^4
(1 - 3 \tanh^2 h_i)(1- \tanh^2 h_i) 
\right.\nonumber\\
&&
+ O(1/N^{3/2}) \Bigg\rangle_m \, .
\end{eqnarray}

Those averages can be easily calculated:
\begin{equation}
\left\langle \frac{1}{4 N}\sum_{\mu \neq \nu} s_{\mu \nu}^2 m_\mu^2 m_\nu^2
\right\rangle_m = \frac{1}{4 N} \sum_{\mu \neq \nu} \frac{s_{\mu \nu}^2}{\chi_\mu \chi_\nu} \, ,
\end{equation}

\begin{eqnarray}
\left\langle \left(\sum_\mu m_\mu \xi_i^\mu \right)^4
\right\rangle_m &=& \left\langle \sum_\mu m_\mu^4 (\xi_i^\mu)^4\right\rangle_m +
3 \left\langle \sum_{\mu \neq \nu} m_\mu^2 m_\nu^2 (\xi_i^\mu)^2 (\xi_i^\nu)^2\right\rangle_m + \text{odd terms} =
\nonumber\\
&=& 3 \sum_\mu \frac{(\xi_i^\mu)^4}{\chi_\mu^2}
+ 3 \sum_{\mu \neq \nu} \frac{ (\xi_i^\mu)^2 (\xi_i^\nu)^2}{\chi_\mu \chi_\nu} \, .
\end{eqnarray}

Finally, we pose
\begin{eqnarray}
r_{\mu \nu} &=& \frac{1}{N} 
\sum_i (\xi_i^\mu)^2 (\xi_i^\nu)^2 (1 - 3 \tanh^2 h_i)(1- \tanh^2 h_i) \, ,
\end{eqnarray}
and
\begin{eqnarray}
s_{\mu \mu} &=& 0 \, .
\end{eqnarray}
which gives an explicit form for $\log Z$:
\begin{eqnarray}
\log Z = 
\sum_i \log( 2 \cosh h_i) - 
\frac{1}{2} \sum_\mu \log \chi_\mu
+ \frac{1}{4 N}\sum_{\mu, \nu} \frac{(1-\delta_{\mu \nu}) s_{\mu \nu}^2 - r_{\mu\nu}}{\chi_\mu \chi_\nu}a
+ O(1/N^{3/2}) \, , \nonumber\\
\end{eqnarray}
where $\delta_{\mu \nu}$ is the Kronecker symbol.

\chapter{Evaluation of \textit{m} for the entropy of the Ising model}
\label{ap_m}

Starting with Eq.~(\ref{eq_m_isi}):
\begin{equation}
  m_l = \frac{1}{N} \sum_i \sigma_i^l \tanh \left(\beta \sum_s m_s \sigma_i^s\right) \, ,
\end{equation}
which under the hypothesis of $m_l=m$ can be rewritten as
\begin{eqnarray}
m &=& 
\moy{\frac{1}{N} \sum_i \sigma_i^l \tanh \left(\beta \sum_s m_s \sigma_i^s\right)} \, ,
\nonumber\\
&=& \sum_{\{\sigma\}} \sigma^1 \tanh\left(\beta m \sum_l \sigma^l  \right)
\prod_{l=1}^L \frac{e^{\beta m^* \sigma^l \tilde{\xi}}}{2 \cosh (\beta m^*)} \, .
\end{eqnarray}

This last equation can be simplified using the variable change
$\sigma^l \rightarrow \sigma^l \tilde{\xi}$:

\begin{eqnarray}
m &=& \sum_{\{\sigma\}} \sigma^1 \tanh\left(\beta m \sum_l \sigma^l  \right)
\prod_{l=1}^L \frac{e^{\beta m^* \sigma^l}}{2 \cosh (\beta m^*)} \\
&=& \frac{1}{L} \sum_{\{\sigma\}} \left(\sum_l \sigma^l\right) \tanh\left(\beta m \sum_l \sigma^l  \right)
\prod_{l=1}^L \frac{e^{\beta m^* \sigma^l}}{2 \cosh (\beta m^*)} \\
&=& \frac{1}{2L \left[2 \cosh(\beta m^*)\right]^L} 
\left[ \sum_{\{\sigma\}} \left(\sum_l \sigma^l\right) 
\tanh\left(\beta m \sum_l \sigma^l  \right)
 \exp\left(\beta m^* \sum_{l=1}^L \sigma^l\right)  \right.\nonumber\\
&&+ \left. \sum_{\{\sigma\}} \left(\sum_l \sigma^l\right) \tanh\left(\beta m \sum_l \sigma^l  \right)
 \exp\left(\beta m^* \sum_{l=1}^L \sigma^l\right)  \right] \, .
\end{eqnarray}
Using the global symmetry $\sigma^l \rightarrow - \sigma^l$:
\begin{eqnarray*}
m &=& \frac{1}{L \left[2 \cosh(\beta m^*)\right]^L}  \sum_{\{\sigma\}} \left(\sum_l \sigma^l\right) 
\tanh\left(\beta m \sum_l \sigma^l  \right)
\cosh \left(\beta m^* \sum_l^L \sigma^l \right)  \, .
\end{eqnarray*}
We verify the solution of this equation is $m= m^*$:
\begin{eqnarray}
m &=& \frac{1}{L \left[2 \cosh(\beta m)\right]^L}  \sum_{\{\sigma\}} \left(\sum_l \sigma^l\right) 
\sinh\left(\beta m \sum_l \sigma^l  \right) \nonumber \\
&=& \frac{1}{\beta L \left[2 \cosh(\beta m)\right]^L} 
\frac{\partial}{\partial m} 
\sum_{\{\sigma\}} \cosh\left(\beta m \sum_l \sigma^l  \right) \nonumber\\
&=& \frac{1}{\beta L \left[2 \cosh(\beta m)\right]^L} 
\frac{\partial}{\partial m} \left[2 \cosh(\beta m)\right]^L \nonumber \\
&=& \tanh(\beta m) \, .
\end{eqnarray}
From which follows the result, since $m^*$ is defined as the solution of
$m^* = \tanh (\beta m^*)$.

\chapter{Evaluation of \textit{\~{N}} in the paramagnetic phase}
\label{ap_entro_para}

We start with Eq.~(\ref{eq_720}):

\begin{eqnarray}
  \tilde{N} &=& \int \prod_{l=1}^L \frac{\di m_l}{\sqrt{2 \pi \beta^{-1} N^{-1}}}
  \sum_{\{\xi\}}
  \exp\left[ -\frac{\beta N}{2} \sum_{l=1}^L m_l^2 +
 \right.\nonumber\\
&&+\left.
   \beta \sum_{l=1}^L m_l \sum_{i} \sigma_i^l \xi_i 
   \right] \, ,
  \end{eqnarray}
which after averaging yields
\begin{eqnarray}
\moy{\tilde{N}^n} &=& e^{-\beta L n/2} \sum_{\{\xi\}} \int \prod_{l=1}^L \prod_{\rho=1}^n
\frac{\di m_l^\rho}{\sqrt{2 \pi}} \times \nonumber\\
&&\times \moy{\exp \left\{\beta N \left[
- \frac{1}{2} \sum_{l, \rho} \left(m^\rho_l\right)^2 + 
\sum_{l,\rho} m_l^\rho \frac{1}{N} \sum_i \xi_i^\rho \sigma_i^l \right] \right\}}
\nonumber \\
 &=& e^{-\beta L n/2} \sum_{\{\xi\}} \int \prod_{l=1}^L \prod_{\rho=1}^n
\frac{\di m_l^\rho}{\sqrt{2 \pi}} \sum_{\{\tilde{\xi}\},\{\sigma\}} \exp \left[
- \frac{\beta N}{2} \sum_{l, \rho} \left(m^\rho_l\right)^2 
+\right.
\nonumber\\
&&+ \left. 
 \beta N \sum_{l,\rho} m_l^\rho \frac{1}{N} \sum_i \xi_i^\rho \sigma_i^l  
+ \frac{\tilde{\beta}}{N} \sum_l \sum_{i<j} \sigma_i^l \sigma_j^l \tilde{\xi}_i \tilde{\xi}_j\right] \, .
\end{eqnarray}

Doing an integral transformation and making the sum over $\sigma$, we obtain
\begin{eqnarray}
\moy{\tilde{N}^n} &=& e^{-\beta L n/2} \sum_{\{\xi\},\{\tilde{\xi}\}} \int \prod_{l, \rho}
\frac{\di m_l^\rho}{\sqrt{2 \pi}}  \frac{\di \tilde{m}_l}{\sqrt{2 \pi} }
 \exp \left[
- \frac{\beta N}{2} \sum_{l, \rho} \left(m^\rho_l\right)^2 
 \right.\nonumber\\
&& \left. 
- \frac{\beta N}{2} \sum_{l} \left(\tilde{m}_l\right)^2
+\sum_{i,l} \ln 2 \cosh \left(\beta \sum_\rho m^\rho_l \xi_i^\rho + 
\tilde{\beta} \tilde{m}_l \tilde{\xi}_i\right) \right] \, .
\end{eqnarray}
Applying the variable change $\frac{\beta}{N} m^2 \rightarrow m^2$,
$\frac{\tilde{\beta}}{N} \tilde{m}^2 \rightarrow \tilde{m}^2$, we have
\begin{eqnarray}
\moy{\tilde{N}^n} &=& e^{-\beta L n/2}
 \sum_{\{\xi\},\{\tilde{\xi}\}} \int \prod_{l, \rho}
\frac{\di m_l^\rho}{\sqrt{2 \pi}}  \frac{\di \tilde{m}_l}{\sqrt{2 \pi}}
 \exp \left[
- \frac{1}{2} \sum_{l, \rho} \left(m^\rho_l\right)^2 
- \frac{1}{2} \sum_{l} \left(\tilde{m}_l\right)^2 
 \right. \nonumber\\
&& \left. 
+\sum_{i,l} 
\ln 2 \cosh \left(\sqrt{\frac{\beta}{N}} \sum_\rho m^\rho_l \xi_i^\rho+ 
\sqrt{\frac{\tilde{\beta}}{N}} \tilde{m}_l \tilde{\xi}_i\right) \right] \, .
\end{eqnarray}

Since that for \emph{small} $x$,
$\ln\left[ 2 \cosh(x)\right] = \ln(2) + \frac{x^2}{2} + O(x^4)$, we can
approximate this expression by
\begin{eqnarray}
\moy{\tilde{N}^n} &\simeq& e^{-\beta L n/2} \sum_{\{\xi\},\{\tilde{\xi}\}} \int \prod_{l, \rho}
\frac{\di m_l^\rho}{\sqrt{2 \pi}}  \frac{\di \tilde{m}_l}{\sqrt{2 \pi}}
 \exp \left[
- \frac{1}{2} \sum_{l, \rho} \left(m^\rho_l\right)^2 
- \frac{1}{2} \sum_{l} \left(\tilde{m}_l\right)^2 + \right. \nonumber\\
&&+ \left. L N \ln 2 + 
\sum_{i,l,\rho,\alpha} m^\rho_l m^\alpha_l \xi_i^\rho \xi_i^\alpha +
\frac{\tilde{\beta}}{2} \sum_l (\tilde{m}_l)^2 + 
\frac{\sqrt{\beta \tilde{\beta}}}{N} \sum_{l, \rho, i} m^\rho_l \tilde{m}_l 
\xi_i^\rho \tilde{\xi}_i 
\right] \nonumber\\
&\simeq& \sum_{\{\xi\},\{\tilde{\xi}\}} \left[ \det M \right]^{-L/2} e^{-\beta L n/2} \, ,
\end{eqnarray}
where $M$ is the matrix
\begin{equation}
\left(
\begin{array}{ccccc}
1- \tilde{\beta} & 
- \frac{\sqrt{\beta \tilde{\beta}}}{N} \sum_i \xi_i^1 \tilde{\xi}_i &
- \frac{\sqrt{\beta \tilde{\beta}}}{N} \sum_i \xi_i^2 \tilde{\xi}_i &
\cdots &
- \frac{\sqrt{\beta \tilde{\beta}}}{N} \sum_i \xi_i^n \tilde{\xi}_i \\
- \frac{\sqrt{\beta \tilde{\beta}}}{N} \sum_i \xi_i^1 \tilde{\xi}_i &
1 - \beta &
- \frac{\beta}{N} \sum_i \xi_i^1 \xi_i^2 &
\cdots &
- \frac{\beta}{N} \sum_i \xi_i^1 \xi_i^n \\
- \frac{\sqrt{\beta \tilde{\beta}}}{N} \sum_i \xi_i^2 \tilde{\xi}_i &
- \frac{\beta}{N} \sum_i \xi_i^1 \xi_i^2 &
1 - \beta &
\cdots &
- \frac{\beta}{N} \sum_i \xi_i^2 \xi_i^n \\
\vdots \raisebox{-3ex}{} & \vdots & \vdots & \multicolumn{1}{l}{\hfill \raisebox{.4ex}{.} \hfill . \hfill \raisebox{-.4ex}{.}} \hfill  & \vdots \\
- \frac{\sqrt{\beta \tilde{\beta}}}{N} \sum_i \xi_i^n \tilde{\xi}_i &
- \frac{\beta}{N} \sum_i \xi_i^1 \xi_i^n &
- \frac{\beta}{N} \sum_i \xi_i^2 \xi_i^n &
\cdots &
1-\beta
\end{array}
\right) \, .
\end{equation}

Fixing $q_{\rho \sigma} = \frac{1}{N}\sum_i \xi_i^\rho \xi_i^\sigma$ and
$t_\rho = \frac{1}{N}\sum_i \xi_i^\rho \tilde{\xi}_i$ with the Lagrange
multipliers $\hat{q}_{\rho \sigma}$ and $\hat{t}_\rho$, we have
\begin{eqnarray}
\moy{\tilde{N}^n} &=& \int \prod_{\rho <\sigma} \di q_{\rho \sigma} 
\di \hat{q}_{\rho \sigma} \di t_\rho \di \hat{t}_\rho 
\sum_{\{\xi\},\{\tilde{\xi}\}} \exp N \cdot \left[
- \frac{\alpha}{2} \log \det M 
-\frac{\alpha}{2}  \sum_{\rho<\sigma} \hat{q}_{\rho \sigma} q_{\rho \sigma}
\right. \nonumber \\
&&- \left. 
\frac{\alpha}{2} \sum_\rho \hat{t}_\rho t_\rho
+\frac{\alpha }{2} \sum_{\rho < \sigma} \frac{1}{N}
\sum_i \hat{q}_{\rho \sigma}  \xi_i^\rho \xi_i^\sigma +
\frac{\alpha}{2} \sum_\rho \hat{t}_\rho \frac{1}{N} 
\tilde{\xi}_i \xi^\rho_i - \frac{\alpha \beta n}{2}
\right] \, .
\label{eq_bla12}
\end{eqnarray}
Since now the sites are completely uncorrelated, we can write this expression
in a form solvable by the saddle-point method:

\begin{eqnarray}
\moy{\tilde{N}^n} &=& \int \prod_{\rho < \sigma} \di q_{\rho \sigma} 
\di \hat{q}_{\rho \sigma} \di t_\rho \di \hat{t}_\rho\, H^N \, ,
\end{eqnarray}
where $H$ is given by
\begin{eqnarray}
H &=& \sum_{\{\xi\},\{\tilde{\xi}\}} \exp \left[
- \frac{\alpha}{2} \log \det M 
-\alpha  \sum_{\rho<\sigma} \hat{q}_{\rho \sigma} q_{\rho \sigma}
\right. \nonumber \\
&&- \left. 
\frac{\alpha}{2} \sum_\rho \hat{t}_\rho t_\rho
+\alpha \sum_{\rho < \sigma}  \hat{q}_{\rho \sigma}  \xi^\rho \xi^\sigma +
\frac{\alpha}{2} \sum_\rho \hat{t}_\rho 
\tilde{\xi} \xi^\rho - \frac{\alpha \beta n}{2}
\right] \, .
\end{eqnarray}

We will look to the replica-symmetric saddle point of $H$.
Posing $q_{\rho \sigma} = q$, $t_\rho=t$, $\hat{q}_{\rho \sigma} = \hat{q}$ and
$\hat{t}_\rho=\hat{t}$, we have
\begin{eqnarray}
H &=& 
\sum_{\{\xi\},\{\tilde{\xi}\}} \exp \left[
- \frac{\alpha}{2} \log \det M 
-\frac{\alpha n (n-1)}{2}   \hat{q} q
\right. \nonumber \\
&&- \left. 
\frac{\alpha n}{2} \hat{t} t
+\frac{\alpha \hat{q}}{2} \sum_{\rho < \sigma} \xi^\rho \xi^\sigma +
\frac{\alpha \hat{t}}{2} \sum_\rho
\tilde{\xi} \xi^\rho - \frac{\alpha \beta n}{2}
\right] \, , \nonumber \\
&=& \int_{-\infty}^\infty \frac{\di z}{\sqrt{2\pi}} 
\sum_{\{\xi\},\{\tilde{\xi}\}} 
\exp \left[ -\frac{z^2}{2}
- \frac{\alpha}{2} \log \det M 
-\frac{\alpha n (n-1)}{4}   \hat{q} q
\right. \nonumber \\
&&- \left. 
\frac{\alpha n}{2} \hat{t} t
+z\sqrt{\alpha \hat{q}} \sum_\rho \xi^\rho  +
\frac{\alpha \hat{t}}{2} \sum_\rho
\tilde{\xi} \xi^\rho - \frac{\alpha \beta n}{2}
\right] \, .
\end{eqnarray}
Since we can do a variable change $\xi^\rho \rightarrow \xi^\rho \tilde{\xi}$,
this expression evaluates to
\begin{eqnarray}
H &=& 
\int_{-\infty}^\infty \frac{\di z}{\sqrt{2\pi}} 
\exp \left\{ -\frac{z^2}{2}
- \frac{\alpha}{2} \log \det M 
-\frac{\alpha n (n-1)}{2}   \hat{q} q
\right. \nonumber \\
&&- \left. 
\frac{\alpha n}{2} \hat{t} t
+n\log\left[2\cosh\left(
\frac{\alpha \hat{t}}{2}
+z\sqrt{\alpha \hat{q}}  
 \right)\right]
- \frac{\alpha \beta n}{2}
\right\} \, .
\end{eqnarray}
Now we need to evaluate explicitly $\det M$ in the replica-symmetric 
hypothesis:
\begin{eqnarray}
\det M &=& 
\left(1-\beta + \beta q\right)^{n-1} \times\nonumber\\
&&\times \left[(1-\tilde{\beta})(1-\beta) 
- (n-1) (1-\tilde{\beta}) \beta q - n \beta \tilde{\beta}t^2\right] \, , \\
\log (\det M) &=& \log(1-\tilde{\beta}) + n\log\left[1-\beta(1- q)\right] 
\nonumber \\
&&- \frac{n \beta}{1-(1- q)\beta}\left(\frac{ t^2 \tilde{\beta} }{1-\tilde{\beta}}+q
\right) + O(n^2) \, .
\end{eqnarray}
After a small calculation, we find the saddle-point equations
\begin{eqnarray}
t &=& \moy{\tanh\left(\frac{\alpha}{2} \hat{t} + z \sqrt{\alpha \hat{q}}  \right)}  \, ,\\
q &=& \moy{\left[\tanh\left(\frac{\alpha}{2} \hat{t} +  z \sqrt{\alpha \hat{q}}  \right)\right]^2} \, ,\\
\hat{t} &=& \frac{2 t \beta^2}{(1-\beta)[1-(1-q)\beta]} \, ,\\
\hat{q} &=& \beta^2 \frac{q (1 - \beta) + t^2 \beta}{(1-\beta)\left[1 - (1-q)\beta\right]^2} \, ,
\end{eqnarray}
where the average $\moy{\cdot}$ is in respect to the gaussian 
variable $z$ of zero mean and standard deviation $\sigma=1$. 

Finally, the entropy can be written as
\begin{eqnarray}
\moy{S} &=& -\frac{\alpha \beta (1-q) 
\left\{1 - \beta\left[2 - 2 q(1-\beta) - (1+t^2)\beta\right]\right\}}
{2 (1-\beta) \left[ 1- (1-q) \beta\right]^2}
- \frac{\alpha}{2} \ln \left[1-(1-q)\beta\right] \nonumber\\
&&+\moy{\ln \left[ 2 \cosh \left( \frac{\alpha}{2} \hat{t} + z \sqrt{\alpha \hat{q}} \right) \right] }
- \frac{\alpha}{2} \left[(1-q)\hat{q} + t \hat{t}\right]  \, .
\end{eqnarray}

\chapter{Entropy calculations details for continuous patterns}
\label{ap_continu}

Our starting points are Eqs.~(\ref{eqn123}) and~(\ref{eq_seila1123}):
\begin{equation}
\begin{split}
\tilde{N}[\{\sigma^l\},\beta] = \int \di x &\int \di m \int \prod_i \di\xi_i 
\exp 
\Bigg(\frac{\beta}{N} \sum_{l=0}^L \sum_{i<j} \xi_i \xi_j \sigma_i^l  \sigma_j^l
\\
&
+ \frac{\beta L N}{2} m^2
- \beta L \sum_i \log\left[ 2 \cosh (m \xi_i)\right]
\\
&
+ \beta \log P_0(\{\xi_i\})
-i N x m
+ i \sum_i \xi_i \tanh(m \xi_i)
\Bigg) \, .
\end{split}
\end{equation}

In the same way as with continuous patterns, we will use the replica trick
to evaluate $\log \tilde{N}$
\begin{equation}
\begin{split}
\tilde{N}\left[\{\sigma^l\}, \beta\right]^n =
\int& \prod_\nu \di x^\nu \int \prod_\nu \di m^\nu
\int \prod_{\nu=1}^n \prod_i 
\di \xi_i^\nu \exp\Bigg\{ \\
&\phantom{+}\frac{\beta}{N} \sum_{i<j} \sum_{\nu=1}^n \sum_{l=1}^L 
\xi_i^\nu \xi_j^\nu \sigma_i^l  \sigma_j^l + \beta \sum_{\nu=1}^n \log P_0(\{\xi^\nu_i\})
 \\
&+ \frac{\beta L N}{2} \sum_{\nu=1}^n (m^\nu)^2 
- \beta L \sum_{\nu=1}^n \sum_i \log\left[2 \cosh (m^\nu \xi_i^\nu)\right]
\\
&-i N \sum_\nu x^\nu m^\nu
+ i \sum_\nu \sum_i \xi_i^\nu \tanh(m^\nu \xi_i^\nu)
\Bigg\} \, ,
\end{split}
\end{equation}
and we will average the entropy with respect to all possible realizations of
the measurements taken from a system where the patterns are given by 
$\{\tilde{\xi_i}\}$:
\begin{equation}
\begin{split}
\moy{\tilde{N}\left[\{\sigma^l\}, \beta\right]^n} = 
 \sum_{\{\sigma\}} \frac{1}{Z(\{\tilde{\xi}_i\})^L}
\tilde{N}[\{\sigma^l\}, \beta]^n \exp\left[
\frac{1}{N} \sum_{l=1}^L \sum_{i<j} \tilde{\xi}_i \tilde{\xi}_j \sigma_i^l \sigma_j^l
\right]\, .
\end{split}
\end{equation}
Finally we average in respect to the underlying patterns $\{\tilde{\xi}_i\}$
\begin{equation}
\begin{split}
\moy{\tilde{N}\left[\{\sigma\}, \beta\right]^n} &=  \\
=\int \prod_i \di& \tilde{\xi}_i \sum_{\{\sigma\}} \frac{1}{Z(\{\tilde{\xi}_i\})^L}
\tilde{N}[\{\sigma^l\}, \beta]^n \exp\left[
\frac{1}{N} \sum_{l=1}^L \sum_{i<j} \tilde{\xi}_i \tilde{\xi}_j \sigma_i^l \sigma_j^l
+ \log P_0(\tilde{\xi_i})
\right]
\end{split}
\end{equation}
Note that we used the same function $P_0(\{\xi_i\})$ both as the prior and as the
distribution of the ``real'' patterns $\{\tilde{\xi}_i\}$. It is equivalent to say
that one knows the statistical distribution of $\{\tilde{\xi}_i\}$, but wants to 
infer one particular realization of this distribution.

Writing the average explicitly and doing a gaussian transform, we have
\begin{eqnarray}
\moy{\tilde{N}\left[\{\sigma\}, \beta\right]^n}
&=& \int \left(
 \prod_{\nu=1}^n \prod_i \di \xi_i^\nu 
\right)
\left(
\prod_i d \tilde{\xi}_i
\right)
\left(
\prod_{\nu=1}^n \prod_{l=1}^L \frac{\di Q_l^\nu}{\sqrt{2 \pi}}
\right)
\left(
\prod_{l=1}^L \frac{\di \tilde{Q}_l}{\sqrt{2 \pi}}
\right)
\times \nonumber\\
&&\times 
\left(
\prod_{\nu=1}^n \di m^\mu 
\right)
\left(
\prod_{\nu=1}^n \frac{\di x^\mu}{\sqrt{2 \pi}} 
\right)
\frac{\di y}{\sqrt{2 \pi}} \di \tilde{m} \sum_{\{\sigma\}} \exp(H) \, ,
\end{eqnarray}
where
\begin{eqnarray}
H &=&  - \frac{\beta N}{2} \sum_{\nu=1}^n \sum_{l=1}^L (Q_l^\nu)^2
- \frac{N}{2} \sum_{l=1}^L (\tilde{Q}_l)^2 
+ \beta \sum_{\nu=1}^n \sum_{l=1}^L \sum_i  Q_l^\nu \xi_i^\nu \sigma_i^l
\nonumber\\
&&
+  \sum_{l=1}^L \sum_i \tilde{Q}_l \tilde{\xi}_i \sigma_i^l 
+ \beta \sum_{\nu=1}^n \log P(\xi^\nu)
+  \log P(\tilde{\xi}) 
\nonumber\\
&&
+ \frac{\beta L N}{2} \sum_{\nu=1}^n (m^\nu)^2
+ \frac{LN}{2} \tilde{m}^2  
- \beta L \sum_{\nu=1}^n \sum_i \log\left[2 \cosh (m^\nu \xi_i^\nu)\right]
\nonumber\\
&&
- L \sum_i \log\left[2 \cosh(\tilde{m} \tilde{\xi}_i)\right]
- i N \sum_{\nu=1}^n x^\nu m^\nu 
- i N y \tilde{m} 
\nonumber\\
&&
+ i \sum_{\nu=1}^n \sum_i x^\nu \xi_i^\nu \tanh(m^\nu \xi_i^\nu)
+ i y \sum_i \tilde{\xi}_i \tanh (\tilde{m} \tilde{\xi}_i) \, .
\end{eqnarray}

We now suppose that the prior probability of the patterns are independent and
identically distributed across the
sites. Mathematically, that means that  $P_0(\{\xi_i\}) = \prod_i p_0 (\xi_i)$. Under 
that supposition we can make the sites decoupled, so we have
\begin{eqnarray}
\moy{\tilde{N}\left[\{\sigma\}, \beta\right]^n}
&=& \int \left(
\prod_{\nu=1}^n \prod_{l=1}^L \frac{\di Q_l^\nu}{\sqrt{2 \pi}}
\right)
\left(
\prod_{l=1}^L \frac{\di \tilde{Q}_l}{\sqrt{2 \pi}}
\right)
\left(
\prod_{\nu=1}^n \di m^\mu 
\right)
\times \nonumber\\
&&\times
\left(
\prod_{\nu=1}^n \frac{\di x^\mu}{\sqrt{2 \pi}} 
\right)
\frac{\di y}{\sqrt{2 \pi}} \di \tilde{m} A^N \exp(N B) \, ,
\end{eqnarray}
where
\begin{eqnarray}
B &=&- \frac{\beta}{2} \sum_{\nu=1}^n \sum_{l=1}^L (Q_l^\nu)^2
- \frac{1}{2} \sum_{l=1}^L (\tilde{Q}_l)^2 
+\frac{\beta L}{2} \sum_{\nu=1}^n (m^\nu)^2
\nonumber\\
&&
+ \frac{L}{2} \tilde{m}^2 
 -i \sum_{\nu=1}^n x^\nu m^\nu - i y \tilde{m} \, ,\\
A &=& \int  \di \tilde{\xi}  p_0(\tilde{\xi}) \sum_{\{\sigma^l\}} 
\exp \Bigg[\sum_{l=1}^L \tilde{Q}_l \tilde{\xi} \sigma^l 
- L \log(2 \cosh(\tilde{m} \tilde{\xi})) 
\nonumber\\
&&
+ i y \tilde{\xi} \tanh(\tilde{m}\tilde{\xi}) 
+ \sum_{\nu=1}^n \log C_\nu
\Bigg] \, ,\\
C_\nu &=& \int \di \xi \exp \Bigg[
\beta \sum_{l=1}^L Q_l^\nu \xi \sigma^l + \beta \log p_0(\xi) 
\nonumber\\
&&
- \beta L \log(2 \cosh (m^\nu\xi)) 
+i x^\nu \xi \tanh(m^\nu \xi)
\Bigg] \, .
\end{eqnarray}

Like we did previously, we look for the replica-symmetric saddle-point of this
integral
\begin{eqnarray}
\tilde{Q}_l &=& \tilde{m} = \int \di \tilde{\xi} p_0(\tilde{\xi}) \tilde{\xi}
\tanh(\tilde{m} \tilde{\xi}) = \moy{\tilde{\xi} \tanh(\tilde{m}\tilde{\xi})}
\, ,\\
m &=& \int \di \tilde{\xi} \sum_{\{\sigma\}} \exp \left[
\sum_{l=1}^L \tilde{m} \tilde{\xi} \sigma^l - L \log 2 \cosh(\tilde{m} \tilde{\xi})
\right] \times \nonumber\\
&&\times \frac{
\int \di \xi p_0(\xi) \xi \tanh(m \xi) \exp \left[\sum_{l=1}^L Q^l \sigma^l \xi 
- L \log 2 \cosh (m \xi)\right]
}{
\int \di \xi p_0(\xi) \exp \left[\sum_{l=1}^L Q^l \sigma^l \xi 
- L \log 2 \cosh (m \xi)\right]
} \, ,\nonumber\\
\\
Q_k &=& \int \di \tilde{\xi} \sum_{\{\sigma\}} \exp \left[
\sum_{l=1}^L \tilde{m} \tilde{\xi} \sigma^l - L \log 2 \cosh(\tilde{m} \tilde{\xi})
\right] \times\nonumber\\
&&\times \frac{
\int \di \xi p_0(\xi) \xi \sigma^l \exp \left[\sum_{l=1}^L Q^l \sigma^l \xi 
- L \log 2 \cosh (m \xi)\right]
}{
\int \di \xi p_0(\xi) \exp \left[\sum_{l=1}^L Q^l \sigma^l \xi 
- L \log 2 \cosh (m \xi)\right]
} \, .
\end{eqnarray}

Finally, we have
\begin{equation}
\begin{split}
\frac{1}{N}& \moy{\log \tilde{N}} = \frac{\beta}{2} \sum_{l=1}^L Q_l^2
+ \frac{\beta L m^2}{2}
\\
&+ \int \di \tilde{\xi} p_0(\tilde{\xi}) \sum_{\{\sigma^l\}} \exp \left[
\tilde{m} \tilde{\xi} \sum_{l=1}^L \sigma^l - L \log (2 \cosh(\tilde{m} \tilde{\xi}))
\right] \times \\
& 
\phantom{\int}
\times \log\left\{ \int \di \xi \exp \left[ 
\beta \sum_{l=1}^L Q_l \xi \sigma^l - \beta L \log(2 \cosh(m \xi)) 
+ \beta \log p_0(\xi) \right]
\right\} \, ,
\end{split}
\end{equation}
from which it is straightforward to evaluate the entropy using
Eq.~(\ref{eq_s_n_deriv}).

\def\thesection{\thechapter.\arabic{section}}
\chapter{Details of the evaluation of the entropy for the Hopfield model}
\label{ap_entrohop}

Starting with Eq.~(\ref{eqnn}), we have
\begin{eqnarray}
\moy{N^n} &=& \frac{1}{2^N} \sum_{\{\xi^1\}} \frac{1}{2^N} \sum_{\{\tilde{\xi}^2\}}
\sum_{\{\sigma\}} \sum_{\{\xi^{2,\nu}\}} \exp \Bigg\{
\frac{\beta}{N} \sum_{\nu=1}^n \sum_{l=1}^L \sum_{i,j} \sigma_i^l \sigma_j^l (\xi_i^1 \xi_j^2+
\xi_i^{2,\nu} \xi_j^{2,\nu})
\nonumber \\
&&- \frac{\beta}{\tilde{\beta}} L \sum_{\nu=1}^n
 \log Z_\text{Hop}\left[\tilde{\beta}, \{\xi^1\}, \{\xi^{2,\nu}\}\right]
+ \tilde{\beta} \sum_{l=1}^L \sum_{i,j} \sigma_i^l \sigma_j^l (\xi_i^1 \xi_j^2+
\tilde{\xi}^2_i \tilde{\xi}^2_j) \nonumber\\
&&-L  \log Z_\text{Hop}\left[\tilde{\beta}, \{\xi^1\}, \{\tilde{\xi}^2\}\right]
\Bigg\} \, , \nonumber\\
&=&
\sum_{\{\xi^1\}} \sum_{\{\tilde{\xi}^2\}}
\sum_{\{\sigma\}} \sum_{\{\xi^{2,\nu}\}} \int 
\left(\prod_l^L \frac{\di m_l^1}
{\sqrt{\frac{2 \pi}{(\beta n + \tilde{\beta})N}}} \right)
\left(\prod_{l,\nu} \frac{\di m_l^{2,\nu}}
{\sqrt{\frac{2 \pi}{\beta N}}} \right)
\left(\prod_{l}^L \frac{\di \tilde{m}_l}
{\sqrt{\frac{2 \pi}{\tilde{\beta} N}}} \right)
e^{N E_1}  \, , \nonumber \\
\end{eqnarray}
where
\begin{eqnarray}
E_1 &=& (\beta n + \tilde{\beta}) \frac{1}{N} \sum_i \sum_l \sigma_i^l \xi_i^1 m_l^1
+ \frac{\beta}{N} \sum_i \sum_l \sum_\nu \sigma_i^l \xi_i^{2,\nu} m_l^{2,\nu}
+ \frac{\tilde{\beta}}{N} \sum_i \sum_l \sigma_i^l \tilde{\xi}_i^2 \tilde{m}_l
\nonumber \\
&&- \frac{\beta n +\tilde{\beta}}{2} \sum_l (m_l^1)^2
- \frac{\beta}{2} \sum_{l,\nu} (m_l^{2,\nu})^2
- \frac{\tilde{\beta}}{2} \sum_l \tilde{m}_l^2 
- 2 \log 2 \nonumber \\
&&- \frac{L}{N}\log Z_\text{Hop}\left[\tilde{\beta}, \{\xi^1\}, \{\tilde{\xi}^2\}\right]
- \frac{\beta}{\tilde{\beta}} \frac{L}{N}\sum_\nu \log Z_\text{Hop}\left[\tilde{\beta}, \{\xi^1\}, \{\xi^{2,\nu}\}\right]
\end{eqnarray}
Making the sum over $\{\sigma\}$, we obtain

\begin{eqnarray}
\moy{N^n} &=&
\sum_{\{\xi^1\}} \sum_{\{\tilde{\xi}^2\}}
\sum_{\{\xi^{2,\nu}\}} \int 
\left(\prod_l^L \frac{\di m_l^1}
{\sqrt{2 \pi}} \right)
\left(\prod_{l,\nu} \frac{\di m_l^{2,\nu}}
{\sqrt{2 \pi}} \right)
\left(\prod_{l}^L \frac{\di \tilde{m}_l}
{\sqrt{2 \pi}} \right)
e^{N E_3}
\end{eqnarray}
where
\begin{eqnarray}
E_3 &=& \frac{L}{2N} \log[(\beta n + \tilde{\beta})N] 
+ \frac{L n}{2 N} \log(\beta N) + \frac{L}{2 N} \log(\tilde{\beta} N)
- 2 \log 2 
 \nonumber \\
&&+
\frac{1}{N} \sum_i \sum_l \log 2 \cosh \left[ 
(\beta n +\tilde{\beta}) \xi_i^1 m_l^1 + \beta \sum_\nu \xi_i^{2,\nu} m_l^{2,\nu}
+ \tilde{\beta} \tilde{\xi}_i^2 \tilde{m}_l \right] 
- \frac{\beta n + \tilde{\beta}}{2} \sum_l (m_l^1)^2 \nonumber \\
&&- \frac{\beta}{2} \sum_{l,\nu} (m_l^{2,\nu})^2 
-\frac{\tilde{\beta}}{2} \sum_l \tilde{m}_l^2
- \frac{L}{N}\log Z_\text{Hop}\left[\tilde{\beta}, \{\xi^1\}, \{\tilde{\xi}^2\}\right]
\nonumber \\
&&- \frac{\beta}{\tilde{\beta}} \frac{L}{N}\sum_\nu\log Z_\text{Hop}\left[\tilde{\beta}, \{\xi^1\}, \{\xi^{2,\nu}\}\right] \, . 
\end{eqnarray}

Using the following change of variables,
\[
\left\{
\begin{array}{ccl}
m_l^1 & \rightarrow & m^* + \delta m_l/\sqrt{N} \\
\tilde{m}_l & \rightarrow & \tilde{m}_l/\sqrt{N} \\
m^{2,\nu} & \rightarrow & m^{2,\nu}/\sqrt{N}
\end{array}
\right.
\]
we have
\begin{eqnarray}
\moy{N^n} &=&
\sum_{\{\xi^1\}} \sum_{\{\tilde{\xi}^2\}}
\sum_{\{\xi^{2,\nu}\}} \int 
\left(\prod_l^L \frac{\di \delta m_l}
{\sqrt{2 \pi}} \right)
\left(\prod_{l,\nu} \frac{\di m_l^{2,\nu}}
{\sqrt{2 \pi}} \right)
\left(\prod_{l}^L \frac{\di \tilde{m}_l}
{\sqrt{2 \pi}} \right)
e^{N E_4} \, , \\
E_4 &=& C_1 + \frac{1}{N} \sum_i \sum_l \log 2 \cosh\Bigg[
(\beta n + \tilde{\beta}) \xi_i^1 m^* + \frac{\beta n + \tilde{\beta}}{\sqrt{N}}
\xi_i^1 \delta m_l
\nonumber\\
&&\phantom{C_1 + \frac{1}{N} \sum_i \sum_l \log 2 \cosh\Bigg[}
+ \frac{\beta}{\sqrt{N}}\sum_\nu \xi_i^{2,\nu} m_l^{2,\nu}
+ \frac{\tilde{\beta}}{\sqrt{N}} \tilde{\xi}_i^2 \tilde{m}_l
\Bigg] \nonumber \\
&&- L \log 2 \cosh\left[(\beta n + \tilde{\beta}) m^*\right]
- \frac{\beta n + \tilde{\beta}}{2} \frac{1}{N} \sum_l (\delta m_l)^2
- \frac{(\beta n + \tilde{\beta}) m^*}{\sqrt{N}} \sum_l \delta m_l \nonumber \\
&&- \frac{\beta}{2} \frac{1}{N} \sum_{l,\nu} (m_l^{2,\nu})^2 
- \frac{\tilde{\beta}}{2} \frac{1}{N} \sum_l \tilde{m}_l^2
- \frac{L}{2 N} \frac{\tilde{\beta} {m^*}^2}{1-\tilde{\beta}(1-{m^*}^2)}
\left[ \frac{1}{\sqrt{N}} \sum_i \xi_i^1 \tilde{\xi}_i^2\right]^2 \nonumber  \\
&-& \frac{L}{2 N} \frac{\beta {m^*}^2}{1-\tilde{\beta}(1-{m^*}^2)}
\sum_\nu \left[\frac{1}{\sqrt{N}}\sum_i \xi_i^1 \xi_i^{2,\nu}\right]^2 \, ,
\end{eqnarray}
where
\begin{eqnarray}
C_1 &=& \frac{L}{2 N} \log\left[(\beta n + \tilde{\beta}) N \right] +
\frac{L n}{2 N} \log(\beta N) + \frac{L}{2 N} \log(\tilde{\beta} N)
- 2 \log 2 + L (\beta n + \tilde{\beta}) {m^*}^2 \nonumber \\
&&-  L \left(1 + \frac{\beta n}{\tilde{\beta}}\right)
\log 2 \cosh(\tilde{\beta} m^*) + \frac{L}{N} 
\left(1 + \frac{\beta n}{\tilde{\beta}}\right)
\log \left[ 1 - \tilde{\beta}(1-{m^*}^2)\right] \nonumber \\
&&+ L \log 2\cosh \left[(\beta n + \tilde{\beta}) m^*\right] \, .
\end{eqnarray}
Posing $\tilde{m^*} = \tanh\left[(\beta n + \tilde{\beta}) m^*\right]$ and
expanding the first term for large $N$, we get
\begin{eqnarray}
E_4 &=& C_1 - \frac{\beta n - \tilde{\beta}}{2}
\left[1 - (1-{\tilde{m^*}}^2)(\beta n +\tilde{\beta})\right]
\frac{1}{N} \sum_l (\delta m_l)^2 \nonumber \\
&&- \frac{\beta}{2} \left[1 - \beta(1-\tilde{m^*}^2)\right] \frac{1}{N}
\sum_l \sum_\nu (m_l^{2,\nu})^2\nonumber \\
&-& \frac{\tilde{\beta}}{2} \left[1 - \tilde{\beta}(1-\tilde{m^*}^2)\right]
\frac{1}{N} \sum_l \tilde{m}_l^2
+ \frac{1}{\sqrt{N}} 
(\beta n + \tilde{\beta})(\tilde{m^*} - m^*)\sum_l \delta m_l \nonumber \\
&&+ \tilde{m^*} \beta \frac{1}{N} \sum_{l,\nu} m_l^{2,\nu} \frac{1}{\sqrt{N}}
\sum_i \xi_i^1 \xi_i^{2,\nu} + \tilde{m^*} \tilde{\beta} \frac{1}{N}
\sum_l \tilde{m}_l \frac{1}{\sqrt{N}} \sum_i \xi_i^1 \tilde{\xi}^2_i\nonumber \\
&&+ (1-\tilde{m^*}^2) \frac{\beta^2}{2} \frac{1}{N^2} \sum_{i,l} 
\sum_{\nu\neq \sigma} m_l^{2,\nu} m_l^{2,\sigma} \xi_i^{2,\nu} \xi_i^{2,\sigma}
\nonumber \\ &&
+ (1-\tilde{m^*}^2) \beta \tilde{\beta} \frac{1}{N^2} \sum_{i,l,\nu} 
\tilde{m}_l m_l^{2,\nu} \tilde{\xi}^2_i \xi_i^{2,\nu}
\nonumber \\ &&
+ \beta (\beta n + \tilde{\beta}) (1- \tilde{m^*}^2) \frac{1}{N^2}
\sum_{i, l, \nu} m_l^{2,\nu} \delta m_l \xi_i^1 \xi_i^{2,\nu}
\nonumber \\ &&
+ (1-\tilde{m^*}^2) \tilde{\beta} (\beta n + \tilde{\beta})
\frac{1}{N^2} \sum_{i,l} \delta m_l \tilde{m}_l \xi_i^1 \tilde{\xi}_i^2 \nonumber \\
&&- \frac{L}{2 N} \frac{\tilde{\beta} {m^*}^2}{1-\tilde{\beta}(1-{m^*}^2)}
\left[ \frac{1}{\sqrt{N}} \sum_i \xi_i^1 \tilde{\xi}_i^2\right]^2   
\nonumber \\ &&
- \frac{L}{2 N} \frac{\beta {m^*}^2}{1-\tilde{\beta}(1-{m^*}^2)}
\sum_\nu \left[\frac{1}{\sqrt{N}}\sum_i \xi_i^1 \xi_i^{2,\nu}\right]^2  \, .
\end{eqnarray}

One may note that this expression is linear on $l$, so it can be rewritten as

\begin{equation}
\label{int1}
\moy{N^n} =
\sum_{\{\xi^1\}} \sum_{\{\tilde{\xi}^2\}}
\sum_{\{\xi^{2,\nu}\}} \left[\int 
\frac{\di \delta m_l}{\sqrt{2 \pi}}
\left( \prod_\nu
\frac{\di m_l^{2,\nu}}{\sqrt{2 \pi}} \right)
 \frac{\di \tilde{m}_l}{\sqrt{2 \pi}}
e^{E_5}
\right]^L \, .
\end{equation}

We define
\begin{eqnarray*}
u = \frac{1}{\sqrt{N}} \sum_i \xi_i^1 \tilde{\xi}_i^2
& \quad \quad \quad &
t_\mu = \frac{1}{N} \sum_i \tilde{\xi}_i^2 \xi_i^{2,\mu}\\
s_\mu = \frac{1}{\sqrt{N}} \sum_i \xi_i^1 \xi_i^{2,\mu} &&
q_{\mu, \nu} = \frac{1}{N} \sum_i \xi_i^{2,\mu} \xi_i^{2,\nu}
\end{eqnarray*}
and we do the gaussian integral in (\ref{int1}):
\begin{eqnarray}
\moy{N^n} &=& 
\sum_{\{\xi^1\}} \sum_{\{\tilde{\xi}^2\}}
\sum_{\{\xi^{2,\nu}\}}
\exp \left[
C_1 - \frac{L}{2} \frac{\tilde{\beta} \tilde{m^*}^2 u^2}{1 - \tilde{\beta}(1-\tilde{m^*}^2)}
 - \frac{L}{2} \frac{\beta \tilde{m^*}^2}{1 - \tilde{\beta}(1-\tilde{m^*}^2)}
\sum_\nu s_\nu^2
\right. \nonumber \\
&&- \left.  \frac{L}{2} \log \det M + \frac{L}{2} A^t M^{-1} A
\right] \, ,
\end{eqnarray}
where
\begin{equation}
M= \left(
\begin{array}{ccccccc}
d_1 & \tilde{u} & \tilde{s}_1 & \tilde{s}_2 & \tilde{s}_3 & \cdots & \tilde{s}_n\\
\tilde{u} & d_2 & \tilde{t}_1 & \tilde{t}_2 & \tilde{t}_3 & \cdots & \tilde{t}_n\\
\tilde{s}_1 & \tilde{t}_1 & d_3 & \tilde{q}_{1,2} & \tilde{q}_{1,3} & \cdots & \tilde{q}_{1,n}\\
\tilde{s}_2 & \tilde{t}_2 & \tilde{q}_{1,2} & d_3  & \tilde{q}_{2,3} & \cdots & \tilde{q}_{2,n} \\
\vdots \raisebox{-3ex}{} & \vdots & \vdots & &\multicolumn{2}{l}{\hfill \raisebox{.8ex}{.} \hfill . \hfill \raisebox{-.8ex}{.}} \hfill  & \vdots \\
\tilde{s}_n & \tilde{t}_n & \tilde{q}_{1,n} &   \multicolumn{2}{l}{\dotfill} & \tilde{q}_{n-1,n} & d3 \\
\end{array}
\right) \, ,
\end{equation}
\begin{equation}
A = \left(
\begin{array}{cccccc}
\sqrt{N} (\tilde{m^*}-m^*)(\beta n + \tilde{\beta}) &
\tilde{m^*} \tilde{\beta} u &
\tilde{m^*} \beta s_1 &
\tilde{m^*} \beta s_2 &
\cdots &
\tilde{m^*} \beta s_n
\end{array}
\right) \, ,
\end{equation}
and
\begin{eqnarray}
d_1 &=& (\beta n + \tilde{\beta}) \left[ 1 -(\beta n + \tilde{\beta})
(1-\tilde{m^*}^2)\right]  \, ,\\
d_2 &=& \tilde{\beta} \left[ 1 - \tilde{\beta}(1-\tilde{m^*}^2)\right] \, ,\\
d_3 &=& \beta \left[ 1 - \beta(1-\tilde{m^*}^2)\right] \, ,\\
\tilde{u}_\nu &=& -(1-\tilde{m^*}^2)\tilde{\beta}(\beta n+\tilde{\beta})u_\nu \, ,\\
\tilde{t}_\nu &=& - (1-\tilde{m^*}^2) \beta \tilde{\beta} t_\nu \, , \\
\tilde{s}_\nu &=& - (1-\tilde{m^*}^2) \beta(\beta n+\tilde{\beta})s_\nu \, ,\\
\tilde{q}_{\nu,\sigma} &=& - (1-\tilde{m^*}^2) \beta^2 q_{\nu,\sigma} \, .
\end{eqnarray}

We can now add the Lagrange multipliers
\begin{eqnarray}
\moy{N^n} &=&  
\sum_{\{\xi^1\}} \sum_{\{\tilde{\xi}^2\}}
\sum_{\{\xi^{2,\nu}\}}
\int \prod_{\rho<\sigma} \di q_{\rho \sigma}
\prod_{\rho < \sigma}  \di \hat{q}_{\rho \sigma}
\prod_\rho \di s_\rho \prod_\rho  \di \hat{s}_\rho
\prod_\rho \di t_\rho \prod_\rho \di \hat{t}_\rho
\di u \di \hat{u} \, e^{N F}  \, ,\\
F &=& 
\frac{C_1}{N} - \frac{\alpha}{2} \frac{\tilde{\beta} \tilde{m^*}^2 u^2}{1 - \tilde{\beta}(1-\tilde{m^*}^2)}
 - \frac{\alpha}{2} \frac{\beta \tilde{m^*}^2}{1 - \tilde{\beta}(1-\tilde{m^*}^2)}
\sum_\nu s_\nu^2
-  \frac{\alpha}{2} \log \det M + \frac{\alpha}{2} A^t M^{-1} A \nonumber \\
&-& \frac{1}{2} \sum_{\rho,\sigma} q_{\rho \sigma} \hat{q}_{\rho \sigma} 
-\frac{1}{2 \sqrt{N}} \sum_\rho s_\rho \hat{s}_\rho 
-\frac{1}{2} \sum_\rho t_\rho \hat{t}_\rho - \frac{1}{2\sqrt{N}} u \hat{u} 
+ \frac{1}{2N} \sum_{i,\rho,\sigma} \hat{q}_{\rho \sigma} \xi_i^{2,\rho} \xi_i^{2,\sigma} \nonumber \\
&&+ \frac{1}{2N} \sum_{i,\rho} \hat{s}_\rho \xi_i^1 \xi_i^{2,\rho} 
+ \frac{1}{2N} \sum_{i,\rho} \hat{t}_\rho \tilde{\xi}_i^2 \xi_i^{2,\rho}
+ \frac{1}{2N} \sum_i \hat{u} \xi_i^1 \tilde{\xi}_i^2 \, ,
\end{eqnarray}
and we have finally the expression presented in chapter~\ref{chap_hop}.

\chapter{Details of the evaluation of $\bm{\langle}$\textit{\~{N}}$\bm{\rangle}$ for the Hopfield model
with large \textit{\protect\iih}}
\label{app_B}

We start with
\begin{equation}
S = - \sum_{\{\xi^2\}} P[\{\xi^2\},\{\xi^1\}|\{\sigma^l\}] \log P[\{\xi^2\},\{\xi^1\}|\{\sigma^l\}] \, ,
\end{equation}
and average this entropy in respect to both the first pattern and the measured
configurations. The probability is given by

\begin{eqnarray}
P[\{\xi^2\},\{\xi^1\}|\{\sigma^l\}] &=& 
\frac{P[\{\sigma^l\}|\{\xi^1\},\{\xi^2\}] P[\{\xi^1\},\{\xi^2\}]}{\mathcal{N}[\{\sigma_l\}]} \, ,\\
P[\{\sigma^l\}|\{\xi^1\},\{\xi^2\}] &=&
\frac{1}{Z_\text{Hop}[\beta, \{\xi^1\},\{\xi^2\}]^L} 
\exp\left[\frac{\beta}{N}\sum_{\mu=1,2} \sum_{l=1}^L \sum_{i<j}
\sigma_i^l \sigma_j^l \xi_i^\mu \xi_j^\mu  \right.
\nonumber\\
&&+ \beta \left. \sum_i \sum_{l=1}^L h_i \sigma_i^l \right] \, .
\end{eqnarray}

The normalization $\mathcal{N}$ is given by
\begin{eqnarray}
\mathcal{N}[\{\sigma^l\}] &=& \sum_{\{\xi^1\}} \sum_{\{\xi^2\}} 
\frac{ P[\{\xi^1\},\{\xi^2\}]}{Z_\text{Hop}[\beta, \{\xi^1\},\{\xi^2\}]^L}
\exp\Bigg[
\frac{\beta}{N} \sum_{\mu=1,2} \sum_l^L \sum_{i<j} \sigma_i^l \sigma_j^l \xi_i^\mu \xi_j^\mu
\nonumber\\
&&+\beta\sum_i \sum_{l=1}^L h_i \sigma_i^l
\Bigg]
\end{eqnarray}
As done in Section~\ref{sec_mattis}, we write our entropy as a 
derivative of a modified normalization $\tilde{N}$:
\begin{eqnarray}
\label{eqNtil2}
\tilde{N}[\{\sigma^l\}] &=& \sum_{\{\xi^1\}} \sum_{\{\xi^2\}} 
\exp\Bigg[
\frac{\beta}{N} \sum_{\mu=1,2} \sum_l^L \sum_{i<j} \sigma_i^l \sigma_j^l \xi_i^\mu \xi_j^\mu + \beta \sum_i \sum_{l=1}^L h_i \sigma_i^l
\nonumber\\
&&- \frac{\beta}{\tilde{\beta}} L 
\log Z_\text{Hop}[\tilde{\beta},\{\xi^1\},\{\xi^2\},\{\sigma^l\}]
\Bigg] \, ,\\
\label{eqS002}
S &=& \left. \log \tilde{N} \right|_{\beta=\tilde{\beta}} 
- \left.\frac{\partial \log \tilde{N}}{\partial \beta}\right|_{\beta=\tilde{\beta}} \, ,
\end{eqnarray}
where we have supposed $P[\{\xi^1\},\{\xi^2\}]=2^{-2N}$.

\section{Determination of log \textit{Z}${}_\text{Hop}$}
To write an explicit
expression for $\tilde{N}$, we have to evaluate $Z_\text{Hop}$:
\begin{eqnarray}
  Z_\text{Hop} &=&
  \int \frac{\di m_1}{\sqrt{2 \pi}} \frac{\di m_2}{\sqrt{2 \pi}}
  \exp\Bigg\{
  -\frac{\beta}{2} m_1^2 -\frac{\beta}{2} m_2^2
  \nonumber \\
&&  + \sum_i \log\left[2 \cosh \left(\beta h_i
    + \frac{\beta}{\sqrt{N}}m_1\xi_i^1
    + \frac{\beta}{\sqrt{N}}m_2\xi_i^2
    \right)\right]
  \Bigg\} \nonumber \\
   &=&
  \int \frac{\di m_1}{\sqrt{2 \pi}} \frac{\di m_2}{\sqrt{2 \pi}}
  \exp\Bigg\{
  -\frac{\beta}{2} m_1^2 -\frac{\beta}{2} m_2^2
  + \sum_i \log\left[2 \cosh (\beta h_i)\right]
  \nonumber \\
&&  + \sum_i   \left[ \frac{\beta}{\sqrt{N}}m_1 \xi_i^1 +
                     \frac{\beta}{\sqrt{N}}m_2 \xi_i^2 \right] \tanh h_i
  \nonumber \\
&&+ \frac{1}{2}\sum_i\left[ \frac{\beta}{\sqrt{N}}m_1 \xi_i^1 +
                  \frac{\beta}{\sqrt{N}}m_2 \xi_i^2 \right]^2 (1- \tanh^2 h_i)
  +O(1/\sqrt{N})\Bigg\}
  \nonumber \\
  &=&
  \int \frac{\di m_1}{\sqrt{2 \pi}} \frac{\di m_2}{\sqrt{2 \pi}}
  \exp\Bigg\{
  -\frac{\beta \chi}{2} m_1^2 -\frac{\beta \chi}{2} m_2^2 + \beta m_1 q_1
  +\beta m_2 q_2 \nonumber \\
  &&+ \sum_i \log\left[2\cosh(\beta h_i)\right] +O(1/\sqrt{N})\Bigg\} \, ,
\end{eqnarray}
\noindent where
\begin{eqnarray}
  q_\mu &=& \frac{1}{\sqrt{N}} \sum_i \xi_i^\mu \tanh (\beta h_i) \, ,\\
  \chi &=& 1 - \frac{\beta}{N} \sum_i [1-\tanh^2 (\beta h_i)) \, .
\end{eqnarray}
Finally
\begin{equation}
  \log Z_\text{Hop} = \sum_i \log\left[2 \cosh(\beta h_i)\right]
   - \log\left(\beta \chi\right) + \frac{\beta}{2 \chi} (q_1^2 + q_2^2)
   + O(1/\sqrt{N}) \, .
\end{equation}





\section[Determination of $\moy{\protect\log \tilde{N}}$]{Determination of $\bm\langle$log \textit{\~{N}}$\bm\rangle$}

To determinate $\moy{\log \tilde{N}}$, we use again the replica
trick,
\begin{equation}
\begin{split}
\moy{\tilde{N}^n} = &\frac{1}{2^{2N}} \sum_{\{\tilde{\xi}^1\}} \sum_{\{\tilde{\xi}^2\}}
\sum_{\{\sigma\}} \sum_{\{\xi^{1,\nu}\}}\sum_{\{\xi^{2,\nu}\}}
\exp \Bigg\{ (\beta n + \tilde{\beta}) \sum_i \sum_{l=1}^L  h_i \sigma_i^l
\\
&+\frac{\beta}{2 N} \sum_{\nu=1}^n \sum_{l=1}^L \sum_{i,j} \sigma_i^l \sigma_j^l (\xi_i^{1,\nu} \xi_j^{1,\nu}+
\xi_i^{2,\nu} \xi_j^{2,\nu})
-  \frac{\beta}{\tilde{\beta}} L \sum_{\nu=1}^n
 \log Z_\text{Hop}\left[\tilde{\beta}, \{\xi^{1,\nu}\}, \{\xi^{2,\nu}\}\right]
 \\
&+ \tilde{\beta} \sum_{l=1}^L \sum_{i,j} \sigma_i^l \sigma_j^l (\tilde{\xi}_i^1 \tilde{\xi}_j^1+
\tilde{\xi}^2_i \tilde{\xi}^2_j)
- L  \log Z_\text{Hop}\left[\tilde{\beta}, \{\tilde{\xi}^1\}, \{\tilde{\xi}^2\}\right]
\Bigg\} 
\end{split}
\label{eqnn2}
\end{equation}
After doing an integral transform, we obtain
\begin{eqnarray}
\moy{\tilde{N}^n} =
\sum_{\{\tilde{\xi^1}\}} \sum_{\{\tilde{\xi}^2\}}
\sum_{\{\sigma\}}\sum_{\{\xi^{1,\nu}\}} \sum_{\{\xi^{2,\nu}\}}
\int 
\left(\prod_{l,\nu} \frac{\di m_l^{1,\nu}}{\sqrt{\frac{2 \pi}{\beta}}} \right)
\left(\prod_{l,\nu} \frac{\di m_l^{2,\nu}}{\sqrt{\frac{2 \pi}{\beta}}} \right)
\left(\prod_{l}^L \frac{\di \tilde{m}_l^1 \di \tilde{m}_l^2}{\frac{2 \pi}{\tilde{\beta}}} \right)
e^{N E_1} \, , \nonumber \\
\end{eqnarray}
where
\begin{eqnarray}
E_1 &=&
 \frac{\beta}{\sqrt{N}} \sum_i \sum_l \sum_\nu \sigma_i^l \xi_i^{1,\nu} m_l^{1,\nu}
+ \frac{\beta}{\sqrt{N}} \sum_i \sum_l \sum_\nu \sigma_i^l \xi_i^{2,\nu} m_l^{2,\nu}
\nonumber \\
&&
+ \frac{\tilde{\beta}}{\sqrt{N}} \sum_i \sum_l \sigma_i^l \tilde{\xi}_i^1 \tilde{m}_l^1 
+ \frac{\tilde{\beta}}{\sqrt{N}} \sum_i \sum_l \sigma_i^l \tilde{\xi}_i^2 \tilde{m}_l^2 \nonumber \\
&&
- \frac{\beta}{2} \sum_{l,\nu} (m_l^{1,\nu})^2
- \frac{\beta}{2} \sum_{l,\nu} (m_l^{2,\nu})^2
- \frac{\tilde{\beta}}{2} \sum_l (\tilde{m}_l^1)^2 
- \frac{\tilde{\beta}}{2} \sum_l (\tilde{m}_l^2)^2 
\nonumber \\
&&- \frac{L}{N}\log Z_\text{Hop}\left[\tilde{\beta}, \{\tilde{\xi}^1\}, \{\tilde{\xi}^2\}\right]
- \frac{\beta}{\tilde{\beta}}  \frac{L}{N}\sum_\nu\log Z_\text{Hop}\left[\tilde{\beta}, \{\xi^{1,\nu}\}, \{\xi^{2,\nu}\}\right]\nonumber \\
&&+(\beta n + \tilde{\beta}) \sum_i \sum_{l=1}^L  h_i \sigma_i^l
\end{eqnarray}

Summing over the spin variables $\sigma$ yields
\begin{eqnarray*}
\tilde{N} &=&
\sum_{\{\tilde{\xi^1}\}} \sum_{\{\tilde{\xi}^2\}}
\sum_{\{\xi^{1,\nu}\}} \sum_{\{\xi^{2,\nu}\}}
\left[
\int 
\left(\prod_{\nu} \frac{\di m_{1,\nu}}{\sqrt{2 \pi}} \right)
\left(\prod_{\nu} \frac{\di m_{2,\nu}}{\sqrt{2 \pi}} \right)
\left(            \frac{\di \tilde{m}_1 \di \tilde{m}_2}{2 \pi} \right)
e^{N E_3}
\right]^L\\
E_3 &=& 
 \frac{n}{N} \log(\beta N) + \frac{1}{ N} \log(\tilde{\beta} N)
- \frac{\beta}{2} \sum_{\nu} (m_{1,\nu})^2 
- \frac{\beta}{2} \sum_{\nu} (m_{2,\nu})^2 
-\frac{\tilde{\beta}}{2} (\tilde{m}_1)^2
-\frac{\tilde{\beta}}{2} (\tilde{m}_2)^2
\\
&&
- \frac{1}{N}\log Z_\text{Hop}\left[\tilde{\beta}, \{\tilde{\xi}^1\}, \{\tilde{\xi}^2\}\right]
- \frac{\beta}{\tilde{\beta}} \frac{1}{N} \sum_\nu \log Z_\text{Hop}\left[\tilde{\beta}, \{\xi^{1,\nu}\}, \{\xi^{2,\nu}\}\right]\\
&&+
 \sum_i \log 2 \cosh \left[ 
(\beta n + \tilde{\beta}) h_i + \frac{\beta}{\sqrt{N}} \sum_\nu \xi_i^{1,\nu} m_{1,\nu} + \frac{\beta}{\sqrt{N}} \sum_\nu \xi_i^{2,\nu} m_{2,\nu}
\right.\\
&&
\left.
\phantom{+\frac{1}{N} \sum_i  \log 2 \cosh \Bigg[ }
+ \frac{\tilde{\beta}}{\sqrt{N}} \tilde{\xi}_i^1 \tilde{m}_1
+ \frac{\tilde{\beta}}{\sqrt{N}} \tilde{\xi}_i^2 \tilde{m}_2
\right]
\end{eqnarray*}

Doing a Taylor expansion and using the result for $\log Z_\text{Hop}$ we have
\begin{eqnarray*}
  E_3 &=& 
 \frac{n}{N} \log(\beta N) + \frac{1}{ N} \log(\tilde{\beta} N)
- \frac{\beta}{2} \sum_{\nu} m_{1,\nu}^2 
- \frac{\beta}{2} \sum_{\nu} m_{2,\nu}^2 
-\frac{\tilde{\beta}}{2}  \tilde{m}_1^2
-\frac{\tilde{\beta}}{2} \tilde{m}_2^2
\\
&&
- \frac{\beta n + \tilde{\beta}}{\tilde{\beta}}
\sum_i \log\left[2 \cosh (\tilde{\beta} h_i)\right]
+ \frac{\beta n + \tilde{\beta}}{\tilde{\beta}}\log(\tilde{\beta} \tilde{\chi})
+ \frac{\tilde{\beta}}{2 \tilde{\chi}} (\tilde{q}_1^2 + \tilde{q}_2^2)
+ \frac{\tilde{\beta}}{2 \tilde{\chi}} \sum_\nu\left(q_{1,\nu}^2 + q_{2,\nu}^2\right) \\
&& + \sum_i \log\left[2 \cosh[(\beta n + \tilde{\beta})h_i]\right]
+ \tilde{\beta} \sum_{\mu=1,2} \tilde{m}_\mu \frac{1}{\sqrt{N}} \sum_i \tilde{\xi}_i^\mu  \tanh [(\beta n + \tilde{\beta}) h_i]
\\
&&
+ \beta \sum_{\nu} \sum_{\mu=1,2} m_{\mu,\nu} \frac{1}{\sqrt{N}} \sum_i \xi_i^{\mu,\nu}  \tanh [(\beta n + \tilde{\beta}) h_i]
+ \frac{\tilde{\beta}^2}{2 N} \sum_{\mu=1,2} \tilde{m}_\mu^2 \sum_i \left[1-\tanh^2[(\beta n + \tilde{\beta})h_i]\right]
\\
&&
+ \frac{\beta^2}{2 N} \sum_{\mu=1,2} \sum_\nu m_{\mu,\nu}^2 \sum_i \left[1-\tanh^2[(\beta n + \tilde{\beta})h_i]\right]
\\
&&
+ \frac{\beta^2}{N} \sum_{\nu, \rho} m_{1,\nu} m_{2,\rho} \sum_i  \xi_i^{1,\nu} \xi_i^{2,\rho}\left[1-\tanh^2[(\beta n + \tilde{\beta})h_i]\right]
\\
&&
+ \frac{\beta^2}{N} \sum_{\mu=1,2} \sum_{\nu \neq \rho} m_{\mu,\nu} m_{\mu,\rho} \sum_i  \xi_i^{\mu,\nu} \xi_i^{\mu,\rho}\left[1-\tanh^2[(\beta n + \tilde{\beta})h_i]\right]
\\
&&
+ \frac{\beta^2}{N} \sum_{\mu=1,2} \sum_{\mu'=1,2} \sum_{\nu} m_{\mu,\nu} \tilde{m}_{\mu'} \sum_i  \xi_i^{\mu,\nu} \tilde{\xi}_i^{\mu'}\left[1-\tanh^2[(\beta n + \tilde{\beta})h_i]\right]
\\
&&
+ \frac{\beta^2}{N} \tilde{m}_{1} \tilde{m}_{2} \sum_i  \tilde{\xi}_i^{1} \tilde{\xi}_i^{2}\left[1-\tanh^2[(\beta n + \tilde{\beta})h_i]\right]
\end{eqnarray*}

Several of these terms can be ignored: first of all,
$ \frac{1}{N} \sum_i  \tilde{\xi}_i^{1} \tilde{\xi}_i^{2}\left[1-\tanh^2[(\beta n + \tilde{\beta})h_i]\right]$ can be considered of order $O(1/\sqrt{N})$, since
we suppose that the patterns are orthogonal in the leading order.
The same thing can be said to the inferred patterns, so
$ \frac{1}{N} \sum_i  \xi_i^{1,\nu} \xi_i^{2,\rho}\left[1-\tanh^2[(\beta n + \tilde{\beta})h_i]\right] = O(1/\sqrt{N})$. Finally, since we can permute $\xi^1 \leftrightarrow \xi^2$, we will only lose one unity of entropy if we suppose
that $\xi^{1}$ is an approximation to $\tilde{\xi}^1$ and not $\tilde{\xi}^2$.
We can consider thus
$ \frac{1}{N} \sum_i  \xi_i^{\mu,\nu} \tilde{\xi}_i^{\mu'}\left[1-\tanh^2[(\beta n + \tilde{\beta})h_i]\right] = O(1/\sqrt{N})$ for $\mu \neq \mu'$.

After neglecting these terms, we can see that the two different patterns are
completely decorrelated in $\tilde{N}$, with no extra term accounting for an
effective influence of one pattern over the other. We can hence conclude that
the results taken for the Mattis model can be applied for this system.

\bibliography{these}

\backmatter
\end{document}